\documentclass[12pt]{report}
\usepackage[left=1in,top=1in,right=1in,bottom=1in]{geometry} 
\usepackage{fancyheadings}
\usepackage{multirow}          % for tabular
\usepackage{graphicx}          % for diagrams

\raggedright                   % do not right justify
\usepackage{amsmath}
\usepackage{amsfonts}
\usepackage{amssymb}

\newcommand{\bra}[1]{\langle #1|}
\newcommand{\ket}[1]{|#1\rangle}
\newcommand{\braket}[2]{\langle #1|#2\rangle}

\newcommand{\set}[1]{\{#1\}}

\newcommand{\E}{\operatorname{{\mathrm E}}}
\newcommand{\Var}{\operatorname{{\mathrm Var}}}

\begin{document}         

\title{\bf Quantum Operator Design for\\
Lattice Baryon Spectroscopy}
\author{Adam C. Lichtl\\[10mm]
Advisor: Colin Morningstar\\[10mm]
Submitted in partial fulfillment of the\\[2mm]
requirements for the degree\\[2mm]
of Doctor of Philosophy\\[2mm]
in the Mellon College of Science at\\[2mm]
Carnegie Mellon University}
\date{September 7, 2006}
\maketitle

\pagenumbering{roman} \setcounter{page}{2}

\chapter*{Abstract}
A previously-proposed method of constructing spatially-extended 
gauge-invariant three-quark operators for use in Monte Carlo
lattice QCD calculations is tested, and a methodology for 
using these operators to extract the energies of a large number
of baryon states is developed.  This work is part of a long-term
project undertaken by the Lattice Hadron Physics Collaboration
to carry out a first-principles calculation of the low-lying
spectrum of QCD.

The operators are assemblages of smeared and
gauge-covariantly-displaced 
quark fields having a definite flavor structure.  The importance
of using smeared fields is dramatically demonstrated.  It is found 
that quark field smearing greatly reduces the couplings to the unwanted
high-lying short-wavelength modes, while gauge field smearing
drastically
reduces the statistical noise in the extended operators.  
Group-theoretical projections onto the irreducible representations of
the
symmetry group of a cubic spatial lattice are used to endow the
operators
with lattice spin and parity quantum numbers, facilitating
the identification of the $J^P$ quantum numbers of the corresponding 
continuum states.

The number of resulting operators is very large; consequently
a key aspect of
this work is the development of a selection method for finding 
a sufficient subset of operators for accurately extracting the lowest
seven or eight energy levels in each symmetry channel.  A procedure in
which the diagonal elements of the correlation matrix of the operators 
are first evaluated to remove noisy operators, followed by the selection
of 
 sixteen operators whose renormalized correlation matrix at a fixed 
small time separation has a low condition number for both the even-
and odd-parity channels,
is found to work well.

These techniques are applied in the construction of nucleon operators.  
Correlation matrix elements between these operators are estimated using 
$200$ configurations on a $12^3\times 48$ anisotropic lattice in the 
quenched approximation with unphysically heavy $u,d$ quark masses (the
pion mass is approximately 700 MeV).  After a change of basis operators 
using a variational method is applied, the energies of up to eight 
states are extracted in each symmetry channel.  
 Although comparison
with experiment is not justified, the pattern of levels obtained
qualitatively agrees with the observed spectrum.  
A comparison with quark model predictions 
is also made; 
the quark model predicts more low-lying even-parity
states than this study yields, but both the quark model and
this study predict more odd-parity states near 2 GeV than
currently observed in experiments.

\chapter*{Dedication}
This work is dedicated to my family and friends for their unwavering
support and love as I pursue my dreams.  They are a continuing source of
strength as I climb, and they are always there to catch me if I fall.

\chapter*{Acknowledgements}
This project would not have been possible without the support of many people.
Special thanks goes to my advisor Colin Morningstar who had the
confidence to put me on such an important project, and who had 
the patience to let me learn from
my mistakes.  I am also grateful to
 Keisuke (Jimmy) Juge, Matthew Bellis, David Richards, 
Robert Edwards, and George Fleming who took the time to answer my 
questions and to show me some of the nuances of their respective specialties.  

The configurations 
used in this work were provided by David Richards and were generated using
the computing resources at Thomas
Jefferson National Laboratory.  
The bulk of the computational work for this project was performed
on the Medium Energy Group's computing cluster at Carnegie Mellon University.
I am grateful to the Medium Energy Experimental Group for their flexibility
concerning the allocation of computing resources, and am especially grateful to 
Curtis Meyer for keeping the cluster running as smoothly as possible (even
on the weekends).

\newpage

\setcounter{tocdepth}{1}
\tableofcontents
\listoffigures
\listoftables
\newpage
\pagenumbering{arabic} \setcounter{page}{1}

\chapter{Introduction}
\label{chap:intro}

Particle physicists seek to identify the elementary building blocks of the
universe, and the mechanisms by which these elements interact.  This view, 
known as reductionism, is seen by many as the necessary starting point for the 
discovery of a unified theory which quantitatively describes all 
natural phenomena.  If we can identify the building blocks and the rules which
hold them together, then we can begin to
explore which features of our universe arise from the complex interactions
of these fundamental components.

\section{The atom and nucleus}

In 1803, John Dalton presented his atomic theory which stated that all
compounds were composed of and reducible to collections of 
atoms.  This sparked
a great effort to use any chemical means available to separate compounds into
their fundamental atoms, the so called elements.  As scientists discovered 
new elements and observed their properties in reactions, they began to notice
that entire groups of elements behaved in similar ways.  In 1869, Dmitri 
Mendeleev~\cite{halzen:quarks} introduced his now ubiquitous periodic table of the elements,
which placed elements with similar reaction properties into columns, ordered by
atomic mass.  Mendeleev's table classified the 63 known elements at the time,
and he predicted the existence and characteristics of three elements which
were later discovered: gallium, scandium, and germanium.

The similar behavior of the different elements in each column was a 
reassurance that there was indeed
an order hidden beneath the diversity of the everyday world.  Today,
the 118 known elements fit neatly into the eighteen columns of the modern
periodic table (there are also the lanthanoid and actinoid classifications).
These elements include stable atoms such as hydrogen, carbon,
nitrogen, and oxygen; and unstable atoms, such as radium and uranium which 
last long enough to be characterized, but which eventually break down into 
lighter elements through radioactive decay.

The large size and redundancy in the table were also significant hints that 
the atom possessed substructure.  A breakthrough on this front was the 
identification of 
the electron in 1897 by J.J. Thomson~\cite{griffiths:quantum}.  At this time,
scientists were fascinated by the cathode ray tube, a glass
chamber with a metal electrode at each end and a port for a vacuum pump.  When the air was pumped out of the tube and a voltage
difference was applied to the electrodes, the tube would glow with `rays'
emanating from the negative electrode (the cathode).  This effect
was later enhanced by J.B. Johnson who added a heating element to the cathode.
Before 1897
scientists suspected, but were not certain, that the rays were composed of 
electrically charged particles.

J.J. Thomson's cathode ray experiments built on those of his contemporaries, and involved the deflection of these rays by known electric and magnetic fields.  
By
balancing the electric force $\vec{F}_E=q\vec{E}$ with the magnetic force $\vec{F}_B=q\vec{v}\times
\vec{B}$, he determined the velocity of the particles to be roughly one tenth
 of the speed of light from the relation
$$v=\frac{E}{B}.$$

  He then switched off the electric field, measured
the radius of curvature $R$ of the beam, and used the condition of uniform 
circular motion $F=mv^2/R$ to determine the charge-to-mass ratio
from the relation
$$\frac{q}{m}=\frac{v}{RB}.$$

Thomson found that the charge-to-mass ratio was abnormally large when compared
to the known ions.  He assumed that the mass was very small (as opposed to
assuming that the charge was very large), and concluded that the 
ray was composed of negatively
charged elementary 
particles, and identified the particles as constituents of the atom.  Later, 
they were named `electrons'\footnote{From the Greek {\em
elektron}, meaning amber.  Early experiments in electricity used amber 
rods rubbed with fur to build up a static charge.}, taken from a term coined 
in 1894 by 
electro-chemist G. Johnstone Stoney.  
Thomson had discovered that cathode ray electrons came from the atoms composing
the cathode; these electrons were `boiled' off of the cathode as it heated up.

The charge of the electron (and mass, from Thomson's ratio) was measured in
1909 by Robert Millikan in his oil-drop 
experiment.
He used an atomizer to spray a mist of oil droplets above two plates.  The
top plate had a small hole through which a few droplets could pass.  Millikan
would vary the electric field between the plates until he had one (charged)
drop of oil suspended.  Then he turned off the electric field and allowed the 
drop to fall until it had reached its terminal velocity $v_1$.  
The velocity-dependent drag force on the drop was given by
Stoke's Law:
$$F_{D1} = 6\pi r \eta v_1,$$
where $r$ is the radius of the (assumed spherical) drop and $\eta$ is the
viscosity of air.
At terminal velocity, the drag force
 balanced the the `apparent weight' of the oil drop 
$F_{D1}=W$, which is
$$W = (4/3)\pi r^3(\rho_{\mathrm{oil}}-\rho_{\mathrm{air}})g,$$
where $\rho_{\mathrm{oil}}$
is the density of the oil, and $\rho_{\mathrm{air}}$ is the density of air.
His measurement of $v_1$ allowed him to infer $r$, and thus the apparent
weight of the drop $W$.  He then turned the electric field back on before
the drop reached the bottom plate and measured the terminal velocity $v_2$ of 
the drop's upward motion.
At this point, Millikan knew that $qE = F_{D2} + W = F_{D1}v_2/v_1 + W$ and 
could therefore determine the charge $q$ on the drop:
$$q = \frac{W}{E}\left(1+\frac{v_2}{v_1}\right).$$
After repeating the experiment for many droplets,
Millikan confirmed that the quantity of charge on a drop was always a multiple
 of the same number, the charge on a 
single electron ($1.602 \times 10^{-19}$ Coulomb, 
in SI units).

Electrically neutral atoms must 
possess enough positive charge to compensate for the negative charge of the
electrons.  The modern model of the atom was born from Geiger and Marsden's~\cite{geiger:alpha}
alpha scattering experiments performed under Ernest Rutherford~\cite{rutherford:alpha}, a former student of Thomson's, in 1909.  The goal of these experiments was to probe the positive charge
distribution of the atom.  Charged alpha particles (doubly ionized helium atoms) emitted from a radioactive
 radium 
source were directed at a gold foil.  A zinc sulfide screen was placed at
various positions to detect the scattered alpha particles.  Rutherford, 
Geiger, and Marsden found
that most of the alpha particles passed through the foil with little 
deflection but some deflected through large angles.  This suggested that the
positive charge in each atom was concentrated at the center and occupied 
just a fraction of the total atomic volume.

The nucleus of the lightest element (Hydrogen) was named the 
proton\footnote{From the Greek {\em proton}, meaning first.}.  Its charge is
exactly equal to the magnitude of the electron charge, but its mass is 
roughly 2000 times greater.  In the atomic model, a hydrogen atom consisted of
a proton and electron bound together via the electromagnetic force.

\section{Hydrogen spectroscopy}
Once scientists had a model of the hydrogen atom, they were in a position to
discuss the hydrogen spectrum.
When an
electrical current is passed through pure hydrogen gas, the atoms absorb
energy and then radiate at specific discrete wavelengths, which can
 be observed by passing the emitted light through a prism or diffraction
grating.  Spectroscopy was pioneered by people such as A.J. Angstrom and was in widespread use in chemistry for the classification of elements.
The first observed series of emission lines was the Balmer series,
named after J.J. Balmer who in 1885 first developed the empirical 
relationship for the spacing of the lines.  The relationship was generalized
in 1888 by Johannes Rydberg for the complete emission line spectrum:
$$\kappa = \mathcal{R}\left(\frac{1}{n_1^2}-\frac{1}{n_2^2}\right),$$
where $\kappa=1/\lambda$ is the discrete emission wavenumber and $\mathcal{R}=10967757.6\pm 1.2\, m^{-1}$ is the Rydberg constant for the hydrogen spectrum~\cite{eisberg:quantum}.  The Balmer series lines correspond to 
$$n_1=2\mbox{ and }n_2=3,4,\ldots.$$
  A burning question was: could the atomic
model reproduce the observed hydrogen spectrum?

In 1913, Niels Bohr combined the Rutherford model of the atom with Einstein's
quantum theory of the photoelectric effect\footnote{Einstein proposed that radiant
energy comes in quanta known as photons with the energy frequency relation $E=h\nu$.} introduced in 1905 by introducing
his quantization condition.  Bohr's model related the electromagnetic radiation
emitted by an atom to electron transitions from states of definite angular
momentum:
$$L=n\hbar.$$

The quantization of atomic energy states was experimentally verified in 1914
by Frank and Hertz who accelerated electrons through a potential difference 
in a tube filled with mercury
vapor~\cite{eisberg:quantum}.  They measured the current as a function of applied voltage, 
which was an indicator of the number
of electrons which passed unimpeded through the gas.  When the kinetic energy
of the electrons reached a threshold level, the current abruptly dropped.  
Frank and Hertz interpreted this to mean that the electrons were exciting
the mercury atoms and being scattered.  As the voltage was increased, the 
current would increase until successive thresholds were reached.  This showed that the mercury atoms possessed discrete energy levels.  The mercury atoms would
 only absorb energy from the electrons by transitioning from one energy level to another. 

The urge to understand the underpinnings of Bohr's (and in 1916 Sommerfeld and Wilson's) quantization conditions 
sparked the rapid development of an entirely new field of physics: quantum
mechanics.  Two notable contributions were Werner Heisenberg's 1925 matrix 
mechanics paper, and Erwin Schr\"odinger's 1926 paper on quantization
as an eigenvalue problem (which introduced `Schr\"odinger's equation').  
Quantum mechanics succeeded not only in accurately describing 
atomic spectra, but also lead to revolutionary technological advances such
as the semiconductor used in computers and nuclear magnetic resonance (NMR)
used in medical imaging.

The hydrogen spectrum not only provided a test of the atomic model, it also
led to refinements in our understanding of the atom and opened up an entirely
new branch of physics.  The importance of spectroscopy cannot be overstated.

\section{Nucleons and the strong nuclear force}

The neutron, the electrically neutral partner of the 
proton in the nucleus, was identified
in 1932 by James Chadwick~\cite{chadwick:neutron} from previous experiments
in which a polonium source was used to bombard beryllium with alpha particles.

In modern notation~\cite{cahn:experiment}, the nuclear reaction was:
$$\mbox{He}_2^4+\mbox{Be}_4^9\to \mbox{C}_6^{12}+\mbox{n}_0^1,$$
where the subscript denotes the atomic number (number of protons), and the 
superscript denotes the atomic weight (number of protons$+$neutrons).

The observation that every atom contains a nucleus in which protons and 
neutrons are confined within $0.01\%$ of the volume of the atom raises
the question: what keeps the protons from flying apart from electrostatic
repulsion?  Physicists inferred the existence of a new force which could
overpower the electromagnetic force, but which had a range on the order
of the nuclear radius.  In 1934, Hideki Yukawa worked out the quantization
of the strong nuclear force field and predicted a new particle, the
pion~\cite{griffiths:quantum}.
In Yukawa's theory, nucleons interact with each other via the pion field.
When the field is quantized according to the formalism of quantum field
theory, the potential felt by the nucleons goes as
$$\frac{1}{r}\exp\left\{-\frac{m_\pi r}{\hbar c}\right\},$$
where $r$ is the inter-nucleon separation, and $m_\pi$ is the mass of the
pion, the quantum which mediates the strong nuclear force\footnote{Note that the
electromagnetic force is mediated by the massless photon, giving the usual
$1/r$ Coulomb potential.}.  To get a force with a range of 1 fm, the order
of the typical nuclear radius
$$m_\pi\approx\frac{\hbar c}{1 \mbox{ fm}}\approx200\mbox{ MeV}.$$

It was then up to particle physicists to use all of the tools at their disposal
to confirm the existence of the pion and complete the atomic model.

\section{Particle sources, particle detectors, and the particle zoo}

\subsection{Sources of subatomic particles}
We have already discussed the cathode ray tube, a source of electrons when
the cathode is heated enough to boil them off of the metal.  A wide variety
of experiments were performed
 which involved accelerating the electrons through a potential
difference.

In 1895, Henri Becquerel discovered that uranium was radioactive from the
darkening of photographic film~\cite{cahn:experiment}.  The radioactive
decay of heavy elements such as polonium produced light nuclei such as 
that of helium (two protons and two neutrons), the so called alpha
radiation.  When a neutron decays into a proton, it emits an electron, the so
called beta radiation.  Electromagnetic radiation is called gamma radiation.
Sources of radiation would be placed in front of collimators, which would
only let through thin beams of the particles.  These beams could then be
directed onto various targets and the reactions observed.

In 1912, Victor Hess used a hot air balloon to take measurements of ionizing
radiation at varying altitudes.  He discovered that the rate of ionization
was roughly four times greater at an altitude of 5,300 meters than it was
at ground level, thus showing that the radiation which ionizes the atmosphere
is cosmic in origin.  The discovery of these so called cosmic rays led
to the discovery of a wide range of subatomic particles.

Particle accelerators (`atom smashers') use electromagnetic
fields to accelerate
beams of charged particles to velocities comparable with the speed of light.
Accelerators may be linear or circular\footnote{The Large Hadron Collider (LHC) is a proton-proton collider with a ring 27 km in circumference.}.  Circular 
colliders use
magnetic fields to curve the paths of particles as they are accelerated around
the ring.
These high energy beams of particles can be directed upon stationary targets
or made to collide with other beams of particles.  Particles of one type could
also
be directed onto targets, producing particles of a different type which can
themselves be collimated into a beam and focused with magnets.

\subsection{Detectors of subatomic particles}
A conspicuous signature of a charged cosmic ray is the ion trail left behind as 
electrons are stripped off of atoms in the ray's path.  Some particle
detectors turn these ion trails into visible paths, but this is not the only
way to detect a subatomic particle.  Here is a brief description of some
of the tools experimentalists use to detect subatomic particles.

\textbf{Nuclear emulsions} are a special mixture of gelatin and 
silver bromide salts and work under the same principles as chemical
photography.
The track of a charged particle creates sites of silver atoms on the salt
grains.  Photographic development chemicals reduce the silver bromide salts to
silver, but are most effective when there are already silver atoms present.
Early cosmic ray researchers would stack plates of nuclear emulsions, and then
develop them into pictures of subatomic paths.

\textbf{Cloud chambers}, also called Wilson chambers after their inventor
C.T.R. Wilson, contain a supersaturated vapor (such as isopropyl alcohol) 
that forms droplets
along the trail of ionization.
The trails are illuminated with a bright light source and photographed.  

\textbf{Bubble chambers}
work under the same principle as cloud chambers, but use a superheated
transparent liquid (such as liquid hydrogen).  When a 
charged particle passes through the liquid, it
interacts with the molecules and deposits enough energy to boil the liquid
around the interaction points.  The result is the formation of a string
of small bubbles along the particle's path which can then be illuminated
and photographed.

\textbf{Drift chambers}
are described by~\cite{online:cleo_drift}:
\begin{quote}Charged particles passing through the chambers ionize the gas in the chambers (DME or an argon/ethane mixture), and the resulting ionization is collected by wires maintained at high voltage relative to the surrounding "field" wires. The electrical signals from these wires are amplified, digitized, and fed into a computer which reconstructs the path of the original particle from the wire positions and signal delay times.
\end{quote}
 
\textbf{Silicon detectors} operate under the same principle as drift chambers
but utilize a semiconducting material instead of a gas.  This allows for
much a higher energy resolution and a much higher spatial resolution.  
However, they are more expensive and
more sensitive to the degradating effects of radiation than drift chambers.

\textbf{Scintillators}
are compounds which absorb energy from interactions with charged particles, 
and then re-emit that energy in the form of electromagnetic radiation at a
longer wavelength (fluorescence).  Scintillators have short decay times
and are optically transparent to the flashes.  Thus, particle tracks register
as a series of rapid flashes within the material which can be seen by light
detectors (such as photomultiplier tubes).
%usually you dope a plastic with scintillating material

\textbf{Calorimeters}
measure the energy content of the particle shower which occurs when a
subatomic particle strikes a dense barrier in the detector.  Often 
calorimeters are segmented into different chambers, and the energy deposited
by the particle showers in each chamber is used to infer the particle's identity
and direction of travel.  Calorimeters can be designed to detect either
electromagnetic showers or hadron showers, which occur when a subatomic
particle interacts with the barrier.

\subsection{The particle zoo}
Particle physicists use detectors to examine the contents of cosmic
rays and the products of beam-target and beam-beam scattering experiments.
They measure the trajectories either directly in the detector, or
through reconstruction via the principle of
conservation of mass and momentum.  A particularly useful quantity measured
is the
{\em differential cross section}, the reaction rate per unit incident flux
as a function of angle, energy, and any 
other parameters of interest~\cite{perkins:experiment}.
Other properties of the particles such as mass, spin, parity, form factors
(describing the structure of the particle), life-time, and 
branching ratios (describing the relative likelihood of decay into each of
several final states), can then be deduced
from interaction cross section data.

There are three classes of outcomes when
two subatomic particles $A$ and $B$ approach each other:
\begin{enumerate}
\item Nothing happens: the particles pass right by each other without interacting
\item Elastic scattering ($A+B \to A+B$): the particles interact through the
exchange of a `messenger' particle.  The outgoing particles are of the same
type as the incoming particles, but may have a different energy and trajectory.

\item Inelastic scattering ($A+B \to C+D+\cdots$): Einstein's mass-energy relation\footnote{Einstein's original formulation of special relativity made a distinction between the mass of a body at rest and the perceived mass of a body in motion.  Modern convention takes mass to be a characteristic of an object (e.g. on equal footing with charge), and explicitly includes the relativistic dilatation factor $\gamma\equiv 1/\sqrt{1-v^2/c^2}$.} $E=\gamma mc^2$ allows for the transmutation the combined energy and mass of the input particles into a completely different set of output particles, subject to quantum transition rules and  kinematic constraints. 
\end{enumerate}

It was expected that the above sources and detectors of particles would
lead to the discovery of Yukawa's pion, and the completion of the atomic
model.  Some scientists anticipated that the proton, neutron, electron, photon,
and pion would constitute the fundamental building blocks of all matter.
It was a great surprise, then, when detailed observations of cosmic rays and 
scattering experiments uncovered a plethora of subatomic particles.  
One of the early cosmic ray candidates
for the pion ended up being the muon, a more massive relative
of the electron.  I.I. Rabi put it best when he said "Who ordered that?"~\cite{cahn:experiment}

The pion was experimentally confirmed in 1947 when D.H. 
Perkins~\cite{perkins:pion}
observed the explosion of a nucleus after capturing a cosmic-ray pion.
He saw the ion trails created by the incoming pion and outgoing nuclear debris 
in a photographic
emulsion.  Nuclear disintegration by pion capture had been predicted in
 1940 by Tomonaga and Araki.  

Modern estimates~\cite{hagiwara:pdg} of the mass of the charged and neutral pions are:
$$\begin{array}{l|r@{.}l}
\mbox{Pion}&\multicolumn{2}{c}{\mbox{Mass (MeV)}}\\ 
\hline
\pi^{\pm}&139&57018(35)\\
\pi^0&134&9766(6)
\end{array}
$$
which is close to the very rough estimate of $200$ MeV made by Yukawa.

The muon, now identified as a distinct particle from the pion, was only the 
beginning of a long revolutionary series of discoveries. 
Positrons, kaons, antiprotons, tau leptons, neutrinos, and more took their place
with the proton, neutron, and electron in the rapidly expanding `particle
zoo.'

Most of these particles were unstable resonances.  In analogy with radioactive
elements in the periodic table which decay into simpler elements, a resonance
is defined as an object of mass $M$ with a lifetime $\tau$ much longer than 
the period 
associated with its `characteristic frequency' 
$\nu = M/h$:
$$\tau >> h/M.$$
For very massive resonances, this lifetime can be very short and is quoted
in terms of the decay width $\Gamma = \hbar/\tau$, a natural spread in the 
energy of the decaying
state induced by the uncertainty principle~\cite{perkins:experiment}.  
Resonances are then defined by the property (neglecting factors of $2\pi$ and working in energy units)
$$\Gamma << M.$$
For example the 
$Z^0$ resonance, discovered in 1983 at CERN, has a mass of approximately
91 GeV~\cite{hagiwara:pdg}
and a decay width of approximately 2.5 GeV, which corresponds to a
lifetime of roughly $2.6 \times 10^{-25}$ seconds, enough time for light
to travel about one-tenth of a Fermi, much less than than the spatial extent
of a proton.  Such resonances cannot, therefore, be observed directly in
particle detectors.  Rather, resonances in the elastic 
channel,\footnote{A {\em channel} refers to the way a reaction can proceed.
The channel concept is also applied to initial, intermediate, or final states 
in a reaction,
e.g. the `nucleon channel' contribution for a process has a nucleon resonance
as an intermediate state, and the $\Delta^{++}$ can be created in the 
$p\pi$ channel.}
$$A+B\to X\to A+B,$$
show up as enhancements
in the differential and total cross sections compared 
to what would be expected from
simple kinematics alone.  Resonances in the inelastic channel
$$A+B\to X\to C+D+\ldots$$
can be identified from trajectory reconstruction and decay products.

The overwhelming number of subatomic particles discovered
(including resonances) lead
 Wolfgang Pauli to exclaim in the 1950s ``Had I foreseen this, I would have 
gone into botany.''

\section{A new periodic table}

\subsection{Classification of particles}
As new particles were discovered, they were classified by their values of
conserved quantities, such as mass and electric charge.  The existence of
`forbidden' reactions, which were never observed, led to the introduction
of two more conserved quantum numbers: lepton number and baryon number.
Leptons (such as the electron and muon) have lepton number = 1, 
baryon number = 0, and do
not participate in any strong interactions.  Hadrons have lepton number = 0,
participate in
strong interactions, and either have baryon number = 1 for baryons (such
as the proton and neutron), or baryon number = 0 for mesons\footnote{From the
Greek {\em mesos}, meaning middle.  Mesons are so named because the first mesons had masses between that
 of the electron (0.5 MeV) and the proton (938 MeV).} (such as the pion).
Because an antiparticle annihilates its corresponding particle, it must
have the opposite quantum numbers (except for mass, which is always
positive and which is conserved in the radiation emitted from an
annihilation reaction).

The discovery of kaons, unstable mesons which were created easily via the
strong interaction but which decayed slowly via a different process
(the weak interaction, which also governs radioactive beta decay), lead
to the introduction of `strangeness', a new quantum number.  The strong
interaction conserves strangeness, but the weak interaction violates it.
The time-scale over which the strong interaction acts is very short, thus
giving rapid creation and decay rates.  On the other hand, weak interactions
have long time-scales which cause strangeness-changing reactions to proceed
slowly.  For example, a kaon (meson with strangeness = +1) and lambda 
(baryon with strangeness = -1) can be produced easily by the strong 
reaction:
$$\pi^-+p^+\to K^0+\Lambda^0,\qquad |\Delta S| = 0,$$
but must decay separately via the much slower weak reactions
$$\begin{array}{cclll}
K^0&\to&\pi^+\pi^-,&\qquad |\Delta S| = 1,&\nonumber\\
\Lambda^0&\to& p^+\pi^-,&\qquad |\Delta S| = 1.
\end{array}$$

\begin{figure}[hb!]
\centering
\includegraphics[height=1.8in]{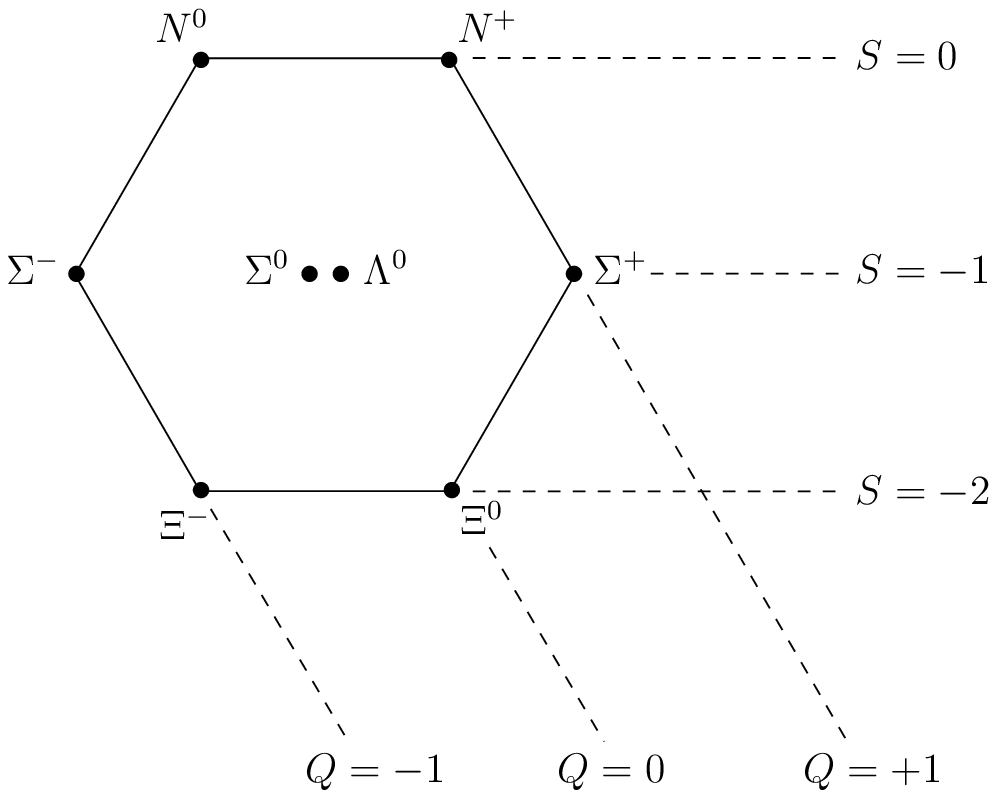}
\hfill
\includegraphics[height=1.8in]{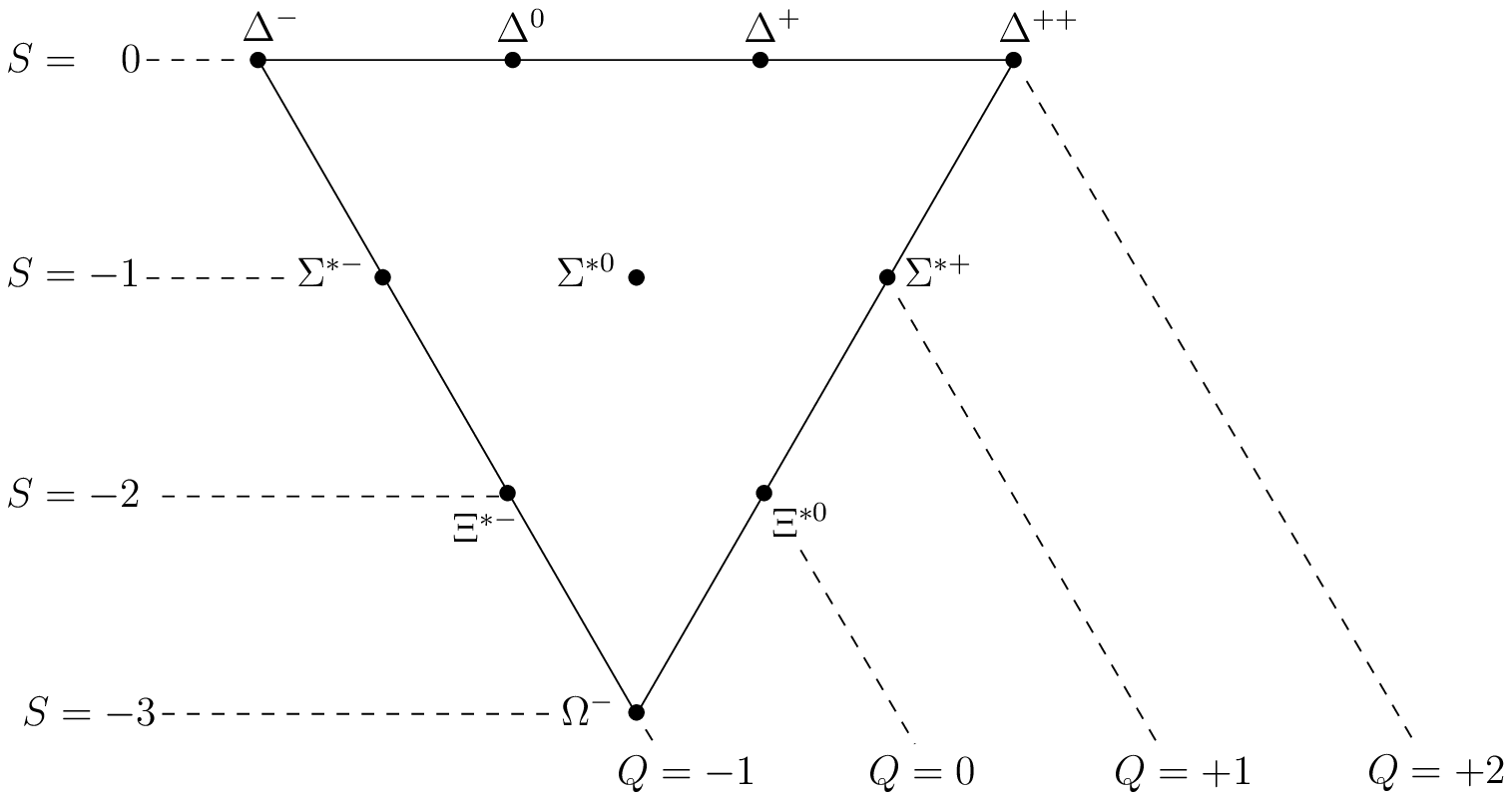}
\caption{The baryon octet (left) and decuplet (right).  The $N^0$ is the 
neutron and the $N^+$ is the proton.}
\label{fig:baryon_octet}
\end{figure}

In the tradition of Mendeleev, the discovered hadrons were
placed into tables according to their masses and reaction properties 
(quantum numbers) independently by Murray Gell-Mann and Yuval Ne'eman 
in 1961~\cite{gell-mann:eightfold}. 
In this arrangement scheme, dubbed the `Eightfold Way' by Gell-Mann, 
particles with similar masses were placed into hexagonal and triangular
arrays which were labeled by strangeness $S$ and electric
charge $Q$.  Two examples of these tables are shown in 
Figure~\ref{fig:baryon_octet}.

Just as Mendeleev had predicted the existence and properties of gallium, 
scandium, and germanium from holes in his periodic table, so also did
Murray Gell-Man predict the existence, mass, and quantum numbers of a 
new particle, the $\Omega^-$, a strangeness = -3 heavy baryon, which was
later discovered at Brookhaven National Laboratory in 1964~\cite{barnes:omega}.

The existence of such a large number of hadrons along with their 
classification according to a comparably small number of quantum numbers
hinted, once again, that there was substructure yet to be discovered.  What
was needed was a model of nucleon substructure which could
explain the patterns in Gell-Man and Ne'eman's tables.

\subsection{The constituent quark model}
 The constituent quark model describes hadrons as being built up from
combinations of point-like particles.  
This model was proposed independently in 1964
by Gell-Mann and Zweig, and Gell-Mann named the constituents `quarks'\footnote{From
 James Joyce's {\em Finnigan's Wake}, referring to the sound a seagull makes.}.

Quarks are spin-1/2 objects (fermions) with baryon number = 1/3
and come in different flavors.
Originally three flavors were introduced: `up', `down', and `strange':

\begin{center}\begin{tabular}{c|cc}
& Charge & Strangeness\\
\hline
u & $+2/3$ & $\phantom{-}0$\\
d & $-1/3$ & $\phantom{-}0$\\
s & $-1/3$ & $-1$
\end{tabular}\end{center}

\begin{figure}[ht!] 
  \centering 
  \includegraphics[width=1.0in]{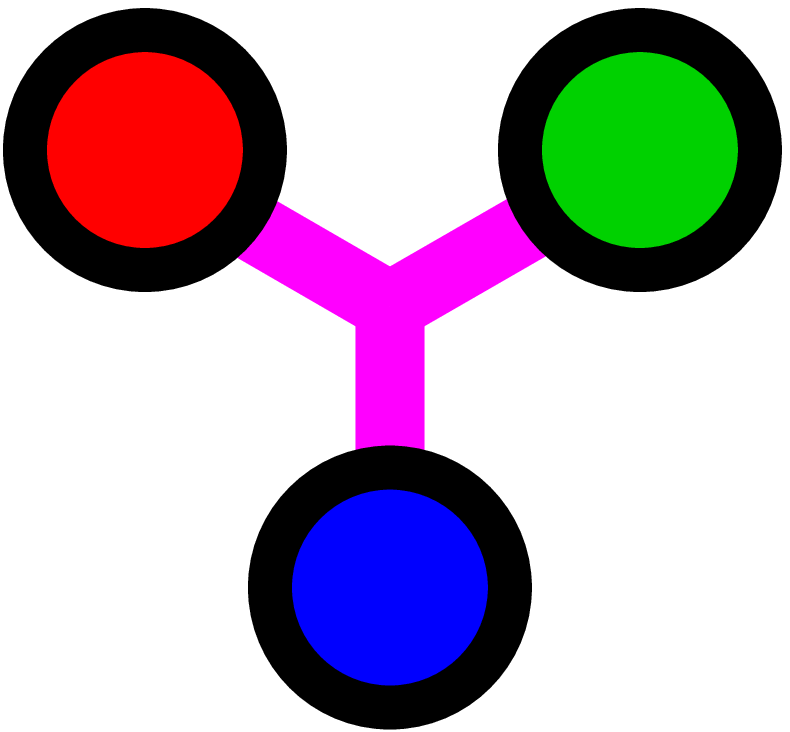}
  \hspace{2.0in}
  \includegraphics[width=1.0in]{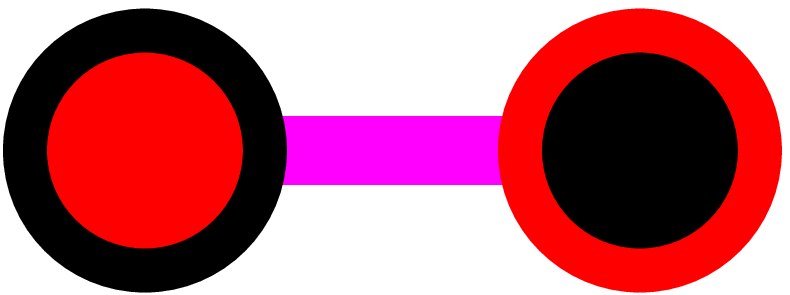}
  \caption{The quark model view of hadrons.  Left: a baryon consisting of three quarks joined by flux-tubes of glue.  Right: a meson consisting of a quark-antiquark pair joined by a flux-tube of glue.} 
  \label{fig:baryon} 
\end{figure}

\begin{figure}[ht!] 
  \centering 
  \includegraphics[width=4.0in]{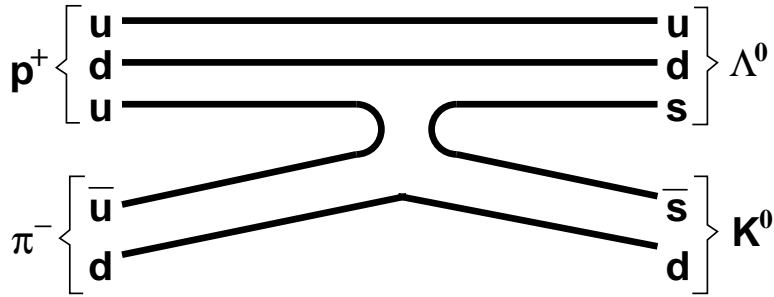}
  \caption{The quark model view of the reaction $\pi^-+p^+\to K^0+\Lambda^0$.
    Time increases from left to right. 
    The up quark and antiquark annihilate, and a new strange 
    quark-antiquark pair appears.}
  \label{fig:lambda-k} 
\end{figure}

The quarks combine into baryon and meson multiplets following the 
well-established rules of addition of angular momenta.  According to the
quark model, baryons are
three-quark states, and mesons are quark-antiquark states.
All of the hadrons known in 1964 (and most discovered since) could be 
described using this model. 
Figure \ref{fig:baryon} shows baryons and mesons as viewed in the quark model,
and figure \ref{fig:lambda-k} shows pictorially the reaction
$$\pi^-+p^+\to K^0+\Lambda^0.$$

\begin{table}[ht!]
\begin{center}\begin{tabular}{c|c|c|c|c}
Hadron & Quark Content & Baryon Number & Charge & Strangeness\\
\hline
$\mbox{p}^+$ & $uud$ & $1$ & $+1$ & $\phantom{+}0$\\
$\mbox{n}^0$ & $udd$ & $1$ & $\phantom{+}0$ & $\phantom{+}0$\\
$\Lambda^0$ & $uds$  & $1$ & $\phantom{+}0$ & $-1$ \\
$\Omega^{-}$ & $sss$ & $1$ & $-1$ & $-3$\\
$\pi^+$ & $u\bar{d}$ & $0$ & $+1$ & $\phantom{+}0$\\
$\pi^-$ & $d\bar{u}$ & $0$ & $-1$ & $\phantom{+}0$\\
$\pi^0$ & $\frac{1}{\sqrt{2}}(u\bar{u}+d\bar{d})$ & 0 & $\phantom{+}0$ & $\phantom{+}0$\\
$\mbox{K}^0$ & $d\bar{s}$ & $0$ & $\phantom{+}0$ & $+1$
\end{tabular}\end{center}
  \caption{The quark flavor content of some sample hadrons.  The properties
of the hadrons are determined by the properties of the quarks from which
they are composed.}
  \label{table:quark_content}
\end{table}

The quark model correctly described the baryon octet and baryon decuplet.
It also predicted that the light meson octet and singlet would mix, forming
a meson nonet.  This corrected the Eightfold Way which had 
treated a newly discovered meson, 
the $\eta'$, as a singlet with no relation to the existing meson octet.

Observed states such as the $\Delta^{++} (uuu)$ produced from
$$\pi^++p^+\to\Delta^{++},$$
and
the $\Omega^- (sss)$ appeared to violate the Pauli exclusion principle, 
because all three
fermions were in a symmetric spin-flavor-spatial wavefunction.  This led
Greenberg to postulate the existence of a new quantum number in 
1964~\cite{greenberg:paraquarks} which could take on one of three values.
This additional quantum number later evolved into the concept of `color,' the 
charge associated with the force holding the quarks together.  Thus, every
flavor of quark comes in three `colors': `red', `green', or `blue.'  The color 
hypothesis holds that all hadrons are colorless states (combinations
of all three colors in the case of baryons, or a color-anticolor
combination in the case of mesons).  A profound and disturbing prediction
is that the fundamental building blocks themselves, the quarks, could never
be observed in isolation.  This phenomenon is known as `color confinement.'

Experimental verification for quarks came from the SLAC (Stanford Linear
Accelerator Center) experiments
performed under ``Pief'' Panofsky~\cite{cahn:experiment} in the late 1960s 
using a beam of electrons directed onto a hydrogen target. 
The goal of the experiments was to repeat the Rutherford experiment,
but at energies high enough (up to about 18 GeV) 
to probe the structure of the proton by deep inelastic 
scattering.  They found
that the proton did indeed contain concentrations of charge which were
`point-like' in comparison to its spatial extent.  High energy electrons
 were unable to knock isolated quarks out of hadrons as expected
from the color confinement hypothesis.  On the other hand, it appeared that
quarks behaved as free particles within the hadrons.  This vanishing of
the color force at high scattering energies (short distance resolution) 
is known as {\em asymptotic freedom}.

In 1974, the $J/\Psi$, a meson containing a new heavier flavor of quark
dubbed the `charm' was discovered independently at Brookhaven and SLAC.
Currently, a total of six quark flavors has been identified.

Other experiments measured the ratio of cross sections
$$\frac{\sigma(e^+e^-\to \mbox{hadrons})}{\sigma(e^+e^-\to \mu^+\mu^-)},$$
as a function of center of mass energy.  At lower energies, only $u\bar{u}$ and
$d\bar{d}$ quark pairs can be created\footnote{The dynamics of the strong
interaction is independent of quark flavor, but the kinematics does in general
depend on the masses of the different quark flavors.}. 
  However, as the center of mass
energy increases, it reaches the threshold where the more massive $s\bar{s}$ 
pairs can be
produced, then $c\bar{c}$ pairs and so on.  Because each quark comes in one
 of three colors, there is an overall factor of three in the reaction rate.
These experiments provided an impressive confirmation of the existence of 
exactly three colors~\cite{povh:experiment}.

\section{The Standard Model}

Currently, the interactions among objects can be understood in terms of
three basic forces: the force of gravity, the electroweak 
force, and the strong nuclear force.

The electroweak and strong interactions are treated in a theoretical framework
known as the Standard Model, which successfully predicts virtually all observed
phenomena in particle physics.  In the Standard Model, matter consists of
 quarks and leptons (and associated antiparticles),
 which are described by quantum field theories 
possessing a local gauge symmetry.  This local gauge symmetry gives rise to
the interactions among the quarks and leptons mediated by `gauge bosons.'
The electroweak gauge bosons are the massless $\gamma$ (photon), and the
massive $W^\pm$ and $Z^0$ which acquire their mass (it is 
believed\footnote{The Higgs boson has not yet been observed experimentally.})
 from the
Higgs boson $H$ via the `Higgs mechanism.'
The strong gauge bosons are the massless gluons
(see Table~\ref{tbl:standard_model}).

 It is widely
recognized that this model is incomplete (most conspicuously, it fails to
cleanly integrate gravity with the other forces), and there is active research
into extensions and revisions.  For this work, we will focus on extracting
predictions from quantum chromodynamics, that
component of the Standard Model which defines the dynamics of the quarks and
which gives rise to the strong nuclear force.

\begin{table}[ht!] 
  \centering 
  $\begin{array}{|c|c|c|}
  \hline
  \mbox{Generation} & \mbox{Leptons} & \mbox{Quarks}\\
  \hline
  \mbox{I} & e, \nu_e & d, u \\
  \mbox{II} & \mu, \nu_\mu & s, c\\
  \mbox{III} & \tau, \nu_\tau & b, t\\
  \hline
  \multicolumn{3}{c}{}\\
  \hline
  \multicolumn{3}{|c|}{\mbox{Gauge Bosons}}\\
  \hline
  \multicolumn{3}{|c|}{(\gamma, W^\pm, Z^0), g}\\
  \hline
  \end{array}$
  \caption{The standard model of particle interactions.  This model describes
all matter as made up of quarks and leptons, held together by the interactions
of the force-carrying mediators. The Higgs boson is the only particle of 
the Standard Model which has not been observed, and is not shown.}
  \label{tbl:standard_model} 
\end{table}

\section{Quantum chromodynamics}
\subsection{Quantum field theory}

Quantum field theory (QFT) is the extension of quantum mechanics to 
relativistic
systems, allowing for a unified description of matter and radiation
fields.
The first foundations of QFT were laid by Dirac in his 1927 
paper~\cite{dirac:radiation} which treated 
the emission and absorption of electromagnetic radiation by atoms.  

A quantum field theory is based on a Hilbert space of possible
physical states of the system.  The fundamental degrees of freedom are fields
of operators which act on the Hilbert space and 
which are defined over a space-time manifold.  
 
 Natural units are used in which
$\hbar=c=1$.  In Minkowski (flat) space-time, points are labeled by
$(x^0, x^1, x^2, x^3)\equiv(t, \vec{x})$ with respect to some basis with
associated metric $\eta_{\mu\nu}=\mbox{diag}(1,-1,-1,-1)$.  Quantities
of the form\footnote{Summation over repeated indices is implied unless noted otherwise.
}\begin{eqnarray*}
\eta_{\mu\nu}(x^\mu-y^\mu)(x^\nu-y^\nu)&\equiv& (x^\mu-y^\mu)(x_\mu-y_\mu)=
(x^0-y^0)^2-(\vec{x}-\vec{y})\cdot(\vec{x}-\vec{y})
\end{eqnarray*}
are invariant under
the Poincar\'e group of translations, rotations, and relativistic boosts.
The fields in QFT may have indices which label different components, and usually
these components are required to transform irreducibly under the 
Poincar\'e group~\cite{ramond:quantum}.  This means that an arbitrary
 Poincar\'e transformation $R$ (e.g. a rotation) transforms the field 
$\Phi_a$ as:
$$\Phi_a(x)\to\Phi'_a(x)=\Phi_b(R^{-1}x)D_{ba}(R),$$
where $D(R)$ is an irreducible\footnote{A representation of a group is a set
of matrices $\{D\}$ which satisfies $D(R)D(R')=D(RR')$ for all $R$ and $R'$ in 
the group.  The representation is irreducible if there is no invariant subspace
which only mixes with itself under the group operations, i.e. if the representation 
matrices cannot all be simultaneously block diagonalized.} representation
matrix representing the effect of $R$ on the components of the field.  In 
general, there are many different irreducible representations allowing for
the definition of a different type of field for each.

The Poincar\'e group has ten generators.  Two operators which commute
with all of the generators of the Poincar\'e group are called Casimir operators
and consist of the mass and the relativistic spin.  Thus, we label the
different types of fields by their mass and spin. 

In the Heisenberg operator picture, we single out the time direction and 
treat the degrees of freedom at a given time as operators with 
eigenvectors and associated eigenvalues:
$$\Phi(\vec{x};t)\ket{\phi; t}=\phi(\vec{x}; t)\ket{\phi; t}.$$
For fixed $t$, all of the spatial points $\vec{x}$ are space-like separated,
which allows us to define $\ket{\phi; t}$, the simultaneous eigenstate
for all of the operators at different spatial locations at time $t$.
The eigenvectors represent the possible states of the field at time $t$ and 
form a complete (Hilbert) space:
$$\braket{\phi; t}{\phi'; t}=\delta(\phi-\phi')$$
$$\int d\phi\, \ket{\phi; t}\bra{\phi; t} = 1$$
with the appropriate definition of the inner product and integration measure.
A general state of the system at time $t$ is specified by the state-vector 
$\ket{\psi; t}$,
and the probability amplitude\footnote{In quantum theory, the probability 
of a particular observed outcome is given by the absolute square
of the sum of the probability amplitudes for all of the ways that
outcome could occur.  The fields may not be directly observable, but we may 
always speak of the probability {\em amplitude} for a particular 
configuration's contribution to a process.} of the system to be in state
$\ket{\psi'; t'}$ if it was known that it was in the state $\ket{\psi; t}$
at $t<t'$ is given by the inner product
$$\braket{\psi'; t'}{\psi; t}.$$

A field at a spatial location $\vec{x}$ evolves in time via the
Heisenberg time-evolution equation:
$$\Phi(\vec{x}; t')=e^{iH(t'-t)}\Phi(\vec{x};t)e^{-iH(t'-t)}$$
where $H$ is the Hermitian Hamiltonian operator, the Poincar\'e 
generator of temporal 
translations.  The Hamiltonian operator governs the dynamics 
of the theory, and its
spectrum consists of the steady states of the theory, including all
stable single- and multi-particle energy states.  
In order for the particle content
of the field theory to be well-defined, the Hamiltonian must be bounded from
below.  Subtracting a suitable constant from $H$, 
we may define the `vacuum' state
$\ket{\Omega}$, the time-independent state of lowest energy:
$$H\ket{\Omega}=0$$

By the action of a suitable operator, any state may be excited from the 
vacuum:
$$\ket{\phi; t}=\mathcal{O}_\phi(t)\ket{\Omega}$$
Because all operators of interest can be expressed as analytic functions of 
the fundamental degrees of freedom of the theory,
 all of the information about the 
properties and dynamics of the theory is
contained in the so called Green's functions, the vacuum expectation
values of products of fields, such as:
$$\bra{\Omega}T\left[\Phi(\vec{x}_b; t_b)
\Phi(\vec{x}_a; t_a)\right]\ket{\Omega},$$
where $T[\cdots]$ is the right-to-left time ordering operator which arranges the
fields in time order from earliest on the right to latest on the 
left.

Inspired by the work of Dirac, Richard Feynman developed 
the path integral approach to quantum mechanics in
1948~\cite{feynman:path_integral}, 
which was quickly extended to quantum field theory.  His work made 
possible
the evaluation of vacuum expectation values (VEVs) by use of the Feynman 
functional integral~\cite{peskin:quantum}:
\begin{eqnarray*}
\bra{\Omega}T\left[\Phi(\vec{x}_b; t_b)
\Phi(\vec{x}_a; t_a)\right]\ket{\Omega} &=& \lim_{\epsilon\to 0} 
\lim_{T\to \infty(1+i\epsilon)}\frac{\int \mathcal{D}\phi\, 
\phi(\vec{x}_b; t_b)\phi(\vec{x}_a; t_a) \exp\left\{iS[\phi]\right\}
}{\int \mathcal{D}
\phi\, \exp\left\{iS[\phi]\right\}}\nonumber\\
 &=& \lim_{\epsilon\to 0}\lim_{T\to\infty(1+i\epsilon)}\frac{1}{Z[0]}
\frac{\delta}{\delta J(\vec{x}_b; t_b)}\frac{\delta}{\delta J(\vec{x}_a; t_a)}
Z[J]\left|_{J=0}\right.\nonumber
\end{eqnarray*}
where
$$Z[J]\equiv\int \mathcal{D}
\phi\, \exp\left\{iS[\phi]+iJ\phi\right\}$$
is a generating functional of the fields, and
$$S[\phi]\equiv\int_{-T/2}^{T/2}dt\int_V d^3x\, \mathcal{L}(\phi, \partial\phi)$$
is the {\em action} functional, the 
space-time integral of the Lagrangian density
$\mathcal{L}$ which determines the dynamics of the theory.\footnote{
When the magnitude of the action is large (in units of $\hbar$), then the theory
becomes classical and the expectation values are dominated by 
values of the field around which the phase remains stationary.  This principle 
of stationary phase (often referred to as the `principle of least action') gives
the Euler-Lagrange equations which govern all of classical mechanics: 
$$\frac{\delta S}{\delta \phi}=0\to
 \partial_\mu\left(\frac{\delta \mathcal{L}}{\delta(\partial_\mu \phi)}
\right)-\frac{\delta \mathcal{L}}{\delta \phi}=0$$.}

In practice, the infinite-dimensional functional integral over field
configurations $\phi$ is ill-defined.
In order to calculate quantities in quantum field theory, we work
in a finite box and introduce a regulator which makes the integrals convergent.  The regulator is then
removed by using the method of renormalization, which allows the 
`bare' parameters in the original Lagrangian (such as mass, coupling, and
field normalization) to vary as functions of the
regulator parameter.  These bare parameters are not observable, and can be
used to absorb the divergences encountered in the theory.  It is not always
possible to do this; in non-renormalizable theories the divergences
cannot all be absorbed into the small number of original bare parameters.

\subsection{Reference frame covariance}
The concept of reference frame covariance~\cite{moriyasu:gauge, schutz:gr} 
was developed in Einstein's Special and
General Theories of Relativity and states simply that coordinate 
systems at 
different space-time
points may have different orientations.  
A physical quantity represented by a vector $v$ can always be 
written as a linear combination of some basis
vectors $\{\hat{\eta}_{(a)}\}$ with coefficients $v^a$:
$$v=v^{a}\hat{\eta}_{(a)}$$
where summation over repeated indices is implied.  The basis vectors are
abstract entities which define some reference frame, and the components are 
simply numbers which express the orientation of the vector within that
 reference frame.  The vector $v$ is an abstract entity and is independent of 
the choice of reference frame.  A different reference frame corresponds to
a different set of basis vectors $\{\hat{\eta}'_{(a)}\}$ and coefficients
$\{v'^a\}$, but the vector $v$ remains unchanged:
$$v=v'^a\hat{\eta}'_{(a)}$$

General relativity postulates that a reference frame can only be 
defined locally.  If we want to make meaningful comparisons of
the components of a vector field $\{v^a(x)\}$ 
 at two different points,  we need a way to specify the relative orientations
of the basis vectors at those points.  Manifolds are locally flat, which
means that we can define vectors connecting the points in a small neighborhood
of $x$ using a basis $\{\hat{e}_{(\mu)}\}$.  Because the neighborhood is flat,
we can always write the basis vectors at 
$x+\epsilon \hat{e}_{(\mu)}$ 
as linear combinations of the basis vectors at $x$:
$$\hat{\eta}_{(a)}(x+\epsilon\hat{e}_{(\mu)})=\hat{\eta}_{(a)}(x)+ig\epsilon  \hat{\eta}_{(b)}(x)A_{\mu b a}(x)+O(\epsilon^2)$$
which gives the correct behavior as $\epsilon\to 0$.  We have pulled out a 
factor of $ig$ for later convenience and have not yet specified the form of 
the {\em connection} $A_{\mu a b}(x)$.
We now have a way of expressing the components of a vector at $x+\epsilon\hat{e}_{(\mu)}$ on the same basis as the components of a vector at $x$:
\begin{eqnarray*}
v(x+\epsilon\hat{e}_{(\mu)})&=&v^a(x+\epsilon\hat{e}_{(\mu)})
\hat{\eta}_{(a)}(x+\epsilon\hat{e}_{(\mu)})\\
&=&v^a(x+\epsilon\hat{e}_{(\mu)})(
\hat{\eta}_{(a)}(x)+ig\epsilon \hat{\eta}_{(b)}(x)A_{\mu b a}(x)+O(\epsilon^2))\\
&=&\left[(\delta_{ab}+ig\epsilon A_{\mu ab}(x))v^b(x+\epsilon\hat{e}_{(\mu)})\right]
\hat{\eta}_{(b)}(x)+O(\epsilon^2)\\
&\equiv& \left[U_{a b}(x, x+\epsilon\hat{e}_{(\mu)})v^b(x+\hat{e}_{(\mu)})\right]
\hat{\eta}_{(a)}(x),
\end{eqnarray*}
where we have introduced the infinitesimal linear {\em parallel transporter}
$$U(x,x+\epsilon \hat{e}_{(\mu)})=1+ig\epsilon A_\mu(x)+O(\epsilon^2)$$ which
expresses the components of a vector at $x+\hat{e}_{(\mu)}$ with respect
to the basis at $x$. Alternatively, we may say that the parallel transporter
moves a vector from $x+\epsilon \hat{e}_{(\mu)}$ to $x$ while keeping it
(locally) parallel to its original orientation.
$$U(x,x+\epsilon \hat{e}_{(\mu)}):\qquad (x)\leftarrow(x+\epsilon \hat{e}_{(\mu)}).$$
We may build up a finite
parallel transporter along a directed curve $\mathcal{C}_{yx}$ from $x$ to $y$
by repeated application of the above.
Break $\mathcal{C}_{yx}$ up into an $N+1$ point mesh $\{z_0, z_1, \cdots, z_N\}$
where $z_0=x$, $z_N=y$, and the distance between $z_k$ and $z_k-1$ 
is $\epsilon$. Letting $dz_k=z_k-z_{k-1}$ we may define the 
left-to-right {\em path ordered exponential} by~\cite{rothe:lgt}:
\begin{eqnarray*}
U(\mathcal{C}_{yx})&=&\mathcal{P}\exp\left\{ig\int_{\mathcal{C}_{yx}} dz^\mu\,
 A_\mu(z)\right\}\\
&\equiv&
\lim_{\epsilon\to 0}(1+ig dz_1^\mu A_\mu(z_0))(1+ig dz_2^\mu 
A_\mu(z_1))\cdots\left(1+ig dz_N^\mu A_\mu(z_{N-1})\right)
\end{eqnarray*}

We may also define the covariant derivative $D_\mu$, which takes into 
account both the 
spatial change and basis change of $v^a(x)$ under an infinitesimal 
displacement $\epsilon$ in the $\mu^{th}$ direction $\hat{e}_{(\mu)}$
\begin{eqnarray*}
D_\mu v^a(x) &\equiv& \lim_{\epsilon\to 0}\frac{U_{ab}(x, x+\epsilon\hat{e}_{(\mu)})
v^b(x+\epsilon\hat{e}_{(\mu)})-
v^a(x)}{\epsilon}\\
&=& \lim_{\epsilon\to 0}\frac{(\delta_{ab}+ig\epsilon A_{\mu ab}(x))v^b(x+\epsilon 
\hat{e}_{(\mu)})-v^a(x)}{\epsilon}\\
&=& (\delta_{ab}\partial_\mu + igA_{\mu ab}(x))v^b(x)
\end{eqnarray*}

\begin{figure}[ht!] 
  \centering 
  \includegraphics[width=5.0in]{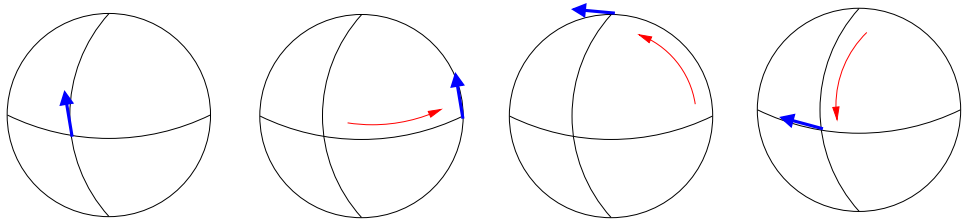}
  \caption{An example of parallel transport around a non-trivial manifold.
The blue arrow ends up in a 
different orientation even though it made only locally
parallel moves.}
  \label{fig:sphere} 
\end{figure}

The significance of this approach is two-fold.  First, the quantity
$(D_\mu v^a(x))\hat{\eta}_{(a})$ is a proper abstract vector; the components
transform `covariantly' under operators on the vector space.  Second, 
the connection $A(x)$ encapsulates
 any non-trivial topological properties of the space of different coordinate
systems on which the vector is 
defined.
For example, the surface of a sphere is locally flat, but has a path-dependent
parallel transporter $U(\mathcal{C})$ as can be seen in figure 
\ref{fig:sphere}.

\subsection{Quantum electrodynamics}
Hermann Weyl had attempted to apply the concept of local gauge symmetry
to electrodynamics in 1919.  He suggested that
one could choose a different scale, or gauge, at each point in space, but this
idea was quickly proven incorrect (e.g. the atom had an observable 
characteristic scale).  He succeeded, however,
in the late 1920's along with Fock and London to identify the
the `coordinate system' of relevance in electrodynamics.  An electron's 
quantum mechanical wavefunction $\psi(x)$ can be rotated by a global phase
without changing the dynamics of the theory.  They postulated that
the phase of an electron's wavefunction was a local quantity and required
that comparisons of the phase at two different points $x$ and $y$ 
required the use of the parallel transporter 
$$U(\mathcal{C}_{yx})=\mathcal{P}\exp\left\{ig\int_{\mathcal{C}_{yx}} dz^\mu\,
 A_\mu(z)\right\}$$
where the {\em gauge} field $A_\mu(x)$ is a real-valued field.  

Dirac had already found a quantum field theory describing a free 
electron when he sought to write a relativistic form of
Schr\"odinger equation (which is first order
in the time derivative) by factoring the relativistic 
energy-momentum expression~\cite{griffiths:quantum}:
$$(p^\mu p_\mu - m^2)=(\gamma^\mu p_\mu+m)(\gamma^\nu p_\nu-m)$$
where 
the four Dirac matrices $\gamma^\mu$ must satisfy
\begin{equation}
\{\gamma^\mu,\gamma^\nu\} =
\gamma^\mu\gamma^\nu+\gamma^\nu\gamma^\mu=2g^{\mu\nu}
\label{eqn:mink_gamma1}
\end{equation}
The Dirac equation of motion 
$(i\gamma^\mu \partial_\mu -m)\psi=0$
comes from the classical limit of the 
quantum field
theory described by the Dirac Lagrangian~\cite{peskin:quantum}:
$$\mathcal{L}_{\mathrm{Dirac}}=\overline{\psi}(i\gamma^\mu \partial_\mu -
m)\psi$$
where $\psi(x)$ is the electron field and
$\overline{\psi}(x)\equiv\psi^\dagger(x)\gamma^0$.  In order to have a real
action (leading to a Hermitian Hamiltonian), we must have 
$(\gamma^0\gamma^\mu)^\dagger=(\gamma^0\gamma^\mu)$ and 
$\gamma^{0\dagger}=\gamma^0$, or equivalently:
\begin{equation}
\gamma^{0\dagger}=\gamma^0,\quad\gamma^{j\dagger}=-\gamma^j
\label{eqn:mink_gamma2}
\end{equation}

Replacing the regular derivative in $\mathcal{L}_{\mathrm{Dirac}}$
 with a covariant derivative gives rise to an electromagnetic interaction term:
\begin{eqnarray*}
\overline{\psi}(x)(i\gamma^\mu \partial_\mu - m)\psi(x)&\to&\overline{\psi}(x)(i\gamma^\mu D_\mu - m)\psi(x)\\
&=&\overline{\psi}(x)(i\gamma^\mu \partial_\mu - m)\psi -  g A_\mu(x)\overline{\psi}(x)\gamma^\mu\psi(x)\\
&=&\mathcal{L}_{\mathrm{Dirac}}+\mathcal{L}_{\mathrm{Int}}
\end{eqnarray*}
Where we see that the gauge field $A_\mu(x)$ is to be identified with the 
electromagnetic potential, $\overline{\psi}\gamma^\mu\psi$ is the electron
probability `current' (which transforms as a Poincar\'e 4-vector), and 
$g=-|e|$ is the electron charge, the coupling constant for the 
electromagnetic interaction between the electron field $\psi$ and 
electromagnetic field $A$.

This theory possesses an important symmetry: local gauge invariance.  The 
Lagrangian is unchanged under local gauge transformations 
(picking a new phase at each 
space-time point, and updating the connection accordingly):
\begin{eqnarray*}
\psi(x) &\to& e^{i\alpha(x)}\psi(x)\\
\overline{\psi}(x) &\to& \overline{\psi}(x)e^{-i\alpha(x)}\\
A_\mu(x) &\to& A_\mu(x)-\frac{1}{g}\partial_\mu \alpha(x).
\end{eqnarray*}
Quantum field theories which are based on a local gauge symmetry are known
as {\em gauge} theories.  
In 1971, Gerard 'tHooft~\cite{hooft:gauge_renormalization}
 showed that all gauge theories were renormalizable.
This was a critical success for gauge theory, because it means that 
calculations are guaranteed to give finite results for physical processes
as the regulator is removed.

In analogy with General Relativity and figure \ref{fig:sphere}, the field strength tensor is 
associated with the
curvature of the gauge field, the amount the phase changes around an
infinitesimal square loop~\cite{schutz:gr, montvay:lgt} in the $\mu-\nu$ plane:
$$F_{\mu\nu}\equiv\frac{1}{ig}[D_\mu, D_\nu]$$
Considering first
\begin{eqnarray*}
D_\mu D_\nu &=& \frac{1}{ig}\left(\partial_\mu + ig A_\mu(x)\right)\left(\partial_\nu + ig A_\nu(x)\right)\\ 
&=&\frac{1}{ig}\partial_\mu\partial_\nu+\partial_\mu A_\nu(x)+ A_\nu(x)\partial_\mu
A_\mu(x)\partial_\nu+ig^2 A_\mu A_\nu
\end{eqnarray*}
we see that 
\begin{eqnarray}
F_{\mu\nu}\equiv\frac{1}{ig}[D_\mu, D_\nu]&=&\partial_\mu A_\nu-\partial_\nu A_\mu +
  ig[A_\mu, A_\nu]\label{eqn:commutator}\\
&=&D_\mu A_\nu - D_\nu A_\mu\nonumber
\end{eqnarray}
The gauge group of electrodynamics is the group of all unimodular complex
phases $U(1)$.
Because $U(1)$ is Abelian (commutative), the commutator
in \ref{eqn:commutator} vanishes and we are left with the familiar expression 
$$F_{\mu\nu}=\partial_\mu A_\nu - \partial_\nu A_\mu$$

The kinetic term in quantum electrodynamics gives Maxwell's equations in the
classical limit, and is obtained by considering the
 simplest gauge-invariant quantity involving the field strength tensor:
$$\mathcal{L}_{\mathrm{Maxwell}}=-\frac{1}{4}F_{\mu\nu}F^{\mu\nu}.$$
Putting it all together gives the Lagrangian describing quantum electrodynamics
(QED),
one of the most successful theories in the history of physics:
\begin{eqnarray*}
\mathcal{L}_{\mathrm{QED}}&=&\mathcal{L}_{\mathrm{Dirac}}+
\mathcal{L}_{\mathrm{Int}}+\mathcal{L}_{\mathrm{Maxwell}}\\
&=&\overline{\psi}(i\gamma^\mu D_\mu-m)\psi-\frac{1}{4}F_{\mu\nu}F^{\mu\nu}\\
D_\mu&\equiv& \partial_\mu+igA_\mu\\
F_{\mu\nu}&\equiv&D_\mu A_\nu- D_\nu A_\mu=\partial_\mu A_\nu - \partial_\nu A_\mu
\end{eqnarray*}

QED is formulated on a flat space-time manifold, but may have curvature
in `phase space' described by $A_\mu(x)$.  If the phase space is flat, 
then the parallel transporter will be path independent, meaning that 
$A$ is a total derivative of some function. 
In such a case, we can perform a local gauge transformation which eliminates
the electromagnetic field: $A\to0$.  Thus we see that if we treat
the electron
phase as an internal degree of freedom, it can give rise to
electromagnetic interactions as a result of curvature in this 
internal coordinate space.

The gauge field can be quantized to give the mediating boson of the gauge 
force.  
$U(1)$ has only one parameter $A_\mu(x)$, and thus there is only one
gauge boson of the quantized theory, the photon.

\subsection{From pions to the Standard Model}
In 1936, Breit~\cite{breit:charge_independence} discovered that the strong nuclear force was blind to electric
charge.  That is, the strong nuclear force between two protons was the same
as the strong nuclear force between a proton and neutron, or 
between two neutrons.  Heisenberg suggested that the proton and neutron 
were two states in the `isospin' doublet:
$$N=\binom{p}{n}.$$
Yang and Mills promoted isospin to a local gauge symmetry by introducing
a connection based on the $SU(2)$ gauge group, the set of all complex
$2\times2$ matrices with unit determinant.  This allowed for the identity
of the proton and neutron to vary as a function of position.
The Yang-Mills $SU(2)$ gauge field was shown to describe the pion in the same
way that the $U(1)$ field describes the photon.  In the case of $SU(2)$, there
are three generators yielding three gauge bosons, the $\pi^-$, $\pi^0$, and
$\pi^+$.

Physicists then searched for the `correct' gauge theories describing the
interactions of the standard model.  Work on the electroweak Lagrangian 
describing
the interactions of leptons and quarks via the $W^\pm$, $Z^0$, and $\gamma$
gauge bosons was started in 1961 by 
Glashow~\cite{glashow:electroweak}, 
completed in 1967 by Weinberg~\cite{weinberg:electroweak}
 and Salam~\cite{salam:electroweak}, and is based on the $SU(2)\times U(1)$
gauge group.

\subsection{Quantum Chromodynamics}
Quantum chromodynamics (QCD) is the gauge theory describing the interactions
of quarks and is the primary focus of this dissertation.
Fritzsch and Gell-Mann~\cite{fritzsch:qcd1, fritzsch:qcd2} identified
the quark color as the fundamental charge associated with QCD.  Quark fields
come in six flavors, possess four spin components 
(which transform irreducibly under 
Poincar\'e transformations), and have three color components.  Let the quark
field be denoted by
$\psi^A_{\alpha a}(x)$, 
where $A$ is the flavor index, $\alpha$ is the spin index,
 and $a$ is the color index.  In this section
we will focus on the behavior of the color index, and will suppress indices
where possible for simplicity of notation.

The color
reference frame (defining the directions of `red', `green', and `blue') is
defined locally.
The infinitesimal linear parallel transporter defining the covariant derivative
is now an $SU(3)$ matrix instead of a $U(1)$ phase:
$$U_{ab}(x,x+\epsilon \hat{e}_{(\mu)})=\exp(i g\epsilon A_\nu^c(x)\frac{\lambda^c}{2})_{ab}$$
where the $\lambda^a$ are the eight $3\times3$ traceless Hermitian Gell-Mann 
matrices which generate the $SU(3)$ group.  $SU(3)$ is the
non-Abelian (non-commutative) group of $3\times 3$ complex unitary
matrices with unit determinant.

We get the covariant derivative $D_\mu=\partial_\mu+igA_\mu$ by defining
$$A_\mu(x) \equiv \frac{1}{2}A_\mu^a(x)\lambda^a.$$
This gives us eight real gauge fields $A^a_\mu(x)$ which transform as vectors
under Poincar\'e transformations.  They represent eight spin-1 gluons,
the gauge bosons which mediate the QCD interactions among quarks.  The 
spin-1/2 quarks
are represented by Grassmann (anticommuting) fields 
which transform as
Dirac spinors under Poincar\'e transformations. 
In a hypothetical free field theory where $A=0$ (all the links are unity), 
the quantum operator $\overline{\Psi}$ creates a quark state
and
annihilates an anti-quark state while $\Psi$ annihilates a quark and
creates an anti-quark.  In the general interacting theory
($A\neq0$), we will still refer to $\overline{\Psi}$ as a 
{\em quark source} and $\Psi$ as a {\em quark sink}.

The gluon field strength tensor is given again by the curvature of the 
gauge field, this time with a non-vanishing gauge field commutator $[A_\mu, A_\nu]$:
\begin{eqnarray*}
F_{\mu\nu}\equiv \frac{1}{ig}[D_\mu, D_\nu] &=& D_\mu A_\nu - D_\nu A_\mu\\
&=&\partial_\mu A_\nu - \partial_\nu A_\mu + ig[A_\mu,A_\nu],
\end{eqnarray*}

The new commutator term can be expressed in terms of the gluon fields by 
using the fact that the 
Gell-Mann matrices $\left\{\lambda_a\right\}$ 
obey the $SU(3)$ Lie Algebra 
$$[\lambda_a, \lambda_b]=i2f_{abc}\lambda_c,$$
where the $f_{abc}$ are the completely antisymmetric $SU(3)$ structure 
constants defining the group.
\begin{eqnarray*}
ig[A_\mu, A_\nu] &=& \frac{ig}{4}A_\mu^a A_\nu^b[\lambda^a, \lambda^b]\\
&=&-\frac{g}{2}f_{abc}A_\mu^a A_\nu^b\lambda^c=-\frac{g}{2}f_{abc}A_\mu^b A_\nu^c \lambda^a
\end{eqnarray*}
Therefore
\begin{eqnarray*}
F_{\mu\nu}&=&\frac{1}{2}F_{\mu\nu}^a\lambda^a,\\
F_{\mu\nu}^a&\equiv&\partial_\mu A_\nu^a - \partial_\nu A_\mu^a
 -gf_{abc}A^b A^c
\end{eqnarray*} 

Using the fact that $\mbox{Tr}(\lambda^a\lambda^b)=2\delta^{ab}$, we may put
these gauge-invariant pieces together to define the QCD Lagrangian density, the 
functional defining the dynamics of the interaction of quarks and gluons:
\begin{eqnarray*}
\mathcal{L}_{\mathrm{QCD}}&\equiv&\overline{\psi}(i\gamma^\mu D_\mu - M)\psi
-\frac{1}{2}\mbox{Tr}(F_{\mu\nu} F^{\mu\nu}),\\
D_\mu&\equiv&\partial_\mu+igA_\mu,\\
A_\mu &\equiv& \frac{1}{2}A_\mu^a\lambda^a\\
F_{\mu\nu}&\equiv&\frac{1}{2}F^{a}_{\mu\nu}\lambda^a,\\
F_{\mu\nu}^a&\equiv&\partial_\mu A_\nu^a - \partial_\nu A_\mu^a
-gf_{abc}A^b A^c\\
M&\equiv&\mbox{diag}(m_d, m_u, m_s, m_c, m_b, m_t).
\end{eqnarray*}

In the above, $M$ in a $6\times 6$ matrix in flavor space, $\gamma^\mu$ is
a $4\times 4$ matrix in spin space, and $\lambda^a$ is a $3\times 3$ matrix in
color space.  In other spaces, the matrices appear as the identity (e.g. 
$\gamma^\mu$ acts as the identity in color space).  The trace is taken with
respect to color. 
The non-Abelian nature of QCD introduces three- and four gluon interactions
through the kinetic term $-\mbox{Tr}(F_{\mu\nu} F^{\mu\nu})/2$.
These gauge field self-interaction terms are believed to reproduce the 
desirable properties of asymptotic freedom and color confinement. Also,
the strong nuclear force binding nucleons together in the nucleus is a 
 seen to be a van der Waals-type residual interaction among the
quarks in different nucleons.  

\subsection{Calculating in quantum chromodynamics}

The Lagrangian density $\mathcal{L}_{\mathrm{QCD}}$ completely determines 
the dynamics and properties of quantum chromodynamics.
However, theorists still need to calculate complicated integrals of the 
form\footnote{The 
exact integrals {\em we} will need will be discussed in the next chapter.}
$$\frac{1}{Z}\int\mathcal{D}\overline{\psi}\mathcal{D}\psi \mathcal{D}A \,
f(\overline{\psi},\psi,A)\exp\left\{iS[\overline{\psi},\psi, A]\right\},$$ 
where
$$Z=\int\mathcal{D}\overline{\psi}\mathcal{D}\psi \mathcal{D}A \, 
\exp\left\{iS[\overline{\psi},\psi, A]\right\}.$$

At high energies, asymptotic freedom tells us the the interactions among
quarks are weak, and can thus be described using perturbation theory.
Perturbative QCD treats the exponential as a power series expansion in
the strong coupling constant $g$.  At high energies, or equivalently short length scales (such
as those probed in deep inelastic scattering experiments), the renormalized
coupling constant becomes small and the power series is convergent.  By
calculating the `tree level' and the first few `correction' terms for various
processes, theorists have shown that the predictions of QCD are in striking
agreement with the results of high-energy experiments~\cite{muta:qcd}.

At lower energies (e.g. below 2 GeV in e+p scattering, where Bjorken 
scaling is violated), we enter the domain of `nonperturbative phenomena,' such
as color confinement (hadronization).  At these energies, the renormalized
coupling constant becomes large, and perturbation expansions diverge badly.

\subsection{The lattice regulator}
In 1974, Kenneth Wilson introduced a regularization 
method~\cite{wilson:lattice} which made rigorous low-energy 
calculations in QCD
possible for the first time.
Wilson defined an action on
a discrete space-time lattice with spacing $a$.
This serves two purposes: first, if the lattice is introduced in a finite
volume, the number of integration variables becomes finite and the integrals
become well defined.  Second, the lattice regulator introduces an 
ultraviolet cutoff at the scale of the inverse lattice spacing $1/a$, thus
controlling divergences.
The introduction of a regulator necessarily breaks symmetries of the theory, 
and it is hoped that these symmetries will be recovered in the continuum 
limit.  For instance, the lattice regulator explicitly breaks Poincar\'e 
invariance.

Let the elementary lattice vectors which give the minimal
displacement in each direction be denoted as: 
$$\hat{\mu}\equiv a\hat{e}_{(\mu)}.$$
The {\em quark fields} are defined at each site as
$$\psi^A_{\alpha a}(x),$$ 
where $x$ denotes the discrete site on the lattice, $A$ is the flavor index, 
$\alpha$ is the Dirac spin index, and $a$ is the color index.
Rather than simply replacing
the derivatives in the continuum action with finite differences, Wilson
preserved local gauge invariance by working directly with 
{\em gauge links}, the $SU(3)$ parallel 
transporters defined on the elementary links between neighboring sites:
$$U_{\mu ab}(x)\equiv U_{ab}(x, x+\hat{\mu})$$
where $a$ and $b$ are color indices.
For simplicity of notation, we will continue to suppress indices where
possible.  $SU(3)$ matrices 
$U$ and $\Omega$ will always act on the color degrees of freedom, Dirac
gamma matrices $\gamma^\mu$ will always act on the spin degrees of freedom,
and the mass matrix $M$ will always be $\mbox{diag}(m_d, m_u,\cdots)$ and will
act on the flavor degrees of freedom.

\begin{figure}[ht!] 
  \centering 
  \includegraphics[width=2.5in]{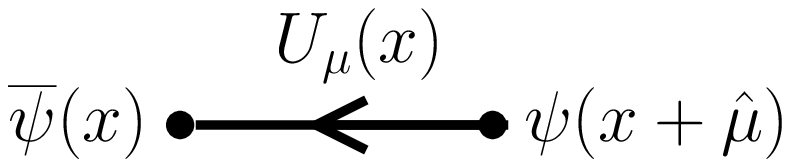}
  \hspace{0.5in}
  \includegraphics[width=2.5in]{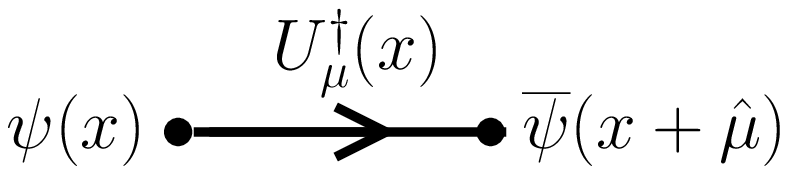}
  \caption{Two gauge-invariant quark-antiquark operators.  The quark and
antiquark fields at neighboring sites $x$ and $x+\hat{\mu}$ may be combined
by use of a gauge link at $x$.  The gauge link $U_\mu(x)$ parallel
transports a color vector from $x+\hat{\mu}$ to $x$ and the Hermitian
conjugate gauge link $U^\dagger(x)$ transports a color vector in the opposite
direction.  Any gauge-invariant quark-antiquark
operator can be formed by connecting the fields at any two sites by a suitable
product of gauge links.}
  \label{fig:q-q_bar-diagram} 
\end{figure}

Under a local gauge transformation represented by $SU(3)$ matrices 
$\Omega(x)$ at each site, the quark fields and gauge links 
transform as:
\begin{eqnarray*}
\psi(x)&\to& \Omega(x)\psi(x)\\
\overline{\psi}(x)&\to& \overline{\psi}(x)\Omega^\dagger(x)\\
U_\mu(x) &\to& 
\Omega(x)U_\mu(x)\Omega^\dagger(x+\hat{\mu})\\
U^\dagger_\mu(x) &\to& 
\Omega(x+\hat{\mu})U_\mu(x)\Omega^\dagger(x)
\end{eqnarray*}
This formulation allowed Wilson to form gauge-invariant expressions such as
$$\overline{\psi}(x) U_\mu(x)\psi(x+\hat{\mu})$$
and
$$\overline{\psi}(x+\hat{\mu}) U^\dagger_\mu(x)\psi(x)$$
denoted pictorially in Figure \ref{fig:q-q_bar-diagram}.

If we replace $D_\mu\psi(x)$ with the symmetric difference
$$\frac{1}{a}\nabla_\mu\psi(x)\equiv \frac{1}{2a}\left(U_\mu(x)\psi(x+\hat{\mu})
-U^\dagger_\mu(x-\hat{\mu})\psi(x-\hat{\mu})\right)$$
we get the fermion part of the action (in Minkowski space-time):
\begin{eqnarray*}
S^M_F&\equiv&\sum_x a^4\overline{\psi}(x)(i\gamma^\mu \frac{1}{a}\nabla_\mu - M)\psi(x)
\end{eqnarray*}
The exact form of fermion action we will use will be discussed in chapter
\ref{chap:overview}.  It differs from the above because we will
work in Euclidean (not Minkowski) space-time, use an anisotropic lattice,
absorb the factors of $a$ to work in `lattice units,'
and add the `Wilson' term which fixes a discretization problem (the
`fermion doubling' lattice artifact).  

\begin{figure}[ht!] 
  \centering 
  \includegraphics[width=2.5in]{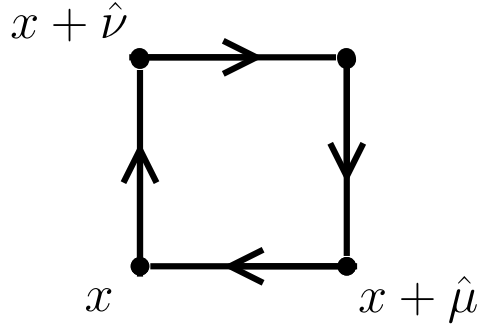}
  \caption{The gauge-invariant plaquette operator $U_{\mu\nu}(x)$.}
  \label{fig:plaquette} 
\end{figure}

Another gauge-invariant quantity involving just the links is the 
{\em plaquette}, the trace (with respect to color) of a product of links
around an elementary loop on the lattice:
$$U_{\mu\nu}\equiv \mbox{Tr}(U_\mu(x)U_\nu(x+\hat{\mu})U^\dagger_\mu(x+\nu)U^\dagger(x)),$$
denoted in figure \ref{fig:plaquette}.

The plaquette is the trace of the parallel transporter around an
elementary loop.  It is reasonable to expect that the plaquette is related
to the field strength tensor, and this can be verified by writing 
$U_{\mu\nu}(x)$ in terms of the $A$ fields using the Campbell-Baker-Hausdorff
formula:
$$\exp(A)\exp(B)=\exp(A+B+\frac{1}{2}[A,B]+\mbox{(terms higher in A and B)}).$$

Because we are interested in the continuum limit where $a\to 0$, we may
write
$$U_\mu(x)\stackrel{a\to 0}{\to} \exp(igaA_\mu(x))$$

Then expanding the exponentials and keeping terms only up to order $a^2$
gives:
\begin{eqnarray*}
U_{\mu\nu}(x)&\equiv&U_\mu(x)U_\nu(x+\hat{\mu})U^\dagger_\mu(x+\nu)U^\dagger(x)\\
&=&\exp(igaA_\mu(x))\exp(igaA_\nu(x+\hat{\mu}))\exp(-igaA_\mu(x+\hat{\nu}))
\exp(-igaA_\nu(x))\\
&=&\exp\left( ig a^2(\nabla^f_\mu A_\nu(x) - \nabla^f_\nu A_\mu(x)
+ig\left[A_\mu(x),A_\nu(x)\right])+O(a^3)\right)\\
&\stackrel{a\to 0}{\to}&\exp\left(iga^2 F_{\mu\nu}(x)+O(a^3)\right)
\end{eqnarray*}
where
$$\nabla^f_\mu A_\nu(x) \equiv \frac{1}{a}(A_\nu(x+\hat{\mu})-A_\nu(x))
\stackrel{a \to 0}{\to}\partial_\mu A_\nu(x)$$
is the forward lattice difference when applied to gauge fields,
and we have evaluated all of the commutator terms at $x$, for example:
$$-\frac{g^2a^2}{2}[A_\mu(x),A_\nu(x+\hat{\mu})]
=-\frac{g^2a^2}{2}[A_\mu(x),A_\nu(x)]+O(a^3)$$
Thus 
\begin{eqnarray*}
\mbox{Re}\,\mbox{Tr}(U_{\mu\nu})
&=&\frac{1}{2}(\mbox{Tr}(U_{\mu\nu})+\mbox{Tr}(U^\dagger_{\mu\nu}))\\
&=&\frac{1}{2}(\mbox{Tr}(U_{\mu\nu})+\mbox{Tr}(U_{\nu\mu}))\\
&=&\mbox{Tr}(1)-g^2a^4\frac{1}{2}\mbox{Tr}(F_{\mu\nu}F^{\mu\nu})+O(a^5)
\qquad\mbox{(no sum)}
\end{eqnarray*}
because $F_{\nu\mu}=-F_{\mu\nu}$ (also because Tr$(F_{\mu\nu})=0$ for $SU(3)$).
In the gauge action, we will write $N\equiv \mbox{Tr}(1)$.

The pure gauge (also called {\em Yang-Mills}) part of the lattice action 
(in Minkowski space-time) is 
thus given by~\cite{montvay:lgt}:
\begin{eqnarray*}
S^M_G
&\equiv&-\frac{\beta}{N}\sum_x\sum_{\mu<\nu}\mbox{Re\,Tr}(1-
U_{\mu\nu}(x)),\qquad \beta\equiv \frac{2N}{g^2},\\
&=&-\frac{2}{g^2}\sum_x\sum_{\mu<\nu}\left\{N-\mbox{Re\,Tr}(U_{\mu\nu}))
\right\},\\
&=&-2\sum_{x}\sum_{\mu<\nu} a^4\, \frac{1}{2}\mbox{Tr}(F_{\mu\nu}F^{\mu\nu})+O(a^5).\\
&=&-\sum_{x}\sum_{\mu,\nu} a^4\, \frac{1}{2}\mbox{Tr}(F_{\mu\nu}F^{\mu\nu})+O(a^5).\\
\end{eqnarray*}
For $SU(3)$, $\beta =\frac{6}{g^2}$.  In the case of Abelian gauge
groups, the expression for $\beta$ is divided by two in order to
reproduce the continuum action to leading order in $a$~\cite{montvay:lgt}.

\subsection{Renormalization and integration}
 Wilson's lattice regulator breaks Poincar\'e invariance, but it maintains
local gauge invariance which is crucial to ensuring that the regularized
theory is renormalizable.  In the continuum limit ($\lim a\to 0$), 
it is easy to see that
Poincar\'e invariance is restored.  Any quantity calculated using
the lattice regulator will equal the continuum quantity plus an infinite
number of $O(a)$ terms.  There is no way to guarantee that the total
effect of these terms will vanish as $a\to0$ (a series of terms may 
combine to give an $O(1/a)$ contribution which will remain in the continuum
limit).  Thus it is critical to make a theory which is renormalizable at 
finite $a$.

\label{sec:gauge_fixing}
Another advantage is that the Wilson action is a functional of
the gauge links $U_\mu(x)$, which are compact gauge fields.  This means
that the parameter space we must integrate over is closed.  In contrast,
the underlying gauge fields $A^a_\mu(x)$ may take any value in the
open set $(-\infty, \infty)$.  There are an infinite number of values of
$A^a_\mu(x)$, related by gauge transformations, that correspond to a 
single value of the gauge link $U_\mu(x)$. 
 This gauge freedom implies that gauge fixing must be
used in integrals over the $A$ fields to avoid divergences.  No gauge fixing
is required when integrating over the gauge links 
directly~\cite{wilson:lattice}. 

The coupling constant $g$ is renormalized by tuning $\beta=6/g^2$ and 
calculating some physical 
quantity to get $a$.  In this way, we may send $a$ to zero by increasing
$\beta$.  When all ratios become independent of $a$, then
we have reached the {\em scaling regime}.  Similarly, we may renormalize
the quark masses by tuning the bare quark masses while measuring the pion
mass $m_\pi$ and watching the ratios of different calculated masses.
It is hoped that these parameters will `flow' to a critical point
where the theory will reproduce the predictions of continuum QCD.

The appearance of the {\em inverse} power of the coupling constant in
$\beta=6/g^2$ allowed
Wilson to calculate some, but not all, non-perturbative quantities using
the so called strong coupling expansion.  For instance, he showed that
the static quark-antiquark potential increases linearly with separation in 
the Yang-Mills (pure gauge) $SU(3)$ theory, which is evidence
that $QCD$ may be confining in the continuum limit. 

The lattice regulator is also perfectly suited for calculation on a computer.
The framework
of calculating QCD quantities using a lattice regulator on a computer is 
called {\em lattice QCD}.  
Starting from the work of Michael Creutz~\cite{creutz:abelian_lgt} and others
in 1979, lattice gauge theorists have applied the techniques
of Monte Carlo integration discussed in chapter \ref{chap:monte_carlo}
to calculate increasingly ambitions quantities.  
This work is a contribution
to one such effort: quantum operator design for the low-lying baryon spectrum.

\section{Unresolved mysteries in hadronic physics}

The high energy predictions of QCD obtained from perturbation theory
do a spectacular job of describing the measurements of high-energy physics 
experiments.  On the other hand, the low-energy behavior of QCD has not
 been thoroughly tested.   This is due to the fact, mentioned above, that
such calculations are nonperturbative with respect to the coupling constant
of the theory.  Fortunately, theoretical
methods and computational power have now evolved to the point where the 
extraction of low-energy predictions from QCD has become feasible.

Predictions from lattice QCD will come at an opportune time; there are several 
`mysteries'
concerning the hadron spectrum which the constituent quark 
model, along with other QCD 
inspired models, have failed to resolve in a definitive manner.

\subsection{Dynamics in the constituent quark model}
The constituent quark model is able to predict the masses of many observed
low-energy hadron resonances.  It does this by treating the hadron states
as either a quark-antiquark state (for mesons) or as a three-quark state
(for baryons).  The quark interactions lead to predictions of the total
energy and quantum numbers of the state.  Different quark models
specify different interaction mechanism, but most consist of 
non-relativistic dynamics
of the form (adapted from ~\cite{hagiwara:pdg}):
\begin{enumerate}
\item A confining interaction, which is generally spin-independent.
\item A spin-dependent interaction, modeled after the effects of gluon exchange 
in QCD. For example, in the S-wave states, there is a spin-spin hyperfine
interaction of the form
$$H_{HF}=-g M \sum_{i>j}(\vec{\sigma}\lambda_a)_i(\vec{\sigma}\lambda_a)_j$$
where $M$ is a constant with units of energy.  Spin-orbit interactions, 
although allowed, seem to be small.  This may be due to relativistic 
QCD-inspired  
corrections such as a Thomas precession term~\cite{capstick:quark_model}.
\item A strange quark mass somewhat larger that the up and down quark masses,
in order to split the $SU(3)$ flavor multiplets.
\item In the case of isoscalar mesons, an interaction for mixing 
$q\overline{q}$ configurations of different flavors (e.g. $u\bar{u}
\leftrightarrow d\bar{d} \leftrightarrow s\bar{s}$), in a manner which is
generally chosen to be flavor independent.
\end{enumerate}

\subsection{Missing baryon resonances}
The constituent quark model predicts many more baryon
 resonances than have been observed experimentally.  Because most of the 
observed
baryon resonances have been produced from electron beams (electroproduction)
or photon beams (photoproduction), it is possible that
these `missing resonances' are created in channels 
which have not yet been sufficiently examined.  Experiments are currently
underway using different types of beams (such as linearly polarized photons)
to excite resonances through different processes.  

An alternative explanation suggests that the missing resonances 
are due to an excess of degrees of freedom in the quark model.  One proposed 
fix is to reduce the 
degrees of freedom by 'locking' two of the quarks of a baron together in a 
di-quark configuration.  In this picture, the higher resonances would
be various single-quark excitations of the three-quark bound state.

Direct QCD calculations are needed to determine the properties of these
missing resonances.  Lattice QCD is in a position to identify
the relevant degrees of freedom in a baryon.  Simulations can probe the 
spatial structure of the fields which compose the baryon state thus leading
to a better understanding.

\subsection{The Roper resonance}
The simplest confining interaction in the quark model
is the linear oscillator potential.  
If that potential is used, then the baryon spectrum can be divided
into {\em oscillator bands}, which correspond to increasing excitations
of the spatial wavefunctions with alternating parity.  Within each 
band, the levels split according to spin-spin and spin-orbit interactions.
Consider the five lowest known spin-1/2 nucleons (J=1/2, I=1/2) 
 and their corresponding levels in the oscillator 
approximation~\cite{hagiwara:pdg} shown in table \ref{table:nucleons}.
\begin{table}
$$
\begin{array}{|c|c|c|c|c|}
\hline
\mbox{Name} & L & S & P & \mbox{Oscillator Band}\\
\hline
N(939)  & 0 & 1/2 &+&0\\
N(1440) & 0 & 1/2 &+&2\\
N(1535) & 1 & 1/2 &-&1\\
N(1650) & 1 & 3/2 &-&1\\
N(1710) & 0 & 1/2 &+&2\\
\hline
\end{array}
$$
\label{table:nucleons}
\caption{The lowest lying spin-1/2 nucleons.  The numbers in parenthesis
are the masses of the resonances in MeV, $L$ is the orbital
angular momentum of the quarks, $S$ is the total spin
of the quarks, and $P$ is the parity of the state.  The quark model
treats the nucleon resonances as increasing excitations in an oscillator
potential.  The $N(1440)$, or Roper, is measured to be lower than predicted
by the quark model.}
\end{table}

The $N(1440)$, known as the `Roper' resonance, is the first excited even-parity 
nucleon state, but it is {\em below} the lowest odd-parity spin-1/2 
nucleon state, 
the $N(1535)$.  This is surprising~\cite{mathur:roper}, because most quark models treat
the Roper as a radially-excited `breathing mode' of the $N(939)$, 
which belongs to
the second oscillator band having two nodes in the spatial wavefunction.
On the other hand, the odd-parity state $N(1535)$ is modeled as an orbitally
excited state in which the quarks have orbital angular momentum $L=1$ with
only one node in the spatial wavefunction.
 
Alternative models have proposed that the Roper resonance is a $qqqg$
hybrid state
with excited glue or that it is a $qqqq\bar{q}$ 
five-quark molecular state.  The mystery is complicated by the facts that the 
Roper
has a wide width and 
has only been detected from phase shift analysis, never directly as an 
enhancement in the cross section. 

\subsection{The $\Lambda(1405)$}
The $\Lambda(1405)$ having $I(J^P)=0(\frac{1}{2}^-)$ is 
another puzzle in baryon 
spectroscopy~\cite{nemoto:lambda1405}.
In the quark model, the $\Lambda$ series of resonances are described as 
different excitations of a three-quark $uds$ flavor singlet system.  
Because the $\Lambda$ contains a strange quark, the first odd-parity
state would be expected to be above the first odd-parity nucleon, which
contains only the lighter up and down quarks.
However we see from table \ref{table:nucleons} that
 the $\Lambda(1405)$ is below the $N(1535)$, the first
negative-parity nucleon.
It has been proposed that the $\Lambda(1405)$ is a linear combination the
three-quark $uds$ state with  two
baryon-meson bound states 
$N\bar{K}$ and $\pi\Sigma$, where the $\Sigma$ baryons have even parity,
strangeness=-1, and $I(J^P)=1(\frac{1}{2}^+)$.
 
Because the molecular states will mix with the 
three-quark state, lattice calculations are needed to determine the 
amount $qqqq\bar{q}$ contribution vs the amount of $qqq$ contribution in
the $\Lambda(1405)$.  Understanding molecular hadronic states will lead to 
a better understanding of hadron structure and hadron reactions.

\subsection{Exotic hadron states} 
The quark model treats hadron resonances as excitations of quarks in a 
confining potential with inter-quark interactions.  
It does not have anything to say about excitations
of the gluon field.
Laboratories are dedicating significant resources to the exploration of 
these `hybrid' states of quarks bound by excited glue.  
QCD predicts that the gluon field can carry quantum numbers such as angular 
momentum, resulting
in the possibility of {\em exotic} states, with quantum numbers inexplicable
 within the quark model.  If such a state was found, it would be a 
significant low-energy validation of QCD.  Two exotic $J^{PC}=1^{-+}$ 
hybrid meson candidates
have been tentatively identified at 1.6 GeV and 1.4 GeV in the $\rho\pi$ 
and $\eta\pi$ channels respectively by the E852 collaboration at 
BNL~\cite{adams:hybrid_meson, chung:hybrid_meson}.

Lattice QCD theorists are assisting in this effort by calculating
the properties of exotic mesons and baryons.  Understanding hybrid states
will lead to a better understanding of when and how the gluonic degrees 
of freedom play a role in the baryon spectrum.  

\subsection{Current experimental efforts}
The spectrum is a fundamental quantity of QCD, yet remains incompletely 
understood by experimentalists and theorists alike.  There are a number of 
high-precision baryon spectroscopy initiatives at laboratories including
Jefferson Laboratory (JLab), BNL, Mainz, Graal, and BES.

CLEO-c at Cornell is committed to exploring heavy quark 
systems~\cite{online:cleo_c}, Hall B at JLab is dedicated to 
mapping out the low-lying baryon spectrum, 
and the planned Hall D at JLab will search for exotic hadrons
~\cite{manley:baryon_status}.  Each one of these efforts will rely on 
lattice QCD calculations for valuable theoretical input.
Such calculations will tell the experimentalists the quantum numbers of the 
states they may find, and will help them decide on the optimal 
energy ranges to explore.

\section{Goal of this work}
The need has never been greater for a systematic exploration of the 
hadron spectrum using the best theory we have, quantum chromodynamics.  
Unfortunately, low-energy features of QCD cannot be 
treated by perturbation expansions
in the coupling constant.  As the world's experimental dataset grows,
so does the number of resonances which cannot be adequately described
by QCD-inspired phenomenological models.
The lattice approach allows us to numerically solve QCD, not just model it, 
and is currently the best tool for low-energy
QCD calculations available.

The Lattice Hadron Physics
Collaboration (LHPC) was formed in 2000, with the stated goal of using lattice
QCD to understand the structure and interactions of
 hadrons~\cite{isgur:lhpc_proposal}.
One major goal of the collaboration is to calculate a significant portion
of the low-lying spectrum of baryon resonances in QCD by developing and using
the best possible methods in lattice QCD.

Spectrum calculations serve four main purposes.  First, such calculations help
answer the question: can QCD reproduce the observed hadron 
spectrum?  Second, predictions of the energy and quantum numbers of 
new states will help the experimental community as they plan and design the 
next generation of experiments.  Third, spectrum calculations help
identify the important degrees of freedom in the theory.
Specifically, theorists hope to identify subsets of configurations which 
dominate the functional integral.
In thermodynamics, systems with an unmanageably large number of degrees
of freedom can be described in terms of a few bulk properties such as 
pressure and specific heat.  Similarly, lattice QCD may find a way to describe 
systems in quantum field theory in terms of a few dominant subsets
of configurations.
Such a description would facilitate the construction of simpler models and 
calculation schemes which would serve as an analytical complement to a
fundamentally numerical approach.
 Among other things, such models may help us better understand the process 
of hadron formation in the early universe. 
Fourth, just as hydrogen spectroscopy led to the rapid development of quantum
mechanics, so too may hadron spectroscopy lead to new theoretical
breakthroughs. 

This work presents the operator design, tuning, and pruning techniques 
developed during the course of my research.  These methods have been used
to achieve unprecedented signals for spectral states in the 
nucleon channel, and are readily extended to the other
baryon channels ($\Delta$, $\Lambda, \Sigma, \Xi, \cdots$). 
This work describes both 
the methodology used in the operator design and the computational steps 
involved in tuning and pruning the operators.  Specific operators are 
presented which can be used to extract the low-lying nucleon spectrum
in future high-statistics, low-quark mass calculations.  Preliminary 
heavy-quark quenched spectrum results are given which, while unsuitable
for comparison with experiment, do show the signal quality achieved by 
the operators.
It is expected that these operators will play a valuable role in the
LHPC's  ongoing
high-precision baryon spectrum calculations.

\section{Organization of this dissertation}
The formalism of extracting the spectrum of a quantum field theory is
reviewed in chapter \ref{chap:overview}.  
We describe how QCD is formulated on a 
computer, how the Monte Carlo method is used to calculate expectation
values of correlation functions of quantum operators, and what these functions
can tell us about the spectrum of QCD.  
We discuss the role of 
Euclidean space-time correlation functions, a method for reliably
extracting multiple excited states, and the importance of the lattice
symmetry group in identifying the spin of the created states.

In chapter \ref{chap:monte_carlo}, we introduce the lattice-regulated
formulation of QCD, and the review the Monte Carlo method for evaluating 
functional integrals.  We also discuss our error analysis methodology which 
uses bootstrap and jackknife resampling.

In chapter \ref{chap:construction}, we discuss construction of nucleon operators
from the building blocks we have at our disposal on the lattice.  We introduce
smeared quark- and gauge-fields which provide a closer link between our
operators and phenomenological models.  Covariant displacement is discussed as a
natural way to build up radial structure.  We treat in
detail the classification of our quantum operators based on their behavior
under groups of symmetry transformations, and how this behavior relates to the
continuum spin values of the states created by our operators. 

In chapter \ref{chap:correlators}, we discuss the combination of our operators
into correlation matrices.  We integrate out the quark degrees of freedom and
reduce the problem to that of evaluating three-quark propagators.
We then provide a detailed 
walk-through of the computational steps performed on the Carnegie Mellon
University Medium Energy Physics computer cluster which allowed us to
tune and select the best quantum operators for extracting the low-lying QCD
spectrum.

The tuning of the operators is discussed in chapter \ref{chap:smearing}, where 
stout-link gauge-field smearing is shown to reduce the
 noise of extended baryon operators, while Laplacian 
quark field smearing is shown to reduce the contamination in
 of our baryon operators due to high-frequency modes.  The systematic approach
we developed to tune the smearing parameters is presented, and 
the optimal smearing parameter values are reported.

The construction of nucleon operators utilizing the symmetries of the lattice
discussed in chapter \ref{chap:construction} leads to a unmanageably large
number of operators for our correlation matrices.  Chapter \ref{chap:pruning}
discusses the systematic approach we developed to prune these operators down
to an optimal linear combination of sixteen operators.

These sixteen operators are then used to extract the low-lying nucleon spectrum
on 200 quenched configurations.  In chapter \ref{chap:results} the methodology 
and analysis code written for this task are discussed, 
and fit results are presented.  Although the configurations are quenched, 
in a small volume, and have an unrealistic quark mass, we still get reasonable
results for the nucleon spectrum.  The spectrum extracted in this work
is discussed in comparison to the experimentally measured nucleon excitation
spectrum, and the predictions of the relativistic constituent quark model.
We conclude by summarizing the main results of this work, and by
discussing the ongoing efforts of the Lattice Hadron Physics Collaboration.

\chapter{Calculation overview}
\label{chap:overview}

\section{Spectral states and resonances}
\subsection{Hilbert spaces}
In quantum field theory, all physical states of a system are
represented by rays in a Hilbert space $\mathcal{H}$.  A ray is an equivalence
class of vectors differing only by a non-zero scale factor.  Thus $\ket{\phi}$
and $\alpha\ket{\phi}$ ($\alpha\neq0$) represent the same state of the system.

A Hilbert space is defined as a normed (or metric) space in which all 
Cauchy sequences are convergent.
The norm $\parallel \cdot \parallel$ is defined for any vector
$\ket{\phi}$ as
$$\parallel \ket{\phi} \parallel \equiv \sqrt{\braket{\phi}{\phi}}.$$
A sequence of vectors $\{\ket{\phi_n}\}$ is a Cauchy sequence if for every
real number $\epsilon>0$ there is a non-negative integer $N$ such that
$$\parallel \ket{\phi_n}-\ket{\phi_m}\parallel < \epsilon\qquad 
\forall\;\;m,n>N.$$
A normed space is a Hilbert space if all Cauchy sequences in the space converge
to an element in the space.
This property is essential to justify many assumptions we will make about the
convergence of integrals over elements in this space.

\subsection{The Hamiltonian}
Allowed transformations on the system are represented by unitary operators 
which 
map Hilbert space states to Hilbert space states.  Examples of 
transformations include Poincar\'e transformations (space-time translations, 
rotations, and boosts), reflections, and charge conjugation.
Continuous transformations can be written as the exponentiation of
Hermitian operators, known as generators.
For example, spatial translations are generated by the momentum operator 
$\vec{p}$:
\begin{eqnarray*}
f(\vec{x})\to f'(\vec{x})&=&f(\vec{x}-\vec{a})\\
&=&\exp(- a^j\partial_j)f(\vec{x})\\
&\equiv& \exp(ia^j p_j)f(\vec{x}),\quad p_j\equiv i\partial_j
\end{eqnarray*}
temporal translations are generated by the Hamiltonian, and rotations are
generated by the angular momentum operator.

The spectrum of a theory consists of those states $\ket{k}$ in the Hilbert
space which are eigenstates of the Hermitian generator of temporal translations 
$H$, the Hamiltonian operator:
$$H\ket{k}=E_k\ket{k}.$$
These are the {\em steady-states} of the theory which remain stable
if allowed to freely propagate.  Because the Hamiltonian operator is
Hermitian, the eigenvalues $E_k$ are real and represent the energy of the
states $\ket{k}$.  States which are eigenvectors of any Hermitian generator
 which commute with the Hamiltonian (such as linear and angular momentum)
 may also be labeled by their associated eigenvalue.  Thus, it is possible
to speak of a state having a particular value of both energy and momentum.

An additional constraint placed on a quantum field theory is that the 
Hamiltonian must be bounded from below.  This means that there is a stable
state of minimum energy, the vacuum $\ket{\Omega}$.  
By adding a suitable constant to the Hamiltonian (which does 
not change the spectrum), we may require that
$$H\ket{\Omega}=0.$$ 
The existence of a state of minimum energy is essential in
a quantum field theory to ensure that the field degrees of freedom in a finite
spatial region don't radiate away an infinite amount of energy as they drop
to lower and lower energy states.

The stable energy states $\ket{k}$ form a complete basis for the Hilbert space:
$$\sum_k \ket{k}\bra{k} = 1,$$
where the sum is over both discrete and continuous states (with the appropriate
measure).

\subsection{Resonances}
In a scattering experiment, a resonance is characterized by a dramatic increase
in cross-section with a corresponding sudden variation in the scattering phase
shift.  A resonance observed in the scattering of hadrons can often be 
interpreted as an unstable particle lasting long enough to be recognized as
having a particular set of quantum numbers.  Such a state is not an
eigenstate of the Hamiltonian, but has a large overlap onto a single 
eigenstate. Note that if the quark mass is taken to be large in the simulation, then
it may happen that a normally unstable state would be kinematically 
forbidden to decay due to energy conservation.  Because its would-be
decay products are too heavy, such a resonance would appear as a stable state
 in
the spectrum. 

Finite spatial volumes provide a second way to infer the existence and 
properties of resonances.
A method of using the lattice approach to probe resonances
is presented in Refs.~\cite{luscher:scattering, 
luscher:torus, luscher:unstable}.  In this method,
the dependence of a two-particle spectrum on the spatial
box size is used to determine elastic scattering phase shifts, which
are in turn related to the energies and widths of any resonances in 
that channel.
For this work, we will extract the spectrum in a single volume.  Future
studies will be needed to determine the single and multi-particle content
of the spectrum we find.

\section{Euclidean space-time} 
Having the Lagrangian describing QCD, our next step is to
find a method to extract the spectrum.
In ordinary quantum mechanics, we would simply solve the Schr\"odinger 
equation for the stationary states in time: $i\partial_t \psi = E\psi$.
Unfortunately, it is extremely difficult to work directly
with the QCD Schr\"odinger 
equation due to the number of variables and the self-interaction terms.  
It is more practical to extract
the energy levels of the theory from correlation functions of fields in 
Euclidean space-time. 

The following discussion will deal with the theory of a single real
scalar field $\phi$.  Afterward, we will discuss subtleties that arise
when applying the results to the case of quark fields and gauge links.

The Hilbert space and the original field can be reconstructed from the
Green's functions.
Osterwalder and Schrader~\cite{osterwalder:euclidean1, osterwalder:euclidean2}
showed that under general conditions, we may analytically continue the 
Minkowski Green's 
functions to imaginary 
time.
Doing so will make the integrals explicitly convergent and will allow
us to use the computational tools of statistical mechanics to calculate
quantities in this system.
We will summarize the main features of this approach in the following.  A 
more detailed treatment is presented in~\cite{montvay:lgt}.
For the rest of this work, we will work in Euclidean space-time, and will
denote Minkowski space-time quantities with a subscript or superscript $M$.

The points on the Euclidean space-time manifold are related to those on the
Minkowski space-time manifold as:
\begin{equation}
x^4=x_4=ix^0_M=ix^M_0,\qquad x^j=x_j=x^j_M=-x_j^M,
\end{equation}
\begin{equation}
\partial_4=\partial^4=-i\partial_0^M=-i\partial^0_M,\qquad \partial_j
=\partial^j=\partial_j^M=-\partial^j_M.
\end{equation}
With these definitions, the new metric therefore becomes Euclidean:
$\delta_{\mu\nu}=\mbox{diag}(+1,+1,+1,+1)$, and there is no distinction between 
covariant (lower) and contravariant (upper) indices.  Summation over repeated
indices will be assumed unless specified otherwise.

Making the above replacements, the continuum fermionic action becomes:
\begin{eqnarray}
iS_F^M&=&i\int dx^0_M d^3 x_M\,\overline{\psi}_M(x_M)(i\gamma^0_M\partial_0+ i\gamma^j_M\partial^M_j-m)\psi_M(x_M)\\
&\to&\int dx_4 d^3x \,\overline{\psi}(x)(-\gamma^0_M\partial_4+ i\gamma^j_M\partial_j-m)
\psi(x)
\end{eqnarray}
If we define the Euclidean space-time Dirac $\gamma$ matrices as
\begin{equation}
\gamma^4=\gamma_4=\gamma^0_M,\qquad \gamma_k=\gamma^k=-i\gamma^k_M
\label{eqn:eucl_gamma}
\end{equation}
then we have:
\begin{eqnarray}
iS_F^M&\to&-\int d^4x\,\overline{\psi}(x)(\gamma_\mu\partial_\mu+m)\psi(x)\\
&\equiv&-S_F.
\end{eqnarray}
Using Eqn. \ref{eqn:mink_gamma1} and Eqn. \ref{eqn:mink_gamma2}, we
see that the Euclidean space-time Dirac $\gamma$ matrices defined
in \ref{eqn:eucl_gamma} satisfy:
\begin{equation}
\{\gamma_\mu,\gamma_\nu\}=2\delta_{\mu\nu},\qquad
\gamma_\mu^\dagger=\gamma_\mu
\end{equation}

We will use the standard Dirac-Pauli representation for the $\gamma$ matrices:
\begin{equation}
\gamma_4=\left(
\begin{array}{cc}I & 0\\ 0 & -I\end{array}\right),\qquad
\gamma_j=\left(\begin{array}{cc} 0 & -i\sigma_j\\
i\sigma_j & 0\end{array}\right),
\end{equation}
where $I$ is the $2\times2$ identity matrix and the Pauli spin matrices 
are
\begin{equation}
\sigma_1=\left(\begin{array}{cc} 
0 & 1\\ 
1 & 0\end{array}\right),\quad 
\sigma_2=\left(\begin{array}{cc} 
0 & -i\\
i & 0\end{array}\right),\quad
\sigma_3=\left(\begin{array}{cc}
1 & 0\\
0 & -1\end{array}\right).
\label{eqn:pauli_spin}
\end{equation}

The fermion fields $\psi(x)$ and $\overline{\psi}(x)$ transform irreducibly
under $SO(4)$ rotations in Euclidean space-time, and represent the Minkowski
fields analytically continued to imaginary time.  Unlike in Minkowski 
space-time, $\psi$ and $\overline{\psi}$ must be treated as independent
fields: $\overline{\psi}\neq\psi^\dagger \gamma_4$.  This is required to
simultaneously satisfy Euclidean covariance of the fields, the canonical
anticommutation relations of the fermion fields, and the equality
of the Euclidean two-point function with the relativistic Feynman 
propagator continued to imaginary times~\cite{osterwalder:euclidean_fermions,
williams:euclidean_fermions}.

\section{Euclidean gauge links and gauge action}
The fermion action may now be discretized as before, using gauge links 
which are now
defined between neighboring quark field sites on the Euclidean
manifold.
\begin{eqnarray*}
U^M_\mu(x)&=&\mathcal{P}\exp\left\{ig\int_x^{x+\hat{\mu}}dz_M^\nu\,A^M_\nu(z)
\right\}\\
&\to& \mathcal{P}\exp\left\{ig\int_x^{x+\hat{\mu}}dz_\nu\,A_\nu(z)\right\}\\
&\equiv& U_\mu(x)
\end{eqnarray*}
with the Euclidean gauge field defined by
\begin{equation}
A_4=A^4=-iA_0^M=-iA^0_M,\qquad A_j
=A^j=A_j^M=-A^j_M.
\end{equation}
This gives
$$F_{44}=-F^{00}_M=-F_{00}^M,\qquad F_{4j}=iF^{0j}_M=-iF^M_{0j},\qquad 
F^{jk}=F^{jk}_M=F^M_{jk}.$$

In order to find the Euclidean gauge action, we must keep track of the temporal
vs spatial directions.  Consider the anisotropic Minkowski lattice spacings
defined by:
$$a^M_t\equiv\int_x^{x+\hat{0}}dz_M^0,\qquad
a^M_s=\int_x^{x+\hat{\jmath}}dz_M^j,\qquad
\xi_0^M \equiv a^M_s/a^M_t+O(a)$$
where $\xi_0^M$ is the bare anisotropy\footnote{The bare anisotropy parameter 
$\xi^M_0$ in the action is note the same as $a_s/a_\tau$ due to quantum
corrections.}
 of the lattice.  
The corresponding 
Euclidean lattice spacings are given by:
$$a_\tau=ia^M_t,\qquad a_s=a_s^M,\qquad \xi_0=-i\xi_0^M.$$

The anisotropic gauge action becomes
\begin{eqnarray*}
iS^M_G
&=&-i\frac{\beta}{N}\sum_x\left[\sum_{j<k}\frac{1}{\xi^M_0}\mbox{Re\,Tr}(1-
U^M_{jk}(x))+\sum_j \xi^0_M\mbox{Re\,Tr}(1-U^M_{j4}(x))\right],\\
&=&-i\sum_x\sum_{\mu,\nu} a^M_t(a_s^M)^3\, \frac{1}{2}\mbox{Tr}(F^M_{\mu\nu}F^{\mu\nu}_M)+O(a^5)\\
&\to&-\sum_x\sum_{\mu,\nu} a_s^3 a_\tau \frac{1}{2}\mbox{Tr}(F_{\mu\nu}F_{\mu\nu})+O(a^5)\\
&=&-S_G
\end{eqnarray*}
where the anisotropic Euclidean gauge action is given 
by~\cite{klassen:aniso_gauge_action}:
\begin{equation}
S_G
=\frac{\beta}{N}\sum_x\left[\sum_{j<k}\frac{1}{\xi_0}\mbox{Re\,Tr}(1-
U_{jk}(x))+\sum_j \xi^0\mbox{Re\,Tr}(1-U_{j4}(x))\right].
\label{eqn:aniso_gauge_action}
\end{equation}

\section{The Euclidean fermion lattice action}
For clarity, we will show all sums explicitly in this section and begin
by describing the fermion action on an isotropic lattice.
On the lattice, we will replace first and second covariant derivatives 
by symmetric differences using gauge links as parallel transporters:
\begin{eqnarray*}
\partial_\mu\to\frac{1}{a}(\nabla_\mu)_{x,y} &\equiv& \frac{1}{2a}\left(U_\mu(x)\delta_{x+\hat{\mu},y}-U^\dagger(x-\hat{\mu})\delta_{x-\hat{\mu},y}\right)\\
\partial_\mu\partial_\mu\to\frac{1}{a^2}(\Delta_\mu)_{x,y} &\equiv& \frac{1}{a^2}\left(U_\mu(x)\delta_{x+\hat{\mu},y}+
U^\dagger_\mu(x-\hat{\mu})\delta_{x-\hat{\mu},y}-2\delta_{x,y}\right)
\end{eqnarray*}
We use symmetric differences in the action in order to guarantee 
a Hamiltonian which satisfies {\em
reflection positivity}:
$$H\stackrel{\tau\to-\tau}{\to}H.$$
Minkowski time evolution $e^{iHt}\mathcal{O}(0)e^{-iHt}$ requires 
$H^\dagger=H$, but Euclidean time evolution $e^{H\tau}\mathcal{O}(0)e^{-H\tau}$
requires reflection positivity.

Making the appropriate replacements, we may write the fermionic part of the
lattice action as
\begin{eqnarray*}
S_F &=& \int d^4x\,\overline{\psi}(x)\left(\sum_\mu\gamma_\mu\partial_\mu+m
\right)\psi(x),\\
&\to& \sum_{x,y} a^4 \overline{\psi}(x)\left(\sum_\mu\gamma_\mu \frac{1}{a}(\nabla_\mu)_{x,y}+m\delta_{x,y}\right)\psi(y),\\
&=& \sum_{x,y}(a^{3/2}\overline{\psi}(x))\left(\sum_\mu\gamma_\mu (\nabla_\mu)_{x,y}+(am)\delta_{x,y}\right)(a^{3/2}\psi(y)),\\
&\equiv& \sum_{x,y}\overline{\psi}_x \left(\sum_\mu\gamma_\mu(\nabla_\mu)_{x,y}+a m\delta_{x,y}\right)\psi_y,\\
&\equiv&\overline{\psi}\left(\sum_\mu\gamma_\mu\nabla_\mu+a m\right)\psi,
\end{eqnarray*}
where we have defined the dimensionless quark fields $\overline{\psi}_x\equiv a^{3/2}\overline{\psi}(x)$ and $\psi_x\equiv a^{3/2}\psi(x)$ and have grouped a factor
of the lattice spacing $a$ with the bare mass $m$ to make the dimensionless
quantity $am$.

\subsection{The Wilson term}
The na\"ive replacement of the derivative by the symmetric difference in the
above leads to a serious lattice artifact known as {\em fermion 
doubling}~\cite{rothe:lgt} in which there are 15 additional massless
fermion `doublers' which contribute to the calculations.  These doublers
can be made harmless by introducing the {\em Wilson term}, an $O(a)$ term 
(irrelevant in the na\"ive continuum limit) into the action
which gives the doublers a mass of $1/a$:
\begin{eqnarray*}
S_F&\to& S_F-a\frac{1}{2}\sum_{x,y,\mu}a^4\psi(x)\partial_\mu\partial_\mu\psi(y)\\
&=&\overline{\psi}(\sum_\mu\gamma_\mu\nabla_\mu+a m)\psi-\frac{1}{2}\sum_\mu \overline{\psi} \Delta_\mu \psi\\
&\equiv&\overline{\psi}Q\psi\\
Q&\equiv&a m+\sum_\mu(\gamma_\mu\nabla_\mu-\frac{1}{2}\Delta_\mu)\\
Q_{x,y}&\equiv&(a m+4)\delta_{x,y}-\frac{1}{2}\sum_\mu\left[
(1-\gamma_\mu)U_\mu(x)\delta_{x+\hat{\mu},y}
+(1+\gamma_\mu)U^\dagger_\mu(x-\hat{\mu})\delta_{x-\hat{\mu},y}\right]
\end{eqnarray*}

\subsection{Anisotropic lattices}
We would like to use large spatial volumes in order to mitigate finite
volume effects in our calculations.  This requires a coarse spatial mesh in 
order
to have a manageable number of spatial lattice sites. On the other hand, our
temporal correlation functions decay exponentially
with the baryon mass, motivating us to work with a fine mesh in the temporal
direction.  Accordingly, we will work with an anisotropic 
lattice~\cite{klassen:aniso_gauge_action, morningstar:glueball, klassen:aniso_improvement} with 
spatial lattice spacing $a_s$, and temporal lattice spacing $a_\tau$.
The anisotropic gauge action is given by Eqn.~\ref{eqn:aniso_gauge_action}.
Repeating the derivation above for an anisotropic fermionic lattice action
is straightforward 
if we introduce the elementary lattice vector $a_\mu=(a_s,a_s,a_s,a_\tau)$.
We now absorb factors of $a_s^{3/2}$ into $\overline{\psi}$ and $\psi$, and
associate a factor of $a_\tau$ with $m$.  In addition, we put in an explicit
factor of the bare speed of light $\nu$ which takes into account
the fermion anisotropy effects.  This gives us
\begin{eqnarray}
S_F &=&\overline{\psi}Q\psi\nonumber\\
Q&=&a_\tau m+\frac{\nu}{\xi_0}\sum_j(\gamma_j\nabla_j-\frac{1}{2}\Delta_j)+
(\gamma_4\nabla_4-\frac{1}{2}\Delta_4)\nonumber\\
Q_{x,y}&=&(a_\tau m+\frac{3\nu}{\xi_0}+1)\delta_{x,y}\nonumber\\
&&-
\frac{\nu}{2\xi_0}\sum_j\left[
(1-\gamma_j)U_j(x)\delta_{x+\hat{\jmath},y}
+(1+\gamma_j)U^\dagger_j(x-\hat{\jmath})\delta_{x-\hat{\jmath},y}\right]\nonumber\\
&&-\frac{1}{2}\left[
(1-\gamma_4)U_4(x)\delta_{x+\hat{4},y}
+(1+\gamma_4)U^\dagger_4(x-\hat{4})\delta_{x-\hat{4},y}\right]
\label{eqn:aniso_fermion_action}
\end{eqnarray}

\subsection{Setting the scale}
In order to get back to energy units (i.e. MeV and fm) from lattice
units ($a_\tau=1$), we
will relate a measured quantity such as the heavy quarkonium string tension 
(which is 
not very sensitive to mass tuning) to an experimentally measured quantity.

The heavy quark string tension for large separation $R$ is
given experimentally by:
$$V(R)\approx V_0+\sigma R,$$ where $\sigma\approx (465\mbox{ MeV})^2$.
On the lattice, we can measure the static quark-antiquark potential
and fit to the dimensionless form:
\begin{eqnarray*}
a_\tau V(R/a_s)&=&(a_\tau V_0)+(a_\tau a_s \sigma)(R/a_s)\\
&=&(a_\tau V_0)+(\xi a_\tau^2 \sigma)(R/a_s)\\
&=&(a_\tau V_0)+(\frac{a_s^2}{\xi} \sigma)(R/a_s).
\end{eqnarray*}

The anisotropy can be determined by comparing the behavior of the
static quark-antiquark potential as a function of both spatial
separation (regular) and temporal separation (sideways).

Once the anisotropy has been determined, the string tension can be used
to find the temporal (or spatial) lattice spacing in 
units of $\mbox{MeV}^{-1}$.
This fixes the energy scale of all quantities in the 
simulation~\cite{edwards:setting_scale,lepage:lat_pert_theory}.

\section{Tuning the lattice action}
The general steps involved in tuning the lattice paramters typically involve:
\begin{itemize}
\item Tune the bare anisotropy $\xi_0$ and compare the regular 
and sideways static quark
potential to determine the renormalized anisotropy 
$\xi=a_s/a_\tau$~\cite{klassen:aniso_gauge_action}.
\item Tune the bare coupling $\beta$ and use the string 
tension ($\sqrt{\sigma}=465$ MeV) to determine 
the temporal lattice spacing $a_\tau$ or 
spatial lattice spacing $a_s$~\cite{edwards:setting_scale}.
\item Tune the bare fermion anisotropy $\nu$ (the `bare velocity of light')
and measure the pion dispersion 
relation, which will converge to $E^2=m^2+p^2$ as the fermion anisotropy
approaches
$\xi$~\cite{klassen:aniso_improvement, su:aniso_study}.
\item Tune the quark masses and measure the spectrum.  As quark masses
decrease, spectral quantities should scale according to chiral perturbation
theory.
\end{itemize}

For the current work, we use unrealistically high quark masses in order to
have rapid convergence of our propagator inversions.
The pion mass can be estimated by a back-of-the-envelope calculation:
\begin{eqnarray*}
\xi\equiv \frac{a_s}{a_\tau} \approx 3.0&\qquad&\mbox{(from sideways potential tuning)}\\
a_s\approx 0.1 \mbox{ fm}&\qquad&\mbox{(from string tension)}\\
a_\tau m_\pi=0.1125(26)&\qquad&\mbox{(from pion effective mass)}\\
m_\pi =\frac{\xi a_\tau m_\mu}{a_s}\approx 700\mbox{ Mev}&\qquad&(\hbar c \approx 200 \mbox{ Mev-fm})
\end{eqnarray*}
where we have rounded to the most significant digit due to the crudeness
of our estimate.

\subsection{Lattice parameters for this study}
\label{sect:lattice_params}
Our study uses the anisotropic Wilson action with the following parameters.
\begin{itemize}
\item Lattice size: $N_s = 12$, $N_\tau = 48$
\item $\xi_0 = 2.464 \leftrightarrow \xi \approx 3.0$
\item $\beta=6.1\leftrightarrow a_s \approx 0.1$ fm
\item $a_\tau m_{ud}=-0.305 \leftrightarrow m_\pi \approx 700$ MeV
\item $\nu = 0.902$
\end{itemize}

The configurations used for this work were quenched (discussed in the next
chapter) and the spatial volume considered was relatively small ($\approx
1.2$ fm).  This was acceptable because the purpose of the present work was
to develop and test a new operator design methodology, specifically with the
goal of 
designing and tuning good nucleon operators to be used in 
later `production'
runs to extract the low-lying baryon spectrum in the nucleon sector.
It is
expected that the operator design methodology described in this work
will remain relevant as
we work to larger spatial volumes, lower quark masses, and finer lattice
spacings.  Production runs will use the anisotropic Wilson action
with dynamical $u$, $d$, and $s$ quarks and a clover improvement 
term~\cite{alford:lqcd,alford:improved_actions,klassen:schrodinger,
edwards:aniso_clover_action}.

\section{The spectral representation of correlation functions}
The Euclidean formulation of a quantum field theory provides a simple way
to calculate vacuum expectation values $\bra{\Omega}A\ket{\Omega}$~\cite{montvay:lgt}.
Consider
\begin{eqnarray*}
\frac{\mbox{Tr}(e^{-H\tau}A)}{\mbox{Tr}(e^{-H\tau})} &=&
\frac{\displaystyle{\sum_{k=0}^\infty} \bra{k}e^{-H\tau}A\ket{k}}{\displaystyle{\sum_{l=0}^\infty}\bra{l}e^{-H\tau}\ket{l}}\\
&=&\frac{\displaystyle{\sum_{k=0}^\infty}\bra{k}A\ket{k}e^{-E_k\tau}}{
\displaystyle{\sum_{l=0}^\infty}e^{-E_l\tau}}\\
&\stackrel{\tau\to\infty}{\to}&\bra{\Omega}A\ket{\Omega}
\end{eqnarray*}
To see how this Hamiltonian approach connects to the functional integral
approach with a lattice regulator, we will use a complete set of field
eigenstates
$\{\ket{\phi}\}$ in the trace, rather than the complete set of 
energy eigenstates $\{\ket{k}\}$.
For clarity we will consider the case of a complex scalar
field operator $\Phi$ for the rest of this section.  Let the integration
measure over field configurations be denoted by $d\mu(\phi)$.  Then for a 
fixed time slice we have:
\begin{eqnarray*}
\Phi\ket{\phi}&=&\phi\ket{\phi},\\
\bra{\phi}\Phi^\dagger&=&\bra{\phi}\phi^*,\\
\int d\mu(\phi)\, \ket{\phi}\bra{\phi}&=&
\int d\phi^*d\phi \,e^{-\phi^*\phi}\,\ket{\phi}\bra{\phi},\\
&=&1.
\end{eqnarray*}
where the factor of $e^{-\phi^*\phi}$ arises due to the definition of the
field eigenstates.  This will be illustrated in the case of fermions in
Subsection~\ref{subsec:fermion_bc}.  A rigorous treatment for both bosons and fermions is
given in~\cite{brown:quantum}.
Consider the following partition function with the trace taken with respect
to a complete basis of field configurations:
\begin{eqnarray*}
Z&\equiv& \mbox{Tr}(e^{-HT}),\\
&=&\int d\mu(\phi_0)\,\bra{\phi_0}e^{-HT}\ket{\phi_0},\\
&&\mbox{(inserting $N_\tau-1$ complete sets of states, with $T=N_\tau a_\tau$)},\\
&=&\int d\mu(\phi_{N_\tau-1})d\mu(\phi_{N_\tau-2})\cdots d\mu(\phi_0)\,\bra{\phi_0}e^{-Ha_\tau}
\ket{\phi_{N_\tau-1}}\bra{\phi_{N_\tau-1}}e^{-H a_\tau}\ket{\phi_{N_\tau-2}}\cdots
\bra{\phi_1}e^{-H a_\tau}\ket{\phi_0},\\
&=&\int_{\phi_{N_\tau}=\phi_0} d\mu(\phi_{N_\tau-1})\cdots d\mu(\phi_0)\,
\bra{\phi_{N_\tau}}e^{-Ha_\tau}\ket{\phi_{N_\tau-1}}\cdots
\bra{\phi_1}e^{-H a_\tau}\ket{\phi_0}.
\end{eqnarray*}

It can be shown~\cite{brown:quantum} that: 
\begin{equation}
\bra{\phi_{\tau+1}}e^{-Ha_\tau}\ket{\phi_\tau}\equiv 
\exp\{\phi_{\tau+1}^*(\vec{x})\phi_{\tau+1}(\vec{x})-a_\tau L(\phi_{t+1}(\vec{x}),\phi_t(\vec{x}))\}.
\label{eqn:transfer_matrix}
\end{equation}
The quantity on the left is the matrix element of the {\em
transfer} operator $e^{-Ha_\tau}$ 
between two states in the Hilbert space.  The quantity on the 
right is a function of two different field configurations 
$\phi_{\tau+1}(\vec{x})$ and $\phi_\tau(\vec{x})$.  We have shown
the spatial indices as a reminder that we are dealing with the field
degrees of freedom defined over all of space (the Hamiltonian is 
time-independent).
If we regularize the individual integrals over the $\phi_\tau$
by placing the system in a finite spatial volume with periodic boundary
conditions and 
by discretizing the spatial coordinates we have:
\begin{eqnarray*}
\int_{\phi_{N_\tau}=\phi_0} \prod_{\tau}\,d\mu(\phi_\tau)&=&
\int \prod_\tau\, \prod_{\vec{x}} \,d\mu(\phi_\tau(\vec{x})),\\
&=&\int_{\mathrm{p.b.c}}\prod_\tau\,\prod_{\vec{x}}\, d\phi^*_\tau(\vec{x})d\phi_\tau(\vec{x})\,e^{-\phi_\tau^*(\vec{x})
\phi_\tau(\vec{x})},\\
&\equiv&\int_{\mathrm{p.b.c.}}\mathcal{D}\phi\,e^{-\phi^*\phi}.
\end{eqnarray*}
The choice of periodic boundary conditions (p.b.c.) for
the spatial volume eliminates edge effects, but does does not eliminate 
finite volume effects due to images.  The periodic boundary conditions
in the `temporal' direction ($\tau$) arise naturally as a consequence of the
trace.  It is not necessary to use a trace to extract the spectrum, but the 
trace has the useful property of temporal translation invariance.  The 
anti-periodic temporal fermion boundary conditions to be used with the trace 
method are
discussed in Subsection~\ref{subsec:fermion_bc}.

If we identify the lattice action 
$S[\phi]=\sum_{\tau}a_\tau L(\phi_{\tau+1},\phi_\tau)$, 
we arrive at the 
fundamental relation between the trace of the operator $e^{-HT}$ and the 
(well-defined) functional integral over field configurations $\phi_\tau(\vec{x})\equiv 
\phi(x)$ on our Euclidean space-time lattice (identifying $x_4\equiv \tau$):
\begin{equation}
Z=\mbox{Tr}(e^{-HT})=\int_{\mathrm{p.b.c.}} \mathcal{D}\phi\,e^{-S[\phi]}
\label{eqn:partition}
\end{equation}
For finite lattice spacings $a_\tau,a_s$ and a given lattice action $S$, 
the lattice Hamiltonian 
$H$ defined via Eqn.~\ref{eqn:partition} will differ from the continuum
Hamiltonian of the theory by terms of order $a_\tau$ and $a_s$.
The goal of action improvement programs is to reduce these discretization
errors by adding offsetting $O(a)$ terms to the lattice action $S$.

We can access the spectrum of the theory by inserting an arbitrary creation
operator
$\overline{\mathcal{O}}$ at `time slice' $0$ and an arbitrary annihilation 
operator $\mathcal{O}$ at `time slice' $\tau$ into the functional integral
expression for $Z$:
\begin{eqnarray}
\langle\mathcal{O}(\tau)\bar{\mathcal{O}}(0)\rangle&\equiv&
\frac{1}{Z}\int_{\mathrm{p.b.c.}} \mathcal{D}\phi\, \mathcal{O}(\tau)
\overline{\mathcal{O}}(0)e^{-S[\phi]}\\
&=&\frac{\mbox{Tr}\left(e^{-H(T-\tau)}\mathcal{O}e^{-H\tau}\bar{\mathcal{O}}\right)}{\mbox{Tr}\left(e^{- HT}\right)}\nonumber\\
&=& \frac{\sum_n \bra{n}e^{-H(T-\tau)}\mathcal{O}e^{-H\tau }\bar{\mathcal{O}}\ket{n}}{\sum_m\bra{m}e^{- HT}\ket{m}}\nonumber\\
&=& \frac{\sum_n e^{-E_nT}\bra{n}e^{+ H\tau}\mathcal{O}e^{-H\tau}\bar{\mathcal{O}}\ket{n}}{\sum_m e^{-E_mT}}\nonumber\\
&\stackrel{T\to\infty}{\to}&\bra{\Omega} e^{+ H\tau}\mathcal{O}e^{- H\tau}\bar{\mathcal{O}}\ket{\Omega}\label{eqn:imag_time}\\
&=&\sum_{k=0}^\infty\bra{\Omega} \mathcal{O}e^{-H\tau}\ket{k}\bra{k}\bar{\mathcal{O}}\ket{\Omega}\nonumber\\
&=&\sum_{k=1}^\infty|\bra{k}\bar{\mathcal{O}}\ket{\Omega}|^2 e^{-E_k\tau}
\label{eqn:spectrum_sum}
\end{eqnarray}
where in \ref{eqn:imag_time} we have taken the limit $T \to \infty$ (large
temporal extent of our lattice).
If we require our creation operators $\bar{\mathcal{O}}$ to have a vacuum
expectation value of zero ($\bra{\Omega}\bar{\mathcal{O}}\ket{\Omega}=0$), then we
can extract the first (non-zero) energy state of the theory by fitting to
the asymptotic form of Eqn.~\ref{eqn:spectrum_sum}.

The form of \ref{eqn:imag_time} looks like a Heisenberg time-evolution equation if we replace $t \to -i\tau$ (giving diffusion dynamics, not wave dynamics).  
This was to be expected because we constructed our Euclidean theory such that
the correlation functions would be related to
the Minkowski Green's functions analytically continued to imaginary time.

\subsection{Temporal boundary conditions for fermions}
\label{subsec:fermion_bc}
There is some subtlety involved with taking the trace using
eigenstates of fermion fields due to their Fermi-Dirac 
statistics~\cite{brown:quantum}.  For
clarity in this
section, we consider the mathematics of a single fermion field carrying no
indices.

The fermion fields in the functional integral are Grassmann (anticommuting)
numbers, not c-numbers (which commute).
Consider the case of two independent Grassmann variables: $z$ and $z^*$
(not related by complex conjugation).  These
objects satisfy:
$$zz^*=-z^*z,\qquad z^2=z^{*2}=0.$$
Consequently, any polynomial in these variables has a finite (exact)
power series expansion:
$$P(z,z^*)=p_0+p_1z+p_2z^*+p_{12}zz^*.$$
Specifically:
$$e^z\equiv 1+z.$$
We can define integration over Grassmann variables formally by
$$\int dz  = 0\quad \int dz\,z = 1\quad \int dz^* = 0\quad \int dz^*\,z^*=1$$
with the requirement that $\int dz^*\,dz=-\int dz\,dz^*$ for consistency.
Note that the above definition defines a translationally invariant integral
$$\int dz\, (z+d)=\int dz\,z$$
where $d$ is any c-number or Grassmann number.

The pair of fermionic field operators $\Psi$ and $\Psi^\dagger$ satisfy
the fermion anticommutation relations:
$$\Psi^2=\Psi^{\dagger 2}=0\qquad \Psi\Psi^\dagger+\Psi^\dagger\Psi=1.$$
These relations are sufficient to show that these operators act upon a two-dimensional Hilbert space spanned
by the states $\ket{0}$ and $\ket{1}$.  The behavior of the
operators $\Psi$ and $\Psi^\dagger$ on this space is completely specified
by
$$
\begin{array}{ccccc}
\Psi\ket{0}&=&\bra{0}\Psi^\dagger&=&0,\\
\Psi^\dagger\ket{1}&=&\bra{1}\Psi&=&0,\\
\braket{0}{0}&=&\braket{1}{1}&=&1,\\
\braket{0}{1}&=&\braket{1}{0}&=&0
\end{array}
$$
and
\begin{eqnarray*}
\Psi^\dagger\ket{0}=\ket{1},\qquad \bra{0}\Psi=\bra{1},\\
\Psi\ket{1}=\ket{0},\qquad \bra{1}\Psi^\dagger=\bra{0}.
\end{eqnarray*}

In order to take a trace over field configurations, we will need the eigenstates of the field operators.  These are given by:
$$\ket{z}=e^{\Psi^\dagger z}\ket{0}\qquad \bra{z}=\bra{0}e^{z^*\Psi}$$
where the Grassmann numbers $z$ and $z^*$ anticommute with the operators 
$\Psi$ and $\Psi^\dagger$
\begin{eqnarray*}
\Psi z = -z\Psi,&\qquad& z\Psi^\dagger = -\Psi^\dagger z,\\
\Psi z^* = -z^*\Psi,&\qquad& z^*\Psi^\dagger = -\Psi^\dagger z^*.
\end{eqnarray*}
Using the rules we have just defined, we can verify that 
$$\Psi\ket{z}=z\ket{z},\qquad\bra{z}\Psi^\dagger=\bra{z}z^*.$$
Notice that (by construction) this {\em looks} like complex conjugation.
To define the measure over the set of field eigenstates, we need
\begin{eqnarray*}
\braket{z_1}{z_2}&=&\bra{0}e^{z_1^*\Psi}e^{\Psi^\dagger z_2}\ket{0}\\
&=&\bra{0}(1+z_1^*\Psi)(1+\Psi^\dagger z_2)\ket{0}\\
&=&\bra{0}(1+z_1^*\Psi+\Psi^\dagger z_2+z_1^*\Psi\Psi^\dagger z_2\ket{0}\\
&=&\braket{0}{0}+z_1^*\bra{0}\Psi\ket{0}+\bra{0}\Psi^\dagger\ket{0}
+z_1^*z_2\bra{0}\Psi\Psi^\dagger\ket{0}\\
&=&1+z_1^*z_2=e^{z_1^*z_2}
\end{eqnarray*}
If we define the measure over coherent states $d\mu(z)$ by
$$d\mu(z)=dz^*dz\, e^{-z^*z}$$
then we have the resolution of the identity in terms of a complete set 
of coherent field states:
$$\int d\mu(z)\, \ket{z}\bra{z}=1.$$
Checking this:
\begin{eqnarray*}
\int d\mu(z)\braket{z_1}{z}\braket{z}{z_2}&=&\int dz^*dz\, e^{-z^*z}e^{z_1^*z}e^{z^*z_2},\\
&=&\int dz^*dz\, (1-z^*z)(1+z_1^*z)(1+z^*z_2),\\
&=&\int dz^*\,z^*\,dz\, z\, (1+z_1^*z_2),\\
&=&e^{z_1^*z_2}=\braket{z_1}{z_2}.
\end{eqnarray*}
Alternatively:
\begin{eqnarray*}
\int d\mu(z)\, \ket{z}\bra{z}&=&\int dz^* dz\,e^{-z^*z}(1+\psi^\dagger z)\ket{0}
\bra{0}(1+z^*\psi)\\
&=&\ket{0}\bra{0}+\ket{1}\bra{1}.
\end{eqnarray*}

Let $F(\Psi,\Psi^\dagger)$ be an analytic function of an even number of fermion
operators $\Psi^\dagger$ and an even number of antifermion operators $\Psi$.
The trace of $F$ is:
\begin{eqnarray*}
\mbox{Tr}(F)&=&\int d\mu(z)\, \bra{-z}F\ket{z}\\
&=&\int dz^* dz\,(1-z^*z)\bra{0}(1-z^*\Psi)F(1+\Psi^\dagger z)\ket{0}\\
&=&\bra{0}F\ket{0}+\bra{1}F\ket{1}
\end{eqnarray*}
Notice that the trace must be taken between $\bra{-z}\cdots\ket{z}$, not
$\bra{z}\cdots\ket{z}$.  In the functional integral, this translates to
using periodic temporal ($x_4$) boundary conditions for the bosonic
fields (i.e. gauge links) but {\em anti-periodic} temporal boundary conditions
for the fermionic fields (i.e. quark fields).  
We use periodic spatial boundary conditions
for both bosonic and fermionic fields.  These boundary conditions will be
 implied when we write `p.b.c.' on our functional integrals.

\section{Effective mass plots}
Returning to the spectral representation of the Euclidean correlation function:
\begin{equation}
C(\tau)=\langle \mathcal{O}(\tau)\overline{\mathcal{O}}(0)\rangle
\stackrel{T\to\infty}{\to}\sum_{k=1}^\infty|c_k|^2e^{-E_k\tau},\qquad 
c_k \equiv \bra{k}\overline{\mathcal{O}}\ket{\Omega},
\label{eqn:correlation_function}
\end{equation}
we can get an idea of how well the temporal extent of our lattice is
approximating the $T\to\infty$ limit
by examining how calculated physical quantities vary with the number
of lattice sites in the temporal direction for a fixed temporal lattice
spacing $a_\tau$.

We wish to extract as many of the energy levels $E_k$ as possible from
the exponentially decaying signal in Eqn.~\ref{eqn:correlation_function}.
It will be a recurring theme in this work to examine plots of {\em effective 
mass functions}
to judge the quality of our signals and to tune our parameters.
The effective mass function is defined by:
\begin{equation}
a_\tau M(\tau)=\ln\left(C(\tau)/C(\tau+a_\tau)\right).
\label{eqn:effective_mass}
\end{equation}
This definition of the effective mass is not unique, but it is one of the 
simplest, involving the value of the correlation function at only two neighboring 
values of time.  

For large values of $\tau$ in the $T\to\infty$ limit, 
the effective mass may be expanded as:
\begin{eqnarray*}
a_\tau M(\tau) &\equiv& \ln\left[\frac{C(\tau)}{C(\tau+a_\tau)}\right]\\
&=&\ln\left[\frac{|c_1|^2e^{-E_1 \tau}+|c_2|^2 e^{-E_2 \tau}+\cdots}{|c_1|^2e^{-E_1(\tau+a_\tau)}+|c_2|^2 e^{-E_2(\tau+a_\tau)}+\cdots}\right]\\
&=&a_\tau E_1\left(1 + O\left(|c_2|^2/|c_1|^2e^{-(E_2-E_1)\tau/a_\tau}\right)\right)
\end{eqnarray*}
and thus $a_\tau M(\tau)$ approaches a plateau of $a_\tau E_1$ for sufficiently
large $\tau/a_\tau$.
We emphasize that effective mass plots are simply visualization tools; 
final fits are best made to the correlation matrix elements directly.

Because the lattice is periodic (anti-periodic) in the temporal direction for
bosons (fermions), we can expect significant contamination from the 
{\em backward propagating} states for large correlation lengths $\tau$.  In
the case of meson operators, the backward propagating state contamination comes
from a time-reversed mirror image of the meson itself.  We can explicitly take 
this contamination into account by defining the effective mass 
$a_\tau M_{\mathrm{cosh}}(\tau)$
as the solution to:
$$\mbox{ln}[\mbox{cosh}((a_\tau M_{\mathrm{cosh}})(\tau/a_\tau))] = 
\mbox{ln}[\mbox{cosh}((a_\tau M_{\mathrm{cosh}})(\tau/a_\tau))] + \mbox{const}.$$

\begin{figure}[ht!] 
  \centering 
  \includegraphics[width=5.0in]{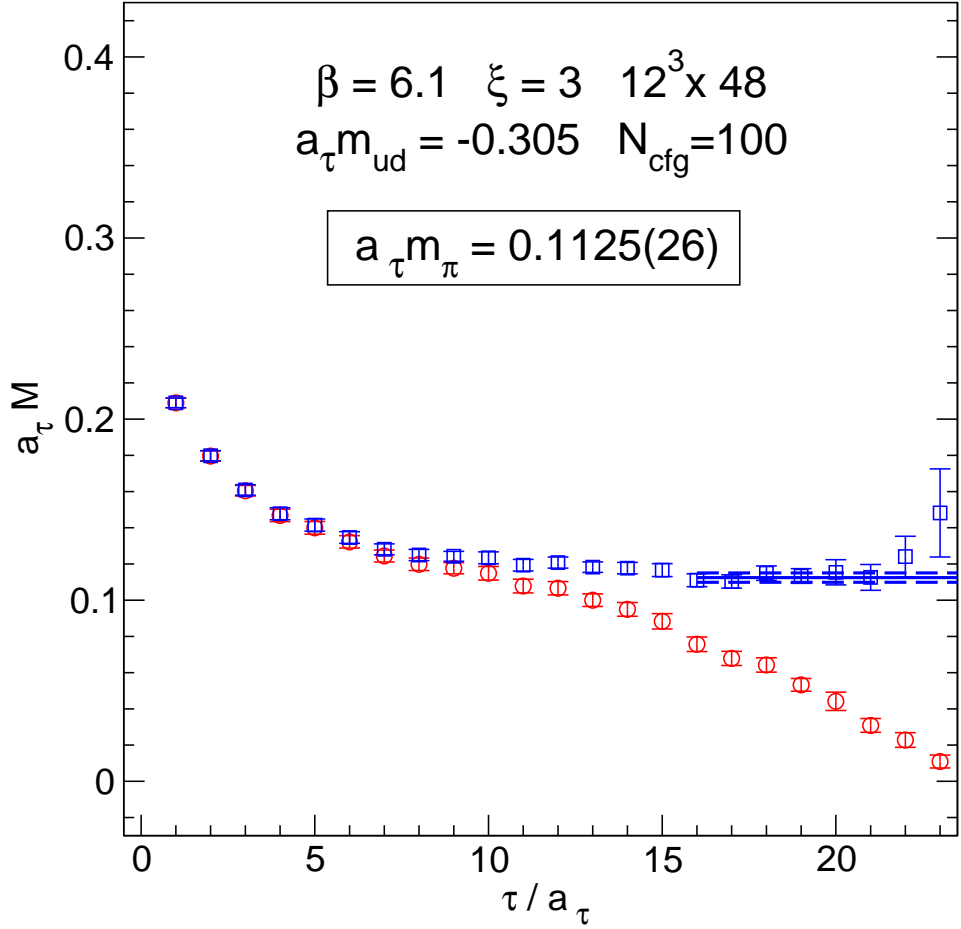}
  \caption{The pion effective mass plots.  The standard effective mass 
$a_\tau M(\tau)$ (red) fails to plateau due to backward
channel contamination.  This contamination is explicitly treated by 
the meson effective mass function $a_\tau M_{\mathrm{cosh}}(\tau)$ (blue).
A $\mathrm{cosh}$ fit to the correlation function yields a pion mass in
lattice units of $a_\tau m_\pi=0.1125(26)$.}
  \label{fig:pion} 
\end{figure}

An example of both types of effective mass plot is given in Figure~\ref{fig:pion}, which gives the pion effective mass as measured on our configurations.

In the case of baryon operators, the backward propagating state
 contamination comes
from states in the opposite-parity channel, which will generally have
different masses due to spontaneous chiral symmetry breaking.  As a result,
we are unable to model the correlation function using a simple $\mathrm{cosh}$ form.
In this work, we do not correct for
this contamination, but limit our effective mass plots to the first or twenty
values of $\tau/a_\tau$ (we are working on a lattice with $N_\tau=48$).
  Because our baryon states are much heavier than
mesons, the backward propagating states decay significantly by the
time their signal reaches early values of $\tau/a_\tau$.  This is treated in
detail in Chapters~\ref{chap:correlators} and~\ref{chap:results}.

\section{Extracting excited states}

In order to extract higher states, we may try to fit multiple exponentials to the decaying correlation function.  
In practice, the correlation function is estimated using numerical techniques 
(c.f. Chapter~\ref{chap:monte_carlo}),
which results in an uncertainty associated with each data point.
The method of fitting many exponentials to a numerically estimated
 correlation function is ineffective due to the rapid decay of the signal to 
noise ratio. 

We introduce a method for extracting multiple excited states which is
based on the approach of L\"uscher and Wolff~\cite{luscher:excited}.
In this approach we utilize a set of $n$ `candidate' operators 
$\{\overline{\mathcal{O}}_1, 
\overline{\mathcal{O}}_2,\ldots,\overline{\mathcal{O}}_n\}$ and construct
 linear combinations 
$$\overline{\Theta}_i = \sum_{a=1}^n\overline{\mathcal{O}}_a v_{ai},$$
with associated correlation functions:
\begin{eqnarray*}
C_i(\tau)&=&\langle \Theta_i(\tau)\overline{\Theta}_i(0)\rangle,\\
&=&\sum_{a,b}v^*_{ia}\langle \mathcal{O}_a(\tau)\overline{\mathcal{O}}_b(0)
\rangle v_{bi},\\
&\equiv&v_i^\dagger C(\tau)v_i,
\end{eqnarray*}
where we have introduced the correlation matrix 
$C_{ab}(\tau) \equiv \langle \mathcal{O}_a(\tau)\overline{\mathcal{O}}_b(0)\rangle$
and the fixed coefficient vector $v_i$.  $C(\tau)$ consists of correlation functions between every possible pair
of candidate operators in our original set, and therefore contains more
information about the spectrum than the correlation function of a single
operator.  Note that all of the $C_i(\tau)$ and $C_{aa}(\tau)$ 
(diagonal elements) are real and positive for all $\tau\neq 0$.
Our task is to extract as much information as possible about the spectrum
from the correlation matrix $C(\tau)$ by finding suitable coefficient
vectors $\{v_i\}$.

\subsection{The variational method}
\label{section:variational}
Let the state $\ket{v_i}$ be defined by:
\begin{eqnarray*}
\ket{v_i}&\equiv& e^{-H\tau_0/2}\overline{\Theta}_i\ket{\Omega}=
e^{-H\tau_0/2}\sum_a \overline{\mathcal{O}}_a\ket{\Omega}v_{ai}\\
\bra{v_i}&\equiv& \bra{\Omega}\Theta_i e^{-H\tau_0/2}=
\sum_a v^*_{ai}\bra{\Omega}\mathcal{O}_a e^{-H\tau_0/2}
\end{eqnarray*}
for some reference time $\tau_0>0$.
In inclusion of a reference time in the normalization of $\ket{v_i}$ 
is necessary to ensure that we never have to 
evaluate a correlation function at zero time separation:
\begin{eqnarray*}
\braket{v_i}{v_i}&=&\bra{\Omega}\Theta_i e^{-H\tau_0}\overline{\Theta}_i\ket{\Omega}\\
&=&C_i(\tau_0)\\
&=&v^\dagger_iC(\tau_0)v_i
\end{eqnarray*}
$C(0)$ will generally involve products of multiple fields evaluated at 
a single space-time point.  Such expressions lead to subtle difficulties 
which arise during the
regularization process~\cite{weinberg:vol1}.
Specifically, $C(0)$ will often receive nonzero contributions from 
`Schwinger terms'~\cite{schwinger:commutators}.

In the space of all possible states $\{\ket{v_i}\}$ parameterized by 
the vectors $v_i$, we would like to find states which are as close to
the stationary states $\{\ket{k}\}$ as possible.
The variational method~\cite{cohen-tannoudji:quantum} is a powerful technique 
that is used to diagonalize some Hermitian operator $F$ within
some subspace spanned by a set of parameterized trial states $\{\ket{v}\}$.
To illustrate this method, let $f(v)$ be the expectation value
of $F$ in the state $\ket{v}$:
\begin{eqnarray*}
f(v)&\equiv&\frac{\bra{v}F\ket{v}}{\braket{v}{v}},\\
\bra{v}F\ket{v}&=&f(v)\braket{v}{v}.
\end{eqnarray*}
Taking the first-order variation with respect to the parameter $v$
gives
\begin{eqnarray*}
\bra{\delta v}F\ket{v}+\bra{v}F\ket{\delta v}&=&(\delta f(v))\braket{v}{v}
+f(v)\braket{\delta v}{v}+f(v)\braket{v}{\delta v}\\
\bra{\delta v}(F-f(v))\ket{v}+\bra{v}(F^\dagger-f^*(v)\ket{\delta v}&=&
(\delta f(v))\braket{v}{v}
\end{eqnarray*}
where we have used the Hermitian property $F^\dagger=F$.
As long as $\braket{v}{v}\neq 0$, we see that
$$\delta f(v) = 0\quad \leftrightarrow \quad F\ket{v}=f(v)\ket{v}.$$
The variational method diagonalizes the operator $F$ {\em restricted} to 
the subspace spanned by the $\{\ket{v}\}$.  If any eigenstates of the 
unrestricted operator $F$ are within the subspace, this method will find them.
Otherwise, the method will produce the `closest states' to the eigenstates
of the full operator $F$ it can find in the subspace. 

For our purposes, we will vary the parameters $v_i$ until the following 
expression is
at an extremum:
\begin{eqnarray*}
\frac{\bra{v_i}e^{-H(\tau-\tau_0)}\ket{v_i}}{\braket{v_i}{v_i}}
&=&\frac{v_i^\dagger C(\tau)v_i}{v^\dagger_i C(\tau_0)v_i}.
\end{eqnarray*}

The infinitesimal change
$$v_i \to v_i+\delta v_i$$
gives
\begin{eqnarray}
\delta\left[\frac{v_i^\dagger C(\tau) v_i}{v_i^\dagger C(\tau_0) v_i}\right]=\frac{(v_i+\delta v_i)^\dagger C(\tau) (v_i+\delta v_i)}{(v_i+\delta v_i)^\dagger C(\tau_0) (v_i+\delta v_i)}-\frac{v_i^\dagger C(\tau) v_i}{v_i^\dagger C(\tau_0) v_i}&=&0,\nonumber\\
(v_i+\delta v_i)^\dagger C(\tau) (v_i+\delta v_i)-\left(\frac{v_i^\dagger C(\tau) v_i}{v_i^\dagger C(\tau_0) v_i}\right)(v_i+\delta v_i)^\dagger C(\tau_0) (v_i+\delta v_i)&=&0,\nonumber\\
\delta v_i^\dagger \left[C(\tau)v_i - \left(\frac{v_i^\dagger C(\tau) v_i}{v_i^\dagger C(\tau_0) v_i}\right)C(\tau_0)v_i)\right]+\mbox{(c.c.)} + O(\delta v_i^2)&=&0,\label{eqn:var1}
\end{eqnarray}
where (c.c.) stands for the complex conjugate of the first term,
 and we have used the fact 
that $C^\dagger(\tau)$ = $C(\tau)$.

Equation (\ref{eqn:var1}) is satisfied for all infinitesimal variations $\delta v_i$ if:
\begin{equation}
C(\tau)v_i = \lambda_i(\tau,\tau_0)C(\tau_0)v_i\label{eqn:geneigen}
\end{equation}

The eigenvalues $\lambda_i(\tau,\tau_0)$ 
(ordered by increasing value at each $\tau$) 
of this generalized eigenvalue 
problem are called the {\em principal correlation functions}.

To find the asymptotic behavior of the principal correlation functions expressed
in terms of the desired energies $E_k$, consider
the quantity $v_i^\dagger C(\tau)v_j$.  From the orthogonality property of the
generalized Hermitian eigenvalue problem, we can always normalize our 
eigenvectors $v_i$ such that:
\begin{equation*}
v_i^\dagger C(\tau_0) v_j = \braket{v_i}{v_j} = \delta_{ij},
\end{equation*}
which gives:
\begin{eqnarray*}
v_i^\dagger C(\tau) v_j &=& \lambda_j(\tau,\tau_0)
v^\dagger_i C(\tau_0)v_j,\\
\bra{v_i}e^{-H(\tau-\tau_0)}\ket{v_j}&=&0 \qquad i\neq j.
\end{eqnarray*}
We can write any state in the 
Hilbert space as a linear combination
of the energy basis states
$$\ket{v_j}=\sum_k \ket{k}\alpha_{kj},$$
with the associated time evolution (or {\em relaxation}) given by
$$e^{-H(\tau-\tau_0)}\ket{v_j}=\sum_k\ket{k}\alpha_{kj}e^{-E_k(\tau-\tau_0)}.$$
Here we see that 
$\ket{v_j}$ evolves to a state which is orthogonal to $\ket{v_i}$.  Because $\ket{v_i}$ has no overlap with the large-time relaxation state of $\ket{v_j}$, it is forced to relax to a {\em different} excited state.
Thus, the $n$ principal correlation functions $\lambda_i(\tau)$ behave as:
\begin{equation}
\lim_{\tau\to\infty}\lambda_k(\tau,\tau_0)=e^{-E_k(\tau-\tau_0)}(1+O(e^{-\Delta_k(\tau-\tau_0)})
\label{eqn:prin_corr}
\end{equation}
$$\Delta_k\equiv \min_{l\neq k}|E_l-E_k|$$
where the $E_k$ are the energies of the first $n$ states accessible from the 
vacuum by the action of our creation operators $\overline{\mathcal{O}}$.
An energy state $\ket{k}$ is accessible if there exists some 
$a\in 1,\cdots,n$ such that
$$\bra{k}\overline{\mathcal{O}}_a\ket{\Omega}\neq0.$$
  The contamination $\Delta_k$
arises due to the fact we are restricted to the 
subspace of states spanned by $\{\overline{\mathcal{O}}_a\ket{\Omega}\}$.

Looking at Eqn.~\ref{eqn:prin_corr}, we may define the {\em principal effective mass functions}:
\begin{eqnarray*}
a_\tau M_k(\tau)&\equiv&\ln\left[\frac{\lambda_k(\tau,\tau_0)}{\lambda_k(\tau+a_\tau,\tau_0)}
\right]\\
\lim_{\tau\to\infty} a_\tau M_k(\tau)&\to& a_\tau E_k
\end{eqnarray*}

We may turn the generalized Hermitian eigenvalue problem (\ref{eqn:geneigen}) 
into a regular Hermitian eigenvalue problem:
\begin{eqnarray}
C(\tau)v_i &=& \lambda_i(\tau,\tau_0)C(\tau_0)v_i,\nonumber\\
C(\tau)C^{-1/2}(\tau_0)C^{+1/2}(\tau_0)v_i &=& \lambda_i(\tau,\tau_0)C(\tau_0)v_i,\nonumber\\
C^{-1/2}(\tau_0)C(\tau)C^{-1/2}(\tau_0)C^{+1/2}(\tau_0)v_i &=& \lambda_i(\tau,\tau_0)C^{+1/2}(\tau_0)v_i,\nonumber\\
\left[C^{-1/2}(\tau_0)C(\tau)C^{-1/2}(\tau_0)\right]u_i(\tau,\tau_0) &=& \lambda_i(\tau,\tau_0)u_i(\tau,\tau_0),\label{eqn:eigen}
\end{eqnarray}
where we assume a positive-definite correlation matrix at $\tau=\tau_0$ and
have added an explicit $\tau$ dependence to the eigenvector
solutions $u_i(\tau,\tau_0)$ to remind the
reader that fluctuations in the Monte Carlo estimates of the correlation
matrix elements at different values of $\tau$ will result in corresponding
fluctuations in the solution of Eqn.~\ref{eqn:eigen}.

In practice, degeneracies or near-degeneracies in 
the energy levels combine with 
numerical uncertainties
to introduce ambiguities into the estimate of a consistent set of 
fixed-coefficient vectors $v_i$ and 
the consistent ordering of the principal correlation functions 
$\lambda_i$
 across different values of $\tau$.  In order to avoid these ambiguities
and to fix a basis in degenerate subspaces,
we will adopt the approach of solving for the fixed coefficient vectors $v_i$
at a single time $\tau^*$:
$$v_i= C^{-1/2}(\tau_0)u(\tau^*,\tau_0),$$
where $\tau^*$ is chosen to be as small as possible (to avoid noise) but large
enough to ensure that the estimates of the $v_i$ have stabilized.
The diagonal elements of the {\em rotated} correlation matrix $\tilde{C}(\tau)$
are called {\em fixed-coefficient} correlation functions, and are defined by
$$\tilde{C}_{kk}(\tau)\equiv v^\dagger_k C(\tau)v_k,$$
and can be used to define {\em fixed-coefficient} effective mass functions:
$$a_\tau M_k(\tau)\equiv\ln\left[\frac{\tilde{C}_{kk}(\tau)}{\tilde{C}_{kk}(\tau+a_\tau)}\right]=\ln\left[\frac{v^{\dagger}_k C(\tau)v_k}{v^{\dagger}_k C(\tau+a_\tau)v_k}\right].$$

The fixed-coefficient correlation functions $\tilde{C}_{kk}(\tau)$ differ
from the principal correlation functions $\lambda_k(\tau,\tau_0)$ because we 
are only solving the eigenvalue problem once, at $\tau^*$.  Thus we are
only guaranteed orthogonality at $\tau^*$:
$$\tilde{C}_{ij}(\tau^*)=0\qquad i\neq j.$$
The lack of orthogonality for $\tau > \tau^*$, no matter how small, means that
our fixed-coefficient correlation functions will eventually relax to the
lowest excited state
$$\tilde{C}_{kk}(\tau) \stackrel{\tau \gg \tau^*}{\to} |c_1|^2\exp(-E_1\tau).$$

In Chapter~\ref{chap:results}, we will seek (and find) a range of 
$\tau$ values over which orthogonality holds to good approximation.  This 
will allow us to extract the values of the excited states using exponential
fits to the fixed-coefficient correlation functions.

\subsection{Discussion}
The variational method allows the physics to tell us which combination of the 
candidate operators best interpolates for the stationary states of the 
Hamiltonian.
Because the coefficients $\{v_{ai}\}$ are time-independent, we may use them
 to 
 infer something about the physical structure of the state 
$\ket{k}$ from the way in which the operators $\overline{\mathcal{O}_a}$ 
combine to form $\overline{\Theta}_k$.

Our need to perform the inversion and square root in Eqn.~\ref{eqn:eigen} underscores the importance of choosing candidate operators which give a correlation matrix with a good condition number.
We will keep this in mind in Chapter~\ref{chap:pruning} when we prune our large
operator set down to manageable size.

Our task is to construct a `good' set of candidate
operators $\overline{\mathcal{O}}_a$ which we believe capture a representative 
set of features of the baryon spectrum we seek (in order to
construct a `good' trial subspace for the variational method).

\chapter{The Monte Carlo method}
\label{chap:monte_carlo}

Our goal is to compute the low-lying spectrum of lattice QCD.
We do this by calculating the elements of Euclidean space-time correlation 
matrices, each of which is a ratio of functional integrals:
\begin{eqnarray}
C_{ab}(\tau)&=&\langle \mathcal{O}_a(\tau)\overline{\mathcal{O}}_b(0)\rangle,\\
&=&\frac{1}{Z}\int_{\mathrm{p.b.c.}} \mathcal{D}U\, \mathcal{D}\overline{\psi}\,
\mathcal{D}\psi\, \mathcal{O}_a(\tau)
\overline{\mathcal{O}}_b(0)e^{-S_F[\overline{\psi},\psi,U]-S_G[U]},
\label{eqn:corr_mat}
\end{eqnarray}
where
\begin{eqnarray}
Z&=&\int_{\mathrm{p.b.c.}} \mathcal{D}U\,\mathcal{D}\overline{\psi}\,
\mathcal{D}\psi\, e^{-S_F[\overline{\psi},\psi,U]-S_G[U]},\label{eqn:z}\\
S_F[\overline{\psi},\psi,U]&=&\overline{\psi}\,Q[U]\,\psi,\\
Q_{x,y}[U]&=&(a_\tau m+\frac{3\nu}{\xi_0}+1)\delta_{x,y}\\
&&-\frac{\nu}{2\xi_0}\sum_j\left[
(1-\gamma_j)U_j(x)\delta_{x+\hat{\jmath},y}
+(1+\gamma_j)U^\dagger_j(x-\hat{\jmath})\delta_{x-\hat{\jmath},y}\right]\nonumber\\
&&-\frac{1}{2}\left[
(1-\gamma_4)U_4(x)\delta_{x+\hat{4},y}
+(1+\gamma_4)U^\dagger_4(x-\hat{4})\delta_{x-\hat{4},y}\right],\\
S_G[U]&=&\frac{\beta}{N}\sum_x\left[\sum_{j<k}\frac{1}{\xi_0}\mbox{Re\,Tr}(1-
U_{jk}(x))+\sum_j \xi^0\mbox{Re\,Tr}(1-U_{j4}(x))\right].
\end{eqnarray}
Before we turn to the design of the operators $\overline{\mathcal{O}}_a$, we
briefly discuss how we will evaluate the ratio of functional integrals defined
by Eqns.~\ref{eqn:corr_mat} and~\ref{eqn:z}.

\section{The need for the Monte Carlo method}
These integrals
are regularized by the space-time lattice, but still consist of an
unmanageable number of integration variables.
The Feynman diagram method is not applicable to QCD functional integrals
 because expansions parameterized by the coupling constant do not converge
at the energy scales in which we are interested.

We turn instead to the Monte Carlo method, a powerful technique 
which requires us to formulate the problem on a computer.
Our quark fields satisfy Fermi-Dirac statistics, 
and are therefore represented by anticommuting Grassmann variables which
are not suitable for direct numerical integration on a computer.  
Fortunately, the quark
fields appear quadratically in the action and in the operator pairs
$\mathcal{O}(\tau)\overline{\mathcal{O}}(0)$, which means that the 
Gaussian integral over the fermion fields is symbolically integrable.  
Therefore, we will integrate out the Grassmann fields by hand before we 
translate the functional integral to computer language.

\section{Integrating the quark fields}
We wish to perform integrals of the form
\begin{eqnarray}
&&\frac{1}{Z_F}\int_{\mathrm{p.b.c}}\mathcal{D}\overline{\psi}\,\mathcal{D}\psi\, 
\psi_{x_1}\overline{\psi}_{x_2}\cdots \psi_{x_{n-1}}\overline{\psi}_{x_n}
\exp(-\sum_{x,y}\overline{\psi}_xQ_{x,y}[U]\psi_y)\label{eqn:ferm_int}\\
&=&
\frac{\delta}{\delta J_{x_2}}\frac{\delta}{\delta \bar{J}_{x_1}}
\cdots \frac{\delta}{\delta J_{x_n}}\frac{\delta}{\delta \bar{J}_{x_{n-1}}}
\ln(Z_F[\bar{J},J])|_{\bar{J},J=0}\nonumber
\end{eqnarray}
where we have introduced the fermion generating functional $Z_F[\bar{J},J]$:
\begin{equation}
Z_F[\bar{J},J]\equiv\int_{\mathrm{p.b.c.}}\mathcal{D}\overline{\psi}\,\mathcal{D}\psi\,
\exp\left(-\sum_{x,y}\overline{\psi}_xQ_{x,y}[U]\psi_y+
\sum_x [\bar{J}_x\psi_x+\overline{\psi}_xJ_x]\right),
\label{eqn:z_j}
\end{equation}
which will allows us to insert arbitrary factors of
$\psi_x\overline{\psi}_y$ into the functional integral by taking
functional derivatives of $Z_F[\bar{J},J]$ with respect to the source
fields $J_y$ and $\bar{J}_x$ and then setting $J_y=0$, $\bar{J}_x=0$,
which eliminates the source terms and gives $Z_F=Z_F[0,0]$.

The sources $\bar{J}_x$, $J_x$ and the functional derivatives 
$\frac{\delta}{\delta \bar{J}_x}$, $\frac{\delta}{\delta J_x}$
 obey the Grassmann algebra just like
the quark variables $\overline{\psi}_x$ and $\psi_x$, and satisfy:
$$\frac{\delta}{\delta \bar{J}}_x\,\bar{J}_y = \delta_{x,y}\quad \frac{\delta}{\delta \bar{J}}_x\, J_y = 0\quad \frac{\delta}{\delta J}_x\,J_y=\delta_{x,y}
\quad \frac{\delta}{\delta J}_x\,\bar{J}_y = 0$$
We can integrate $Z_F[\bar{J},J]$ by shifting the quark field variables
to complete the square in the exponent:
\begin{eqnarray*}
\overline{\psi'}_x&\equiv&\overline{\psi}_x-\bar{J}_y Q^{-1}_{y,x}[U]\\
\psi_x'&\equiv&\psi_x-Q^{-1}_{x,y}[U]J_y\\
\end{eqnarray*}
\begin{eqnarray}
Z_F[\bar{J},J]&=&\int_{\mathrm{p.b.c.}}\mathcal{D}\overline{\psi'}\,\mathcal{D}\psi'\,
\exp\left(-\sum_{x,y}\overline{\psi'}_xQ_{x,y}[U]\psi'_y+
\sum_{x,y}\bar{J}_x Q^{-1}_{x,y}[U]J_y\right),\nonumber\\
&=&\int \prod_{x'} d\overline{\psi}_{x'}\,d\psi_{x'}\,
\exp\left(-\sum_{x,y}\overline{\psi}_xQ_{x,y}[U]\psi_y\right)
\exp\left(\sum_{x,y}\bar{J}_x Q^{-1}_{x,y}[U]J_y\right)
\label{eqn:grass_int}
\end{eqnarray}
where we have dropped the primes, written the integration measure explicitly,
 and have used the
fact that the product of two Grassmann numbers behaves as a c-number.
The product of integrations extends over all lattice sites $x'$, and the 
fermionic boundary 
conditions are applied to expressions such as $\delta_{x,x+\hat{\mu}}$ within
the quark matrix $Q[U]$ and inverse quark matrix $Q^{-1}[U]$.

The Gaussian integral over Grassmann variables in Eqn.~\ref{eqn:grass_int}
can be evaluated~\cite{montvay:lgt}
 by expanding the integral and keeping only terms 
which do not vanish when integrated:
$$\int \prod_{x'} d\overline{\psi}_{x'}\,d\psi_{x'}\,
\exp\left(-\sum_{x,y}\overline{\psi}_xQ_{x,y}[U]\psi_y\right)
=\mbox{det}(Q[U])\int\prod_{x'} \left(d\overline{\psi}_{x'}\,\overline{\psi}_{x'}\right)\,
\left(d\psi_{x'}\,\psi_{x'}\right)=\mbox{det}(Q[U])
$$
Thus
$$Z_F[\bar{J},J]=\mbox{det}(Q[U])\exp\left(\sum_{x,y}\bar{J}_x Q^{-1}_{x,y}[U]J_y\right)$$

At this point it is essential to address the correct treatment of the
suppressed color, flavor, and Dirac indices.  The matrix $Q[U]$ carries
those indices, as do the variables $\overline{\psi}$,
$\psi$, $\bar{J}$ and $J$.  
The quark matrix $Q[U]$ can be thought of as a tensor product
of matrices acting in position space, color space, flavor space, and spin space:
\begin{itemize}
\item $\delta_{x,x+\hat{\mu}}$ acts in position space
\item $(U_\mu)_{ab}(x)$ acts in color space
\item $Q[U]$ is diagonal in flavor space (QCD conserves flavor)
\item $(\gamma_\mu)_{\alpha\beta}$ acts in spin space
\end{itemize}
Consequently, when we take a matrix determinant or inversion, we must
do so with respect to all indices.

We can use the fermion generating functional to define the full generating
functional $Z[\bar{J},J]$:
\begin{equation}
Z[\bar{J},J]\equiv \int_{\mathrm{p.b.c.}}\mathcal{D}U\,\mbox{det}(Q[U])
e^{-S_G[U]+\bar{J}Q^{-1}[U]J}
\label{eqn:gen_func}
\end{equation}
where now the functional integral is only over gauge links, with the
fermionic boundary conditions applied to $Q[U]$, $Q^{-1}[U]$, and the bosonic
boundary conditions applied to $S_G[U]$.

{\em Any} correlation matrix element of the form in Eqn.~\ref{eqn:corr_mat}
can be evaluated with the use of the generating functional in Eqn.~\ref{eqn:gen_func}.  For every fermion pair $\psi_x\overline{\psi}_y$, we apply
$\frac{\delta}{\delta J_y}\frac{\delta}{\delta \bar{J}_x}$ to 
$\ln(Z[\bar{J},J])$.
Any gauge link factors are left as they are.  We then set the
Grassmann
source fields $\bar{J}$ and $J$ to zero, and evaluate the remaining 
regularized
and well-defined integral over the gauge links $U$.

For example, the single quark propagator is:
\begin{eqnarray*}
\langle \psi_x\overline{\psi}_y\rangle &=&\frac{1}{Z}\int_{\mathrm{p.b.c.}}
\mathcal{D}U\,Q^{-1}_{x,y}[U]\,\mbox{det}(Q[U])e^{-S_G[U]}\\
Z&=&\int_{\mathrm{p.b.c.}}\mathcal{D}U\,\mbox{det}(Q[U])e^{-S_G[U]}.
\end{eqnarray*}
In the case of two quark/antiquark fields, an additional
term is generated by the derivatives:
$$\langle \psi_{x_1}\overline{\psi}_{x_2}\psi_{x_3}\overline{\psi}_{x_4}\rangle 
=\frac{1}{Z}\int_{\mathrm{p.b.c.}}
\mathcal{D}U\,\left(Q^{-1}_{x_1,x_2}Q^{-1}_{x_3,x_4}-
Q^{-1}_{x_1,x_4}Q^{-1}_{x_3,x_2}\right)\,\mbox{det}(Q)e^{-S_G}$$
The general way to evaluate correlation functions of arbitrary numbers
of quark-antiquark field pairs is to add up all possible permutations the fields 
which are in
alternating $\psi$-$\overline{\psi}$ form (this can be done graphically
using Wick's theorem~\cite{brown:quantum}).  For each term we can read off
the inverse quark matrix indices from the indices of the fields, and include
a minus sign if we performed an odd number of anticommutations
to get the quark fields into their final ordering.

In Chapter~\ref{chap:correlators}, we will need correlation matrix elements of 
the form
$$\langle \psi_{x_1}\psi_{x_2}\psi_{x_3}
\overline{\psi}_{x_4}\overline{\psi}_{x_5}\overline{\psi}_{x_6}\rangle$$
The six relevant permutations and signs are given by:
$$
\begin{array}{|c|c|}
\hline
\mbox{Permutation} & \mbox{Sign}\\
\hline
14\,25\,36 & -\\
14\,26\,35 & +\\
15\,24\,36 & +\\
15\,26\,34 & -\\
16\,24\,35 & -\\
16\,25\,34 & +\\
\hline
\end{array}
$$

Thus
\begin{eqnarray}
\langle \psi_{x_1}\psi_{x_2}\psi_{x_3}
\overline{\psi}_{x_4}\overline{\psi}_{x_5}\overline{\psi}_{x_6}\rangle=
\frac{1}{Z}\int_{\mathrm{p.b.c.}}\mathcal{D}U\,(
&-&Q^{-1}_{x_1,x_4}Q^{-1}_{x_2,x_5}Q^{-1}_{x_3,x_6}\nonumber\\
&+&Q^{-1}_{x_1,x_4}Q^{-1}_{x_2,x_6}Q^{-1}_{x_3,x_5}\nonumber\\
&+&Q^{-1}_{x_1,x_5}Q^{-1}_{x_2,x_4}Q^{-1}_{x_3,x_6}\nonumber\\
&-&Q^{-1}_{x_1,x_5}Q^{-1}_{x_2,x_6}Q^{-1}_{x_3,x_4}\nonumber\\
&-&Q^{-1}_{x_1,x_6}Q^{-1}_{x_2,x_4}Q^{-1}_{x_3,x_5}\nonumber\\
&+&Q^{-1}_{x_1,x_6}Q^{-1}_{x_2,x_5}Q^{-1}_{x_3,x_4})\,\mbox{det}(Q)e^{-S_G}
\end{eqnarray}
The integrand now contains only pure functions of the gauge link variables:
$Q[U]$, $Q^{-1}[U]$, and $S_G[U]$.

\section{The Monte Carlo method of integration}
Now that we have a well defined integral and have integrated out the quark
degrees of freedom, we would like to evaluate expressions such as:
\begin{eqnarray}
\langle f[U]\rangle &\equiv& \frac{1}{Z}\int_{\mathrm{p.b.c.}} \mathcal{D}U\,f[U] \mbox{det}(Q[U])e^{-S_G[U]}\nonumber\\
&=& \frac{\int\prod_{x,\mu}dU_\mu(x)\,f[U]\mbox{det}(Q[U])e^{-S_G[U]}}
{\int\prod_{x,\mu}dU_\mu(x)\,\mbox{det}(Q[U])e^{-S_G[U]}}\label{eqn:mc_integral}\end{eqnarray}
where each $dU_\mu(x)$ is the appropriate Haar measure over the $SU(3)$ 
group~\cite{creutz:quarks}.  There are $(2\times N_s^3\times N_\tau \times 4)$ $SU(3)$
integration variables, where $N_s$ and $N_\tau$ are the number of
lattice sites in the spatial and temporal directions, respectively.
$SU(3)$ has eight generators, so a brute force integration over our small
$12^3\times 48$ lattice would consist of a
$2\times 12^3\times 48\times 4\times 8 = 5,308,416$ dimensional
real integral over a compact domain.

The Monte Carlo method was specifically developed to estimate integrals over a 
large number of degrees of freedom~\cite{press:numerical}:
$$\frac{1}{V}\int_V d\mu(x)\, f(x),$$
where
$$V\equiv\int_V d\mu(x).$$ 

The Monte Carlo method uses the normalized integration 
measure $d\mu(x)/V$  at each point
in the integration domain to define the measure $p_F(x)dx$ over a probability 
space $F$.  An ensemble of $N$ configurations 
$\{x_1,x_2, \cdots,x_n\}$ is generated according to the probability measure
over $F$, 
and the integral is estimated by:
$$\frac{1}{V}\int_V d\mu(x)\, f(x) \approx \frac{1}{N}\sum_{n=1}^N f(x_n)\equiv \bar{f}$$
The expectation value of the ensemble average $\bar{f}$ over $F$ is the same
as the expectation value of the function evaluated on any single member
of the ensemble $x_i$, and equals the desired integral:
\begin{eqnarray}
\E_F[\bar{f}]&=&\frac{1}{N}\sum_{n=1}^N E_F[f(x_n)]=E_F[f(x)]\\
&=&\int_V p(x)d\mu(x)\, f(x)\\
&=&\frac{1}{V}\int_V d\mu(x)\, f(x)
\end{eqnarray}

From the central limit theorem~\cite{billingsley:probability} we know that if $N$ is sufficiently 
large\footnote{Finite ensemble error analysis involves subtle issues and is
treated in detail in section~\ref{sect:error_analysis}.}, then 
$\bar{f}$ will be normally distributed with:
\begin{equation}
\Var_F[\bar{f}]\propto \frac{1}{N}\label{eqn:var}
\end{equation}
The variance of the estimate $\Var_F[\bar{f}]$ is a measure of the uncertainty,
or expected error, caused by using a finite number $N$ of configurations in
 the Monte Carlo estimate of the integral.  We will return to the task of
estimating 
$\Var_F[\bar{f}]$ in Section~\ref{sect:error_analysis}.

\subsection{Importance sampling}
For our present purposes, we would like to perform high-dimensional integrals
of the form in Eqn.~\ref{eqn:mc_integral}.
Because the Euclidean
gauge action $S_G[U]$ is real and bounded from below, we can absorb the 
exponential damping term
$\exp(-S_G[U])$ into our definition of the measure.
Weighting the probability of generating a configuration with the 
Boltzmann-type factor $\exp(-S_G[U])$ greatly reduces the variance of our 
estimates by suppressing configurations
which give exponentially small contributions to the functional integral.
The fermion determinant $\mbox{det}(Q[U])$ is not 
real for all configurations because we treat $\overline{\psi}$ and $\psi$
as independent variables in the functional integral.  This unfortunate
situation, known as the {\em fermion sign problem}, makes it difficult
to include the fermion determinant into our definition of the 
measure.

After absorbing the exponential factor into the measure, we can estimate
the functional integral in Eqn.~\ref{eqn:mc_integral}
by first generating a ensemble of 
lattice-wide gauge link configurations 
$\{\{U_\mu(x)\}_1,\cdots,\{U_\mu(x)\}_N\}\equiv \{U^{(1)},
\cdots, U^{(N)}\}$ according to the joint probability 
density:
$$p(U^{(n)}) = \frac{e^{-S_G[U^{(n)}]}}{\int\prod_{x,\mu}dU_\mu(x)\,\mbox{det}(Q[U])e^{-S_G[U]}}$$
and then approximating:
\begin{equation}
\langle f[U] \rangle\approx\frac{\frac{1}{N}\sum_{n=1}^N f[U^{(n)}]\mbox{det}(Q[U^{(n)}])}
{\frac{1}{N}\sum_{m=1}^N\mbox{det}(Q[U^{(m)}])}.
\label{eqn:mc_expectation}
\end{equation}

\section{Markov updating}
We now turn to the task of generating the gauge configurations $U^{(n)}$.
Given some configuration, we may generate a new
configuration by cycling through the lattice and proposing random
(valid) changes to the individual link variables.
The new proposed change is accepted or rejected based on some
probabilistic rule which considers the Boltzmann factor of the
 configuration that would result if the link change was accepted.  
The process is repeated for several sweeps through the entire lattice in order
to produce an updated configuration. 
This is an example of a {\em Markov chain}, a class of processes in which the current state 
depends probabilistically on the previous 
state.  We require our Markov chains to be {\em ergodic}: any state in
the configuration space can be reached from any other state in a finite
number of updating steps.  This is necessary to ensure that we are 
adequately exploring the integration domain during our Monte Carlo integration.

A Markov chain is initialized to some initial state $U^{(0)}$.  It is then 
updated until it `thermalizes.'  At this point, all of the statistical 
properties of the states being generated in the chain have stabilized.  For
example, one can plot a moving average of the plaquette value to
determine when the Markov chain has thermalized.  The plot will drift
rapidly at first, and then fluctuate around some average value.  If
the configuration at which this begins to happen is denoted $U^{(m)}$, then
we take the configurations from $U^{(m+1)}$ to $U^{(m+N)}$ for our ensemble.
$U^{(0)}\to U^{(1)}\to U^{(2)} \to \cdots \to U^{(m)} \to \underbrace{U^{(m+1)} \to
\cdots\to U^{(m+N)}}_{\mbox{keep}}$

Examining the moving average of the plaquette is just a heuristic; 
in practice thermalization is a subtle issue requiring careful treatment.  
Whenever possible, it is preferred to initialize a Markov chain with
a thermalized configuration from a previous instance of the Markov chain.

\subsection{Quenching}
Because the fermion determinant $\mbox{det}(Q[U])$ is a complex quantity,
it is an extremely difficult quantity to estimate numerically due
to cancellations.  
There exist tricks, such as the Weingarten pseudo-fermion 
method~\cite{weingarten:pseudofermion},
 which enable us to include the fermion determinant 
in the updating algorithm as part of the integration measure.  Such methods
are computationally expensive because they involve the inversion
of a matrix having space-time, color, and spin indices.
The inversion of such matrices is currently
the single largest bottleneck in Lattice QCD calculations.
  
For this work, we will work
in the {\em quenched `approximation'} where we simply set the fermion
determinant in Eqn.~\ref{eqn:mc_expectation} to one:
$$\mbox{det}(Q[U])\to 1.$$
This is an unparameterized `approximation' which breaks the unitarity
of the theory.  Nonetheless, it is expected to to be mostly harmless
 for tuning studies 
with unphysically heavy quark masses (such as this work)
 because quenching is expected to affect the levels of the spectrum, but 
not
the coupling of the operators to the spectral states.
The fermion determinant is a highly non-local object containing
the complete contribution to the action of the fermionic degrees
of freedom.  From the 
(non-convergent) Feynman diagram point of view, quenching is equivalent
to removing all diagrams containing sea-quark loop contributions.

\subsection{Configurations used in this work}
Although updating is an important and subtle topic, it is not the focus
of this work.  
We briefly describe the method used to generate our gauge configurations.

The fact that Wilson's formulation of lattice QCD
 uses compact gauge links removes
the need for gauge fixing (see the discussion in section 
\ref{sec:gauge_fixing}).  Thus, it is possible to update the gauge links
freely according only to the probability distribution
$\exp(-S_G[U])$.  There is no need to require that the new configuration
adhere to any fixed gauge prescription.

The gauge configurations used in this work were
 generated using the Chroma QCD library developed at JLab, and 
use the parameters reported in Subsection~\ref{sect:lattice_params}.
That code used
the Cabibbo-Marinari heatbath algorithm~\cite{cabibbo:gauge_updating,
kennedy:gauge_updating} in conjunction with 
Creutz's overrelaxation technique~\cite{creutz:gauge_updating} to update the gauge configurations.

Because the configurations are quenched, the quark masses
did not play a role in the updating.  However, they are used in
the propagator calculations, which involve factors of inverse quark
matrix elements such as $Q^{-1}_{x,y}[U]$.  The operators presented
in this work act on the nucleon sector, thus only the bare light quark mass 
$a_\tau m_u=a_\tau m_d=-0.305$ was reported in 
Subsection~\ref{sect:lattice_params}.

Quenching will not be used in the production runs, and it will be assumed here 
that our operator choices and tuning parameters will not be significantly 
affected by quenching. 
In this study, we find that our quenched configurations still yield a
useful spectrum.
Rather, they serve as an indication of the quality of the operators to be used
in this ongoing research effort. 

\section{Error analysis}
\label{sect:error_analysis}
Monte Carlo integration is a statistical procedure.  It only provides 
stochastic estimates of quantities subject 
to statistical uncertainty.  The use of a Markov chain to 
generate the configurations
in our ensemble introduces autocorrelations (discussed below).  Additionally, we will
be transforming our data in a complicated manner (e.g. $C^{-1/2}(\tau_0)$) 
and need to quantify the statistical uncertainty for such quantities.

\subsection{Estimate variance in the presence of autocorrelation}
We have an ensemble of $N$ (thermalized) configurations $\{x_i\}$ which
looks as though it was drawn from some probability distribution $F$ but was
actually produced in sequential order by a Markov chain.  
The Markov process introduces serial 
autocorrelations~\cite{kennedy:ect06} into the ensemble which
increases the true variance of the estimate $\Var_F[\bar{f}]$.  This
implies that estimates\footnote{In this chapter, we will denote sample
estimates of statistics with a hat.  For example, if we use the sample
mean of some quantity $x$ to estimate $x$, then we may write $\hat{x}=\bar{x}$.}
$\widehat{Var}[\bar{f}]$ of the variance of $\bar{f}$ will {\em 
underestimate} the true variance $\Var_F[\bar{f}]$
of $\bar{f}$.  In other words, our error bars will be too small if our data
is autocorrelated.

To reduce clutter in the following, we define
\begin{eqnarray*}
g(x)&=&f(x)-E_F[f(x)],\\
\bar{g}&\equiv& \frac{1}{N}\sum_{n=1}^N g(x_n)=\bar{f}-E_F[\bar{f}],
\end{eqnarray*} 
which has the properties
\begin{eqnarray*}
E_F[g(x)] &=& 0,\\
E_F[g(x)^2] &=& \Var[f(x)],\\ 
E_F[\bar{g}^2] &=& \Var[\bar{f}].
\end{eqnarray*}
The variance of the ensemble average is
\begin{eqnarray}
\Var_F[\bar{f}]&=&E_F[\bar{g}^2],\nonumber\\
&=& \frac{1}{N^2}\sum_{n=1}^N\sum_{m=1}^N E_F[g(x_m)g(x_n)],\nonumber\\
&=&\frac{1}{N^2}\sum_{n=1}^N E_F[g(x_n)^2] + \frac{2}{N^2}\sum_{n=1}^{N-1} \sum_{m=n+1}^N E_F[g(x_m)g(x_n)],\nonumber\\
&=&\frac{1}{N}E_F[g(x)^2]\left(1+\frac{2}{N}\sum_{n=1}^{N-1}
\sum_{m=n+1}^N \frac{E_F[g(x_m)g(x_n)]}{E_F[g(x)^2]}\right)\label{eqn:autocorr1}
\end{eqnarray}

We can define the {\em autocorrelation function} $C_k[f(x)]$ for a function
$f(x)$ evaluated on an ordered series of observations 
$\{x_1, x_2, \cdots, x_N\}$ by
\begin{eqnarray}
C_k[f(x)] &=& \frac{E_F[(f(x_{n+k})-E_F[f(x_{n+k})])(f(x_n)-E_F[f(x_n)])]}
{\sqrt{E_F[(f(x_{n+k})-E_F[f(x_{n+k})])^2] E_F[(f(x_n)-E[f(x_n)])^2]}}
\nonumber\\
&&\nonumber\\
&=&\frac{E_F[(f(x_{n+k})-E_F[f(x_{n+k})])(f(x_n)-E_F[f(x_n)])]}
{E_F[(f(x)-E_F[f(x)])^2]}
\label{eqn:autocorr}
\end{eqnarray}

There is no $n$ index on the left hand side of Eqn.~\ref{eqn:autocorr}
 because we assume that the process generating the configurations has reached 
a stationary state (i.e. the statistical properties of the sequence are 
invariant under translations in $n$).

We may rewrite equation (\ref{eqn:autocorr1}) in terms of $C_k[f(x)]$:
\begin{eqnarray*}
\Var_F[\bar{f}]&=&\frac{1}{N}\Var_F[f(x)]\left(1+
\frac{2}{N}\sum_{n=1}^{N-1}\sum_{m=n+1}^N C_{m-n}[f(x)]\right),\\
&=& \frac{1}{N}\Var_F[f(x)]\left(1+\frac{2}{N}\sum_{k=1}^{N-1}(N-k) C_k[f(x)]\right),\\
&=& \frac{1}{N}\Var_F[f(x)]\left(1+2\sum_{k=1}^{N-1}C_k-\frac{2}{N}\sum_{k=1}^{N-1}kC_k[f(x)]\right),
\end{eqnarray*}
where we remind the reader that $\Var_F[\bar{f}]$ is the {\em true} variance
of $\bar{f}$.  In the case of a Markov chain, $C_k[f(x)]$ falls exponentially with $k$, and 
 for sufficiently large $N$ we have~\cite{binder:autocorr}
\begin{equation}
\Var_F[\bar{f}] \to \frac{1}{N}\Var_F[f(x)]\left(1+2A[f(x)]\right).\label{eqn:stderr}
\end{equation}
where
$$A[f(x)] = \sum_{k=1}^{\infty}C_k[f(x)]$$
is the `integrated' autocorrelation of the function $f$.  Thus we see that
in the presence of autocorrelation, the simple estimate of variance:
$$\widehat{Var}[\bar{f}]=\frac{1}{N}\Var_F[f(x)]$$
understates the true variance $\Var_F[\bar{f}]$.

A simple unbiased estimate of $\Var_F[f(x)]$ on an ensemble
$\{x_1,x_2,\cdots,x_N\}$ is 
\begin{equation}
\widehat{\Var}[f(x)] \equiv \frac{1}{N-1}\sum_{n=1}^N (f(x_n) - \bar{f})^2
\label{eqn:samplevar}
\end{equation}
with a similar expression for $\widehat{C_k}$.  In practice, the Markov chain 
updating method is tuned to reduce the autocorrelation in the data
to negligible levels.

\subsection{The Jackknife estimate of variance}
The variance estimate in Eqn.~\ref{eqn:samplevar} is applicable to functions
$f$ evaluated separately on each ensemble member.  As the ensemble member
fluctuates, so does $f(x)$.  Eqn.~\ref{eqn:samplevar} is not generally 
applicable to more complicated functions $f$ which involve multiple members
of the ensemble simultaneously.

In the following, each $x_n$ denotes a collection of simple quantities
(such as correlation matrix elements) evaluated on the $n^{th}$ gauge
link configuration 
generated by the Markov chain.
We are interested in a special class of functions $f$ which depend on the 
all of the ensemble members $x_i$ simultaneously through
the ensemble average $\bar{x}$:
$$f(\bar{x}),\qquad \bar{x}\equiv\frac{1}{N}\sum_{n=1}^N x_n.$$
In this work, the $\bar{x}$ correspond to Monte Carlo estimates of
correlation functions, and the $f$ correspond to the analysis methods
applied to these estimates.
Consequently, $f$ will typically be a complicated function of the 
ensemble average $\bar{x}$, e.g. taking the inverse square root of a matrix or
solving a Hermitian eigenvalue problem.
We would like to know how $f(\bar{x})$ varies across ensembles, not across
the members of the ensembles.

The typical variance estimate in Eqn.~\ref{eqn:samplevar} is not
generally applicable to functions of the {\em mean} of an ensemble: 
$f(\bar{x})$.
Using estimates $f(x_i)$ on individual ensemble members in an attempt to
estimate the variance of $f(\bar{x})$ 
is often a poor approach because the function may only be well defined
for certain values of its argument. 
The individual members of the ensemble $x_i$ fluctuate far more than does the
ensemble average $\bar{x}$, and these larger
fluctuations may result in an undefined $f(x_n)$.  
For example, if $f(\bar{x})=\ln(\bar{x})$, then $f(x_n)$ will be undefined 
 any time
$x_n\leq 0$, even if $\bar{x} > 0$.

The {\em jackknife} estimate of the variance~\cite{efron:jackknife} is a 
non-parametric alternative that uses re-sampling to numerically estimate the 
expected squared fluctuation of $f$ across ensembles.  
The jackknife method is applicable
to virtually any function $f$, but we will specialize it here to functions
of ensemble averages $f(\bar{x})$.

Let $\bar{x}_{(n)}$ be the ensemble average excluding the $n^{th}$ ensemble 
member $x_n$:
$$\bar{x}_{(n)}\equiv\frac{1}{N-1}\mathop{\sum_{m=1}^N}_{m\neq n} x_m
=\frac{N\bar{x}-x_n}{N-1}=\bar{x}-\left(\frac{x_n-\bar{x}}{N-1}\right)$$

Accordingly, we can define the $n^{th}$ {\em jackknife replicate} $f_{(n)}$ by:
$$f_{(n)}\equiv f(\bar{x}_{(n)}).$$
In contrast to $f(x_n)$ which relies on the information in {\em only} the 
$n^{th}$ ensemble member, the jackknife replicate $f_{(n)}$ is supported
by the information in all ensemble members {\em except} the $n^{th}$.
The power of this technique is that we can usually choose $N$ large enough
to ensure that both $f(\bar{x})$ and $f(\bar{x}_{(n)})$ are well defined.

The $N$ jackknife replicates $f_{(n)}$ are highly correlated, but it
is reasonable to expect that their mean squared fluctuations  
are proportional to the mean squared fluctuations of $f(\bar{x})$.
The jackknife estimate of the variance of $f$ across ensembles is:
\begin{eqnarray}
\widehat{\Var}_{\mathrm{jack}}[f(\bar{x})]&\equiv& \left(\frac{N-1}{N}\right)\sum_{n=1}^N
(f_{(n)}-f_{(\cdot)})^2,\nonumber\\
f_{(\cdot)}&\equiv&\frac{1}{N}\sum_{n=1}^N f_{(n)}.
\label{eqn:jackknife}
\end{eqnarray}
The {\em jackknife error} is defined as the square root of the jackknife
estimate of the variance.

We can verify the constant of proportionality $(N-1)/N$ 
in Eqn.~\ref{eqn:jackknife} by considering the known case $f(\bar{x}) = \bar{x}$:
\begin{eqnarray*}
\widehat{\Var}_{\mathrm{jack}}[\bar{x}] &=& \left(\frac{N-1}{N}\right)\sum_{n=1}^N
(\bar{x}_{(n)}-\bar{x}_{(\cdot)})^2,\\
&=& \left(\frac{N-1}{N}\right)\sum_{n=1}^N\left(\bar{x}-
\left(\frac{x_n-\bar{x}}{N-1}\right)-\bar{x}\right)^2\\
&=& \left(\frac{1}{N(N-1)}\right)\sum_{n=1}^N (x_n-\bar{x})^2,\\
&=&\frac{1}{N}\widehat{\Var}[f(x)]
\end{eqnarray*}
which is the expected expression for the (uncorrelated) estimate of the 
variance of $\bar{x}$.
We also emphasize that we use $f(\bar{x})$, not $f_{(\cdot)}$, to estimate $f(x)$:
$$\hat{f}=f(\bar{x}).$$

A generalization of the jackknife is the {\em bootstrap}, in which 
a large number $N_B$ (typically $O(1024)$) of bootstrap ensembles are each
created by 
selecting $N$ members (with replacement) from the original ensemble.
$f$ is then evaluated on the bootstrap ensembles to yield $N_B$ bootstrap
replicates $f_{[b]}$.
The bootstrap estimate of the variance of $f$ across ensembles is given by:
\begin{eqnarray*}
\widehat{\Var}_{\mathrm{boot}}[f(\bar{x})]&=&\left(\frac{1}{N_B-1}\right)\sum_{b=1}^{N_B}
(f_{[b]}-f_{[\cdot]})^2\\
f_{[\cdot]}&\equiv&\frac{1}{N_B}\sum_{b=1}^{N_B} f_{[b]}
\end{eqnarray*}

We use the simple jackknife method to estimate the errors on our effective mass
plots and the more thorough bootstrap method to estimate the uncertainties in our
 final fit parameter estimates.

\chapter{Baryon operator construction}
\label{chap:construction}
\section{Operator design goals}
The spectrum is calculated by evaluating correlation matrix elements
of quantum operators:
\begin{equation}
C_{ab}(\tau)=\langle \mathcal{O}_a(\tau)\overline{\mathcal{O}}_b(0)\rangle
=\sum_{k=1}^\infty|c_k|^2e^{-E_k\tau},\qquad 
c_k \equiv \bra{k}\overline{\mathcal{O}}\ket{\Omega},
\label{eqn:corr_mat_element}
\end{equation}
assuming a large temporal extent $T$ of the lattice and that the lattice
action satisfies reflection positivity~\cite{osterwalder:euclidean1,osterwalder:euclidean2}.

This chapter describes the design of
operators which excite the low-lying baryon spectral states.
We seek to find operators such that $c_k$ in Eqn.~\ref{eqn:corr_mat_element}
is as large as possible for the low-lying states of interest while being
as small as possible for higher-lying contaminating states.
Because we are using the Monte Carlo method to estimate Eqn.~\ref{eqn:corr_mat_element}, we also need to improve our operators to reduce the variance in
the estimates of $C_{ab}(\tau)$.
With these issues in mind, our goal is to design a set of operators which
\begin{itemize}
\item are gauge-invariant,
\item have the correct flavor content,
\item facilitate continuum spin identification,
\item rapidly plateau in the effective mass,
\item have minimal noise in the effective mass,
\item are linearly independent, and
\item compose a set of manageable size.
\end{itemize}

This chapter is based on the operator design approach developed by the LHPC
and described in~\cite{basak:clebsch,basak:group, lichtl:smeartune}.  
We begin by 
introducing the basic building blocks of our operators: smeared and 
covariantly-displaced quark fields, and smeared gauge links which are combined
into gauge-invariant three-quark operators.  Next, we discuss the group
theoretic approach adopted by the LHPC. 
The gauge-invariant three-quark operators are then endowed with the 
appropriate flavor structure to make our
three-quark elemental operators.  These elemental operators are then
combined into full baryon operators which transform irreducibly under
spatial lattice symmetry operations (rotations and reflections).
The chapter concludes by showing how the states excited by these 
operators are connected to the continuum $J^P$ baryon states.

\section{Building blocks}
On the lattice, we have quark sources
$\overline{\psi}^A_{a\alpha}(x)$, quark sinks
$\psi^A_{a\alpha}(x)$, and gauge links $U_{\mu ab}(x)$, 
where $A$ is a flavor index, $a,b$ are color indices,
and $\alpha$ is a Dirac spin index.  We have returned to writing the 
space-time index $x$ in parenthesis in order to avoid clutter.  Note that
we are still working with the unitless fields defined previously.

\subsection{Spatial inversion}
Under spatial inversion $I$, the field operators transform as:
$$U_I\overline{\psi}_{\alpha}(\vec{x},\tau)U_I^\dagger
=\overline{\psi}_{\beta}(-\vec{x},\tau)\bar{I}_{\beta\alpha}
\qquad 
U_I\psi_{\alpha}(\vec{x},\tau)U_I^\dagger= \psi_{\beta}(-\vec{x},\tau)I_{\beta\alpha}$$
where the $4\times 4$ parity matrices $\bar{I}$ and $I$ keep track of how
the spin components of the fields change under spatial inversion.  Summation
over repeated indices is implied.
To find the form of $\bar{I}$ and $I$, we consider how the fermion terms
in the continuum Euclidean action transform under the action of spatial 
inversion:
\begin{eqnarray*}
\int d^4x\, \overline{\psi}_\alpha(\vec{x},\tau)\psi_\alpha(\vec{x},\tau)&\to& 
\int d^4x\, \overline{\psi}_\beta(-\vec{x},\tau)\psi_\gamma(-\vec{x},\tau)
\bar{I}_{\beta\alpha}I_{\gamma\alpha}\\
&=&\int d^4x\, \overline{\psi}_\beta(\vec{x},\tau)\psi_\gamma(\vec{x},\tau)
\bar{I}_{\beta\alpha}I_{\gamma\alpha},\\
\int d^4x\, \overline{\psi}_\alpha(\vec{x},\tau) \gamma_{j\alpha\beta}\partial_j 
\psi_\beta(\vec{x},\tau)&\to&
\int d^4x\, \overline{\psi}_\lambda(-\vec{x},\tau) \gamma_{j\alpha\beta}\partial_j \psi_\sigma(-\vec{x},\tau) \bar{I}_{\lambda\alpha}I_{\sigma\beta}\\
&=&-\int d^4x\, \overline{\psi}_\lambda(\vec{x},\tau) \bar{I}_{\lambda\alpha}
\gamma_{j\alpha\beta}I_{\sigma\beta}\partial_j(\vec{x},\tau) \psi_\sigma\\
\int d^4x\, \overline{\psi}_\alpha(\vec{x},\tau) \gamma_{4\alpha\beta}\partial_4 
\psi_\beta(\vec{x},\tau)&\to&
\int d^4x\, \overline{\psi}_\lambda(-\vec{x},\tau) \gamma_{4\alpha\beta}\partial_\tau \psi_\sigma(-\vec{x},\tau) \bar{I}_{\lambda\alpha}I_{\sigma\beta}\\
&=&\int d^4x\, \overline{\psi}_\lambda(\vec{x},\tau) \bar{I}_{\lambda\alpha}
\gamma_{4\alpha\beta}I_{\sigma\beta}\partial_\tau \psi_\sigma(\vec{x},\tau) 
\end{eqnarray*}
To maintain the covariant derivative, we see that:
\begin{eqnarray*}
A_j(x)\to -A_j(-x),\qquad U_\mu(x)\to U^{*}_\mu(-x),\\
A_4(x)\to +A_4(-x),\qquad U_\mu(x)\to U_\mu(-x).\\
\end{eqnarray*}
In order for the action to be invariant under spatial inversion, we must have
$$\bar{I}I^T=1,\quad \bar{I}\gamma_j I^T = -\gamma_j,\quad
\bar{I}\gamma_4 I^T=\gamma_4$$
Using the definitions of the Euclidean Dirac gamma matrices
given in Chapter~\ref{chap:overview}, we see that
$\bar{I}=I=\gamma_4$, and that the fields transform as:
$$\overline{\psi}(\vec{x},\tau)\to \overline{\psi}(-\vec{x},\tau)\gamma_4,\qquad \psi(\vec{x},\tau)\to \psi(-\vec{x},\tau)\gamma_4.$$

\subsection{Gauge invariance}
We require our hadron operators to be gauge-invariant; all
operators must transform as scalars under local $SU(3)$ color transformations.
We may combine three quark sources (or sinks) into a color singlet by
contracting the
free color indices with the totally antisymmetric Levi-Civita tensor:  
$$\epsilon_{abc}=\left\{\begin{array}{cl}
+1 & abc = 123,231,312\\
-1 & abc = 132,213,321\\
\phantom{-}0 & \mbox{otherwise}.\end{array}
\right.$$

As an example, consider the behavior of the local (single-site) operator
under a local gauge transformation $\Omega(x)\in SU(3)$:
\begin{eqnarray*}
\epsilon_{abc}\overline{\psi}_a(x)\overline{\psi}_b(x)\overline{\psi}_c(x)
&\to& \epsilon_{abc}\overline{\psi}_{a'}(x)\overline{\psi}_{b'}(x)\overline{\psi}_{c'}(x)
\Omega^\dagger_{a'a}(x)\Omega^\dagger_{b'b}(x)\Omega^\dagger_{c'c}(x)\\
&=&
\det(\Omega^\dagger(x))\overline{\psi}_{a'}(x)\overline{\psi}_{b'}(x)\overline{\psi}_{c'}(x)
\epsilon_{a'b'c'}\\
&=&
\overline{\psi}_{a'}(x)\overline{\psi}_{b'}(x)\overline{\psi}_{c'}(x)
\epsilon_{a'b'c'}
\end{eqnarray*}

We will always contract our operators with the Levi-Civita tensor, with 
summation over repeated color indices implied.

\subsection{Gaussian quark field smearing}
 To successfully extract the spectrum, operators which couple
strongly to the low-lying states of interest and weakly to the high-lying states must
be used.
Effective phenomenological models tell us that we can improve our operators
by associating wavefunctions with the valence quarks, rather than working
with point quark sources.  This translates to the use of spatial quark field
{\em smearing} in our functional integrals.

\begin{figure}[ht!]
  \centering 
  \includegraphics[width=6.0in]{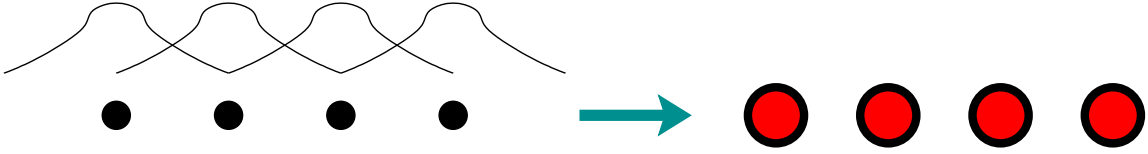}
  \caption{A schematic view of Gaussian quark field smearing.  The quark operators are replaced by `fatter' versions which better mimic the `fuzzy' nature of the quark wavefunctions.}
  \label{fig:gauss_smear} 
\end{figure}

Damping out couplings to the short-wavelength, high-momentum modes is
the crucial feature which any effective smearing 
prescription\cite{gusken:wuppertal,allton:jacobi} must have.  Gaussian suppression of
the high-momentum modes is perhaps the simplest method one can use.  Since a Gaussian in momentum
space remains a Gaussian in coordinate space, we decided to employ a gauge-covariant smearing 
scheme\cite{alford:gaussiansmear} in which the smeared quark field is defined at a given site as
a Gaussian-weighted average of the surrounding sites on the same time slice (see Figure~\ref{fig:gauss_smear}).

The smeared quark operators are `fatter', consisting of 
weighted contributions from nearby values with a (real) Gaussian radius of 
$\sigma$.
We expect that baryon operators using smeared quark fields
will better represent the spatial 
structure of the low-lying QCD baryon states, and will 
therefore interpolate better for those resonances.  

On a given time slice, the smeared continuum quark fields are defined as the 
integral transform of the unsmeared fields using the real, rotationally
invariant, Gaussian kernel 
$\mathcal{Q}(r^2)$:
\begin{equation*}
\tilde{\psi}(\vec{x})\equiv\int d^3r\, \mathcal{Q}(r^2)\psi(\vec{x}+\vec{r}),\qquad
\mathcal{Q}(r^2)\equiv\frac{1}{(\sqrt{2\pi\sigma^2})^3}e^{-r^2/(2\sigma^2)},
\end{equation*}
where $r^2\equiv (\vec{r}\cdot\vec{r})$.  The first three moments of the real 
Gaussian kernel are
$$\int d^3r\,\mathcal{Q}(r^2)=1,\qquad\int d^3r\,\mathcal{Q}(r^2)\vec{r}=0,\qquad\int d^3r\,\mathcal{Q}(r^2)r^2=\sigma^2.$$

We can transform the integral equation into a multiplicative operator equation
by applying the Euclidean spatial Fourier transform relations:
$$\psi(\vec{x})=\int\frac{d^3 k}{(2\pi)^3}\,e^{-i\vec{k}\cdot\vec{x}}\psi(\vec{k}),
\qquad \psi(\vec{k})\equiv\int d^3x\,e^{+i\vec{k}\cdot\vec{x}}\psi(\vec{x})$$
to the expression for the smeared quark field:
\begin{eqnarray}
\tilde{\psi}(\vec{x})&=&\frac{1}{(\sqrt{2\pi\sigma^2})^3}\int d^3r\, 
e^{-r^2/(2\sigma^2)}\psi(\vec{x}+\vec{r})\nonumber\\
&=&\frac{1}{(\sqrt{2\pi\sigma^2})^3}\int d^3r\, 
e^{-r^2/(2\sigma^2)}\int \frac{d^3k}{(2\pi)^3}\,e^{-i\vec{k}\cdot(\vec{x}+
\vec{r})}\psi(\vec{k})\nonumber\\
&=&\frac{1}{(\sqrt{2\pi\sigma^2})^3}\int d^3r\, 
\int \frac{d^3k}{(2\pi)^3}\,e^{-(r^2/\sigma^2+2i\vec{k}\cdot \vec{r})/2}
e^{-i\vec{k}\cdot\vec{x}}\psi(\vec{k})\nonumber\\
&&(\mbox{Letting }\vec{z}\equiv \vec{r}/\sigma+i\sigma\vec{k})\nonumber\\
&=&\left(\frac{1}{(\sqrt{2\pi})^3}\int d^3z\,e^{-z^2/2}\right) 
\int \frac{d^3k}{(2\pi)^3}\,e^{-\sigma^2 k^2/2}
e^{-i\vec{k}\cdot\vec{x}}f(\vec{k}),\qquad k^2\equiv \vec{k}\cdot\vec{k}\nonumber\\
&=&e^{\sigma^2\nabla^2/2}\int \frac{d^3k}{(2\pi)^3}\,e^{-i\vec{k}\cdot\vec{x}}
\psi(\vec{k})\nonumber\\
\tilde{\psi}(\vec{x})&=&e^{\sigma^2\nabla^2/2}\psi(\vec{x})
\label{eqn:cont_gauss_smear}
\end{eqnarray}

Using the relation
$$e^x = \lim_{n\to\infty}\left(1+\frac{x}{n}\right)^n$$
Eqn.~\ref{eqn:cont_gauss_smear} may be approximated on the lattice by
introducing
the quark smearing matrix $S$:
\begin{equation}
\tilde{\psi}_x\equiv \sum_y S_{x,y}\psi_y,\qquad 
S(\sigma,n_\sigma)\equiv\left(1+\frac{\sigma^2}{2n_\sigma}\Delta\right)^{n_\sigma}\
\label{eqn:quark_smear}
\end{equation}
where
\begin{eqnarray}
\Delta_{x,y}&\equiv&\sum_{k=1}^3\left(U_k(x)\delta_{x+\hat{k},y}+
U^\dagger_k(x-\hat{k})\delta_{x-\hat{k},y}-2\delta_{x,y}\right)\nonumber\\
&=&\sum_{k=\pm1,\pm2,\pm3}\left(U_k(x)\delta_{x+\hat{k},y}-\delta_{x,y}\right),
\qquad U_{-k}(x)\equiv U^\dagger_k(x-\hat{k})
\nonumber
\end{eqnarray}
is the three-dimensional 
gauge-covariant Laplace operator\footnote{We have defined $\Delta$ and $\sigma$
to be dimensionless.  The smearing radius in $\mbox{MeV}^{-1}$ or fm is given
by $\sigma a_s$.}.  
It is simple to check that the analogous expression for $\overline{\psi}$
in the continuum 
$$\tilde{\overline{\psi}}(\vec{x})=e^{\sigma^2\nabla^2/2}\overline{\psi}(\vec{x})$$
is approximated on the lattice by:
$$\tilde{\overline{\psi}}\equiv\overline{\psi}S^{n_\sigma\dagger}(\sigma).$$
There are two quark field
smearing parameters in Eqn.~\ref{eqn:quark_smear} that
we will tune in Chapter~\ref{chap:smearing}: the
 real (dimensionless) Gaussian radius $\sigma$ and the integer number of iterations in the 
approximation
of the exponential $n_{\sigma}$.

The smeared quark fields at each site are linear combinations of the 
original lattice quark fields;
we note that the smeared quark fields obey the same transformation 
laws and Grassmann algebra as the unsmeared fields.

\subsection{Gauge link smearing}
We can improve our operators further by replacing the gauge
links in our quark smearing matrix $S$ with smeared
gauge links $\tilde{U}_\mu(x)$.

Let $C_k(x)$ denote the sum of perpendicular spatial {\em staples} 
which
begin at lattice site $x$ and terminate at the neighboring site $x+\hat{\mu}$:
\begin{eqnarray*}
C_k(x)&\equiv&\sum_{j\neq k}\left(U_j(x)U_k(x+\hat{\jmath})
U^\dagger_j(x+\hat{k})\right.\\
&&+\left.U^\dagger_j(x-\hat{\jmath})U_k(x-\hat{\jmath})
U_j(x-\hat{\jmath}+\hat{k})\right).
\end{eqnarray*}
We incorporate the spatial information contained in the
staple sums $\{C_k\}$
with the original links $U^{[0]}\equiv \{U_k\}$ in 
some manner to produce a smeared 
gauge link configuration $U^{[1]}$ (the superscript refers to the 
smearing iteration, and has nothing to do with the Markov chain).  This
process is then iterated $n_\rho$ times to produce the final spatially
smeared link configuration $\tilde{U}$:
$$U^{[0]}\to U^{[1]}\to U^{[2]}\to \cdots \to U^{[n_\rho]}\equiv 
\tilde{U}$$

The APE smearing\cite{albanese:apesmear} iteration rule is commonly used:
$$U^{[n+1]}_k=\mathcal{P}_{SU(3)}\left(U^{[n]}_k(x)+\rho C^{[n]}_k(x)\right)$$
where $\rho$ is some real weight, and $\mathcal{P}_{SU(3)}$ is a projection 
operator into the gauge group.  The projection is
 needed because the sum of two $SU(3)$ matrices is not necessarily
an $SU(3)$ matrix.

To avoid the abrupt projection back into the gauge
group needed by this prescription, we decided instead to use the analytic and
computationally efficient gauge link smearing scheme
known as {\em stout-link}\footnote{The term `stout-link' was coined by
Morningstar and Peardon in a Dublin pub as a nod to Guinness beer.  
Officially, `stout' refers to ``their thick-bodied nature from the large
brew of paths used in their formulation''~\cite{morningstar:stout}} smearing~\cite{morningstar:stout} defined by
\begin{eqnarray}
 U^{[n+1]}_k(x)&=&\exp\left(i\rho\Theta^{[n]}_\mu(x)\right)U^{[n]}_k(x),\\
\Theta_k(x)&=&\frac{i}{2}\left(\Omega^\dagger_k(x)-\Omega_k(x)\right)-\frac{i}{2N}\mbox{Tr}\left(\Omega^\dagger_k(x)-\Omega_k(x)\right)\\
\Omega_k(x)&=&C_k(x) U^\dagger_k(x)\qquad\mbox{(no summation over $k$)}
\end{eqnarray}
where $N=3$ for $SU(3)$.  The two parameters to tune in this 
smearing procedure are the real staple weight $\rho$ 
and the integer number of iterations $n_\rho$.  

To make use of stout links, we replace the quark smearing matrix $S$ in
Eqn.~\ref{eqn:quark_smear} with
the complete smearing matrix $\tilde{S}$:
\begin{equation}
\tilde{\psi}_x\equiv \sum_y \tilde{S}_{x,y}\psi_y,\qquad 
\tilde{S}(\sigma,n_\sigma,\rho,n_\rho)\equiv\left(1+\frac{\sigma^2}{2n_\sigma}\tilde{\Delta}\right)^{n_\sigma}\
\end{equation}
where
$$
\tilde{\Delta}_{x,y}\equiv\sum_{k=\pm1,\pm2,\pm3}\left(\tilde{U}_k(x)\delta_{x+\hat{k},y}-\delta_{x,y}\right)$$

The quark field
and gauge link smearing schemes preserve the gauge invariance of hadron operators.
In Chapter~\ref{chap:smearing}, we systematically tune the smearing parameters 
$\sigma, n_\sigma, \rho$, and $n_\rho$ to give an optimal effective mass 
signal for our operators.

\subsection{Radial structure through covariant displacement}
The need for extended three-quark operators to capture both the radial and 
orbital structure of baryons has been emphasized and described in 
Ref.~\cite{basak:clebsch} and Ref.~\cite{basak:group}. 
We may add radial structure to our operators without breaking gauge invariance
by using the covariant displacement operators $\tilde{D}^{(p)}_j(x)$ which consist of
$p$ smeared parallel transporters in the $j^{th}$ spatial direction 
which `carry the 
color' back to the reference site $x$ from the displaced site
 $x+p\hat{\jmath}$:
$$
\tilde{D}^{(p)}_j(x)\equiv \tilde{U}_j(x)\tilde{U}_j(x+\hat{\jmath})\cdots \tilde{U}_j(x+(p-1)\hat{\jmath}),\qquad j=\pm1,\pm2,\pm3.
$$
For economy of notation in the upcoming expressions, we also define
the additional zero-displacement operator:
$$\tilde{D}^{(p)}_0(x)\equiv 1$$

A quark sink displaced from 
$x$ to $x+p\hat{\jmath}$ is represented by
$$\tilde{D}^{(p)}_j(x)\tilde{\psi}(x+p\hat{\jmath})\equiv [\tilde{D}^{(p)}_j
\tilde{\psi}](x),$$
and a quark source displaced from $x$ to $x+p\hat{\jmath}$ is represented by
$$\tilde{\overline{\psi}}(x+p\hat{\jmath})\tilde{D}^{(p)\dagger}_j(x)\equiv
[\tilde{\overline{\psi}}\tilde{D}^{(p)\dagger}_j](x)$$

For simplicity, the displaced quarks in a baryon operator will always be 
displaced by an equal amount $p$.
In general, operators corresponding to different values of $p$ may be combined
(via the variational method) to capture more complex radial structure.
In this work, we restrict our consideration to a single displacement
length $p$ of three links for our operators.

At this point, we can combine the smeared quark fields and gauge links into
gauge-invariant {\em extended three-quark operators}
 containing the desired color
and radial structure:
\begin{eqnarray}
\overline{\Phi}^{ABC}_{\alpha\beta\gamma,ijk}(x)&=&
\epsilon_{abc}
[\tilde{\overline{\psi}}D^{(3)\dagger}_i\gamma_4]^A_{a\alpha}(x)
[\tilde{\overline{\psi}}D^{(3)\dagger}_j\gamma_4]^B_{b\beta}(x)
[\tilde{\overline{\psi}}D^{(3)\dagger}_k\gamma_4]^C_{c\gamma}(x)\\
\Phi^{ABC}_{\alpha\beta\gamma,ijk}(x)&=&\epsilon_{abc}
[\tilde{D}^{(3)}_i\tilde{\psi}]_{Aa\alpha}(x)
[\tilde{D}^{(3)}_j\tilde{\psi}]_{Bb\beta}(x)
[\tilde{D}^{(3)}_k\tilde{\psi}]_{Cc\gamma}(x)
\end{eqnarray}
where a specific displacement length of $p=3$ has been used. 
An explicit factor of 
$\gamma_4$ is included in the creation operator 
$\overline{\Phi}^{ABC}_{\alpha\beta\gamma,ijk}(x)$
 to
ensure that our correlation matrices are Hermitian. 
 
The displacement direction indices $i,j,k=0,\pm1,\pm2,\pm3$ 
define the radial structure of the operator.
The types of displacements for the three-quark operators
considered in this work are illustrated in Table~\ref{table:operators}.
In particular, comparing the effectiveness of the doubly-displaced-L vs.
the triply-displaced-T operators can shed some light on the much discussed
issue of Y-flux/$\Delta$-flux formation of the gluon field in three-quark
systems (the $\Delta$ is actually a quantum superposition of three V-flux 
forms).

\begin{table}
\centering
\begin{tabular*}{0.75\linewidth}{l@{\extracolsep{\fill}}r}
\hline
\hline
\raisebox{0mm}[5mm]{Operator type} & Displacement indices\\
\hline
\multirow{3}{*}{\raisebox{0mm}{\setlength{\unitlength}{1mm}
\thicklines
\begin{picture}(20,15)
\put(14,7.5){\circle{6}}
\put(13,7){\circle*{2}}
\put(15,7){\circle*{2}}
\put(14,9){\circle*{2}}
\put(6,0){Single-Site}
\end{picture}}} &\\ &$i=j=k=0$\\&\\
\multirow{3}{*}{\raisebox{0mm}{\setlength{\unitlength}{1mm}
\thicklines
\begin{picture}(20,15)
\put(14,7.2){\circle{5}}
\put(14,6){\circle*{2}}
\put(14,8.3){\circle*{2}}
\put(21,7){\circle*{2}}
\put(16.5,7){\line(1,0){4}}
\put(2,0){Singly-Displaced}
\end{picture}}} &\\& $i = j =0, k \neq 0$\\&\\
\multirow{3}{*}{\raisebox{0mm}{\setlength{\unitlength}{1mm}
\thicklines
\begin{picture}(20,15)
\put(14,8){\circle{3}}
\put(14,8){\circle*{2}}
\put(8,8){\circle*{2}}
\put(20,8){\circle*{2}}
\put(8,8){\line(1,0){4.2}}
\put(20,8){\line(-1,0){4.2}}
\put(0,0){Doubly-Displaced-I}
\end{picture}}} &\\& $i=0,j=-k,k\neq 0$\\&\\
\multirow{3}{*}{\raisebox{0mm}{\setlength{\unitlength}{1mm}
\thicklines
\begin{picture}(20,15)
\put(14,6){\circle{3}}
\put(14,6){\circle*{2}}
\put(14,12){\circle*{2}}
\put(20,6){\circle*{2}}
\put(20,6){\line(-1,0){4.2}}
\put(14,12){\line(0,-1){4.2}}
\put(0,0){Doubly-Displaced-L}
\end{picture}}} &\\& $i=0, |j| \neq |k|, jk \neq 0$\\&\\
\multirow{3}{*}{\raisebox{0mm}{\setlength{\unitlength}{1mm}
\thicklines
\begin{picture}(20,15)
\put(14,11){\circle{2}}
\put(8,11){\circle*{2}}
\put(20,11){\circle*{2}}
\put(14,5){\circle*{2}}
\put(8,11){\line(1,0){5}}
\put(20,11){\line(-1,0){5}}
\put(14,5){\line(0,1){5}}
\put(0,0){Triply-Displaced-T}
\end{picture}}} &\\& $i=-j,|j|\neq|k|,jk\neq 0$\\&
\vspace{2mm}\\
\hline
\hline
\end{tabular*}
\caption{The spatial arrangements of the extended three-quark baryon
operators $\bar{\Phi}_{ijk}$ and
 $\Phi_{ijk}$. Quark-fields are
shown by solid circles, line segments indicate
gauge-covariant displacements, and each hollow circle indicates the location
of a Levi-Civita color coupling.  For simplicity, all displacements
have the same length in an operator.}
\label{table:operators}
\end{table}

A sixth possible configuration in which the quarks are displaced in three
orthogonal directions (triply-displaced-O) was not considered because
the connecting gauge links for such an operator do not lie in the plane of the 
quarks.  It is expected that such an operator would couple to the higher-lying
contaminating states much more than the similar triply-displaced-T operator.

\section{Classification of states by transformation behavior}
So far, we have constructed gauge-invariant extended 
three-quark operators $\overline{\Phi}^{ABC}_{\alpha\beta\gamma,ijk}(x)$ and
$\Phi^{ABC}_{\alpha\beta\gamma,ijk}(x)$ 
from smeared gauge links, and smeared and covariantly-displaced 
quark fields.  Our goal is to
use these extended operators to construct full baryon operators which 
possess good quantum
numbers on our lattice.  

Quantum numbers tell us how an object transforms under some symmetry operation.
As an example, consider the spherical
harmonics $Y_l^m(\theta,\phi)$, functions which appear frequently in the 
description of systems
possessing spherical symmetry (e.g. the multi-pole expansions of classical 
electrodynamics, or the orbital wavefunction solutions of the hydrogen
atom in quantum mechanics).  The spherical harmonics $Y_l^m(\theta,\phi)$ 
transform under spatial rotations in specific $l$-dependent ways.
The spherical harmonic $Y_0^0$ is a rotationally invariant
function.  The three spherical harmonics $Y_1^{-1}$, $Y_1^{0}$, $Y_1^1$
transform as a vector $v$ having components:
$$v_x = \frac{1}{\sqrt{2}}(Y^{-1}_1-Y^1_1),\quad v_y=\frac{i}{\sqrt{2}}(Y^{-1}_1+Y^1_1),\quad v_z=Y^0_1.$$
These different transformation behaviors can be used to label the state of the
system (in the example of the hydrogen atom, $l$ is the orbital quantum number).

We turn now to the task of combining our three-quark operators 
into baryon operators which transform irreducibly according to
the symmetries of the lattice.

\subsection{Irreducible representations of groups}
Briefly, we define and summarize the key ideas from group theory we will
need for the remainder of this chapter.  Summation will be denoted
explicitly (summation over repeated indices is not implied).

A group $\mathcal{G}$ is a set of elements $\set{R}$ along with 
multiplication law which satisfies the following properties:
\begin{enumerate}
\item \textit{Closure}:  $RR' \in \mathcal{G} \quad \forall \;\; R, R' \in \mathcal{G}$
\item \textit{Associativity}:  $R_1(R_2R_3) = (R_1R_2)R_3\quad \forall \;\;
 R_1, R_2, R_3 \in \mathcal{G}$
\item \textit{Identity}: $\exists \;\; E \in \mathcal{G} \;\; s.t. 
\;\; ER = RE = R \quad \forall \;\; R \in \mathcal{G}$
\item \textit{Inverse}: $\exists \;\; R^{-1} \in \mathcal{G} \;\; s.t. \;\; RR^{-1}
=R^{-1}R = E \quad \forall \;\; R \in \mathcal{G}$
\end{enumerate}                                                                           
If the multiplication operation is commutative, then the group is Abelian, otherwise, it is non-Abelian.  
A {\em representation} of a group is any set of matrices $\{D(R)\}$
which satisfies the group multiplication rule
$$D(R)D(R')=D(RR') \quad \forall \;\; R,R' \in \mathcal{G}.$$

In the following, we will denote our extended three-quark operators on
each time slice $\tau$ using a single index:
$$\overline{B}_\sigma(\tau)\equiv\overline{\phi}^{ABC}_{\alpha\beta\gamma,ijk}(\vec{x},\tau)
,\qquad B_\sigma\equiv\overline{\phi}^{ABC}_{\alpha\beta\gamma,ijk}(\vec{x},\tau),$$
where the master index $\sigma$ stands for the values of $A,B,C,\alpha,
\beta,\gamma,i,j,k$, and $\vec{x}$.

Under a color, flavor, or spatial transformation $R\in\mathcal{G}$ 
represented by the Hilbert space operator $U_R$,
our operators will transform into one another as 
\begin{eqnarray}
\overline{B}_{\alpha}&\to& U_R\overline{B}_{\alpha} U^\dagger_R=\sum_{\beta=1}^d
\overline{B}_{\beta} W_{\beta\alpha}(R),\nonumber\\
B_\alpha&\to& U_R B_\alpha U^\dagger_R=\sum_{\beta=1}^d\overline{B}_{\beta}
W^{-1}_{\beta\alpha}(R),\label{eqn:op_sum}
\end{eqnarray}
where $d$ is the dimension of the (not necessarily unitary) 
representation matrices $\{W(R)\}$.  

For example, lattice rotations through the angles of $0$,$\pi/2$,$\pi$, 
and $3\pi/2$ about the
axis $O\hat{\jmath}$ transform the different orientations of a 
triply-displaced-T operator into each other (taking $O\hat{\jmath}$ to be 
perpendicular to the plane containing the quarks).
Performing the same operations on a single-site
operator yields only one distinct operator.

We will restrict our discussion to these types of transformations.  Other
types of transformations, such as the charge conjugation operation
considered in Chapter~\ref{chap:correlators}, do not transform
baryon operators into baryon operators.

If the collection of transformations $\{R\}$ being considered composes
a group $\mathcal{G}$, then the coefficient matrices $\{W(R)\}$
are a representation for $\mathcal{G}$ defined on the basis $\{\overline{B}_\alpha\}$.
\begin{eqnarray*}
U_{R'} U_R\overline{B}_{\alpha} U^\dagger_R U^\dagger_{R'}&=&\sum_{\beta=1}^d
U_{R'}\overline{B}_{\beta}U^\dagger_{R'} W_{\beta\alpha}(R),\\
&=&\sum_{\gamma=1}^d\sum_{\beta=1}^d \overline{B}_{\gamma}W_{\gamma\beta}(R') W_{\beta\alpha}(R),\\
U_{R'R}\overline{B}_{\alpha} U^\dagger_{R'R} &=&\sum_{\gamma=1}^d
\overline{B}_{\gamma} W_{\gamma\alpha}(R'R).
\end{eqnarray*}
If the representation is {\em reducible}, then we can define $d'<d$ operators
$$\overline{B}'_\alpha\equiv\sum_{\beta=1}^d \overline{B}_\beta A_{\beta\alpha},
\qquad \alpha=1,\cdots,d'<d,$$
which transform only among themselves:
$$\overline{B}'_\alpha\to U_R\overline{B}'_\alpha U^\dagger_R=\sum_{\beta=1}^{d'} 
\overline{B}'_\beta W'_{\beta\alpha}(R),$$
where the representation matrices $\{W'(R)\}$ have a smaller dimension than
the original representation matrices $\{W(R)\}$.  A representation which
is not reducible is called an {\em irreducible representation}, or an 
{\em irrep}, and is labeled by a superscript in parenthesis.  Such 
representations are significant because they can be used to label
the different sectors of our Hilbert space. 

Let $\{D^{(\Lambda)}(R)\}$ denote
a set of $n_\Lambda$-dimensional unitary\footnote{We will be dealing with
unitary representation matrices $(D^\dagger(R) D(R) =1)$ unless 
specified otherwise. In particular, {\em all} finite groups possess
unitary representations.} matrices forming an 
irreducible representation of $\mathcal{G}$ labeled by $\Lambda$.
If we consider all nonequivalent\footnote{Two representations are equivalent
if they are related by a change of basis.} irreducible unitary representations of a
finite group $\mathcal{G}$ of order\footnote{The order $g$ of a finite group 
$\mathcal{G}$ is simply the number
of elements in $\mathcal{G}$.} $g$, then the quantities $D^{(\mu)}_{ij}(R)$ for 
{\em{fixed}} 
$\mu, i, j$, form a vector in a $g$-dimensional space, such that:
\begin{equation}
\sum_R D^{(\mu)}_{il}(R)D^{(\nu)*}_{jm}(R)=\frac{g}{n_\mu}\delta_{\mu\nu}\delta_{ij}\delta_{lm}
\label{eqn:ortho}
\end{equation}
Eqn.~\ref{eqn:ortho} forms the backbone of our operator construction effort,
and is proven in~\cite{hamermesh:group} as a corollary of Schur's Lemma.
The fundamental theme which will guide the operator construction
is that of constructing operators which transform irreducibly under
symmetry transformations.
  
\subsection{Symmetry operations}
When constructing our baryon operators, we will consider
symmetry groups of the Hamiltonian.
Consider a finite group $\mathcal{G}$ of $g$ unitary operators 
 $U_R$  which act on the Hilbert
space of the system ($U^\dagger_RU_R=1$).  $\mathcal{G}$ is a symmetry group of the Hamiltonian
if each element commutes with the Hamiltonian:
$$[U_R,H]=0.$$
Symmetry operations may be applied to operators at different times 
without disrupting the time evolution relation:
\begin{eqnarray*}
U_R\overline{B}_\alpha(\tau_2)U^\dagger_R
&=&U_R e^{H(\tau_2-\tau_1)}\overline{B}_\alpha(\tau_1)e^{-H(\tau_2-\tau_1)}U^\dagger_R\\
&=&e^{H(\tau_2-\tau_1)}\left(U_R\overline{B}_\alpha(\tau_1)U^\dagger_R \right)
e^{-H(\tau_2-\tau_1)}.
\end{eqnarray*}

We would like to find linear combinations of these operators which transform 
irreducibly under $\mathcal{G}$:
\begin{eqnarray}
\overline{B}^\Lambda_{i,a,\alpha}\equiv\sum_S c^{\Lambda}_{iaS}U_S \overline{B}_{\alpha}U^\dagger_S&\qquad&
U_R\overline{B}^{\Lambda}_{i,a,\alpha}U^\dagger_R = 
\sum_j \overline{B}^{\Lambda}_{j,a,\alpha}D^{(\Lambda)}_{ji}(R)\\
B^\Lambda_{i,a,\alpha}\equiv\sum_S c^{\Lambda*}_{iaS}U_S B_{\alpha}U^\dagger_S&\qquad&
U_RB^{\Lambda}_{i,a,\alpha}U^\dagger_R = 
\sum_j B^{\Lambda}_{j,a,\alpha}D^{(\Lambda)*}_{ji}(R)
\end{eqnarray}
where $\Lambda$ labels the irrep, $i$ labels the row of the irrep,
$a$ labels the instance of the 
irrep\footnote{There may be more than one linear combination of the 
$U_R\overline{B}_\alpha U^\dagger_R$ which transforms according to a 
particular irreducible transformation.}, 
 and $\alpha$ denotes
all other operator information (e.g. displacement values).

The power of this approach may be seen by considering a general
correlation function between two operators:
\begin{eqnarray}
C^{\Lambda\Gamma ab}_{ij\alpha\beta}(t)\equiv
\langle B^{\Lambda}_{i,a,\alpha}(t)\overline{B}^{\Gamma}_{j,b,\beta}(0)\rangle &=& 
\mbox{Tr}\left(B^{\Lambda}_{i,a,\alpha}(t)\overline{B}^{\Gamma}_{j,b,\beta}(0)\right)\nonumber\\
&=&\frac{1}{g}\sum_R \mbox{Tr}\left(U^\dagger_R U_R B^{\Lambda}_{i,a,\alpha}(t)
U^\dagger_R U_R\overline{B}^{\Gamma}_{j,b,\beta}(0)U^\dagger_R U_R\right)\nonumber\\
&=&\frac{1}{g}\sum_R \mbox{Tr}\left(U_R B^{\Lambda}_{i,a,\alpha}(t)
U^\dagger_R U_R\overline{B}^{\Gamma}_{j,b,\beta}(0)U^\dagger_R\right)\label{eqn:ortho_corr1}\\
&=&\sum_{k,l}\left(\frac{1}{g}\sum_R D^{(\Lambda)*}_{ki}(R)
D^{(\Gamma)}_{lj}(R)\right)\mbox{Tr}\left(B^{\Lambda}_{k,a,\alpha}(t)
\overline{B}^{\Gamma}_{l,b,\beta}(0)\right)\nonumber\\
&=&\delta_{\Lambda\Gamma}\delta_{ij}\frac{1}{n_\Lambda}\sum_k 
\langle B^{\Lambda}_{k,a,\alpha}(t)\overline{B}^{\Lambda}_{k,b,\beta}(0)\rangle
\label{eqn:ortho_corr2}
\end{eqnarray}
where the last line came from using the orthogonality relation in Eqn.~\ref{eqn:ortho}.

It is important to notice four things:
\begin{enumerate}
\item The irrep and the row must be the same for the correlation function 
not to vanish.
\item To get \ref{eqn:ortho_corr1}, we invoked the cyclic property of the 
trace.  Thus the operators are not orthogonal on a
configuration-by-configuration basis because equation \ref{eqn:ortho_corr2} is
an expectation value relation.  Anticipating orthogonality by 
calculating only non-vanishing correlation functions, we are effectively 
reducing the noise in our Monte Carlo estimates.
\item The value of the correlation function is independent of the row.  
Thus we can get better 
statistics by correlating operators by row, and then averaging over
the rows.\\
\item Different instances of an irrep may mix.  Because the value of the 
instance label does not affect the transformation properties of the operator,
we may group the instance label with the auxiliary index $\alpha$.  Let
the resulting operators be denoted $\overline{B}^\Lambda_{i,j}$, where 
$\Lambda$ is the irrep, $i$ is the row, and $j$
represents the value of $a$ and $\alpha$.
\end{enumerate}
Given a set of elementary operators $\{\overline{B}_\alpha\}$, we will 
construct
the correlation matrix in a given irrep $\Lambda$ as
$$C^{\Lambda}_{ij}(\tau)\equiv\frac{1}{n_\Lambda}\sum_{m=1}^{n_\Lambda}
\langle B^\Lambda_{m,i}(\tau)\overline{B}^\Lambda_{m,j}(0)\rangle.$$

\subsection{Projection onto irreducible representations of finite groups}
\label{section:projection_method}
For any operators $\overline{B}_\alpha$ and $B_\alpha$, we construct
operators which transform irreducibly under transformations $R\in \mathcal{G}$
by applying the following {\em projection operations}:
\begin{eqnarray}
\overline{B}^{\Lambda}_{i,a,\alpha}\equiv \mathcal{O}^{\Lambda}_{i,a} 
\overline{B}_\alpha&\equiv& \frac{n_\Lambda}{g}\sum_R D^{(\Lambda)*}_{ia}(R)
U_R\overline{B}_{\alpha}U^\dagger_R,\nonumber\\
B^{\Lambda}_{i,a,\alpha}\equiv \mathcal{O}^{\Lambda}_{i,a} 
B_\alpha&\equiv& \frac{n_\Lambda}{g}\sum_R D^{(\Lambda)}_{ia}(R)
U_RB_{\alpha}U^\dagger_R.
\label{eqn:projection}
\end{eqnarray}

The required transformation behavior is easily verified:
\begin{eqnarray*}
U_R\overline{B}^{\Lambda}_{i,a,\alpha}U^\dagger_R&=&\frac{n_\Lambda}{g}
\sum_S D^{(\Lambda)*}_{ia}(S)U_{RS}\overline{B}_\alpha U^\dagger_{RS}\\
&=&\frac{n_\Lambda}{g}
\sum_S D^{(\Lambda)*}_{ia}(R^{-1}S)U_{S}\overline{B}_\alpha U^\dagger_{S}\\
&=&\sum_j\left(\frac{n_\Lambda}{g}
\sum_S D^{(\Lambda)*}_{ja}(S)U_{S}\overline{B}_\alpha U^\dagger_{S}\right)D^{(\Lambda)}_{ji}(R)\\
&=&\sum_j \overline{B}^{\Lambda}_{j,a,\alpha} D^{(\Lambda)}_{ji}(R)
\end{eqnarray*}

$\mathcal{O}^{\Lambda}_{i,a}$ acting on an $\overline{B}_\alpha$ produces
an object which transforms according to the $i^{th}$ row of the irreducible
representation labeled by $\Lambda$.  Consider the effect of
two such operations:
\begin{eqnarray*}
\mathcal{O}^{\Lambda}_{i,a}\mathcal{O}^{\Gamma}_{j,b} \overline{B}_\alpha &=&
\frac{n_\Lambda n_\Gamma}{g^2}\sum_{R,S}D^{(\Lambda)*}_{ia}(R)
D^{(\Gamma)*}_{jb}(S)U_{RS}\overline{B}_\alpha U^\dagger_{RS}\\
&=&\frac{n_\Lambda n_\Gamma}{g^2}\sum_{R,S}D^{(\Lambda)*}_{ia}(R)
D^{(\Gamma)*}_{jb}(R^{-1}S)U_{S}\overline{B}_\alpha U^\dagger_{S}\\
&=&\sum_k\left(\frac{n_\Gamma}{g}\sum_{R}D^{(\Lambda)*}_{ia}(R)
D^{(\Gamma)}_{kj}(R)\right)\frac{n_\Lambda}{g}\sum_SD^{(\Gamma)*}_{kb}(S)U_{S}\overline{B}_\alpha U^\dagger_{S}\\
&=&\delta_{\Lambda\Gamma}\delta_{aj}\mathcal{O}^{\Lambda}_{i,b}\overline{B}_\alpha.
\end{eqnarray*}
In particular, if we define 
$P^\Lambda_i\equiv \mathcal{O}^{\Lambda}_{i,i}$, we have
$$P^\Lambda_i P^\Gamma_j\overline{B}_\alpha=\delta_{\Lambda\Gamma}
\delta_{ij}P^\Lambda_i\overline{B}_\alpha$$
The $P^\Lambda_i$ are therefore projection operators.
\begin{eqnarray*}
\sum_{\Lambda,i}P^\Lambda_i \overline{B}_\alpha&=&\sum_{\Lambda,i}
\frac{n_\Lambda}{g}\sum_S D^{(\Lambda)*}_{ii}(S)U_S\overline{B}_\alpha U^\dagger_S\\
&=&\sum_S\left(\frac{1}{g}\sum_\Lambda \chi^{(\Lambda)*}(S)\,n_\Lambda\right)
U_S\overline{B}_\alpha U^\dagger_S\\
&=&\overline{B}_\alpha
\end{eqnarray*}
where the $\chi^{(\Lambda)}(R)=\mbox{Tr}(D^{(\Lambda)})$ are the characters of 
the
group irrep, and we have used the character orthogonality 
relation~\cite{hamermesh:group}:
$$\frac{1}{g}\sum_\Lambda \chi^{(\Lambda)*}(R) 
\chi^{(\Lambda)}(E)=\delta_{RE}.$$
The identity element\footnote{The identity element is denoted by
$E$, from the German {\em einheit}, meaning {\em unit}.} $D(E)=1$ in 
any representation, and thus $\chi^{(\Lambda)}(E)=n_\Lambda$.

We can use Eqn.~\ref{eqn:op_sum} to define projection matrices 
$P^{\Lambda}_{i,\alpha\beta}$:
\begin{eqnarray}
P^\Lambda_i \overline{B}_{\alpha}&=&
\frac{n_\Lambda}{g}\sum_R D^{(\Lambda)*}_{ii}(R)
\sum_{\beta=1}^d \overline{B}_\beta
W_{\beta\alpha}(R),\nonumber\\
&\equiv&\sum_{\beta=1}^d P^{\Lambda}_{i,\alpha\beta}\overline{B}_\beta\\
P^{\Lambda}_{i,\alpha\beta}&\equiv&\frac{n_\Lambda}{g}\sum_R D^{(\Lambda)*}_{ii}(R)W_{\beta\alpha}
(R)
\end{eqnarray}

In the case that the $\{W(R)\}$ are not unitary, we may define the Hermitian
metric matrix $M$:
$$M_{\alpha\beta}=\frac{1}{g}\sum_R \sum_{\gamma=1}^d W^*_{\gamma\alpha}(R)
W_{\gamma\beta}(R).$$ 
We then compute the projection matrix for row $i=1$: 
$$P^{\Lambda}_{1,\alpha\beta}=\frac{n_\Lambda}{g}\sum_R D^{(\Lambda)*}_{11}(R)
W_{\beta\alpha}.$$
From the rows of this projection matrix, $r$ linearly-independent operators
are obtained:
$$\overline{B}^{\Lambda}_{1,\alpha}=\sum_{\beta=1}^d c^{\Lambda}_{1,\alpha\beta}
\overline{B}_\beta,$$
where $r$ is the rank of the projection matrix and the superposition coefficients 
$c^{\Lambda}_{1,\alpha\beta}$ for each operator are a linear combination of the 
rows of the projection matrix such that these coefficients satisfy
$$\sum_{\gamma,\sigma=1}^d c^{\Lambda*}_{1,\alpha\gamma}M_{\gamma\sigma}c^\Lambda_{1,\beta\sigma}=
\delta_{\alpha\beta},\qquad \alpha=1,\cdots,r.$$

In practice, these linear combinations are obtained using the well-known
Gram-Schmidt procedure, but with a modified inner product to incorporate the
metric matrix $M$.  The choice of these operators is not unique.  For each
of the $r$ operators $B^{\Lambda}_{1,\alpha}$ in the first row $i=1$, we 
obtain the partner operators in all other rows $j>1$ using:
$$c^{\Lambda}_{j,\alpha\beta}=\frac{n_\Lambda}{g}\sum_{\gamma=1}^d
\sum_R D^{(\Lambda)*}_{j1}(R)W_{\beta\gamma}(R)c^{\Lambda}_{1,\alpha\gamma}$$

\subsection{Symmetries of the lattice Hamiltonian}
We will construct baryon operators which transform irreducibly according to
the symmetry groups of the Hamiltonian on each time slice:
\begin{itemize}
\item The $SU(3)$ gauge group (color)
\item The $Z_N\otimes Z_N\otimes Z_N$ cyclic translation group (momentum)
\item The $SU(2)$ isospin group (flavor)\qquad (assumes $m_u=m_d$)
\item The $O^D_h$ crystal point group (spin and parity)
\end{itemize}

\subsection{Color}
By requiring our operators to be gauge-invariant, we have already ensured
that the operators transform irreducibly as color singlets. 
The irreducible representation of the color group by which the operators
transform is the trivial representation where 
$$D(R)=1\quad\forall\;\; R\in SU(3).$$
An the projection operation for this representation acts on the color indices
only, and results in the Levi-Civita coupling $\epsilon_{abc}$.

\subsection{Momentum}
The next index we consider for projection is the spatial reference 
site $\vec{x}$.  
Spatial translations form an Abelian group (the product of two translations
is a translation, each translation possesses an inverse, et cetera).
The displacement operator acting on Euclidean space-time functions is given by 
\begin{eqnarray*}
D(a)&=&e^{ia_j P_j}, \qquad P_j\equiv i\partial_j\\
D(a)f(x)&=&f(x-a).
\end{eqnarray*}
Here $D(a)$ is an operator, not a representation matrix.  However, this
abuse of notation will not lead to trouble because the
irreps $D^{(k)}(a)$ of the group of translations are
all one-dimensional and given by:
$$D^{(k)}(a)=e^{i k_j a_j},$$
where $k_j$ is the irrep label (the momentum).

Functions which transform irreducibly under translations are labeled
by their momentum $k$:
\begin{eqnarray*}
f^{(k)}(x)&=&ce^{-ik_j x_j}\\
D(a)f^{(k)}(x)&=&f^{(k)}(x-a)=f^{(k)}(x)D^{(k)}(a)\qquad \mbox{(no sum on $k$)}\\
P_\mu f^{(k)}(x)&=&k_\mu f^{(x)}(x)
\end{eqnarray*}

For our operators, we have a finite number $N_j$ of
discrete spatial lattice sites in each direction $\hat{\jmath}$
with periodic boundary conditions:
$$D^{(k)}(N_j a_s)=D^{(k)}(0)$$
where $a_s$ is the spatial lattice spacing. 
Thus, our group of spatial translations is a direct product of three
cyclic groups $Z_{N_j}=\{z_0,z_1,\cdots, z_{N_j}\}$, where
\begin{itemize}
\item $z_0$ is the identity element
\item $z_1$ corresponds to a translation from $x$ to $x+\hat{\jmath}$
\item $z_n=z_1^n$
\item $z_1^{N_j}=z_0$
\end{itemize}
The unique allowed momenta on the lattice are given by the irrep labels of
$Z_{N_j}$:
$$k_j=\frac{2\pi}{N_j a_s}n_j,\quad\mbox{(no sum on j)},\quad n_j=0,1,\cdots, N_j-1,
.$$

Some researchers shift the range of $n_j$ to allow negative momentum
values.  Because we are interested in the mass spectrum, 
 we will work in the zero-momentum
sector $k_1=k_2=k_3=0$. 
Applying the projection operation (Eqn.~\ref{eqn:projection}) 
with the trivial representation gives
the average over all spatial sites. 
Therefore, our zero-momentum extended three-quark operators are translationally
invariant and given by:
\begin{eqnarray}
\overline{\Phi}^{ABC}_{\alpha\beta\gamma,ijk}(\tau)&=&\frac{1}{N_1 N_2 N_3}
\sum_{\vec{x}}
\epsilon_{abc}
[\tilde{\overline{\psi}}D^{(3)\dagger}_i]^A_{a\alpha}(\vec{x},\tau)
[\tilde{\overline{\psi}}D^{(3)\dagger}_j]^B_{b\beta}(\vec{x},\tau)
[\tilde{\overline{\psi}}D^{(3)\dagger}_k]^C_{c\gamma}(\vec{x},\tau)\gamma_4\\
\Phi^{ABC}_{\alpha\beta\gamma,ijk}(\tau)&=&\frac{1}{N_1 N_2 N_3}
\sum_{\vec{x}}\epsilon_{abc}
[\tilde{D}^{(3)}_i\tilde{\psi}]_{Aa\alpha}(\vec{x},\tau)
[\tilde{D}^{(3)}_j\tilde{\psi}]_{Bb\beta}(\vec{x},\tau)
[\tilde{D}^{(3)}_k\tilde{\psi}]_{Cc\gamma}(\vec{x},\tau)
\end{eqnarray}

\subsection{Flavor}
For this work we will work with `light baryons,' containing 
$u$, $d$, and $s$ valence quarks only. Correspondingly, our flavor indices 
$A$, $B$, and $C$, can only take on the values $u,d,s$.  
In addition to conserving flavor, our Hamiltonian possesses an $SU(2)$ isospin 
symmetry in which the $u$ and $d$
 quarks are assigned the same mass ($m_u=m_d$) and are treated as components 
of an isospin doublet:
$$l=\left(\begin{array}{c}u\\d\end{array}\right)$$
where $l$ stands for {\em light} quark.
The group of transformations on the isospin doublet is $SU(2)$, the 
continuous group of
all $2\times 2$ unitary matrices with determinant $= +1$.  

In nature isospin symmetry 
is broken by a small $u$-$d$ mass splitting
which arises due to electromagnetic interactions and other effects, but
this splitting is less than $1\%$ mass of the lightest baryon, the proton.  
It is therefore
reasonable to treat $SU(2)$ isospin symmetry as exact in 
lattice QCD simulations
at present levels of precisions.  
We will require our operators to transform irreducibly under isospin
transformations.

On the other hand, the action explicitly breaks $SU(3)$ flavor symmetry 
($m_u=m_d=m_s$) by the designation of a heavier strange quark mass.  In
nature, the strange quark - light quark splitting is about $15-30\%$ of the 
proton mass.
Although $SU(3)$ flavor symmetry is not exact, we may design our
 operators to transform according to $SU(3)$ flavor irreps in addition to
$SU(3)$ isospin irreps, but must expect such operators to mix.

The underlying mechanism causing the different $SU(3)$ irreps to mix is 
illustrated by considering the effect of using the projection operation
(Eqn.~\ref{eqn:projection}) for a group which is not a symmetry of the
Hamiltonian:
\begin{eqnarray*}
\overline{B}^{\Lambda,a}_{i,\alpha}(\tau)&=&
\frac{n_\Lambda}{g}\sum_S D^{(\Lambda)*}_{ia}(S)U_S
\overline{B}_\alpha(\tau)U^\dagger_S\\
&=&\frac{n_\Lambda}{g}\sum_S D^{(\Lambda)*}_{ia}(S)U_S e^{+H\tau}
\overline{B}_\alpha(0)e^{-H\tau}U^\dagger_S
\end{eqnarray*}
If $[U_R,H]\neq0$, then
$$\overline{B}^{\Lambda, a}_{i,\alpha}(\tau)\neq
e^{+H\tau}\overline{B}^{\Lambda,a}_{i,\alpha}(0)e^{-H\tau}$$
and the orthogonality relation \ref{eqn:ortho_corr2} does not hold.  If we
require our baryon operators to transform according to the $SU(3)$ flavor
group, then the explicit symmetry breaking ($m_s\neq m_u,m_d$)
in the Hamiltonian will cause the operators in different symmetry sectors
to mix.  
This mixing will occur among baryon operators with equal isospin $I$, isospin
projection $I_3$, and quark flavor content (e.g. strangeness).
Some different baryon sectors labeled by the good quantum numbers $I$ and $S$
are illustrated in Table~\ref{table:isospin_strangeness}.

\begin{table}[ht!]
\centering
$$\begin{array}{|cc|l|}
\hline
\raisebox{0mm}[5mm]{I} & S & \mbox{Baryon Sector}\\
\hline
\raisebox{0mm}[5mm]{$\frac{1}{2}$} & 0 & N^0, N^+\\
\raisebox{0mm}[5mm]{$\frac{3}{2}$} & 0 & \Delta^-,\Delta^0,\Delta^+,\Delta^{++}\\
\raisebox{0mm}[5mm]{$0$} & -1 & \Lambda^0\\
\raisebox{0mm}[5mm]{$1$} & -1 & \Sigma^-, \Sigma^0, \Sigma^+\\
\raisebox{0mm}[5mm]{$\frac{1}{2}$} & -2 & \Xi^-, \Xi^0\\
\raisebox{0mm}[5mm]{$0$} & -3 & \Omega^-\\
\hline
\end{array}$$
\caption{The isospin and strangeness quantum numbers for the different
baryon sectors.  The isospin projection $I_3$ ranges from $-I$ to $I$
by increments of $1$.  The charge $Q$ of each baryon listed increases
with increasing isospin projection $I_3$.}
\label{table:isospin_strangeness}
\end{table}

$SU(2)$ isospin is a subgroup of $SU(3)$ flavor, and consists of the set
of $SU(3)$ flavor elements which leave the $s$ quarks invariant.
Given an irreducible representation of $SU(3)$ flavor (which is not 
necessarily three-dimensional), the subset of irrep matrices corresponding
to the $SU(2)$ subgroup elements forms a representation of the 
 $SU(2)$ isospin group.  This representation will generally be reducible,
which implies that we can find a basis in the $SU(3)$ irrep such that 
the $SU(2)$ representation (which is not two-dimensional) can be reduced
to block diagonal form.  In this basis, we will have states which transform
irreducibly under both $SU(3)$ flavor and $SU(2)$ isospin.
We will return to this theme in section \ref{sec:subduced}
 when we apply this {\em method of subduced
representations} by  restricting the continuum spin group $O(3)$
to the finite lattice rotation group.

\subsection{Flavor irreps}
The finite group projection operation (Eqn.~\ref{eqn:projection}) 
is readily extended
to compact continuous groups\footnote{Elements of a continuous group, 
such as the
group of three dimensional rotations, are characterized by a set of
real parameters.  If there exists a closed set which covers the complete
range of the parameters, then the group is said to be {\em compact}.  
Elements of $U(1)$ may be written as $\exp(i\theta)$, where the parameter
$\theta\in[0,2\pi)\subset[0,2\pi]$.  Thus $U(1)$ is a compact continuous group.}
such as $SU(2)$, but it is simpler to use Lie algebra methods to construct
operators which transform irreducibly.  
We will write out sums
explicitly in this section (summation over repeated indices is not implied).

An element $R$ of $SU(2)$ can be written as 
$$R = \exp(i\sum_{j=1}^3 a_j(R) \tau_j),\qquad \tau_j\equiv \frac{\sigma_j}{2}$$
where the $\sigma_j$ are the traceless Hermitian Pauli spin matrices as 
defined in
Eqn \ref{eqn:pauli_spin}.
The $SU(2)$  group structure is determined by the commutation relations
among the generators:
$$[\tau_i,\tau_j]=i\sum_k\epsilon_{ijk}\tau_k$$
We can construct a Casimir operator which commutes with all of the group
generators:
$$\tau^2\equiv\tau_1^2+\tau_2^2+\tau_3^2,\qquad [\tau^2,\tau_i]=0\quad(i=1,2,3).$$
By Schur's Lemma, we know that $\tau^2$ is proportional
to the identity matrix and can thus be used to label the different
irreps.
We may also choose the basis of Hilbert space states in each irrep
such that $\tau_3$ is diagonal.  By introducing the ladder operators
$$\tau_\pm\equiv \tau_1\pm i\tau_2$$
which satisfy
$$[\tau_3,\tau_\pm] =\pm \tau_\pm,\qquad \tau_3^2+\frac{1}{2}\tau_+\tau_-+\frac{1}{2}\tau_-\tau_+=\tau^2,$$
it can be shown (c.f.~\cite{mertzbacher:quantum}) that
\begin{eqnarray*}
\tau_\pm\ket{I,I_3}&=&\sqrt{I(I+1)-I_3(I_3\pm 1)}\ket{I,I_3\pm 1}\\
&=&\sqrt{(I\mp I_3)(I\pm I_3+1)}\ket{I,I_3\pm 1}\\
\tau_3\ket{I,I_3} &=& I_3\ket{I,I_3}\\
\tau^2\ket{I,I_3}&=&I(I+1)\ket{I,I_3}
\end{eqnarray*}
where the different irreps are labeled by $I$ and the row is 
labeled by $I_3$.

By definition, the irreducible representation matrices $\{D^{(I)}\}$ (also
known as {\em Wigner rotation matrices} in this context) are the matrix
elements of the $SU(2)$ operators between different irrep basis states in
the Hilbert space:
\begin{eqnarray*}
U_R\ket{I,I_3}&=&\sum_{I_3'}\ket{I,I_3'}D^{(I)}_{I_3'I_3}(R)\\
D^{(I)}_{I_3'I_3}(R)&=&\bra{I,I_3'}U_R\ket{I,I_3}.
\end{eqnarray*}

We would like our baryon operators to transform irreducibly under 
$SU(2)$ 
isospin transformations:
\begin{eqnarray*}
U_R\overline{B}^{I}_{I_3}U^\dagger_R&=&\sum_{I_3'}\overline{B}^{I}_{I_3'}
D^{(I)}_{I_3'I_3}(R)\\
&=&\sum_{I_3'}\overline{B}^I_{I_3'}\bra{I,I_3'}U_R\ket{I,I_3}
\end{eqnarray*}
where we have suppressed all indices except the flavor irrep and row.
Considering an infinitesimal transformation $a_i \to 0$ ($i=1,2,3$) and keeping
terms up to first order in $a_i$ at each step gives:
\begin{eqnarray*}
(1+i\sum_j a_j \tau_j)\overline{B}^I_{I_3}(1-i\sum_k a_k \tau_k)&=&\sum_{I_3'}
\overline{B}^I_{I_3'}\bra{I,I_3'}(1+i\sum_l a_l \tau_l)\ket{I,I_3}\\
\overline{B}^I_{I_3}+i\sum_j a_j[\tau_j,\overline{B}^I_{I_3}]&=&
\overline{B}^I_{I_3}+i\sum_j a_j\sum_{I_3'}\overline{B}^I_{I_3'}
\bra{I,I_3'}\tau_j\ket{I,I_3}\\
{}[\tau_j,\overline{B}^I_{I_3}]&=&
\sum_{I_3'}\overline{B}^I_{I_3'}\bra{I,I_3'}\tau_j\ket{I,I_3}\\
\end{eqnarray*}
where in the last step we used the fact that the infinitesimal
 $a_i$ parameters are arbitrary.  Our baryon operators will transform
under isospin according to the irreducible representation labeled
$I$ if and only if
\begin{eqnarray*}
{}[\tau_3,\overline{B}^I_{I_3}]&=&I_3\overline{B}^I_{I_3},\\
{}[\tau_+,\overline{B}^I_{I_3}]&=&\sqrt{(I-I_3)(I+I_3+1)}\overline{B}^I_{I_3+1},\\
{}[\tau_-,\overline{B}^I_{I_3}]&=&\sqrt{(I+I_3)(I-I_3+1)}\overline{B}^I_{I_3-1}.\\
\end{eqnarray*}
From these relations, we see that the analog of the Casimir operator is:
$$[\tau_3,[\tau_3,\overline{B}^I_{I_3}]]+
\frac{1}{2}[\tau_+,[\tau_-,\overline{B}^I_{I_3}]]+
\frac{1}{2}[\tau_-,[\tau_+,\overline{B}^I_{I_3}]]=I(I+1)\overline{B}^I_{I_3}.$$

We may now proceed to construct baryon operators from the individual quark
fields.  The light quarks are represented by $\bar{u}\equiv\tilde{\overline{\psi}}^u$ 
and
$\bar{d}\equiv\tilde{\overline{\psi}}^d$ smeared source operators having $I=1$ and 
$I_3=+1/2,-1/2$ respectively.
The strange quark is represented the $\bar{s}\equiv\tilde{\overline{\psi}}^s$ 
smeared source operator having
$I=0,I_3=0$.
These operators therefore satisfy
\begin{equation*}
\begin{array}{rclcrclcrcl}
{}[\tau_3,\bar{u}]&=&\frac{1}{2}\bar{u}, &&
{}[\tau_3,\bar{d}]&=&-\frac{1}{2}\bar{d}, &&
{}[\tau_3,\bar{s}]&=&0,\\
{}[\tau_+,\bar{u}]&=&\phantom{\frac{1}{2}}0, &&
{}[\tau_+,\bar{d}]&=&\phantom{-\frac{1}{2}}\bar{u}, &&
{}[\tau_+,\bar{s}]&=&0,\\
{}[\tau_-,\bar{u}]&=&\phantom{\frac{1}{2}}\bar{d}, &&
{}[\tau_-,\bar{d}]&=&\phantom{-\frac{1}{2}}0, &&
{}[\tau_-,\bar{s}]&=&0.
\end{array}
\end{equation*}

Due to isospin symmetry in the Hamiltonian, our particle masses will not depend
on the irrep row $I_3$.  We will therefore construct only one operator in each
isospin $I$, strangeness $S$ sector, the operator corresponding to maximal 
isospin projection $I_3=I$.  We now have {\em three-quark elemental operators}
 having definite isospin $I$, maximal 
$I_3=I$, and strangeness $S$: 
$$\overline{B}^{F}_{\alpha\beta\gamma,ijk} \equiv\sum_{A,B,C}
\overline{\Phi}^{ABC}_{\alpha\beta\gamma,ijk}(\tau)\phi^{(F)}_{ABC},$$
where $\phi^{(F)}_{ABC}$ are the coefficients for the flavor channel $F$.  
The three-quark elemental baryon
operators are given explicitly in terms of the extended three-quark operators
in Table~\ref{table:baryon_flavor}.

\begin{table}[ht!]
\centering
\begin{tabular*}{0.75\linewidth}{l@{\extracolsep{\fill}}ccc}
\hline
\hline
\raisebox{0mm}[5mm]{Baryon} & $I=I_3$ & $S$ & Operators\\
\hline
\raisebox{0mm}[5mm]{$\Delta^{++}$} & $\frac{3}{2}$ & $\phantom{-}0$ & $\overline{\Phi}^{uuu}_{\alpha\beta\gamma,ijk}$\\
\raisebox{0mm}[5mm]{$\Sigma^+$} & $1$ & $-1$ & $\overline{\Phi}^{uus}_{\alpha\beta\gamma,ijk}$\\
\raisebox{0mm}[5mm]{$N^+$} & $\frac{1}{2}$ & $\phantom{-}0$ & $\overline{\Phi}^{uud}_{\alpha\beta\gamma,ijk}-\overline{\Phi}^{duu}_{\alpha\beta\gamma,ijk}$\\
\raisebox{0mm}[5mm]{$\Xi^0$} & $\frac{1}{2}$ & $-2$ & $\overline{\Phi}^{ssu}_{\alpha\beta\gamma,ijk}$\\
\raisebox{0mm}[5mm]{$\Lambda^0$} & $0$ & $-1$ & $\overline{\Phi}^{uds}_{\alpha\beta\gamma,ijk}-\overline{\Phi}^{dus}_{\alpha\beta\gamma,ijk}$\\
\raisebox{0mm}[5mm]{$\Omega^-$} & $0$ & $-3$ & $\overline{\Phi}^{sss}_{\alpha\beta\gamma,ijk}$\\
\hline
\hline
\end{tabular*}
\caption{Elemental three-quark baryon operators $\overline{B}^F_{\alpha\beta\gamma,ijk}$ having 
definite isospin $I$, 
maximal 
$I_3=I$, and strangeness $S$ in terms of the gauge-invariant extended
three quark operators $\overline{\Phi}^{ABC}_{\alpha\beta\gamma,ijk}(\tau)$}
\label{table:baryon_flavor}
\end{table}

\section{Spin and parity}
At this point, we have used smeared quark fields and gauge links to
construct three-quark elemental operators $\overline{B}^F_i$ and $B^F_i$,
which have the desired
momentum, color, flavor, and radial structure.  The superscript $F$ denotes
the flavor structure, and can take on the values $(\Delta)$, $(\Sigma)$,
$(N)$, $(\Xi)$, $(\Lambda)$, or $(\Omega)$.
The index $i$ now consists
of only the Dirac spin indices $\alpha\beta\gamma$ and the displacement
direction indices $ijk$.
Our final task in constructing our operators is to project our operators
onto states of definite spin and parity.

\subsection{Baryon spin in the continuum}
In the continuum, we could construct our baryon operators using the 
irreducible representations of the continuous spin group $SU(2)$, the spinorial
double cover of the rotation group $SO(3)$.  The irreps of $SU(2)$ are labeled
 by the 
spin quantum number $J=\frac{1}{2},\frac{3}{2},\frac{5}{2},\cdots$.  

For example, we can construct a single-site spin-1/2 nucleon baryon 
operator by taking
$$\overline{B}^{(N)}_\alpha(\tau)=\sum_{\beta\gamma}B^{(N)}_{\alpha\beta\gamma,i=j=k=0}(\tau)\Gamma_{\beta\gamma}$$
where $\Gamma$ is some $4\times 4$ matrix of constants.

By combining two of the quarks into a Lorentz scalar or pseudoscalar, the 
resulting three-quark operator transforms like a spin-1/2 Dirac field under 
Lorentz transformations.  If two of the quarks are combined into a four-vector
(or axial four vector), then the resulting interpolating 
three-quark operators produce both spin-3/2 and spin-1/2 states.  

Although this method of building up baryon operators is in widespread use,
it becomes extremely cumbersome when constructing higher spin states or
complicated extended operators.  The difficulty is compounded by the fact that
the lattice regulator explicitly breaks Lorentz covariance.

\subsection{Baryon spin on the lattice}
The discretization of space onto a lattice breaks the rotational symmetry
of the system.  Thus, different continuum spin states will mix, making 
the identification of those states problematic.  We may mitigate this
difficulty, however, by designing our operators to transform under the lattice 
symmetry group $O^D_h$, the double-valued (spinorial) octahedral 
crystallographic point 
group.  The states excited by our operators will have well-defined lattice spin
and parity labels, and will not mix with each other.

Applying the projection operation (Eqn.~\ref{eqn:projection}) gives the full
baryon operators:
\begin{equation}
\begin{array}{rcccl}
\overline{B}^{\Lambda\lambda F}_i(\tau) &\equiv& 
\overline{B}^{F}_{\alpha\beta\gamma,jkl}c_{\alpha\beta\gamma,jkl}^{(\Lambda\lambda i)},\\
B^{\Lambda\lambda F}_i(\tau) &\equiv& 
B^{F}_{\alpha\beta\gamma,jkl}c_{\alpha\beta\gamma,jkl}^{(\Lambda\lambda i)*},
\end{array}
\label{eqn:baryon_operators}
\end{equation}
where $\Lambda$ is the $O^D_h$ irrep and $\lambda$ is the row. 
The coefficients $c(\Lambda\lambda i)_{\alpha\beta\gamma,jkl}$ can be
 determined using the methods of Subsection~\ref{section:projection_method} once
 we have the irreducible representation matrices for $O^D_h$.

The exposition below is adapted from~\cite{basak:group} 
and~\cite{morningstar:baryonic_ops}.

\subsection{Irreducible representation matrices for $O_h^D$}
The basic building blocks used to assemble
our baryon operators transform under the allowed spatial rotations and reflections
of the point group $O^D_h$ according to
\begin{eqnarray}
  U_R \left(\tilde{\overline{\psi}}(x)\tilde{D}^{(p)\dagger}_j\right)^A_{a\alpha} U_R^\dagger &=&
   \left(\tilde{\overline{\psi}}(Rx)
   \tilde{D}^{(p)\dagger}_{R j}\right)^A_{a\beta}\, S_{\beta\alpha}(R),\\
  U_R \left(\tilde{D}^{(p)}_j
  \tilde{\psi}(x)\right)^A_{a\alpha} U_R^\dagger &=&
  S(R)^{-1}_{\alpha\beta} \left(\widetilde{D}^{(p)}_{R j}
  \tilde{\psi}(Rx)\right)^A_{a\beta},\label{eq:transform}
\end{eqnarray}
where the transformation matrices for spatial inversion $I_s$
and proper rotations $C_{nj}$ through angle $2\pi/n$ about axis
$O\hat{\jmath}$ are given by
\begin{eqnarray}
 S(C_{nj}) &=& \exp\Bigl(\textstyle\frac{1}{8}\omega_{\mu\nu}
 [\gamma_\mu,\gamma_\nu]  \Bigr),\\
S(I_s) &=& \gamma_4,
\end{eqnarray}
with $\omega_{kl}=-2\pi\varepsilon_{jkl}/n$ and $\omega_{4k}=\omega_{k4}=0$
($\omega_{\mu\nu}$ is an antisymmetric tensor which parameterizes rotations
and boosts). A rotation by $\pi/2$ about the $y$-axis is conventionally 
denoted by $C_{4y}$, and $C_{4z}$ denotes a rotation by $\pi/2$ about the 
$z$-axis.  These particular group elements are given by
\begin{equation}
 S(C_{4y})=\frac{1}{\sqrt{2}}(1+\gamma_1\gamma_3),\quad 
 S(C_{4z})=\frac{1}{\sqrt{2}}(1+\gamma_2\gamma_1).
\end{equation}
The allowed rotations on a three-dimensional spatially-isotropic cubic lattice
form the octahedral group $O$ which has 24 elements.  Inclusion of spatial
inversion yields the point group $O_h$ which has 48 elements occurring
in ten conjugacy classes. All elements of $O_h$ can be generated from 
appropriate products of only $C_{4y}$, $C_{4z}$, and $I_s$.

Operators which transform according to the irreducible representations of $O_h$
can then be constructed using the group-theoretical projections
given in Eqn.~(\ref{eqn:baryon_operators}).  Orthogonality
relations and hence, projection techniques, in group theory apply only to
single-valued irreducible representations.  However, the fermionic
representations are double-valued representations of $O_h$.  The commonly-used
trick to circumvent this difficulty is to exploit the equivalence of the
double-valued irreps of $O_h$ with the extra single-valued irreps of
the so-called {\em double point group} $O_h^D$.  This group is formed by
introducing a new element $\overline{E}$ which represents a rotation by
an angle $2\pi$ about any axis, such that $\overline{E}^2=E$ (the identity).
By including such an element, the total number of elements in $O_h^D$ is
double the number of elements in $O_h$.  The 96 elements of $O_h^D$ occur
in sixteen conjugacy classes.  

Since baryons are fermions, we need only be concerned with the six double-valued
irreps of $O_h$.  There are four two-dimensional irreps
$G_{1g}, G_{1u}, G_{2g}$, and $G_{2u}$, and two four-dimensional irreps
$H_g$ and $H_u$.  The subscript $g$ refers to even-parity states, whereas the
subscript $u$ refers to odd-parity states\footnote{From the German {\em gerade} meaning `even,' and {\em ungerade} meaning `odd'}. The irreps $G_{1g}$ and $G_{1u}$ 
contain the spin-1/2 states, spin-3/2 states reside in the $H_g$ and $H_u$,
and two of the spin projections of the spin-5/2 states occur in the $G_{2g}$
and $G_{2u}$ irreps, while the remaining four projections reside in the
$H_g$ and $H_u$ irreps.  The spin content of each $O_h$ irrep in the continuum limit
is summarized in Table~\ref{table:spin_table}.  This table lists the number of times
that each of the $O_h$ irreps occurs in the 
$J=\frac{1}{2},\frac{3}{2},\frac{5}{2},\cdots$ representations
of $SU(2)$ subduced to $O_h$. 

To carry out the projections in Eqn.~(\ref{eqn:baryon_operators}),
explicit representation matrices are needed. Our choice of representation matrices
is summarized in Table~\ref{tab:ODreps}.  Matrices for only the
group elements $C_{4y},C_{4z}$, and $I_s$ are given in Table~\ref{tab:ODreps}
since the representation matrices for all other group elements can be obtained
by suitable multiplications of the matrices for the three generating elements.
For baryons, the representation matrix for $\overline{E}$ in each of the $O_h^D$
extra irreps is $-1$ times the identity matrix.

\begin{table}
\centering
\begin{tabular*}{0.75\linewidth}{c@{\extracolsep{\fill}}rr}
\hline
\hline
\raisebox{0mm}[5mm]{$\Lambda$} & $\Gamma^{(\Lambda)}(C_{4y})$
 & $\Gamma^{(\Lambda)}(C_{4z})$ 
 \\ \hline
$G_{1g}$  & \rule[-3ex]{0mm}{8ex}
$\displaystyle\frac{1}{\sqrt{2}}\!\left[\begin{array}{rr}
 1 & -1 \\ 1 &  1
\end{array}\right]$ &
$\displaystyle\frac{1}{\sqrt{2}}\!\left[\begin{array}{cc}
  1\!-\!i & 0 \\ 0 & 1\!+\!i 
\end{array}\right] $ \\
$G_{2g}$   &\rule[-3ex]{0mm}{8ex}
$\displaystyle\frac{-1}{\sqrt{2}}\!\left[\begin{array}{rr}
 1 & -1 \\ 1 &  1
\end{array}\right]$ &
 $\displaystyle\frac{-1}{\sqrt{2}}\!\left[\begin{array}{cc}
  1\!-\!i & 0 \\ 0 & 1\!+\!i 
\end{array}\right]$ \\
$H_g$  &\rule[-7ex]{0mm}{15ex}
$\displaystyle\!\!\frac{1}{2\sqrt{2}}\!\!\left[\begin{array}{rrrr}
  \!1 & \!\!-\sqrt{3} & \sqrt{3} & -1 \\
  \!\!\sqrt{3} & -1 & -1 &  \sqrt{3} \\
  \!\!\sqrt{3} &  1 & -1 & \!\!-\sqrt{3} \\
  1 &  \sqrt{3} & \sqrt{3} &  1 \end{array}\right] $
& $\displaystyle\!\!\frac{1}{\sqrt{2}}\!\!\left[\begin{array}{cccc}
   \!\!-1\!-\!i\!\! & 0 & 0 & 0 \\
   0 & \!\!1\!-\!i\!\! & 0 & 0 \\
   0 & 0 & \!\!1\!+\!i\!\! & 0 \\
   0 & 0 & 0 & \!\!-1\!+\!i\!\! \end{array}\right]$\\
\hline
\hline
\end{tabular*}
\caption{Our choice of the representation matrices for the double-valued 
 irreps of $O_h$.  The $G_{1u},G_{2u},H_u$ matrices for the rotations $C_{4y},C_{4z}$
  are the same as the $G_{1g},G_{2g},H_g$ matrices, respectively, given below.
  Each of the $G_{1g},G_{2g},H_g$ matrices for spatial inversion $I_s$ is the identity
  matrix, whereas each of the $G_{1u},G_{2u},H_u$ matrices for $I_s$ is $-1$ times the identity
  matrix. The matrices for all other group elements can be obtained from appropriate
  multiplications of the $C_{4y},C_{4z}$, and $I_s$ matrices.}
 \label{tab:ODreps}
\end{table}

To give an example of this method, the single-site $N^+$ (nucleon) operators
which transform irreducibly under the symmetry group of the spatial lattice
are given in Table~\ref{tab:SSNucleon}.
\begin{table}[ht!]
\centering
\begin{tabular}{c@{\hspace{1em}}cc@{\hspace{1em}}cc}
\hline
\hline
 Irrep & Row & Operator & Row & Operator\\ \hline 
 $G_{1g}$ & 1 &  $N_{211} $ 
          & 2 &  $N_{221} $ \\
 $G_{1g}$ & 1 &  $N_{413} $ 
          & 2 &  $N_{423} $ \\
 $G_{1g}$ & 1 &  $2N_{332}\!+\!N_{413}\!-\!2N_{431}\!\!\! $
          & 2 &  $2N_{432}\!-\!2N_{441}\!-\!N_{423} $ \\ 
 $G_{1u}$ & 1 &  $N_{433} $ 
          & 2 &  $N_{443} $ \\ 
 $G_{1u}$ & 1 &  $N_{321}-N_{312} $ 
          & 2 &  $N_{421}-N_{412} $ \\ 
 $G_{1u}$ & 1 &  $N_{312}\!+\!N_{321}\!-\!2N_{411} $ 
          & 2 &  $2N_{322}\!-\!N_{412}\!-\!N_{421} $ \\ 
 $H_{g}$ & 1 &  $\sqrt{3}\,N_{331} $
         & 2 &  $N_{332}\!-\!N_{413}\!+\!2N_{431} $ \\
 $H_{g}$ & 3 &  $2N_{432}\!+\!N_{441}\!-\!N_{423} $
         & 4 &  $\sqrt{3}\,N_{442} $ \\ 
 $H_{u}$ & 1 &  $-\sqrt{3}\,N_{311} $ 
         & 2 &  $\!-\!N_{312}\!-\!N_{321}\!-\!N_{411} $ \\
 $H_{u}$ & 3 &  $\!-\!N_{322}\!-\!N_{412}\!-\!N_{421} $ 
         & 4 &  $-\sqrt{3}\,N_{422} $ \\
\hline
\hline
\end{tabular}
\caption{The single-site $N^+$ 
 operators which transform irreducibly under the symmetry group
 of the spatial lattice, defining 
 $N_{\alpha\beta\gamma}=\Phi^{uud}_{\alpha\beta\gamma;000}
-\Phi^{duu}_{\alpha\beta\gamma;000}$ (see Table~(\protect\ref{table:baryon_flavor})) and using the Dirac-Pauli representation for the Dirac gamma matrices.
\label{tab:SSNucleon}}
\end{table}

\subsection{Continuum spin identification}
\label{sec:subduced}
Once we have the lattice spectrum, we may use the method of 
subduced representations to to identify the continuum spin 
values of the extracted states.
  Because $O^D_h$ is a subgroup of the continuum rotation group for spinors
(with inversion),
 the continuum states appear as degenerate states within each of the 
lattice irreps.  Table~\ref{table:spin_table} denotes the number of times
a continuum-limit spin $J$ state will appear in each irreducible representation
of the lattice.
Thus, Table~\ref{table:spin_table} allows us to identify the continuum-limit
spin $J$ corresponding to the masses extracted in our Monte Carlo calculations.  
For example, to identify an even-parity baryon as having $J=1/2$, a level must
be observed in the $G_{1g}$ channel, and there must be no degenerate partners
in either of the $G_{2g}$ or $H_g$.  A level observed in the $H_g$ channel
with no degenerate partners in the $G_{1g}$ and $G_{2g}$ channels (in the
continuum limit) is a $J=3/2$ state.  Degenerate partners observed in the
$G_{2g}$ and $H_g$ channels with no partner in the $G_{1g}$ channel indicates
a $J=5/2$ baryon.  In other words, Table~\ref{table:spin_table} 
details the patterns
of continuum-limit degeneracies corresponding to each half-integral $J$ value.

\begin{table}[ht!]
\centering 
$$\begin{array}{|c|ccc|}  
\hline
\raisebox{0mm}[5mm][3mm]{J} & n^J_{G_1} & n^J_H & n^J_{G_2}\\
\hline
\raisebox{0mm}[5mm]{$\frac{1}{2}$} & 1  & 0 & 0\\
\raisebox{0mm}[5mm]{$\frac{3}{2}$} & 0  & 1 & 0\\
\raisebox{0mm}[5mm]{$\frac{5}{2}$} & 0  & 1 & 1\\
\raisebox{0mm}[5mm]{$\frac{7}{2}$} & 1  & 1 & 1\\
\raisebox{0mm}[5mm]{$\frac{9}{2}$} & 1  & 2 & 0\\
\raisebox{0mm}[5mm]{$\frac{11}{2}$} & 1 & 2 & 1\\
\raisebox{0mm}[5mm]{$\frac{13}{2}$} & 1 & 2 & 2\\
\raisebox{0mm}[5mm]{$\frac{15}{2}$} & 1 & 3 & 1\\
\raisebox{0mm}[5mm][3mm]{$\frac{17}{2}$} & 2 & 3 & 1\\
\hline
\end{array}$$
\caption{Because $O^D_h$ is a subgroup of the continuum spin group, the continuum states appear as degenerate states within each of the lattice irreps.  This table allows us to identify the continuum spin state which corresponds to each lattice state.}
\label{table:spin_table} 
\end{table}

\chapter{Evaluation of baryon correlation matrices}
\label{chap:correlators}

\section{Charge conjugation and backward propagating states}
\subsection{Charge conjugation}
As discussed in chapter \ref{chap:overview}, the 
Euclidean $\overline{\psi}$ and $\psi$ operators are the
analytic continuations of $\overline{\psi}_M$ and $\psi_M$, respectively, and
must be treated as independent fields:
$$\overline{\psi}\neq\psi^\dagger\gamma_4.$$

However, we can define a {\em charge conjugation} operator $U_C$ on the Hilbert
space which exchanges quarks and antiquarks.  This introduces a relation
between the operators $\overline{\psi}$ and $\psi$:
$$U_C\overline{\psi}_{\alpha}(x)U_C^\dagger= \psi_{\beta}\bar{C}_{\beta\alpha},
\qquad
U_C\psi_{\alpha}(x)U_C^\dagger=\overline{\psi}_{\beta}C_{\beta\alpha}, 
$$
where the $4\times 4$ charge conjugation matrices $C$ and $\bar{C}$ keep 
track of how
the spin components of the fields change under charge conjugation and summation
over repeated indices is implied.
To find the form of $C$ and $\bar{C}$, we consider how the fermion terms
in the continuum Euclidean action transform under the action of charge
conjugation:
\begin{eqnarray*}
\overline{\psi}_\alpha\psi_\alpha&\to& \psi_\beta\overline{\psi}_\gamma
\bar{C}_{\beta\alpha}C_{\gamma\alpha},\\
&=&-\overline{\psi}_\gamma\psi_\beta,
\bar{C}_{\beta\alpha}C_{\gamma\alpha},\\
\int d^4x\, \overline{\psi}_\alpha \gamma_{\mu\alpha\beta}\partial_\mu 
\psi_\beta&\to&
\int d^4x\, \psi_\lambda \gamma_{\mu\alpha\beta}\partial_\mu 
\overline{\psi}_\sigma \bar{C}_{\lambda\alpha} C_{\sigma\beta},\\
&=&\int d^4x\, \overline{\psi}_\sigma \bar{C}_{\lambda\alpha}
\gamma_{\mu\alpha\beta}C_{\sigma\beta}\partial_\mu 
\psi_\lambda,
\end{eqnarray*}
where we anticommuted the Grassmann variables, integrated by parts, and 
discarded the surface term. 
In order for the action to be invariant under charge conjugation, we must have
\begin{eqnarray*}\bar{C}C^T=-1,&\qquad& \bar{C}\gamma_\mu C^T=\gamma_\mu^T,\\
A_\mu(x)\to -A^T_\mu(x),&\qquad& U_\mu(x)\to U_\mu^\dagger(x).
\end{eqnarray*}
A common choice for the charge conjugation matrix in the Dirac-Pauli basis
(c.f. Chapter~\ref{chap:overview})
is:
$$\bar{C}=C=\gamma_4\gamma_2,$$
and the fields transform as:
$$\overline{\psi}\to \psi \gamma_4\gamma_2,\qquad \psi
\to\overline{\psi}\gamma_4\gamma_2.$$

As an example, consider the charge conjugation behavior of the simple bilinear
form $\psi\gamma_\mu\overline{\psi}$ (which can be used to 
interpolate for vector mesons):
\begin{eqnarray*} 
\psi\gamma_\mu\overline{\psi}&\to&\overline{\psi} \gamma_4\gamma_2\gamma_\mu
(\gamma_4\gamma_2)^T\psi,\\
&=&\overline{\psi} \gamma_4\gamma_2\gamma_\mu
\gamma_2\gamma_4\psi,\\
&=&-\overline{\psi}\gamma_\mu^T\psi,\\
&=&\psi\gamma_\mu\overline{\psi}.
\end{eqnarray*}

\subsection{Backward propagating states} 
Our operators are constructed using fermion fields $\overline{\psi}(x)$ 
which create
a quark and annihilate an antiquark.  Hence, each of our baryon operators 
creates a three-quark system of a given parity $P$ and annihilates
a three-antiquark system of the {\em same} parity $P$.  This means
that in the baryon propagator, a baryon of parity $P$ propagates forward in
time while an antibaryon of parity $P$ propagates backwards in time.  
If we apply charge conjugation to our baryon operators, we 
arrive at the `backward propagating' partners mentioned in 
Chapter~\ref{chap:overview}.

Unlike
boson fields, a fermion and its antifermion have opposite intrinsic parities,
so that the antibaryon propagating backwards in time is the antiparticle of
the \textit{parity partner} of the baryon propagating forwards in time.  Since
chiral symmetry (vanishing quark mass) is spontaneously broken in QCD, the 
masses of opposite-parity partners may differ.
The forward propagating baryon will have a mass different from that of
the antibaryon propagating backwards in time.  If the even- and odd-parity baryon
operators are carefully designed with respect to one another, it is possible to
arrange a definite relationship between the correlation matrix elements of
one parity for $\tau>0$ and the opposite-parity matrix elements
for $\tau<0$, allowing us to increase our statistics.  

We can get a useful relation between the 
correlation matrix elements $C_{ab}(\tau)$ and $C_{ab}(T-\tau)$ if we
utilize the
projection procedure in Chapter~\ref{chap:construction} for our 
{\em even}-parity operators, and
 utilize charge conjugation 
to construct our {\em odd}-parity operators. 
Consider the correlation matrix element
of two even-parity operators for $\tau\geq 0$.  Suppressing irrep, displacement, and
flavor indices,
one sees that invariance under charge conjugation implies that
\begin{eqnarray*}
 C_{ab}(\tau) 
&=& c^{(i)*}_{\alpha\beta\gamma}c^{(j)}_{\overline{\alpha}
\overline{\beta}\overline{\gamma}}
\langle 0\vert\ B_{\alpha\beta\gamma}(\tau)
\overline{B}_{\overline{\alpha}
\overline{\beta}\overline{\gamma}}(0)
\ \vert 0\rangle,\\
&=& c^{(i)*}_{\alpha\beta\gamma}c^{(j)}_{\overline{\alpha}
\overline{\beta}\overline{\gamma}}
\langle 0\vert\ {\cal C}^\dagger {\cal C}B_{\alpha\beta\gamma}(\tau)
{\cal C}^\dagger {\cal C}\overline{B}_{\overline{\alpha}
\overline{\beta}\overline{\gamma}}(0)
{\cal C}^\dagger {\cal C}\ \vert 0\rangle,\\
&=& c^{(i)*}_{\alpha\beta\gamma}c^{(j)}_{\overline{\alpha}
\overline{\beta}\overline{\gamma}}
\langle 0\vert\ \overline{B}_{\alpha^\prime\beta^\prime\gamma^\prime}(\tau)
B_{\overline{\alpha}^\prime
\overline{\beta}^\prime\overline{\gamma}^\prime}(0)
\ \vert 0\rangle\\
&&\times \gamma^2_{\alpha^\prime\alpha}\gamma^2_{\beta^\prime\beta}
\gamma^2_{\gamma^\prime\gamma}\gamma^2_{\overline{\alpha}^\prime\overline{\alpha}}
\gamma^2_{\overline{\beta}^\prime\overline{\beta}}
\gamma^2_{\overline{\gamma}^\prime\overline{\gamma}},\\
&=& c^{(i)*}_{\alpha\beta\gamma}c^{(j)}_{\overline{\alpha}
\overline{\beta}\overline{\gamma}}
\langle 0\vert\ \overline{B}_{\alpha^\prime\beta^\prime\gamma^\prime}(0)
B_{\overline{\alpha}^\prime
\overline{\beta}^\prime\overline{\gamma}^\prime}(-\tau)
\ \vert 0\rangle\\ &&\times \gamma^2_{\alpha^\prime\alpha}\gamma^2_{\beta^\prime\beta}
\gamma^2_{\gamma^\prime\gamma}\gamma^2_{\overline{\alpha}^\prime\overline{\alpha}}
\gamma^2_{\overline{\beta}^\prime\overline{\beta}}
\gamma^2_{\overline{\gamma}^\prime\overline{\gamma}},
\end{eqnarray*}
using invariance under time translations of the above expectation value
and invariance of the vacuum under charge conjugation.  The last line above represents
the correlation of odd-parity operators propagating temporally backwards.  Hence,
for a given even-parity operator $\overline{B}^g_i(t)$, 
we can define an odd-parity
operator $\overline{B}^u_i(t)$ by rotating the three Dirac indices using the $\gamma_2$ matrix and
replacing the expansion coefficients by their complex conjugates
such that the correlation matrices of the even- and odd-parity operators
are related by
\begin{eqnarray}
C^{G_{1g}}_{ab}(\tau)&=&-C^{G_{1u}}_{ab}(-\tau)^\ast,\nonumber\\
C^{H_{g}}_{ab}(\tau)&=&-C^{H_{u}}_{ab}(-\tau)^\ast,\nonumber\\
C^{G_{2g}}_{ab}(\tau)&=&-C^{G_{2u}}_{ab}(-\tau)^\ast.\nonumber 
\label{eq:parityflip}
\end{eqnarray}
For a lattice of time-extent $T$ with anti-periodic 
temporal boundary conditions, this means that
\begin{equation}
C^{(g)}_{ab}(\tau)= C^{(u)}_{ba}(T-\tau),
\end{equation}
and similarly for the other irreps.
This allows us to appropriately average over forward and backward temporal
propagations for increased statistics.

\subsection{Improved estimators}
\label{section:backward}
In the absence of any external applied fields, the energies of the 
baryons
do not depend on the row $\lambda$ of a given irrep $\Lambda$, so we can 
increase statistics by averaging over all rows:
$$C^{\Lambda F}_{ab}\equiv\frac{1}{n_\Lambda}\sum_{\lambda=1}^{n_\Lambda}\langle 
B^{\Lambda\lambda F}_a(\tau)\overline{B}^{\Lambda\lambda F}_b(0)\rangle.$$

Because $\psi$ and $\overline{\psi}$ are independent fields in the functional 
integral, the correlation matrix
for a single configuration will not be Hermitian due to statistical
fluctuations inherent in the Monte Carlo estimate.  
Thus, we can increase our 
statistics by taking the average:
$$C^{(g)}_{ab}(\tau)\to\frac{1}{2}(C^{(g)}_{ab}(\tau)+C^{(g)*}_{ba}(\tau)).$$

In addition, we can use the backward propagating states in our 
correlation matrix to improve the statistics of the opposite-parity correlation
matrix by exploiting the relation
$$C^{(g)}_{ab}(\tau)=C^{(u)}_{ba}(T-\tau).$$

Putting it together gives us the improved estimator for our correlation matrix
elements:
\begin{eqnarray}
C_{ab}^{(g)}(\tau)\to\frac{1}{4}\left[C^{(g)}_{ab}(\tau)+C^{(u)}_{ba}(T-\tau)
+C_{ba}^{(g)*}(\tau)+C^{(u)*}_{ab}(T-\tau)\right],\\
C_{ab}^{(u)}(\tau)\to\frac{1}{4}\left[C^{(u)}_{ab}(\tau)+C^{(g)}_{ba}(T-\tau)
+C^{(u)*}_{ba}(\tau)+C^{(g)*}_{ab}(T-\tau)\right].
\label{eqn:reverse_avg}
\end{eqnarray}

It is important to note that although we have increased the precision of our
estimates by averaging, we have not increased the number of configurations
produced by the Monte Carlo method.  Thus, when estimating the jackknife errors
according to the method presented in Chapter~\ref{chap:monte_carlo}, we must
form our $n^{th}$ jackknife sample {\em before} calculating the improved
estimators.  
Specifically, we do {\em not} treat the different terms in the improved 
estimator (Eqn.~\ref{eqn:reverse_avg}) as additional configurations which
can be excluded to form a jackknife sample.

\section{Three-quark propagators}
Having constructed our baryon operators, we now have correlation
matrix elements which look like:
\begin{eqnarray*}
C^{\Lambda F}_{ab}(\tau)&=&\frac{1}{n_\Lambda}\sum_{\lambda=1}^{n_\Lambda}
\langle B^{\Lambda\lambda F}_a(\tau)\overline{B}^{\Lambda\lambda F}_b(0)
\rangle\\
 &=&\frac{1}{n_\Lambda}\sum_{\lambda=1}^{n_\Lambda}\sum_{k,l=1}^{M_B}
 c_{ak}^{\Lambda\lambda  F*} c_{bl}^{\Lambda\lambda F}
 \langle B_k^F(\tau)
 \overline{B}_l^F(0) \rangle.
\end{eqnarray*}
where the indices $a,b,k,l$ range from $1,\cdots,M_B$, and encapsulate all of 
the information about 
the type of operator (single-site, singly-displaced, doubly-displaced-L,
doubly-displaced-I, or triply-displaced-T), and the particular linear 
combination over Dirac indices used to make an operator which transforms
irreducibly according to row $\lambda$ of the $\Lambda$ irrep ($G_{1g}$, $H_g$,
$G_{2g}$, $G_{1u}$, $H_u$, or $G_{2u}$)
 in the flavor channel $F$ ($\Delta$, $\Sigma$,
$N$, $\Xi$, $\Lambda$, or $\Omega$).

Since the number of elemental operators is large and the quark propagators
are rather expensive to compute, it is very important to use symmetry
to reduce the number of quark-propagator sources.  Given the cyclic property
of the trace and the unitarity of the symmetry transformation operators, we
know that
\begin{eqnarray*}
 \langle B_k^F(\tau)
 \overline{B}_l^F(0)\rangle&=&  \langle U_R B_k^F(\tau) U_R^\dagger
 U_R\overline{B}_l^F(0) U_R^\dagger\rangle,\nonumber\\
&=&  \displaystyle\sum_{k^\prime,l^\prime=1}^{M_B}
  \!\! W_{k^\prime k}(R) W_{l^\prime l}(R)^\ast
 \langle B_{k^\prime}^F(\tau) \overline{B}_{l^\prime}^F(0)
 \rangle,
\end{eqnarray*}
for any group element $R$ of $O_h$.  Hence, for each source 
$\overline{B}_l^F(0)$, we can choose a group element $R_l$ such that we
minimize the total number of source elemental operators which must be considered. 
For example, consider the singly-displaced operators.  We can choose
an $R_l$ such that the displaced quark in the source is always displaced in
the $+z$ direction.  Similarly, a group element $R_l$ can always be chosen to
rotate each of the other types of operators into a specific orientation.
The canonical source orientations for each type of operator are given
in Table~\ref{table:orientations}

\begin{table}
\centering
\begin{tabular}{l|r}
Operator Type & Displaced Quark Direction(s)\\
\hline
Singly-Displaced   & $+z$\\
Doubly-Displaced-I & $+z$, $-z$\\
Doubly-Displaced-L & $+y$, $+z$\\
Triply-Displaced-T & $+z$, $+y$, $-z$
\end{tabular}
\caption{To minimize the number of sources, thereby reducing the number of
three-quark propagators needed, we rotate our source and sink operators
such that the displaced quarks at the source are always in the same
canonical positions.}
\label{table:orientations}
\end{table}

To make zero momentum operators, we averaged over all spatial
reference sites $\vec{x}$ at both the sink and the source.  However, we know
that correlation functions of operators having different momenta will
vanish (as expectation values), and thus we need only perform the spatial
average at the sink.  The source operator, evaluated at only one reference 
site, will no longer be translationally invariant, but the zero momentum
contribution will be the only one which gives a non-canceling contribution
to the correlation function.  This increases the noise in our
Monte Carlo estimates, but dramatically reduces the number of quark
propagators needed.

The coefficients $c^{\Lambda\lambda F}_{ij}$ in the baryon operators
involve only the, Dirac spin components and the quark displacement
directions and are independent of the color indices and spatial sites.  Thus,
in calculating the baryon correlation functions, it is convenient to first calculate
gauge-invariant three-quark propagators in which all summations over color
indices and spatial sites have been done.  A three-quark propagator is 
defined by
\begin{eqnarray}
&&\widetilde{G}^{(ABC)(p\overline{p})}_{(\alpha i\vert\overline{\alpha}\overline{i})
(\beta j\vert\overline{\beta}\overline{j})
(\gamma k\vert\overline{\gamma}\overline{k})}(t)\nonumber\\
&=&\sum_{\vec{x}}
\varepsilon_{abc}\,\varepsilon_{\overline{a}\overline{b}\overline{c}}
\ \widetilde{G}^{(A)}_{a\alpha i p\vert\overline{a}
   \overline{\alpha}\overline{i}\overline{p}}(\vec{x},t\vert\vec{x}_0,0)\nonumber\\
&\times&\widetilde{G}^{(B)}_{b\beta j p\vert\overline{b}
   \overline{\beta}\overline{j}\overline{p}}(\vec{x},t\vert\vec{x}_0,0)
\ \widetilde{G}^{(C)}_{c\gamma k p\vert\overline{c}
   \overline{\gamma}\overline{k}\overline{p}}(\vec{x},t\vert\vec{x}_0,0),
\quad \label{eq:threequarkprop}
\end{eqnarray}
where $\widetilde{G}^{(A)}_{a\alpha i p\vert\overline{a}
   \overline{\alpha}\overline{i}\overline{p}}(\vec{x},t\vert\vec{x}_0,0)$
denotes the propagator for a single smeared quark field of flavor $A$ from
source site $\vec{x}_0$ at time $t=0$ to sink site $\vec{x}$ at time $t$.
At the sink, $a$ denotes color, $\alpha$ is the Dirac spin index, $i$ is
the displacement direction, and $p$ is the displacement length, and
similarly at the source for $\overline{a},\overline{\alpha},\overline{i},$
and $\overline{p}$, respectively.  Notice that the three-quark propagator is
symmetric under interchange of all indices associated with the same flavor.
As usual, translation invariance is invoked at the source so that summation
over spatial sites is done only at the sink.  These three-quark propagators
are computed for all possible values of the six Dirac spin indices.

Each baryon correlation function is simply a linear superposition of elements of the
three-quark propagators.  These superposition coefficients are calculated
as follows: first, the baryon operators at the source and sink are
expressed in terms of the elemental operators; next, Wick's theorem is
applied to express the correlation function as a large sum of three-quark propagator
components;  finally, symmetry operations are applied to minimize the number
of source orientations, and the results are averaged over the rows of the 
representations.  C++ code was written by Colin Morningstar to perform these 
computations,
and the resulting superposition coefficients are stored in computer files
which are subsequently used as input to the Monte Carlo runs.

Wick's theorem is an important part of expressing the baryon correlation functions
in terms of the three-quark propagators.  To simplify the notation in the
following, let the indices $\mu,\nu,\tau$ each represent a Dirac spin
index and a displacement direction, and suppress the displacement lengths.
Define
$   \overline{c}^{(i)}_{\mu\nu\tau} 
 = c^{(i)\ast}_{\mu^\prime\nu^\prime\tau^\prime}
 \gamma^4_{\mu\mu^\prime} \gamma^4_{\nu\nu^\prime} \gamma^4_{\tau\tau^\prime},$
then the elements of the baryon correlation matrix in the $\Delta^{++}$
channel are given in terms of three-quark propagator components (before
source-minimizing rotations) by
\begin{eqnarray}
 C^{(\Delta)}_{ij}(t) &=& c^{(i)}_{\mu\nu\tau}
\, \overline{c}^{(j)}_{\overline{\mu}\overline{\nu}\overline{\tau}}
\Bigl\{
\,\widetilde{G}^{(uuu)}_{(\tau  \vert \overline{\mu})
(\nu   \vert \overline{\nu})
(\mu   \vert \overline{\tau})}(t) \nonumber\\
&+&\,\widetilde{G}^{(uuu)}_{ (\tau  \vert \overline{\mu})
(\nu   \vert \overline{\tau})
(\mu   \vert \overline{\nu})}(t) 
+\,\widetilde{G}^{(uuu)}_{ (\tau  \vert \overline{\nu})
(\nu   \vert \overline{\mu})
(\mu   \vert \overline{\tau})}(t)\nonumber\\
&+&\,\widetilde{G}^{(uuu)}_{ (\tau  \vert \overline{\nu})
(\nu   \vert \overline{\tau})
(\mu   \vert \overline{\mu})}(t)
+\,\widetilde{G}^{(uuu)}_{ (\tau  \vert \overline{\tau})
(\nu   \vert \overline{\nu})
(\mu   \vert \overline{\mu})}(t) \nonumber\\
&+&\,\widetilde{G}^{(uuu)}_{ (\tau  \vert \overline{\tau})
(\nu   \vert \overline{\mu})
(\mu   \vert \overline{\nu})}(t)
\Bigr\}.
\end{eqnarray}
The $N^+$ correlation functions are expressed in terms of components of
three-quark propagators by
\begin{eqnarray}
 C^{(N)}_{ij}(t) &=& c^{(i)}_{\mu\nu\tau}
\, \overline{c}^{(j)}_{\overline{\mu}\overline{\nu}\overline{\tau}}
\Bigl\{
\widetilde{G}^{(uud)}_{ (\mu  \vert \overline{\mu})
(\nu  \vert \overline{\nu})
(\tau \vert \overline{\tau})} \nonumber\\
&+&\widetilde{G}^{(uud)}_{ (\mu  \vert \overline{\nu})
(\nu  \vert \overline{\mu})
(\tau \vert \overline{\tau})} 
-\widetilde{G}^{(uud)}_{ (\mu  \vert \overline{\tau})
(\nu  \vert \overline{\nu})
(\tau \vert \overline{\mu})} \nonumber\\
&-&\widetilde{G}^{(uud)}_{ (\mu  \vert \overline{\nu})
 (\nu  \vert \overline{\tau})
 (\tau \vert \overline{\mu})}
-\widetilde{G}^{(uud)}_{ (\nu  \vert \overline{\nu}) 
 (\tau \vert \overline{\mu})
 (\mu  \vert \overline{\tau})}\nonumber\\
&-&\widetilde{G}^{(uud)}_{ (\nu  \vert \overline{\mu})
 (\tau \vert \overline{\nu}) 
 (\mu  \vert \overline{\tau})}
+\widetilde{G}^{(uud)}_{ (\tau \vert \overline{\tau})
 (\nu  \vert \overline{\nu})
 (\mu  \vert \overline{\mu})}\nonumber\\
&+&\widetilde{G}^{(uud)}_{ (\tau \vert \overline{\nu})
 (\nu  \vert \overline{\tau})
 (\mu  \vert \overline{\mu})} 
\Bigr\},
\end{eqnarray}
and for the $\Sigma^+$ and $\Lambda^0$ channels, one finds
\begin{eqnarray}
 C^{(\Sigma)}_{ij}(t) &=& c^{(i)}_{\mu\nu\tau}
\, \overline{c}^{(j)}_{\overline{\mu}\overline{\nu}\overline{\tau}}
\Bigl\{
\widetilde{G}^{(uus)}_{ (\mu  \vert \overline{\mu})
(\nu  \vert \overline{\nu})
(\tau \vert \overline{\tau})}(t) \nonumber\\
&+&\widetilde{G}^{(uus)}_{ (\mu  \vert \overline{\nu})
(\nu  \vert \overline{\mu})
(\tau \vert \overline{\tau})}(t)\Bigr\},\\
 C^{(\Lambda)}_{ij}(t) &=& c^{(i)}_{\mu\nu\tau}
\, \overline{c}^{(j)}_{\overline{\mu}\overline{\nu}\overline{\tau}}
\Bigl\{
\widetilde{G}^{(uds)}_{ (\mu  \vert \overline{\mu})
( \nu  \vert \overline{\nu})
( \tau \vert \overline{\tau})} \nonumber\\
&-&\widetilde{G}^{(uds)}_{ (\mu  \vert \overline{\nu})
( \nu  \vert \overline{\mu})
( \tau \vert \overline{\tau})}
-\widetilde{G}^{(uds)}_{ (\nu  \vert \overline{\mu})
( \mu  \vert \overline{\nu})
( \tau \vert \overline{\tau})} \nonumber\\
&+&\widetilde{G}^{(uds)}_{ (\nu  \vert \overline{\nu})
( \mu  \vert \overline{\mu})
( \tau \vert \overline{\tau})}
\Bigr\}.
\end{eqnarray}

\subsection{Smeared, covariantly-displaced quark propagators}
\label{sec:elprop}
We are now ready to compute the
quark propagators for smeared and displaced sinks and sources.  Recall that
we are computing expectation values via path integrals, such as
\begin{eqnarray*}
 \langle 0\vert\ T f(\overline{\psi},\psi,U) \ \vert 0\rangle
 &=& \frac{\displaystyle\int {\cal D}(\overline{\psi},\psi,U)
\ f(\overline{\psi},\psi,U) \exp\left(-\overline{\psi}Q[U]\psi - S_G[U]\right)
}{\displaystyle\int {\cal D}(\overline{\psi},\psi,U)
 \  \exp\left(-\overline{\psi}Q[U]\psi - S_G[U]\right)}\\
 &=& \frac{\displaystyle\int {\cal D}U
\ h(\overline{\psi},\psi,U)\ \det Q[U]
 \ \exp\left(- S_G[U]\right)}{\displaystyle\int {\cal D}U
 \ \det Q[U]\ \exp\left(- S_G[U]\right)}
\end{eqnarray*}
where $U$ represents the gluon field (link variables) and $S_G[U]$ is
the pure gauge action.  The remaining path integration over the link
variables is carried out by the Monte Carlo method.  Results in the
so-called {\em quenched approximation} are obtained if $\det Q[U]$
is ignored in the updating and measurement process.  Full QCD simulations
require the incorporation of the computationally-expensive fermion
determinant $\det Q[U]$ in the updating.

Now define
\begin{eqnarray*}
 &&\displaystyle\int {\cal D}(\overline{\psi},\psi)
 \  (\tilde{D}_i^{(p)}\tilde{\psi})_{Aa\alpha}(\vec{x},t)
 \ \  ( \tilde{\overline{\psi}}\overleftarrow{\tilde{D}_{\overline{i}}^{
   (\overline{p})}}
  )_{\overline{A} \overline{a} \overline{\alpha}}(\vec{x}_0,0)
 \ \ \exp\left(-\overline{\psi}Q[U]\psi\right)\\
&&\hspace*{1cm}\equiv \det Q[U]\  \ \delta_{A\overline{A}}
  \ \ \tilde{G}^{(A)}_{a\alpha i p\vert\overline{a}
   \overline{\alpha}\overline{i}\overline{p}}(\vec{x},t\vert\vec{x}_0,0).
\end{eqnarray*}
To see how to compute the quark propagator 
$\tilde{G}^{(A)}_{a\alpha i p\vert\overline{a}
   \overline{\alpha}\overline{i}\overline{p}}(\vec{x},t\vert\vec{x}_0,0)$,
let us proceed in steps of increasing complexity.  First, we shall
focus on the application of the three-dimensional Laplacian.  If we can
understand how to compute the quark propagators with the three-dimensional
covariant Laplacian\ acting on both the source and sink fields, then it
is straightforward to understand how to compute the smeared-smeared
propagator.  Finally, the $p$-link displacements will be included.

First, consider the application of a single Laplacian on both the
source and sink fields:
\begin{eqnarray}
 \left(\Delta\psi\right)_a(x) &=& \sum_{j=\pm 1,\pm 2,\pm 3}
 \biggl\{ U_j^{ab}(x)\ \psi_b(x\!+\!\hat{j})-\psi_a(x)\biggr\},\\
 \left(\overline{\psi}\overleftarrow{\Delta}\right)_a(x) &=&
  \sum_{j=\pm 1,\pm 2,\pm 3}\biggr\{
  \overline{\psi}_b(x\!+\!\hat{j})\ U^{\dagger ba}_j(x)
  - \overline{\psi}_a(x)\biggr\},
\end{eqnarray}
where the indices $a,b$ are color indices.  To simplify the notation,
both flavor and spin indices are omitted, as well as the gauge-field 
smearing tildes.  Our goal now is to compute
\begin{equation} 
\int{\cal D}(\overline{\psi},\psi)\ \left(\Delta\psi\right)_a(x)
 \left(\overline{\psi}\overleftarrow{\Delta}\right)_b(y)
 \ \exp\left(-\overline{\psi}\ Q[U]\ \psi\right).
\end{equation}
First, write the Laplacian operators in matrix form:
\begin{eqnarray}
 \Delta_{ab}(x\vert y) &=& \sum_{j=\pm 1,\pm 2,\pm 3}
 \biggl\{ U_j^{ab}(x)\ \delta(x\!+\!\hat{j},y)
   -\delta^{ab}\delta(x,y)\biggr\},\nonumber\\
&=& \sum_{j=1}^3
 \biggl\{ U_j^{ab}(x)\ \delta(x\!+\!\hat{j},y)
         +U_j^{\dagger ab}(x\!-\!\hat{j})\ \delta(x\!-\!\hat{j},y)
   -2\delta^{ab}\delta(x,y)\biggr\},\nonumber\\
&=& \sum_{j=1}^3
 \biggl\{ U_j^{ab}(x)\ \delta(x\!+\!\hat{j},y)
         +U_j^{ba}(y)^\ast\ \delta(x\!-\!\hat{j},y) 
   -2\delta^{ab}\delta(x,y)\biggr\},
\label{eq:rightDelta}\\
 \overleftarrow{\Delta}_{ab}(x\vert y) &=&
  \sum_{j=\pm 1,\pm 2,\pm 3}\biggr\{
   U^{\dagger ab}_j(y)\ \delta(x,y\!+\!\hat{j})
  - \delta^{ab}\delta(x,y)\biggr\},\nonumber\\
&=& \sum_{j=1}^3\biggr\{
   U^{\dagger ab}_j(y)\ \delta(x,y\!+\!\hat{j})
   +U^{ab}_{j}(y\!-\!\hat{j})\ \delta(x,y\!-\!\hat{j})
  - 2\delta^{ab}\delta(x,y)\biggr\},\nonumber\\
&=& \sum_{j=1}^3\biggr\{
   U^{ba}_j(y)^\ast\ \delta(x\!-\!\hat{j},y)
   +U^{ab}_{j}(x)\ \delta(x\!+\!\hat{j},y)
  - 2\delta^{ab}\delta(x,y)\biggr\},\nonumber\\
&=& \Delta_{ab}(x\vert y).
\label{eq:leftDelta}
\end{eqnarray}
Grassmann integration quickly gives us the quark propagator:
\begin{eqnarray}
&&\int{\cal D}(\overline{\psi},\psi)\ \left(\Delta\psi\right)_a(x)
 \left(\overline{\psi}\overleftarrow{\Delta}\right)_b(y)
 \ \exp\left(-\overline{\psi}\ Q[U]\ \psi\right)\\
&=& \int{\cal D}(\overline{\psi},\psi)\ 
  \Delta_{ac}(x\vert z)\ \psi_c(z)
 \ \overline{\psi}_d(w)\ \Delta_{db}(w\vert y)
 \ \exp\left(-\overline{\psi}\ Q[U]\ \psi\right)\\
&=& \det Q[U]\ \Delta_{ac}(x\vert z)
 \ Q^{-1}_{cd}(z\vert w) \ \Delta_{db}(w\vert y).
\end{eqnarray}
A straightforward application of the above formula would require
the computation of the inverse of the Dirac matrix $Q$ for several
local sources $d$.  For higher powers of the Laplacian which may
arise in usual quark smearings, the number of values of $d$ needed
may be quite large. Due to the computational expense of this inversion,
it is better to proceed as follows.
The inverse of the fermion Dirac matrix is computed by solving
the linear system of equations
\begin{equation} Q_{fc}(v\vert z)\ Q^{-1}_{cd}(z\vert w)
 =\delta_{fd}\delta(v,w).
\end{equation}
Apply $\overleftarrow{\Delta}$ from the right onto both sides of this
equation to obtain:
\begin{equation} Q_{fc}(v\vert z)\ Q^{-1}_{cd}(z\vert w)
 \ \Delta_{db}(w\vert y)
 =\delta_{fd}\delta(v,w)\ \Delta_{db}(w\vert y).
\end{equation}
Write
\begin{eqnarray}
 P_{cb}(z\vert y)&=&Q^{-1}_{cd}(z\vert w)
 \ \Delta_{db}(w\vert y),\\
 R_{fb}(v\vert y) &=&\delta_{fd}\delta(v,w)
 \ \Delta_{db}(w\vert y),\label{eq:sourceapply}
\end{eqnarray}
then the above equation becomes
\begin{equation}
 Q_{fc}(v\vert z)\ P_{cb}(z\vert y) = R_{fb}(v\vert y).
\end{equation}
This is a linear system of equations.  We can solve for $P_{cb}(z\vert y)$
{\em for a single fixed value of $b$ and $y$} by some variant of the conjugate
gradient method.  This is then repeated for the different colors $b$ and
Dirac spin components $\alpha$, but usually only one site $y$ is involved.
The final propagator is obtained using
\begin{equation}
   G_{ab}(x\vert y) = \Delta_{ac}(x\vert z)\ P_{cb}(z\vert y).
\label{eq:sinkapply}
\end{equation}
Eq.~(\ref{eq:sourceapply}) tells us how we must apply the Laplacian
for the source, and Eq.~(\ref{eq:sinkapply}) tells us how we must apply
the Laplacian for the sink.

The generalization of these results to higher powers of the Laplacian
is straightforward.  One starts at the source by forming the right-hand side
\begin{eqnarray}
  R_{ab}(x\vert y) &=& \delta_{ac_1}\delta(x,z_1)
  \ \Delta_{c_1c_2}(z_1\vert z_2)
  \ \Delta_{c_2c_3}(z_2\vert z_3)
  \dots
  \ \Delta_{c_nb}(z_n\vert y), \nonumber\\
 &=& 
   \Delta_{ac_n}(x\vert z_n)
  \dots
  \ \Delta_{c_3c_2}(z_3\vert z_2)
  \ \Delta_{c_2c_1}(z_2\vert z_1) 
 \ \delta_{c_1 b}\delta(z_1,y),
\label{eq:sourcesmear}
\end{eqnarray}
solves the linear system of equations
\begin{equation}
  Q_{ac}(x\vert z)\ P_{cb}(z\vert y) = R_{ab}(x\vert y),
\end{equation}
then applies the Laplacians at the sink to the result:
\begin{equation}
  G_{ab}(x\vert y) =
  \Delta_{ac_n}(x\vert z_n)  \dots
  \ \Delta_{c_3c_2}(z_3\vert z_2)
  \ \Delta_{c_2c_1}(z_2\vert z_1) \ P_{c_1b}(z_1\vert y).
\label{eq:sinksmear}
\end{equation}
Using Eq.~(\ref{eq:rightDelta}), the application of the Laplacians
at the sink in Eq.~(\ref{eq:sinksmear}) is straightforward.  For
fixed $b,y$, $P_{c_1b}(z_1\vert y)$ can be viewed as a known vector
(since it is a column of a matrix), then the applications of the
Laplacians are equivalent to successive matrix-vector multiplications.
Similarly, using Eq.~(\ref{eq:leftDelta}) in Eq.~(\ref{eq:sourcesmear}) at
the source is straightforward, too.

The computation of the propagator for smeared fields
\begin{equation} 
\int{\cal D}(\overline{\psi},\psi)\ \tilde{\psi}_a(x)
 \tilde{\overline{\psi}}_b(y)
 \ \exp\left(-\overline{\psi}\ Q[U]\ \psi\right)
\end{equation}
is now obvious. Let the smearing kernel be
\begin{eqnarray}
  F_{ab}(x\vert y) &=& \delta_{ab}\delta(x,y)
  +\xi\ \tilde{\Delta}_{ab}(x\vert y),\\
&=& (1-6\xi)\delta^{ab}\delta(x,y) +\xi \sum_{j=1}^3
 \biggl\{ \tilde{U}_j^{ab}(x)\ \delta(x\!+\!\hat{j},y)
         +\tilde{U}_j^{ba}(y)^\ast\ \delta(x\!-\!\hat{j},y) 
   \biggr\}. \label{eq:smearkernel}
\end{eqnarray}
One first forms the source field using
\begin{eqnarray}
  R_{ab}(x\vert y)&=& F^{n_\xi}_{ac}(x\vert z)
\ \delta_{c b}\delta(z,y),\\
&=&F_{ac_{n_\xi}}(x\vert z_{n_\xi})
  \dots
  \ F_{c_3c_2}(z_3\vert z_2)
  \ F_{c_2c_1}(z_2\vert z_1) 
 \ \delta_{c_1 b}\delta(z_1,y),
\label{eq:source_smear}
\end{eqnarray}
solves the linear system of equations
\begin{equation}
  Q_{ac}(x\vert z)\ P_{cb}(z\vert y) = R_{ab}(x\vert y),
\end{equation}
then applies the smearing at the sink to the result:
\begin{eqnarray}
  G_{ab}(x\vert y) &=&
  F^{n_\xi}_{ac}(x\vert z) \ P_{cb}(z\vert y),\\
&=&  F_{ac_{n_\xi}}(x\vert z_{n_\xi})  \dots
  \ F_{c_3c_2}(z_3\vert z_2)
  \ F_{c_2c_1}(z_2\vert z_1) \ P_{c_1b}(z_1\vert y).
\label{eq:sink_smear}
\end{eqnarray}

Lastly, we need to include the $p$-link covariant displacement operators:
\begin{equation} 
\int{\cal D}(\overline{\psi},\psi)
 \ \left(\tilde{D}_j^{(p)}\tilde{\psi}\right)_a(x)
 \left(\tilde{\overline{\psi}}\overleftarrow{D_k^{(p)}}\right)_b(y)
 \ \exp\left(-\overline{\psi}\ Q[U]\ \psi\right).
\end{equation}
In matrix form, these operators are given by
\begin{eqnarray}
 \left(\tilde{D}_j^{(p)}\right)_{ab}(x\vert y)
 &=& \tilde{U}^{ac_2}_j(x)\ \tilde{U}^{c_2 c_3}_j(x\!+\!\hat{j})
  \dots\tilde{U}^{c_pb}_j(x\!+\!(p\!-\!1)\hat{j})
 \ \delta(x\!+\!p\hat{j},y),\label{eq:displace}\\
 \left(\overleftarrow{\tilde{D}_j^{(p)}}\right)_{ab}(x\vert y)
 &=& \tilde{U}^{\dagger ac_p}_j(y\!+\!(p\!-\!1)\hat{j})
  \dots\tilde{U}^{\dagger c_3 c_2}_j(y\!+\!\hat{j})
  \ \tilde{U}^{\dagger c_2b}_j(y)
 \ \delta(x,y\!+\!p\hat{j}),\\
 &=& \left(\tilde{U}^{bc_2}_j(y)
  \ \tilde{U}^{c_2 c_3}_j(y\!+\!\hat{j})\dots
  \tilde{U}^{c_pa}_j(y\!+\!(p\!-\!1)\hat{j})
  \ \delta(y\!+\!p\hat{j},x)\right)^\ast,\nonumber\\
 &=&\left(\tilde{D}_j^{(p)}\right)_{ba}(y\vert x)^\ast.\label{eq:displace2}
\end{eqnarray}
Unlike the Laplacian, this is {\em not} a Hermitian operator.
The source function is now given by
\begin{equation}
 R_{ab}(x\vert y) = F_{ac}^{n_\xi}(x\vert z)
  \ \left(\overleftarrow{\tilde{D}_k^{(p)}}\right)_{cd}(z\vert w)
  \ \delta_{db}\delta(w,y).
\end{equation}
Since the $\overleftarrow{\Delta}$ matrix may also be written as
\begin{eqnarray}
 \left(\overleftarrow{\tilde{D}_j^{(p)}}\right)_{ab}(x\vert y)
 &=& \tilde{U}^{\dagger ac_p}_j(y\!+\!(p\!-\!1)\hat{j})
  \dots\tilde{U}^{\dagger c_3 c_2}_j(y\!+\!\hat{j})
  \ \tilde{U}^{\dagger c_2b}_j(y)
 \ \delta(x,y\!+\!p\hat{j}),\nonumber\\
 &=& \tilde{U}^{ ac_p}_{-j}(y\!+\!p\hat{j})
  \dots\tilde{U}^{ c_3 c_2}_{-j}(y\!+\!2\hat{j})
  \ \tilde{U}^{ c_2b}_{-j}(y\!+\!\hat{j})
 \ \delta(x,y\!+\!p\hat{j}),\nonumber\\
 &=& \tilde{U}^{ ac_p}_{-j}(x)
  \dots  \ \tilde{U}^{ c_2b}_{-j}(x\!-\!(p\!-\!1)\hat{j})
 \ \delta(x\!-\!p\hat{j},y),\nonumber\\
 &=& \left(\tilde{D}^{(p)}_{-j}\right)_{ab}(x\vert y),
\end{eqnarray}
the source function may be better expressed as
\begin{equation}
 R_{ab}(x\vert y) = F_{ac}^{n_\xi}(x\vert z)
  \ \left(\tilde{D}_{-k}^{(p)}\right)_{cd}(z\vert w)
  \ \delta_{db}\delta(w,y).\label{eq:source}
\end{equation}
The two key points to note about Eq.~(\ref{eq:source}) are that
(a) the displacement is applied {\em first} to the $\delta$-function,
and then the smearing is applied, and (b) a displacement in the $-\hat{k}$
direction must be made.

{\bf In summary}, to compute
\begin{equation} 
 \int{\cal D}(\overline{\psi},\psi)
 \ \left(\tilde{D}_j^{(p)}\tilde{\psi}\right)_a(x)
 \left(\tilde{\overline{\psi}}\overleftarrow{D_k^{(p)}}\right)_b(y)
 \ \exp\left(-\overline{\psi}\ Q[U]\ \psi\right)
=G_{ab}(x\vert y)\ \det Q[U] ,
\end{equation}
one first forms the following source field
\begin{eqnarray}
 R_{ab}(x\vert y) &=& F_{ac}^{n_\xi}(x\vert z)
  \ \left(\tilde{D}_{-k}^{(p)}\right)_{cd}(z\vert w)
  \ \delta_{db}\delta(w,y),\\
&=&F_{ac_{n_\xi}}(x\vert z_{n_\xi})
  \dots
  \ F_{c_3c_2}(z_3\vert z_2)
  \ F_{c_2c_1}(z_2\vert z_1) 
\ \left(\tilde{D}_{-k}^{(p)}\right)_{c_1d}(z_1\vert w)
 \ \delta_{db}\delta(w,y),
\label{eq:source_smear_displace}
\end{eqnarray}
solves the linear system of equations
\begin{equation}
  Q_{ac}(x\vert z)\ P_{cb}(z\vert y) = R_{ab}(x\vert y),
\end{equation}
then applies the smearing and displacement at the sink to the result:
\begin{eqnarray}
  G_{ab}(x\vert y) &=&
 \left(\tilde{D}_j^{(p)}\right)_{ad}(x\vert w)
  F^{n_\xi}_{dc}(w\vert z) \ P_{cb}(z\vert y),\\
&=&  \left(\tilde{D}_j^{(p)}\right)_{ad}(x\vert w)
 \ F_{dc_{n_\xi}}(w\vert z_{n_\xi})  \dots
  \ F_{c_3c_2}(z_3\vert z_2)
  \ F_{c_2c_1}(z_2\vert z_1) \ P_{c_1b}(z_1\vert y),
\label{eq:sink_smear_displace}
\end{eqnarray}
where the smearing kernel is
\begin{eqnarray}
  F_{ab}(x\vert y) &=& \delta_{ab}\delta(x,y)
  +\xi\ \tilde{\Delta}_{ab}(x\vert y),\\
&=& (1-6\xi)\delta^{ab}\delta(x,y) +\xi \sum_{j=1}^3
 \biggl\{ \tilde{U}_j^{ab}(x)\ \delta(x\!+\!\hat{j},y)
         +\tilde{U}_j^{ba}(y)^\ast\ \delta(x\!-\!\hat{j},y) 
   \biggr\}.
\end{eqnarray}
In short,\\[3mm]
\framebox{\begin{minipage}{14cm}
\begin{itemize}
\item
 at the sink: smear then displace in desired direction,
\item
 at the source: start with point source, displace in {\em opposite}
   direction, then smear.
\end{itemize}
\end{minipage}}

\subsection{Chroma walk-through}
Our generalized three-quark propagators were generated using the Chroma 
QCD library developed at JLab~\cite{edwards:qdp}. 
We first used Chroma to make four smeared, covariantly-displaced single-quark 
sources:
\begin{enumerate}
\item no displacement
\item $+y$-direction displacement
\item $-z$-direction displacement
\item $+z$-direction displacement
\end{enumerate}

Each source object was then propagated to all sink sites on the lattice,
resulting in a smeared, covariantly-displaced 
source and and unsmeared, undisplaced sink.  

It is in this step that
the computationally expensive quark matrix inversion $Q^{-1}$ is performed. 
This is done using the bi-conjugate gradient
algorithm~\cite{golub:matrix}.
The resulting `one-to-all' propagators take the form of $Q^{-1}_{x,0}$, 
requiring the inversion of only a single column of the quark matrix $Q[U]$.

 The sinks were then smeared and covariantly-displaced.
For each of the four input propagators, there were seven output propagators, one for each of the types of displacement: no displacement, $\pm x$, $\pm y$, $\pm z$.  These $28$ single-quark propagators were then combined, three at
a time, to make all of the needed generalized three-quark propagators.
These propagators were saved to disk for later use.

Combining the indices of the single quark propagators to form the three-quark
 propagators was one of the most computationally intensive steps.
This is because the $28$ single-quark propagators combine into a rich
variety of three quark propagators.  At the source, there can be one of five 
three-quark
elemental operators, one for each type (remembering that we have fixed a
canonical orientation).  At the sink, the number of orientations depends 
on the symmetry of the 
elemental three-quark operator:
\begin{itemize}
\item single-site (SD): 1 orientation,
\item singly-displaced (SD): 6 orientations,
\item doubly-displaced-I (DDI): 3 orientations,
\item doubly-displaced-L (DDL): 12 orientations,
\item triply-displaced-T (TDT): 12 orientations.
\end{itemize}
In addition, we must choose which source quark propagators to each sink
quark, and take into account that the up and down quarks are indistinguishable
(isospin symmetry: $m_u=m_d$).

For the correlation matrix involving a singly-displaced operator at the source 
($\overline{SD}$) and a singly-displaced operator at the sink, there is only
one orientation for the quarks at the source (by convention), there are
six ways to orient the sink quarks, and there are two ways to propagate
the displaced source quark: either to the displaced sink quark, or to one of the 
(indistinguishable) undisplaced quarks.

For example: the number of three-quark propagators required for a full 
correlation matrix run including only single-site and singly-displaced nucleon 
operators can be calculated as follows:

\begin{itemize}
\item $SS-\overline{SS} = 1 \times 1$ possibility +
\item $SS-\overline{SD} = 1 \times 1$ possibility +
\item $SD-\overline{SS} = 1 \times 6$ possibilities +
\item $SD-\overline{SD} = 1 \times 6 \times 2$ possibilities = 20 three-quark propagators
\end{itemize}

We restricted the scope of this project to the $N^+$ (nucleon) channel.  
Our final 
correlation matrix run used all five types of operators at the source and sink 
and required 671 three-quark propagators 
(see Table~\ref{table:qqq_prop_numbers}). 
These three-quark propagators were stored on disk, each file containing
one three-quark propagator measured on one configuration.
 We were able to reuse 175 of those 
three-quark propagators from our diagonal runs.
\begin{table}
\begin{center}
\begin{tabular}{c|ccccc}
      & $\overline{SS}$ & $\overline{SD}$ & $\overline{DDI}$ & $\overline{DDL}$ & $\overline{TDT}$\\
\hline
$SS$  & $1$  & $1$  & $1$  & $1$  & $1$   \\
$SD$  & $6$  & $12$ & $18$ & $18$ & $18$   \\
$DDI$ & $3$  & $9$  & $18$  & $18$  & $18$   \\
$DDL$ & $12$ & $36$ & $72$ & $72$ & $72$   \\
$TDT$ & $12$ & $36$ & $72$ & $72$ & $72$   
\end{tabular}
\end{center}
\label{table:qqq_prop_numbers}
\caption{The number of three-quark propagators needed for each part of the
full nucleon correlation matrix.  The total number of propagators needed 
is 671.  The diagonal correlation matrix elements required only 175 
propagators, which were reused for the full run.}
\end{table}

\subsection{Correlation matrix elements}
The C++  program we wrote to tie the three quark propagators together into
correlation matrix elements was based on the JLab \texttt{ttt} 
program\footnote{The program name
\texttt{ttt} stands for `tie them together.'}.

The program would first read in the pre-calculated coefficients.
For each configuration, or {\em bin}, the program would read in the 
three-quark propagator files for that configuration and combine them into the 
desired correlation matrix elements.

We designed the program to process multiple irreps and to
generate only the user-selected elements of a correlation matrix.  Early in the
operator tuning and selection process, we were only interested in the diagonal
elements of the correlation matrices (for effective mass functions).  
Later, we utilized
the off-diagonal elements of our correlation matrices to extract 
excited states.

\chapter{Quark field and gauge link smearing parameter tuning}
\label{chap:smearing}

The smearing method we use to smear our quark fields and gauge links was
developed in Chapter~\ref{chap:construction} and is summarized here:
\begin{equation}
\tilde{\psi}\equiv \tilde{S}\psi,\qquad 
\tilde{\overline{\psi}}\equiv  \tilde{S}^\dagger\overline{\psi}
\end{equation}
where
\begin{eqnarray}
\tilde{S}(\sigma,n_\sigma,\rho,n_\rho)&\equiv&\left(1+\frac{\sigma^2}{2n_\sigma}\tilde{\Delta}\right)^{n_\sigma}\\
\tilde{\Delta}_{x,y}&\equiv&\sum_{k=\pm1,\pm2,\pm3}\left(\tilde{U}_k(x)\delta_{x+\hat{k},y}-\delta_{x,y}\right)\\
\tilde{U}_k&\equiv&U^{[n_\rho]}_k\\ 
U^{[n+1]}_k(x)&=&\exp\left(i\rho\Theta^{[n]}_\mu(x)\right)U^{[n]}_k(x),\\
\Theta_k(x)&=&\frac{i}{2}\left(\Omega^\dagger_k(x)-\Omega_k(x)\right)-\frac{i}{6}\mbox{Tr}\left(\Omega^\dagger_k(x)-\Omega_k(x)\right)\\
\Omega_k(x)&=&C_k(x) U^\dagger_k(x)\qquad\mbox{(no summation over $k$)}\\
C_k(x)&\equiv&\sum_{j\neq k}\left(U_j(x)U_k(x+\hat{\jmath})
U^\dagger_j(x+\hat{k})\right.\\
&&+\left.U^\dagger_j(x-\hat{\jmath})U_k(x-\hat{\jmath})
U_j(x-\hat{\jmath}+\hat{k})\right).
\end{eqnarray}

The four parameters to tune in this 
smearing procedure are the real (dimensionless) Gaussian radius $\sigma$ and staple 
weight $\rho$, and the integer quark field and gauge link smearing iterations 
$n_\sigma$ and $n_\rho$.  Due to Chroma conventions, the Gaussian radius
values reported here will be denoted by $\sigma_s$ and related to the $\sigma$
in our equations by 
$$\sigma_s\equiv \sqrt{2}\sigma.$$

\section{Criterion for judging the effectiveness of smearing}
Recalling our definition of the effective mass function from chapter ~\ref{chap:overview}, 
a good measure of the signal of an operator $\overline{\mathcal{O}}_i$
 is the effective mass 
associated with the correlation function 
$$C_{ii}(\tau)=\bra{\Omega} \mathcal{O}_i(\tau)\overline{\mathcal{O}}_i(0)\ket{\Omega},$$
given by
\begin{eqnarray*}
a_\tau M(\tau) &\equiv& \ln\left[\frac{C_{ii}(\tau)}{C_{ii}(\tau+a_\tau)}\right],\\
&=&\ln\left[\frac{|c_{i,1}|^2e^{-E_1 \tau}+|c_{i,2}|^2 e^{-E_2 \tau}+\cdots}{|c_{i,1}|^2e^{-E_1(\tau+a_\tau)}+|c_{i,2}|^2 e^{-E_2(\tau+a_\tau)}+\cdots}\right],\qquad
c_{i,k}\equiv \bra{k}\overline{\mathcal{O}}_a\ket{\Omega},\\
&=&a_\tau E_1 \left(1 + O\left(|c_{i,2}|^2/|c_{i,1}|^2e^{-(E_2-E_1)\tau}\right)\right).
\end{eqnarray*}

A reasonable way to proceed is to tune our operators to reduce 
$|c_{i,k}|^2/|c_{i,1}|^2$ for large $k$, because an
objective measure of the quality of an operator is the 
time $\tau$ at which its effective mass function reaches its plateau.
In addition, we would like to reduce the variance of the Monte Carlo
estimate.  We define the {\em noise} of the effective mass function at each
time as the absolute ratio of the signal to jackknife error.

To compare the effectiveness of different values of the quark field smearing parameters,
we compared the effective mass $a_\tau M(\tau=4a_\tau)$ for each of the three operators
at a particular temporal separation $\tau=4a_\tau$.  Results using 50 quenched 
configurations on a $12^3\times 48$ anisotropic lattice using the Wilson action 
with $a_s \sim 0.1$ fm and $a_s/a_\tau \sim 3.0$ are shown in Figure~\ref{fig:qtune}.  

\begin{figure}[ht!]
  \centering 
  \includegraphics[width=3.0in]{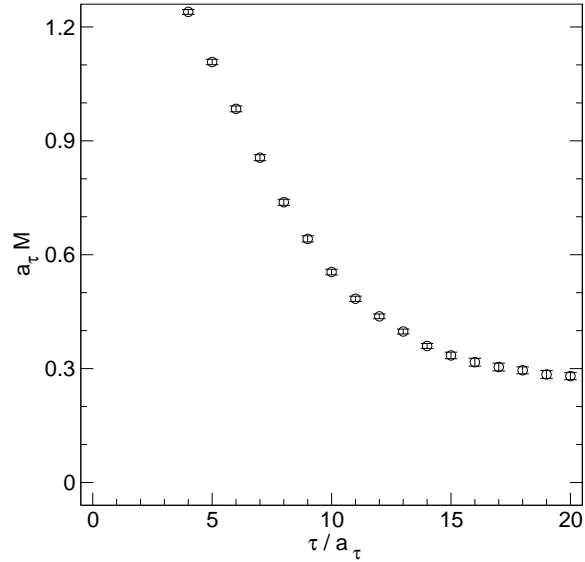}
  \caption{A sample effective mass plot of a single-site operator showing contamination at early times due to operator coupling with higher lying modes.}
  \label{fig:meff_sample} 
\end{figure}

\section{Contamination and noise in extended baryon operators}
\begin{figure}[ht!]
  \centering 
  \includegraphics[width=3.0in]{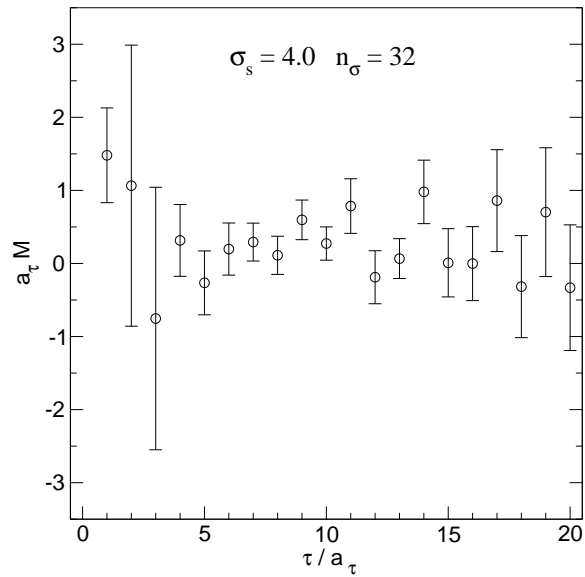}
  \caption{A sample effective mass plot of a triply-displaced operator showing noise due to stochastic gauge link noise.  This noise remains present even after
significant quark field smearing ($\sigma_s=4.0, n_\sigma=32$).}
  \label{fig:meff_noisy} 
\end{figure}

For single-site (local) hadron operators, it is well known that the use
of spatially-smeared quark fields is crucial.  For extended baryon operators,
one expects quark field smearing to be equally important, but the relevance
and interplay of link-field smearing is less well known. Thus, we decided that
a systematic study of both quark field and link-variable smearing was warranted.

From our smearing parameter tuning studies~\cite{lichtl:smeartune} we found that Gaussian quark field
smearing substantially reduces contributions from the short wavelength modes
of the theory, while stout gauge link smearing significantly reduces the 
noise from 
the stochastic evaluations.  The use of gauge link smearing is shown to
be crucial for baryon operators constructed of covariantly-displaced
quark fields. 

\begin{figure}[b]
\includegraphics[width=6.0in]{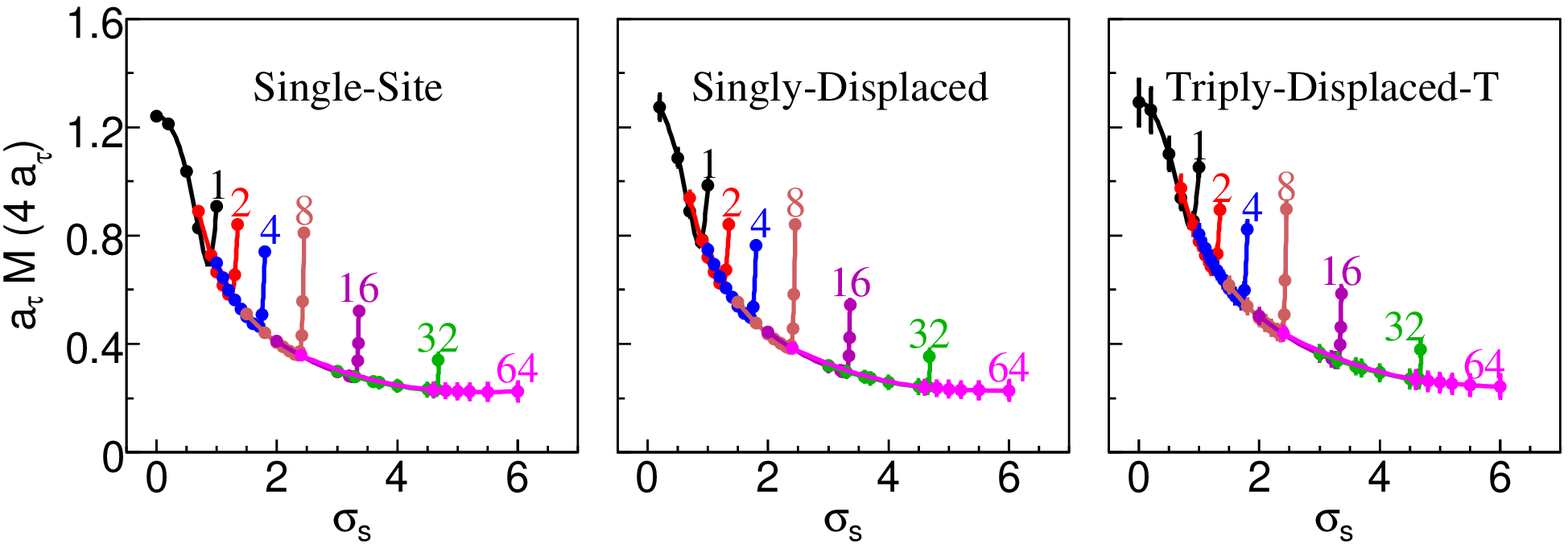}
\caption{The effective mass $a_\tau M(4a_\tau)$ for the operators $\overline{\mathcal{O}}_{SS},\ \overline{\mathcal{O}}_{SD},\ \overline{\mathcal{O}}_{TDT}$
against smearing radius $\sigma_s$ for $n_\sigma=1,2,4,8,16,32,64$.  The gauge field
is smeared using $n_\rho=16$ and $n_\rho\rho=2.5$.
Results are based on 50 quenched configurations on a $12^3\times 48$
anisotropic lattice using the Wilson action with $a_s \sim 0.1$ fm
and $a_s/a_\tau \sim 3.0$.  The quark mass is such that the mass of the pion 
is approximately 700 MeV. 
\label{fig:qtune}}
\end{figure}

Also, the order of approximation of the exponential operator $n_\sigma$ 
determines the maximum value of $\sigma$ we can use before the approximation 
breaks down (see Figure~\ref{fig:qtune}).  Once $\sigma$ becomes too large, the smearing operator behaves as $(1+\frac{\sigma^2}{2n_\sigma}\Delta)^{n_\sigma}$ $\to \; \sim \sigma^{2n_\sigma}\Delta^{n_\sigma}$, which is simply a momentum weighting.  Thus, high momentum modes rapidly start to contribute even more than in the unsmeared case.

\section{Systematic study of the smearing parameter space}

\subsection{Trial operators}
Each choice of smearing parameter values necessitates the generation of a new
set of generalized three-quark propagators.  
In order to maximize the number of different smearing parameter
values we could examine, we constructed our three trial operators out of three
nucleon three-quark propagators having the same operator type at the source
and sink (single-site, singly-displaced, and triply-displaced-T, see 
Figure~\ref{table:operators}).  Our
first trial operator was a single-site operator $\overline{\mathcal{O}}_{SS}$ projected into
the $G_{1g}$ irreducible representation of the cubic point group.
Because 
we did not have different orientations at the sink, we chose a particular
choice of the Dirac indices for the singly-displaced operator 
$\overline{\mathcal{O}}_{SD}$ and triply-displaced-T operator $\overline{\mathcal{O}}_{TDT}$.  In later runs, we found that the choice of smearing parameters was
insensitive to the type of extended baryon operator used.
 
\subsection{Quark field smearing}
Without gauge link smearing, the displaced operators were found to be excessively
noisy, making a meaningful comparison impossible.  For this reason, we first
tuned the link smearing parameters to minimize the noise of the effective
mass at the fourth time slice, using unsmeared quark fields.  We then
fixed the gauge link smearing parameters, and varied the quark field
smearing parameters.  These results are 
shown in Figure~\ref{fig:qtune} and include gauge-field smearing with 
$n_\rho=16$ and 
$n_\rho\rho=2.5$.  One sees that $a_\tau M(\tau=4a_\tau)$ is independent of $n_\sigma$
for sufficiently small $\sigma_s$.  For each value of $n_\sigma$, $a_\tau M(\tau=4a_\tau)$
first decreases as $\sigma_s$ is increased, until the approximation to a Gaussian 
eventually breaks down, signaled by a sudden rapid rise in $a_\tau M(\tau=4a_\tau)$.
This rapid rise occurs at larger values of $\sigma_s$ for larger values of 
$n_\sigma$.

\begin{figure}[t]
\includegraphics[width=6.0in]{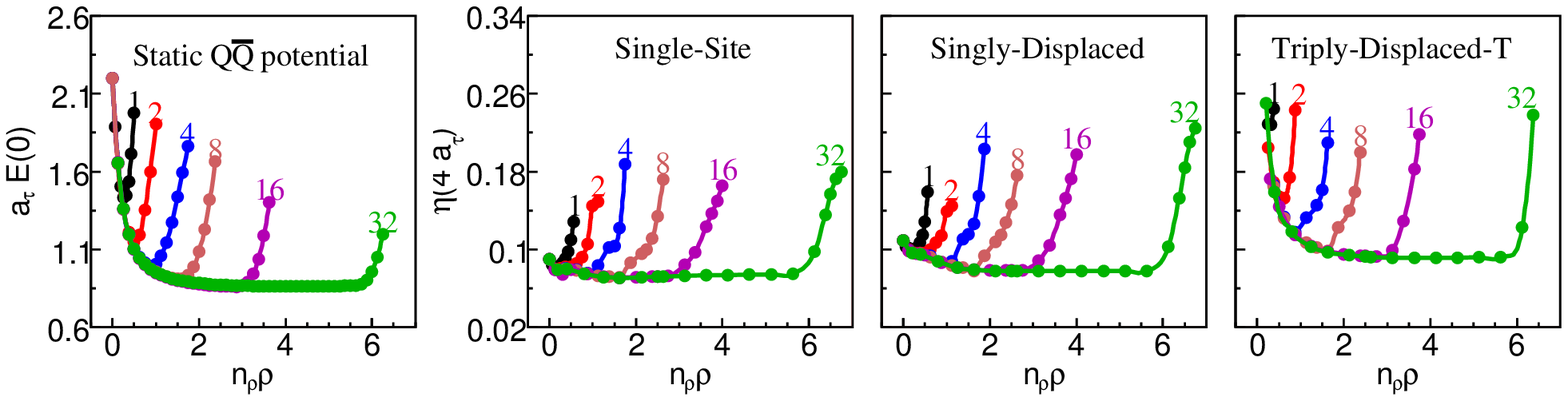}
\caption{
Leftmost plot: the effective mass $a_\tau E(0)$ for $\tau=0$
corresponding to the static quark-antiquark potential at spatial separation
 $R = 5a_s \sim 0.5$ fm against $n_\rho\rho$ for $n_\rho=1,2,4,8,16,32$.
Results are based on 100 configurations on a $16^4$ isotropic lattice using
the Wilson gauge action with $\beta=6.0$.
Right three plots: the relative jackknife error $\eta(4a_\tau)$ of effective 
masses $a_\tau M(4a_\tau)$ of the three nucleon operators $\overline{\mathcal{O}}_{SS},\ \overline{\mathcal{O}}_{SD},\ \overline{\mathcal{O}}_{TDT}$
for $n_\sigma=32,\ \sigma_s=4.0$ against $n_\rho\rho$ for $n_\rho=1,2,4,8,16,32$.
Results are based on 50 quenched configurations on a $12^3\times 48$ anisotropic 
lattice using the Wilson action with $a_s \sim 0.1$ fm, $a_s/a_{\tau} \sim 3.0$.
\label{fig:ltune}}
\end{figure}

\subsection{Gauge link smearing}
Next, we fixed the quark field smearing parameters and 
studied the effect of changing the gauge-field smearing parameters.
Before tuning the trial operators, we examined the the effective mass $a_\tau E(0)$ associated with the static 
quark-antiquark potential at a spatial separation $R=5a_s\sim 0.5$~fm and at a 
particular temporal separation $\tau=0$ to get an idea of the 
effectiveness of different values of $\rho$ and $n_\rho$.  The 
results are shown 
in the leftmost plot in Figure~\ref{fig:ltune}.  The behavior is qualitatively
similar to that observed when we tuned the gauge link smearing parameters for 
our trial operators, shown in Figure~\ref{fig:qtune}.  One sees that the 
$\tau=0$ effective mass is independent of the product $n_\rho \rho$ 
for sufficiently small values of $n_\rho \rho$.  For each value of $n_\rho$, 
$a_\tau E(0)$ decreases as $n_\rho\rho$ increases, until a minimum is
reached and a rapid rise occurs.  The onset of the rise occurs at larger
values of $n_\rho\rho$ as $n_\rho$ increases.  Note that $a_\tau E(0)$ does
not decrease appreciably as $n_\rho\rho$ increases above 2.5.  Hence, 
$n_\rho\rho=2.5$ with $n_\rho=16$ are our preferred values for the link smearing
at lattice spacing $a_s\sim0.1$~fm, based on the static quark-antiquark potential.

Somewhat surprisingly, we found that changing the link-smearing parameters
did not appreciably affect the mean values of the effective masses of our
three nucleon operators.  However, the effect on the variances of the effective
masses was dramatic.   The relative jackknife error $\eta(4a_\tau)$
of $a_\tau M(4a_\tau)$ is shown against $n_\rho\rho$ in the right three
plots in Figure~\ref{fig:ltune}, and amazingly, this error shows the same 
qualitative behavior as in Figure~\ref{fig:qtune} and the leftmost plot
in Figure~\ref{fig:ltune}.  One key point learned here is that the preferred 
link-smearing parameters determined from the static quark-antiquark potential
produce the smallest error in the extended baryon operators.

\begin{figure}[p]
\includegraphics[width=6.0in]{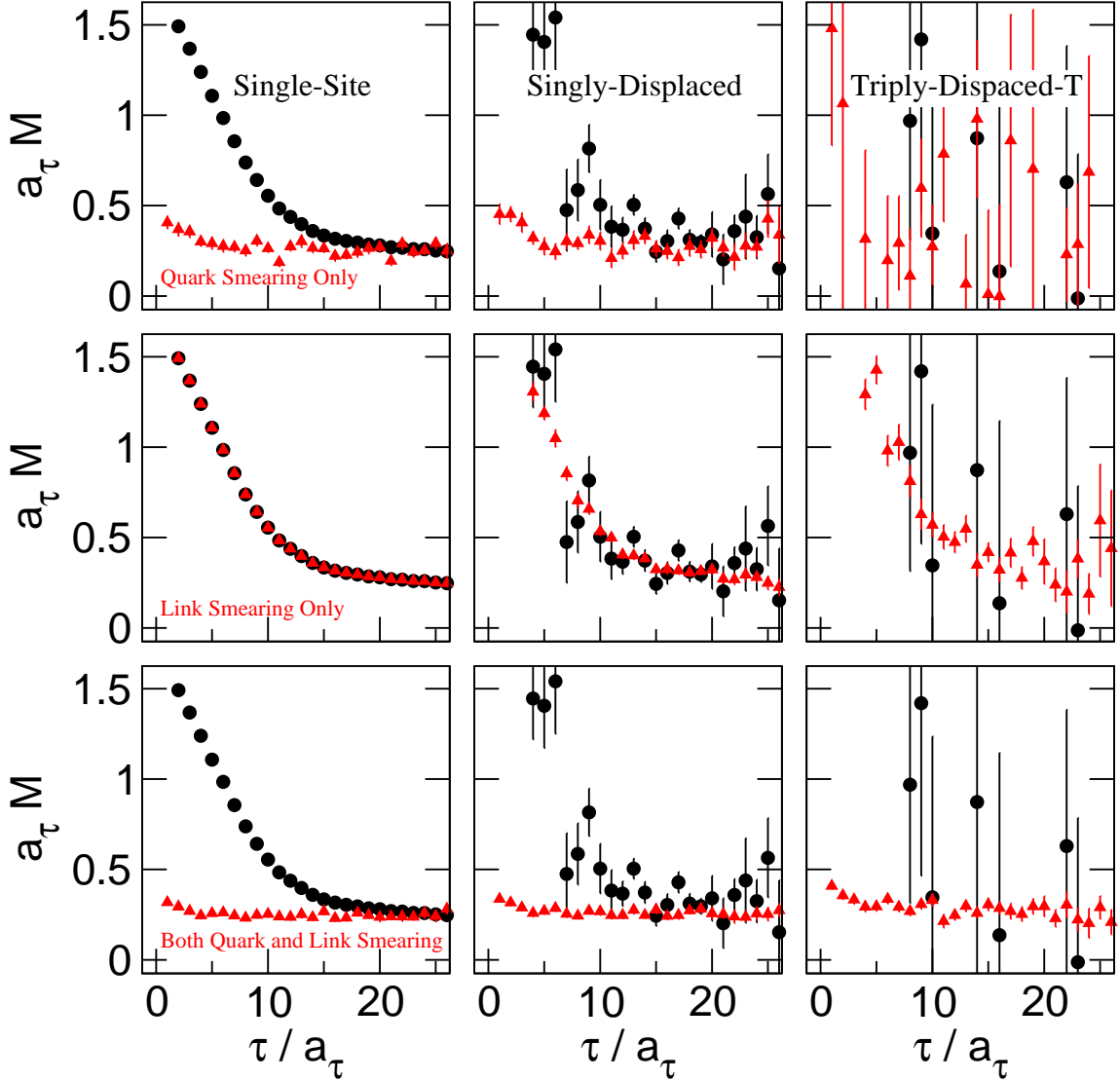}
\caption{Effective masses $a_\tau M(\tau)$ for unsmeared (black circles) and smeared 
(red triangles) operators $\overline{\mathcal{O}}_{SS},\ \overline{\mathcal{O}}_{SD},\ \overline{\mathcal{O}}_{TDT}$. 
Top row: only quark field smearing $n_\sigma=32,\ \sigma_s=4.0$ is used. Middle row: 
only link-variable smearing $n_\rho=16,\ n_\rho\rho=2.5$ is applied.  
Bottom row: both quark and link smearing $n_\sigma=32,\ \sigma_s=4.0, 
\ n_\rho=16,\ n_\rho\rho=2.5$ are used, dramatically improving the signal for all
three operators. Results are based on 50 quenched configurations on a 
$12^3\times 48$ anisotropic lattice using the Wilson action with $a_s \sim 0.1$ fm,
$a_s/a_\tau \sim 3.0$.\label{fig:meff-smear}}
\end{figure}

The effective masses shown in Figure~\ref{fig:meff-smear} also illustrate
these findings.  The top row shows that applying only quark field smearing
to the three selected nucleon operators significantly reduces couplings
to higher-lying states, but the displaced operators remain excessively
noisy.  The second row illustrates that including only link-field smearing
substantially reduces the noise, but does not appreciably alter the effective
masses themselves.  The bottom row shows dramatic improvement from reduced
couplings to excited states and dramatically reduced noise when both
quark field and link-field smearing is applied, especially for the extended
operators. 

\begin{figure} 
  \centering 
  \includegraphics[width=6.0in]{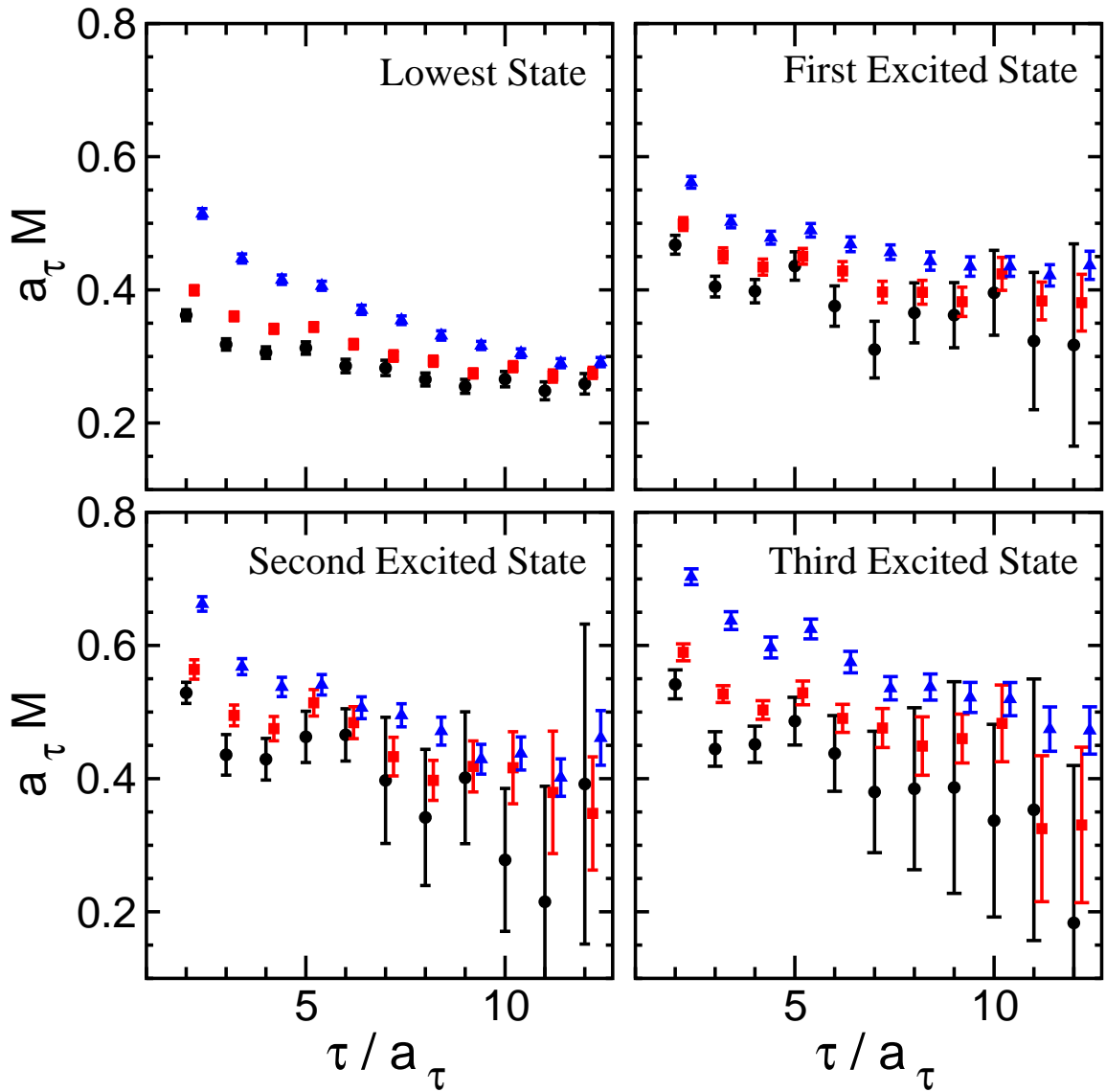}
  \caption{The effects of different values of the quark smearing radius
$\sigma_s$ on the excited states.  Throughout, stout-link smearing is used with $n_\rho\rho=2.5$,$n_\rho=16$, and $n_\sigma=32$ quark smearing interactions are used.  The
black circles have $\sigma_s=4.0$, the red squares have $\sigma_s=3.0$, and
the blue triangles have $\sigma=2.0$.  A Gaussian radius of $\sigma_s=3.0$ was
chosen as a balance between high state contamination vs. stochastic noise
in the excited states.} 
  \label{fig:smear-excited} 
\end{figure}

\subsection{Excited state considerations}
Incorporating both quark field and link-variable smearing is crucial
for extracting the baryon spectrum using gauge-invariant extended three-quark
operators.  Gaussian quark field smearing dramatically diminishes couplings to the
short wavelength modes of the theory, whereas stout-link smearing drastically reduces
the noise in operators with displaced quarks.  Preferred smearing parameters
$\sigma_s=4.0,\ n_\sigma=32,\ n_\rho\rho=2.5,\ n_\rho=16$ were found for a 
lattice spacing $a_s\sim 0.1$~fm and were independent of
the baryon operators chosen.  
After examining the behavior of the excited state signal (using the principal
effective mass method), we decided to reduce the quark smearing in order to decrease the noise
in the excited states (see Figure~\ref{fig:smear-excited}).

The final values of smearing parameters we selected were:
\begin{itemize}
\item Gaussian smearing: $\sigma_s=3.0,\; n_\sigma= 32$ iterations,
\item Stout-link smearing: $n_\rho \rho=2.5,\; n_\rho=16$ iterations.
\end{itemize}

An issue which remains to be addressed in future work is 
the dependence of the preferred smearing parameters on the quark mass.

\chapter{Baryon operator pruning}
\label{chap:pruning}

At this point, we have constructed a set of gauge-invariant single-site and
extended baryon operators endowed with specific momentum,
flavor, and (lattice) spin quantum numbers.  We have tuned the quark field
and gauge link smearing parameters to reduce contamination from high-lying
modes, and to reduce the noise in the extended operators.  These efforts
have rewarded us with a large set of candidate operators which can be
used to extract the low-lying baryon spectrum.
The numbers of baryon operators of each type which project into the different
irreducible representations of $O^D_h$ are given in 
Table~\ref{table:op_numbers}.

\begin{table*}[b!]
\centering
{\small
\begin{tabular}{l|rrr|rrr|rrr|rrr} 
\hline
\hline
 & \multicolumn{3}{c|}{\raisebox{0mm}[5mm]{$\Delta^{++}$}} &  \multicolumn{3}{c|}{$\Sigma^+$} 
 & \multicolumn{3}{c|}{$N^+$} & \multicolumn{3}{c}{$\Lambda^0$} \\
 Operator type &  $G_{1g}$ & $H_g$ & $G_{2g}$ & $G_{1g}$ & $H_g$  &   $G_{2g}$ &  $G_{1g}$  &  $H_g$  &   $G_{2g}$ &  $G_{1g}$  &  $H_g$  &   $G_{2g}$\\ \hline
 Single-Site       &  1 &  2 &  0 &  4 &  3 &  0 &  3 &  1 &  0 &  4 &  1 &  0\\
 Singly-Displaced  & 14 & 20 &  6 & 38 & 52 & 14 & 24 & 32 &  8 & 34 & 44 & 10\\
 Doubly-Displaced-I& 12 & 16 &  4 & 36 & 48 & 12 & 24 & 32 &  8 & 36 & 48 & 12\\
 Doubly-Displaced-L& 32 & 64 & 32 & 96 &192 & 96 & 64 &128 & 64 & 96 &192 & 96\\
 Triply-Displaced-T& 32 & 64 & 32 & 96 &192 & 96 & 64 &128 & 64 & 96 &192 & 96\\
\hline
\raisebox{0mm}[4mm]{\textbf{Total}} & \textbf{91} & \textbf{166} & \textbf{74} & \textbf{270} & \textbf{487} & \textbf{218} & \textbf{179} & \textbf{321} & \textbf{144} & \textbf{266} & \textbf{477} & \textbf{214}\\
\hline
\hline
\end{tabular}}
\caption{The numbers of operators of each type which project
into each row of the $G_{1g}$, $H_g$, and $G_{2g}$ irreps for the 
$\Delta^{++}$, $\Sigma^+, N^+,$ and $\Lambda^0$ baryons.  The numbers
for the $G_{1u}$, $H_u$, and $G_{2u}$ irreps are the same as for the
$G_{1g}$, $H_g$, and $G_{2g}$ irreps, respectively.  The $\Xi^0$ operators
are obtained from the $\Sigma^+$ operators by making the flavor exchange
$u\leftrightarrow s$.  The $\Omega^-$ operators
are obtained from the $\Delta^{++}$ operators by making the flavor replacement 
$u\to s$.
\label{table:op_numbers}}
\end{table*}

For some set of candidate operators $\{\overline{\mathcal{O}}_i\}$,
the Hermitian correlation matrix $C(\tau)$ is the fundamental quantity we
estimate using the Monte Carlo method:
\begin{equation*}
C_{ij}(\tau)=\bra{\Omega}\mathcal{O}_i(\tau)\overline{\mathcal{O}}_j(0)\ket{\Omega},
\end{equation*}
where we require $\tau>0$ (c.f. Subsection~\ref{section:variational})
 and once again assume a sufficiently large spatial extent $T \gg \tau$ 
and an 
action which satisfies reflection positivity.
The variational method for extracting the excited states 
discussed in Chapter~\ref{chap:overview} requires
us to solve:
$$C^{-1/2}(\tau_0)C(\tau)C^{-1/2}(\tau_0)u_i=\lambda_i\, u_i$$
for $u_i$ and $\lambda_i$.
In this study, we will attempt to extract the first eight excited levels
in each irrep channel.  If we were to construct a $179\times 179$ correlation
matrix for the $N^+$ (nucleon) $G_{1g}$ channel, then we would have 
$179$ principal correlation functions $\lambda_i$, of which we would only be interested
in the first eight.  Most of the other $171$ principal correlation functions would
have a negligible signal-to-noise ratio, suggesting that the matrix inversion
step
$$C(\tau_0)\to C^{-1/2}(\tau_0)$$
would most likely fail before we could even solve the eigenvalue problem.
Therefore it is infeasible to use all of the candidate operators to construct
our 
correlation matrices.
We chose instead to prune our set of candidate operators down to
 sixteen operators in each $O^D_h$ irrep before attempting to 
extract the excited states in those channels using the variational method.  
We feel that this number provides a good balance between the need to use a 
large enough
set of operators to reliably extract the states of interest, and the need
to use a small enough set of operators such that the matrix inversions
are not adversely affected by noise.

Additionally, for each irrep we used the same
candidate operators for the even and odd-parity channels ($G_{1g}/G_{1u}$, 
$H_g/H_u$, $G_{2g}/G_{2u}$).  This requirement allows us to improve
our estimates of the correlation matrix elements by averaging with the backward 
propagating state information in the opposite-parity channel (discussed in 
Subsection~\ref{section:backward}).
The two criteria we used to prune our operators were {\em signal quality} and
{\em linear independence}.
For this work, we focused on the $N^+$ (nucleon) channel.

\section{Signal quality}
We first calculated the effective mass function associated with each
nucleon operator $\overline{\mathcal{O}}_i$ using 200 quenched configurations.
  These effective mass functions required Monte Carlo estimates 
of only the diagonal 
elements of the nucleon correlation matrices for each irrep:
\begin{itemize}
\item $N^+\ G_{1g}/G_{1u}$ matrices: $179$ diagonal elements,
\item $N^+\ H_{g}/H_{u}$ matrices: $321$ diagonal elements,
\item $N^+\ G_{2g}/G_{2u}$ matrices: $144$ diagonal elements.
\end{itemize}
The $175$ three-quark propagators we generated to construct these elements were
saved for reuse in our subsequent pruning steps.

\begin{figure}[ht!]
\centering
\includegraphics[width=6in]{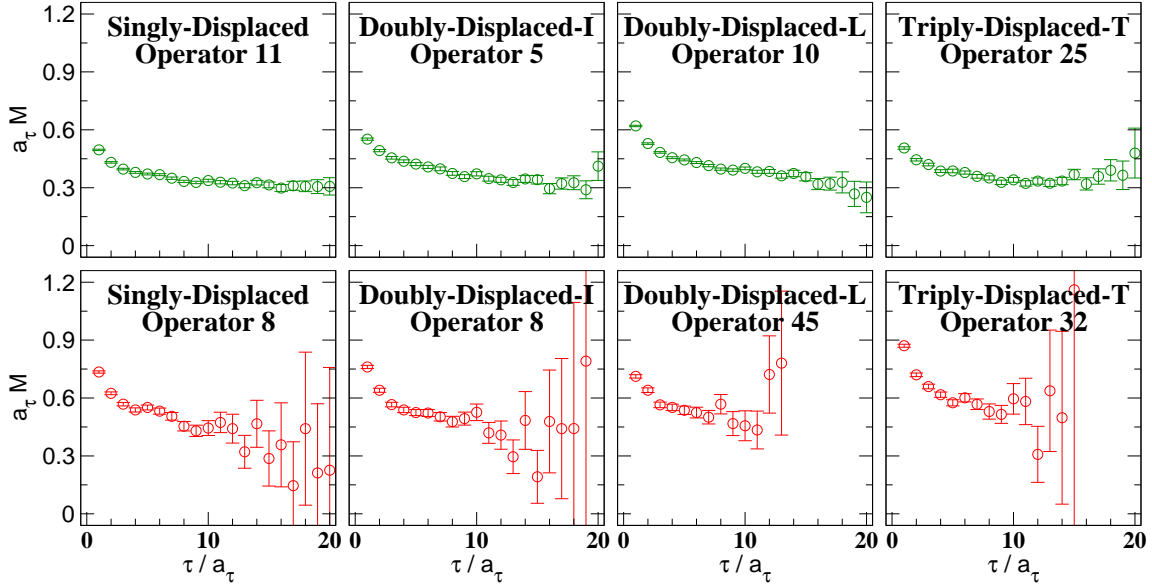}
\caption{Effective mass plots for eight representative operators from the 
complete set of extended baryon operators in the $G_{1u}$ channel.  
We chose up to ten candidate operators of each type 
(SD, DDI, DDL, TDT) based on the average jackknife error over the first
sixteen time slices.  This process helped us to identify `quiet' operators
(top row, in green) and remove some of the noisier operators (bottom row,
in red).}
\label{fig:operator_pruning}
\end{figure}

At this pruning stage, we wanted to select operators of each type which
had a high quality signal. 
It is reasonable to assume that linear combinations of `quiet' operators
would yield superior results over linear combinations of noisy operators.
Examples of quiet vs. noisy operators are shown in Figure~\ref{fig:operator_pruning}.
Our measure of signal quality was the effective mass function's jackknife 
error averaged over $\tau=1,\cdots,16$.  The consideration of average error
rather than the signal-to-noise ratio facilitates a direct
comparison between the different operators.

Because there are three single-site type operators in the $G_{1g}/G_{1u}$ 
channels, one in the $H_g/H_u$ channels, and none in the $G_{2g}/G_{2u}$ 
channels, we selected all single-site operators.
 Within each type of extended operator
(singly-displaced, doubly-displaced-I, doubly-displaced-L, and
triply-displaced-T), we 
sorted the operators in increasing order of the average jackknife error 
over the first
sixteen non-zero time separations: $\tau=1,\cdots, 16$ and selected the first
ten operators.  At this point, we have selected all of the single-site operators
and ten extended operators of each type in each irrep.
For each operator selected in each even-parity channel, we also selected its 
opposite-parity partner in the odd-parity channel, and vice-versa.
We performed the signal quality selection each extended operator type {\em 
separately} to ensure that we still had a representative set of operators
of each type.  We did not want noise to be the only factor in the selection
process; we are primarily interested in the ability of an operator to
interpolate for the excited states of interest.

\section{Linear independence}
We had eliminated our noisiest operators, but still had
over forty candidate operators in each channel.  We pruned our operators
further by looking for subsets of operators which excited Hilbert 
states which were as
distinct as possible.  We considered such operators likely to perform
well in the variational method because they would give a 
more nearly orthogonal
trial state basis.

\subsection{The normalized correlation matrix}
We would like to select a subset of operators such that the corresponding
states are as distinct as possible.  To enable valid comparisons among our 
candidate operators $\{\overline{\mathcal{O}}_i\}$, we considered the
{\em normalized}\footnote{In this section we will use a hat to denote
the normalized correlation matrix.  This is not to be confused with the hat 
notation of Chapter~\ref{chap:monte_carlo}.} Hermitian correlation matrix elements $\hat{C}_{ij}$ defined
in terms of $C_{ij}(\tau=a_\tau)$:
\begin{eqnarray*}
\hat{C}_{ij}&\equiv&\frac{C_{ij}(a_\tau)}{\sqrt{
C_{ii}(a_\tau)C_{jj}(a_\tau)}},\\
&=&\frac{\bra{\Omega}\mathcal{O}_i e^{-Ha_\tau}\overline{\mathcal{O}}_j\ket{\Omega}}
{\sqrt{\bra{\Omega}\mathcal{O}_i e^{-Ha_\tau}\overline{\mathcal{O}}_i\ket{\Omega}
\bra{\Omega}\mathcal{O}_j e^{-Ha_\tau}\overline{\mathcal{O}}_j\ket{\Omega}}}
\end{eqnarray*}
remembering that the diagonal correlation matrix elements $C_{ii}(\tau)$ and 
$C_{jj}(\tau)$ are real and positive for $\tau\neq 0$.  
As in Subsection~\ref{section:variational}, we avoid expressions involving
the correlation matrix evaluated at $\tau=0$ due to the presence of Schwinger
terms.

\subsection{Pruning by condition number}
For any set of $n$ candidate operators
$\{\overline{\mathcal{O}}_1, \overline{\mathcal{O}}_2, \cdots, \overline{\mathcal{O}}_n\}$, 
we have $\hat{C}_{ii}=1\;\;\forall\;\;i$.
In particular:
$$\mbox{Tr}(\hat{C})=n.$$
If all of the states created by our candidate operators 
were orthogonal, then $\hat{C}_{ij}=\delta_{ij}$, which implies that all of
 the eigenvalues of $\hat{C}$ are equal to one.  
In contrast, if two of the states were identical, then $\det(\hat{C})=0$, 
implying that at least one of the eigenvalues of $\hat{C}$ is equal to
zero.  In order for maintain $\mbox{Tr}(\hat{C})=n$, at least one of the other
eigenvalues must therefore be greater than one.

Because our matrix is Hermitian, its eigenvalues are guaranteed to be real.
We therefore choose as our measure of the degree of linear independence
of our operators the {\em condition number} $\kappa$, the ratio of the 
largest to smallest eigenvalue of the normalized correlation matrix $\hat{C}$:
$$\kappa\equiv \lambda_{\mathrm{max}}/\lambda_{\mathrm{min}}\ge 1.$$
For perfectly orthogonal states $\kappa=1$ and for completely degenerate
states $\kappa\to\infty$.
We therefore seek to find a subset of operators in each irrep such 
that the condition
numbers of the associated even and odd correlation matrices satisfy 
some minimization condition.
Selecting a correlation submatrix having a small condition number
has another advantage: small condition numbers lead to the rapid convergence 
of matrix inversion algorithms~\cite{golub:matrix}.

\subsection{Linear independence within each operator type}
We proceeded by calculating all correlation matrix elements having the same
type of extended operator at the source and sink.
We reused the previously generated three-quark propagators for the diagonal 
elements, and generated the additional three-quark propagators needed for
the off-diagonal elements.

For each correlation matrix, we calculated the condition
numbers of all possible $5\times 5$ submatrices using the singular value
decomposition algorithm~\cite{press:numerical}.  Because we wanted to
select the same operators in both the even and odd-parity channels, we proceeded
as follows for each irrep and operator type separately:
\begin{itemize}
\item Let $x$ denote a subset of five extended operators of a specific type 
(i.e. all $SD$ or all $DDL$): 
$$x\equiv\{\overline{\mathcal{O}}_{i_1},\overline{\mathcal{O}}_{i_2},\cdots,\overline{\mathcal{O}}_{i_5}\}.$$
\item Let $\kappa^g(x)$ and $\kappa^u(x)$ be the condition numbers of the even and odd-parity correlation
submatrices corresponding to the operator subset $x$, respectively\footnote{Remember, we had selected
the same operators for both the even and odd-parity channels during our signal quality pruning step.}.
\item Let $\kappa_>(x)\equiv \max(\kappa^g(x),\kappa^u(x))$, the larger of the even and odd-parity condition numbers.
\item We selected the operator set $x^*$ which minimized
$\kappa_>(x)$:
$$\kappa_>(x^*)=\min_x (\kappa_>(x))=\min_x\left(\max(\kappa^g(x), \kappa^u(x)) \right). $$
\end{itemize} 
By minimizing the larger of the even and odd-parity condition numbers, we
found a reasonably distinct set of operators across 
both the even and odd-parity channels.
As a check of our results, we examined the smallest eigenvalue 
$\lambda_{\mathrm{min}}$ of our selected submatrix 
to verify that it was much greater than its jackknife error, avoiding the
possibility of signal-to-noise issues.

A natural question which arises during the application of this method is: 
how much did we lose by requiring that the even
and odd-parity channels of each irrep contain the same operators?  In 
order to ensure that we were not sacrificing too much, we calculated
the percent deviation of the condition number from the 
minimum possible condition number for the even and odd-parity channels 
separately:
\begin{eqnarray*}
\frac{\kappa^g(x^*)-\min_x(\kappa^g(x))}{\min_x(\kappa^g(x))},\qquad
\frac{\kappa^u(x^*)-\min_x(\kappa^u(x))}{\min_x(\kappa^u(x))}.
\end{eqnarray*} 
We found that all of these percentages were below $20\%$, and most were
below $5\%$. The extended operator 
results are shown in Table~\ref{table:condition_numbers}.
Once again, we selected all single-site operators. 

\begin{table}[ht!]
\centering
\small{
\begin{tabular}{l|r@{.}lr@{.}l|r@{.}lr@{.}l|r@{.}lr@{.}l} 
\hline
\hline
 Operator type      &  \multicolumn{2}{c}{$G_{1g}$}  &  \multicolumn{2}{c|}{$G_{1u}$} & 
\multicolumn{2}{c}{$H_g$} & \multicolumn{2}{c|}{$H_u$} & 
\multicolumn{2}{c}{$G_{2g}$}  & \multicolumn{2}{c}{$G_{2u}$} \\ 
\hline
 \multirow{2}{*}{Singly-Displaced}   & $12$&$96(21)$ & $12$&$90(18)$ &
 $1$&$6972(58)$ & $1$&$6920(59)$ &
 $5$&$74(14)$ & $5$&$46(16)$\\
&\multicolumn{2}{c}{$3\%$}&\multicolumn{2}{c|}{$10\%$}&
\multicolumn{2}{c}{$3\%$}&\multicolumn{2}{c|}{$0\%$}&
\multicolumn{2}{c}{$0\%$}&\multicolumn{2}{c}{$0\%$}\\
\hline 
 \multirow{2}{*}{Doubly-Displaced-I} & $3$&$849(30)$ & $3$&$477(28)$ &
 $1$&$2956(32)$ & $1$&$2907(32)$ &
 $2$&$922(18)$ & $2$&$842(21)$\\
&\multicolumn{2}{c}{$1\%$}&\multicolumn{2}{c|}{$7\%$}&
\multicolumn{2}{c}{$0\%$}&\multicolumn{2}{c|}{$0\%$}&
\multicolumn{2}{c}{$1\%$}&\multicolumn{2}{c}{$0\%$}\\
\hline
 \multirow{2}{*}{Doubly-Displaced-L} & $10$&$70(16)$ & $10$&$32(13)$ &
 $4$&$157(27)$ & $3$&$894(27)$ &
 $1$&$697(16)$ & $1$&$712(17)$\\
&\multicolumn{2}{c}{$18\%$}&\multicolumn{2}{c|}{$12\%$}&
\multicolumn{2}{c}{$16\%$}&\multicolumn{2}{c|}{$0\%$}&
\multicolumn{2}{c}{$0\%$}&\multicolumn{2}{c}{$0\%$}\\
\hline 
\multirow{2}{*}{Triply-Displaced-T} & $4$&$209(42)$ & $4$&$048(48)$ &
 $1$&$7796(95)$ & $1$&$9163(89)$ &
 $1$&$744(12)$ & $1$&$856(13)$\\
&\multicolumn{2}{c}{$4\%$}&\multicolumn{2}{c|}{$5\%$}&
\multicolumn{2}{c}{$0\%$}&\multicolumn{2}{c|}{$0\%$}&
\multicolumn{2}{c}{$4\%$}&\multicolumn{2}{c}{$2\%$}\\
\hline
\hline
\end{tabular}}
\caption{The condition numbers $\kappa$ and for the 
$5\times 5$ extended operator
submatrices chosen.  The percentages denote
how much greater each condition number is than its minimum possible value.}
\label{table:condition_numbers}
\end{table}

\subsection{Linear independence across operator types}
In the previous pruning step, we selected all single-site operators and five
extended operators of each type by considering correlation matrix elements
between operators of the same type.  We next considered correlation matrix 
elements between operators of {\em different} types.
The total number of three-quark propagators needed was $671$.
Once again we reused previously generated three-quark propagators when possible,
and generated the additional three-quark propagators needed.

After normalizing the correlation matrices, we took the best 
$16\times 16$ submatrix according to the condition number 
minimization condition presented in the last subsection.
Combining the five operators selected from each of the four types with the 
single-site
operators gave the following correlation matrix sizes and
submatrix numbers for each channel:
\begin{itemize}
\item $N^+\ G_{1g}/G_{1u}$ matrices: $23\times 23$ elements and $245,157$ submatrices,
\item $N^+\ H_{g}/H_{u}$ matrices: $21\times 21$ elements and $20,349$ submatrices,
\item $N^+\ G_{2g}/G_{2u}$ matrices: $20\times 20$ elements and $4,845$ submatrices,
\end{itemize}
where each submatrix has $16\times 16$ elements.

\begin{figure}[ht!] 
  \centering 
  \includegraphics[width=6.0in]{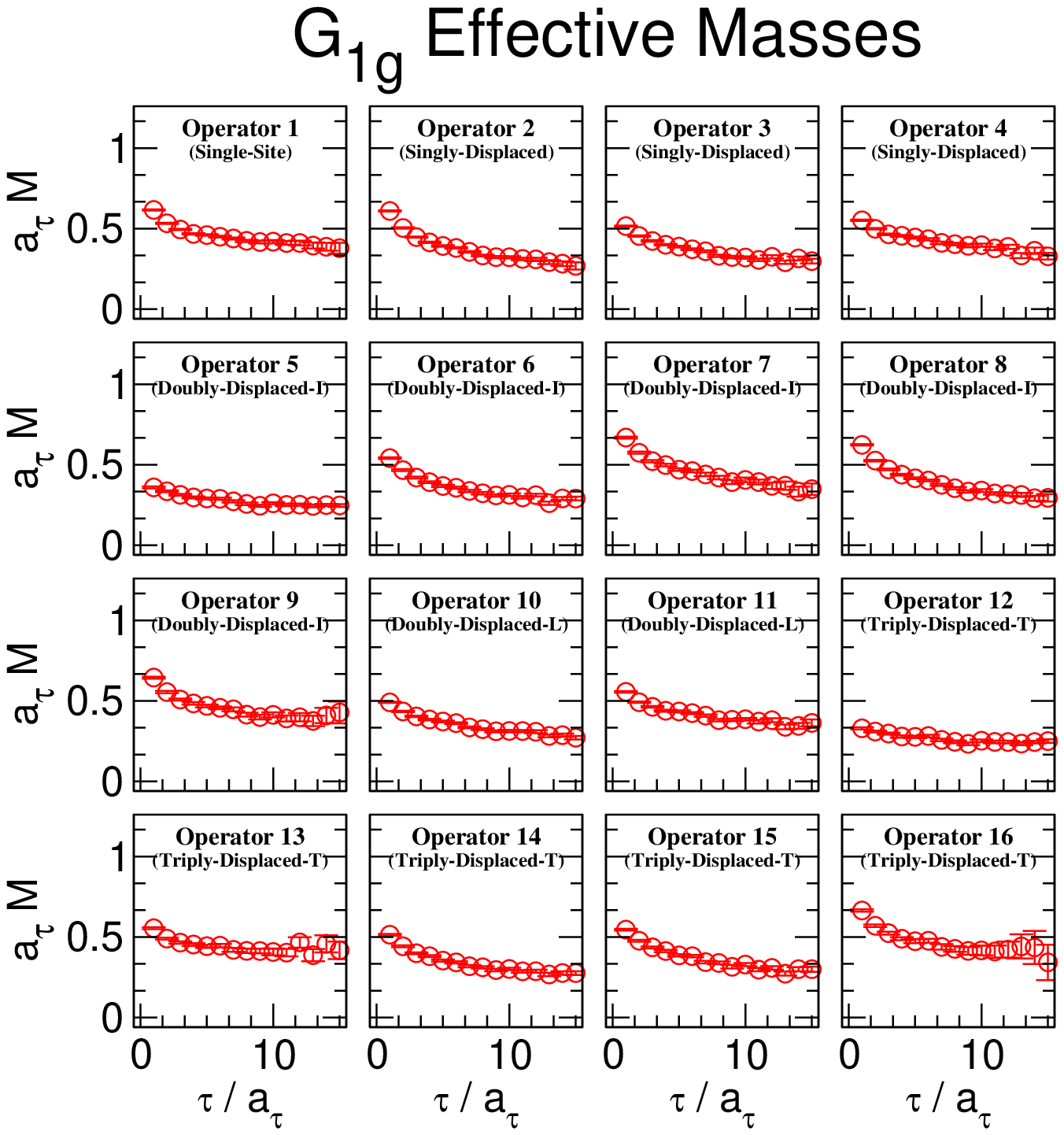}
  \caption{The effective masses for the final sixteen $G_{1g}$ operators selected by our pruning process.} 
  \label{fig:meff_final-G1g} 
\end{figure}

\begin{figure}[ht!] 
  \centering 
  \includegraphics[width=6.0in]{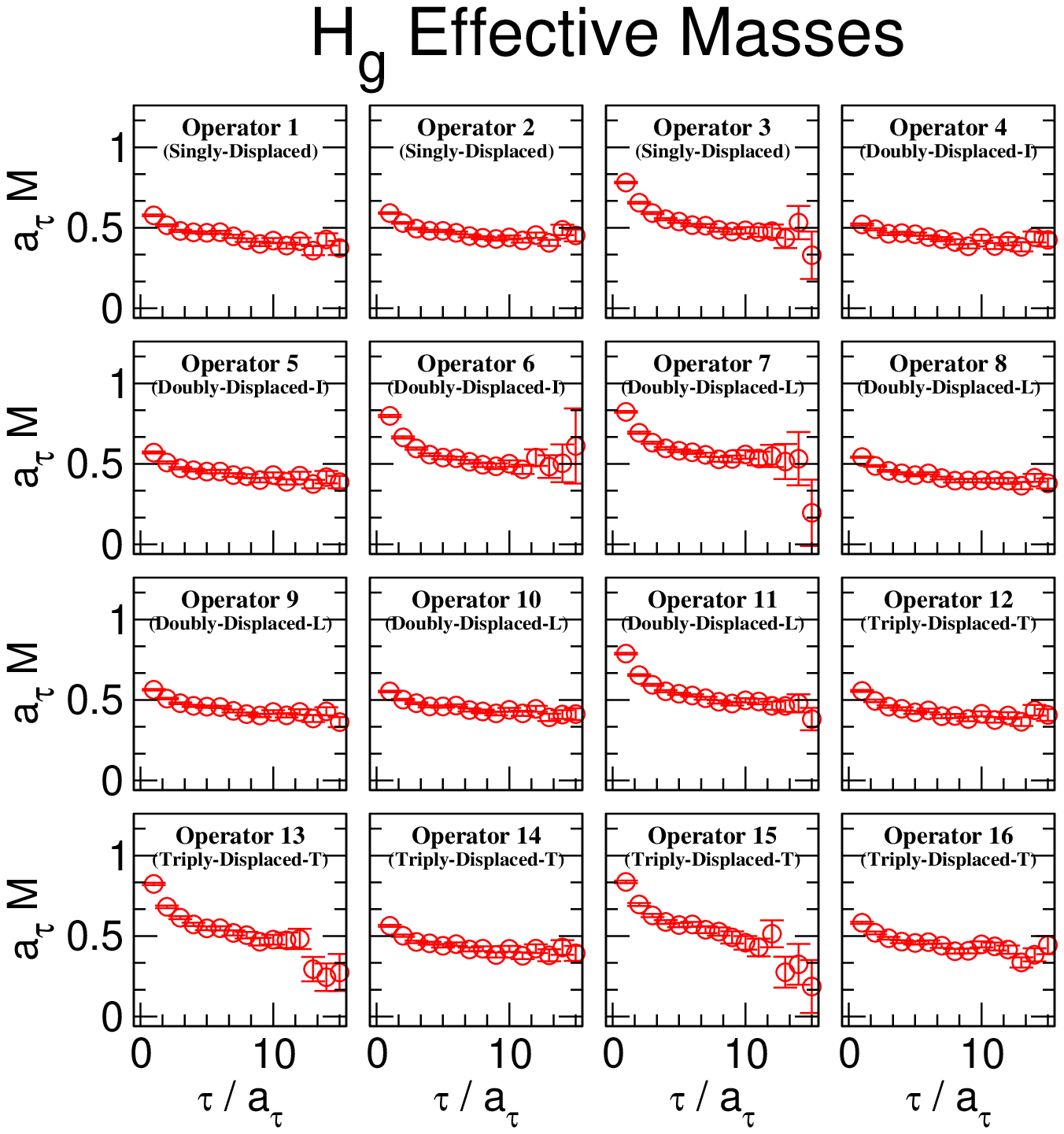}
  \caption{The effective masses for the final sixteen $H_{g}$ operators selected by our pruning process.} 
  \label{fig:meff_final-Hg} 
\end{figure}

\begin{figure}[ht!] 
  \centering 
  \includegraphics[width=6.0in]{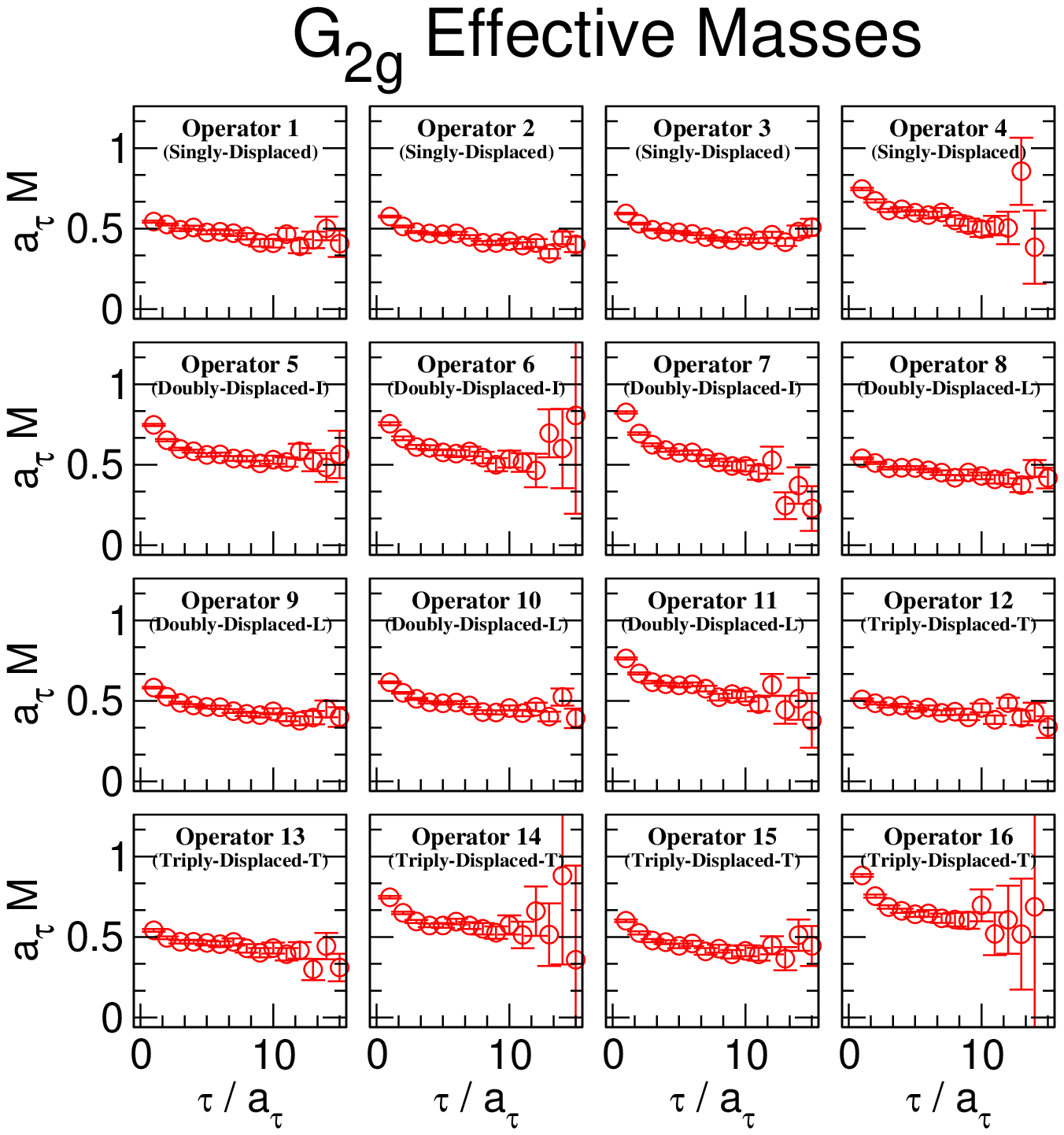}
  \caption{The effective masses for the final sixteen $G_{2g}$ operators selected by our pruning process.} 
  \label{fig:meff_final-G2g} 
\end{figure}

\begin{figure}[ht!] 
  \centering 
  \includegraphics[width=6.0in]{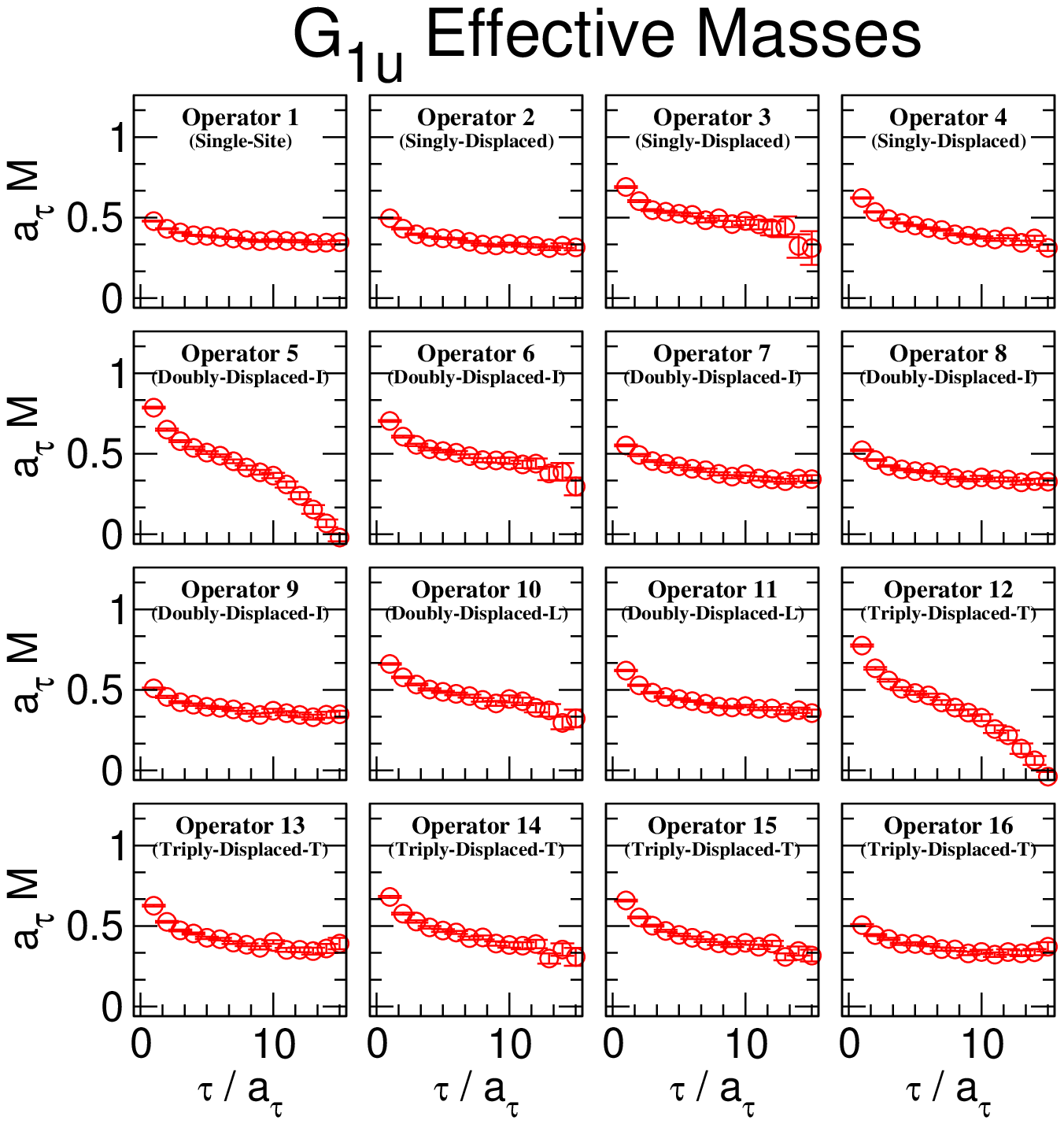}
  \caption{The effective masses for the final sixteen $G_{1u}$ operators selected by our pruning process.} 
  \label{fig:meff_final-G1u} 
\end{figure}

\begin{figure}[ht!] 
  \centering 
  \includegraphics[width=6.0in]{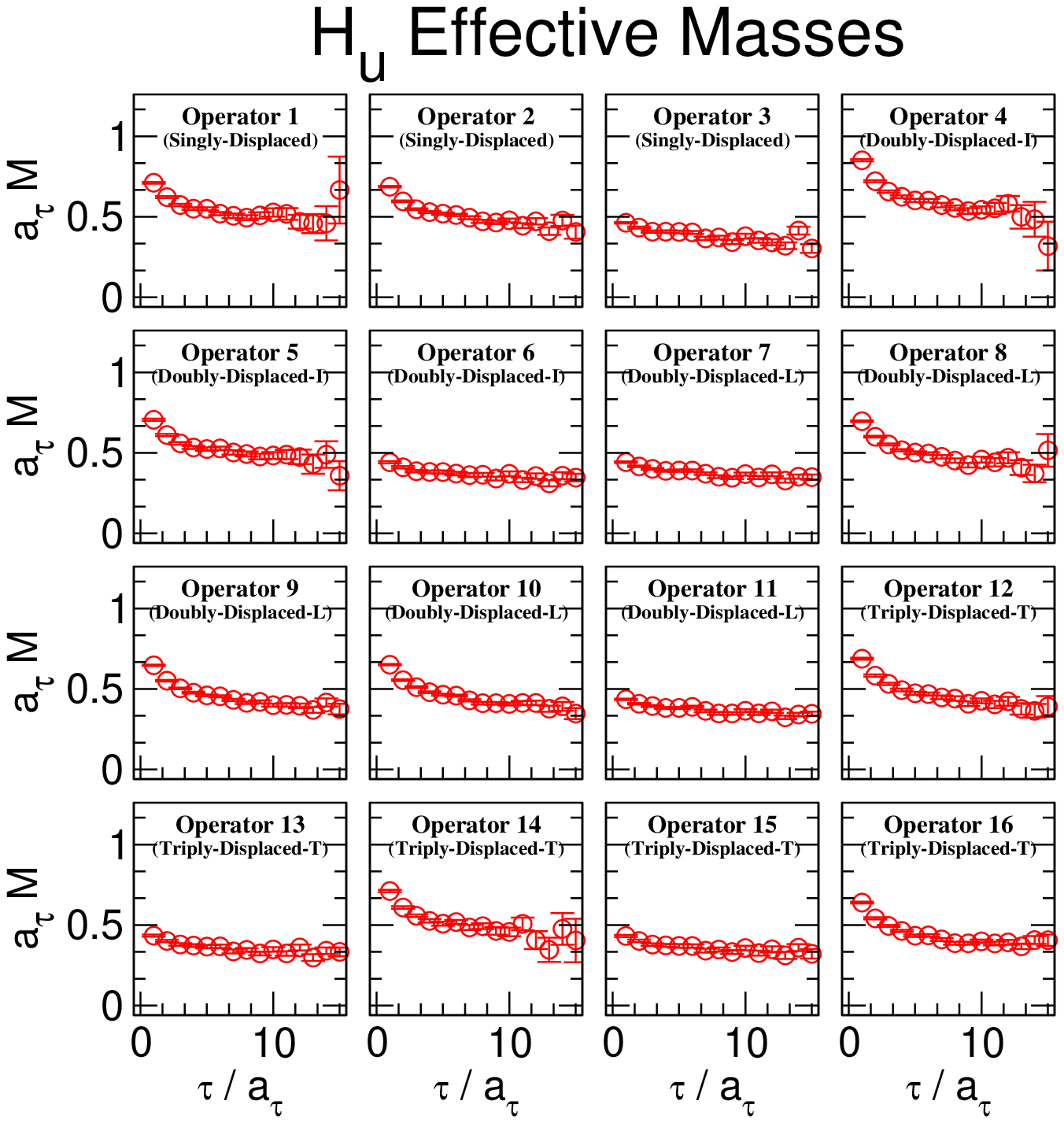}
  \caption{The effective masses for the final sixteen $H_{u}$ operators selected by our pruning process.} 
  \label{fig:meff_final-Hu} 
\end{figure}

\begin{figure}[ht!] 
  \centering 
  \includegraphics[width=6.0in]{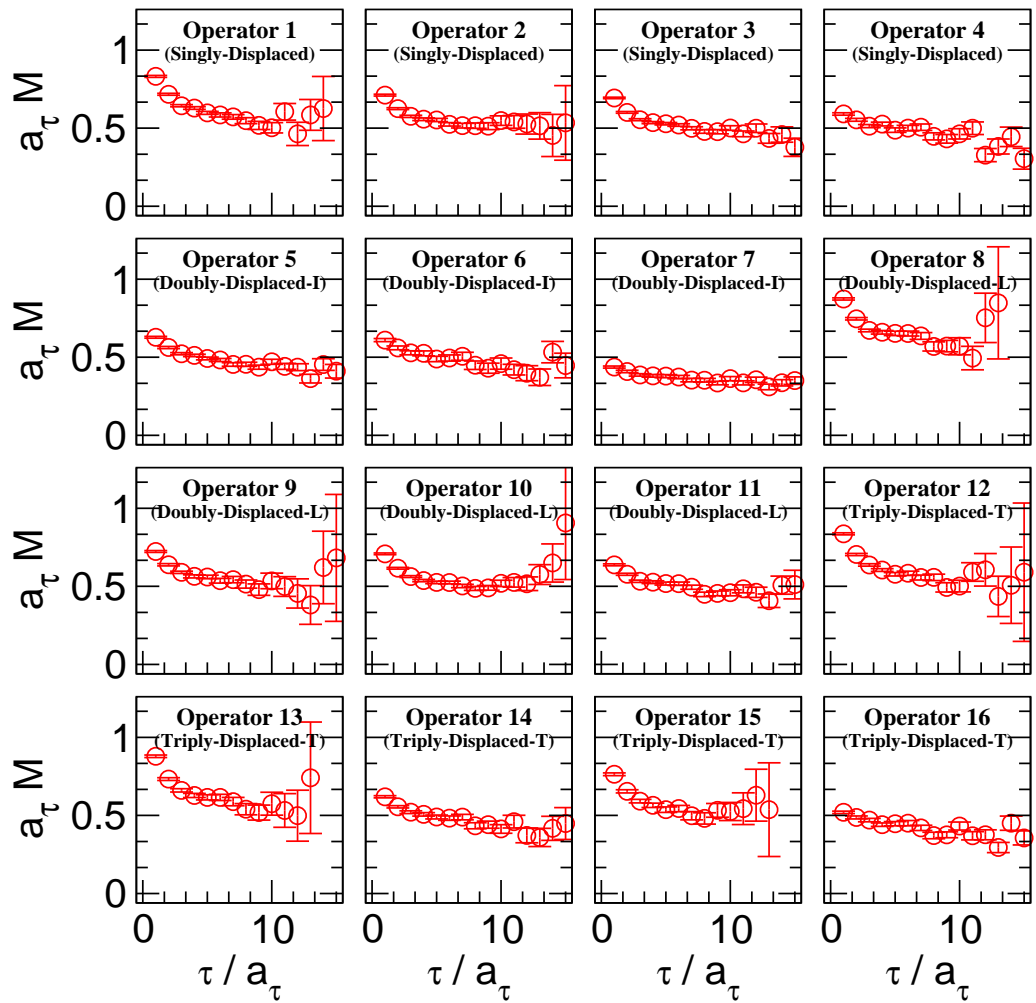}
  \caption{The effective masses for the final sixteen $G_{2u}$ operators selected by our pruning process.} 
  \label{fig:meff_final-G2u} 
\end{figure}

The condition numbers corresponding to the final selected set of sixteen candidate
operators are presented in Table~\ref{table:condition_numbers_final},
and the new operator number information for our final set of sixteen operators
is presented in Table~\ref{table:new_op_numbers}.  The operator identification
numbers used in our programs are given in Appendix~\ref{appx:operators}.
For completeness, the effective mass plots of the operators
as calculated from the diagonal correlation matrix elements are shown
in Figures~\ref{fig:meff_final-G1g},~\ref{fig:meff_final-Hg},
~\ref{fig:meff_final-G2g},~\ref{fig:meff_final-G1u},~\ref{fig:meff_final-Hu},
and~\ref{fig:meff_final-G2u}.  We also examined the principal effective plots
to ensure that our selected operators interpolated well for the excited states
(see Chapter~\ref{chap:results}).

\begin{table}[ht!]
\centering
\begin{tabular}{l|r@{.}l|c} 
\hline
\hline
 Irrep & \multicolumn{2}{c|}{$\kappa$} & $(\kappa-\kappa_{\mathrm{min}})/\kappa_{\mathrm{min}}$\\
\hline
$G_{1g}$ & $68$&$5(1.5)$ & $0\%$ \\
$G_{1u}$ & $63$&$85(68)$ & $4\%$ \\
$H_{g}$  & $53$&$57(84)$ & $6\%$ \\
$H_{u}$  & $50$&$09(67)$ & $0\%$ \\
$G_{2g}$ & $58$&$61(76)$ & $0\%$ \\
$G_{2u}$ & $56$&$5(2.3)$ & $3\%$ \\
\hline
\hline
\end{tabular}
\caption{The condition numbers $\kappa$ for the 
$16\times 16$ final operator
submatrices chosen.  The percentages denote
how much greater each condition number is than its minimum possible value. 
The operators in each odd-parity irrep are the opposite-parity partners
of the operators in the corresponding even-parity irrep.}
\label{table:condition_numbers_final}
\end{table}

\begin{table*}[ht!]
\centering
\begin{tabular}{l|rrr} 
\hline
\hline
 & \multicolumn{3}{c}{\raisebox{0mm}[5mm]{$N^+$}} \\
 Operator type      &  $G_{1g}/G_{1u}$  &  $H_g/H_u$ & $G_{2g}/G_{2u}$\\ 
\hline
 Single-Site        &     1      &     0     &     0\\
 Singly-Displaced   &     3      &     3     &     4\\
 Doubly-Displaced-I &     5      &     3     &     3\\
 Doubly-Displaced-L &     2      &     5     &     4\\
 Triply-Displaced-T &     5      &     5     &     5\\
\hline
\raisebox{0mm}[4mm]{\textbf{Total}} & \textbf{16} & \textbf{16} & \textbf{16}\\
\hline
\hline
\end{tabular}
\caption{
The numbers of $N^+$ operators of each type which are used in each correlation
matrix.
The operators selected in the $G_{1u}$, $H_u$ and $G_{2u}$ irreps 
are the same as those selected for the
$G_{1g}$, $H_g$, and $G_{2g}$ irreps, respectively.
\label{table:new_op_numbers}}
\end{table*}

\newpage

\newpage

\newpage

\newpage

\newpage

\newpage

\chapter{Baryon operator results: The nucleon spectrum}
\label{chap:results}

Now that we have selected sixteen candidate nucleon operators
$\{\overline{\mathcal{O}}_1,\overline{\mathcal{O}}_2,\cdots,
\overline{\mathcal{O}}_{16}\}$ in each $O^D_h$ irreducible representation,
we conclude this work by extracting the lowest eight energy
levels in the $G_{1g}, H_g, G_{2g}, H_{u},$ and $G_{2u}$ nucleon 
channels, and the lowest seven energy
levels in the $G_{1u}$ nucleon channel\footnote{We were unable to find
a fit range which gave a satisfactory estimate for the $8th$ excited state
in the $G_{1u}$ channel}.

\section{Fixed-coefficient correlation functions}
In each $O^D_h$ irrep channel, we evaluated the elements of a 
$16\times 16$ correlation matrix
$C(\tau)$ on $200$ quenched configurations.  
Our next step was to apply
the variational method discussed in Subsection~\ref{section:variational} to
find linear combinations of operators
$$\overline{\Theta}_i\equiv \sum_{a=1}^{16} \overline{\mathcal{O}}_a v_{ai},$$
which couple strongly to single (distinct) states of interest
and weakly to the other states.  In other words, we are looking for fixed
coefficient vectors $v_i$ such that:
\begin{eqnarray}
\langle \Theta_k(\tau)\overline{\Theta}_k(0) \rangle = 
\sum_{a,b=1}^{16}v^*_{ak}C_{ab}(\tau)v_{bk}&\equiv&v_k^\dagger C(\tau)v_k,\label{eqn:fixed_corr_mat}\\
&\approx& |\bra{k}\overline{\Theta}_k\ket{\Omega}|^2e^{-E_k\tau},\qquad
\tau_{\mathrm{min}} \le \tau \le \tau_{\mathrm{max}},\nonumber
\end{eqnarray}
where $[\tau_{\mathrm{min}},\tau_{\mathrm{max}}]$ specifies the range of
$\tau$ values
over which the single-exponential approximation is valid.  It is not 
guaranteed that such a range exists, but this simple approach served
us well in this exploratory study\footnote{We found that two-exponential
fits to our rotated correlation matrix elements $v^\dagger_iC(\tau)v_j$ 
had inferior stability and errors when compared to our single-exponential 
fit results.  The reason for this seems to be our effective removal of 
contamination from the correlation matrix (e.g. see Figure~\ref{fig:fits-G1g})}.  
A more sophisticated approach would 
attempt to take
into account some of the exponential contamination terms neglected by this 
approximation.

  If we solve
\begin{eqnarray*}
\left[C^{-1/2}(\tau_0)C(\tau)C^{-1/2}(\tau_0)\right]u_i(\tau,\tau_0) &=&
\lambda_i(\tau,\tau_0)u_i(\tau,\tau_0),
\end{eqnarray*}
for $u_i(\tau,\tau_0)$ and $\lambda_i(\tau,\tau_0)$, where $\tau_0$ is
a fixed reference time (we chose $\tau_0=a_\tau$ in this work), then
the principal correlation functions $\lambda_i(\tau,\tau_0)$ behave as:
\begin{eqnarray*}
\lim_{\tau\to\infty}\lambda_k(\tau,\tau_0)&=&e^{-E_k(\tau-\tau_0)}(1+O(e^{-\Delta_k(\tau-\tau_0)}),\\
\Delta_k&\equiv& \min_{l\neq k}|E_l-E_k|,
\end{eqnarray*}
where the $E_k$ are the energies of the first $n$ states accessible from the 
vacuum by the action of our trial candidate operators 
$\{\overline{\mathcal{O}}_a\}$.
The eigenvectors $u_i(\tau,\tau_0)$ may be chosen to be orthonormal, in
which case we may write
the principal correlation functions as:
\begin{equation}
\lambda_i(\tau,\tau_0)=u^\dagger_i(\tau,\tau_0)\left[C^{-1/2}(\tau_0)C(\tau)C^{-1/2}(\tau_0)
\right] u_i(\tau,\tau_0).\label{eqn:prin_corr_mat}
\end{equation}
Comparing Eqn.~\ref{eqn:prin_corr_mat} to Eqn.~\ref{eqn:fixed_corr_mat}, we see
that a reasonable choice for the fixed coefficient vector $v_{i}$ is:
$$v_{i}=C^{-1/2}(\tau_0)u(\tau^*,\tau_0),$$
where $\tau^*$ is some fixed time.
We found that our estimates for $v_i$
stabilized around $\tau^*=3a_\tau$, which signaled the reduction of
contamination coming from the higher-lying levels.  
The chosen value of $\tau^*=3a_\tau$ was also small enough to ensure that our
estimate for the fixed coefficient vectors $v_i$ was based on data having
a good signal-to-noise ratio\footnote{The signal-to-noise ratio of Monte
Carlo estimates of $C(\tau)$ decays
exponentially with $\tau$ for baryon correlation
functions~\cite{lepage:signal_to_noise}}.

Our {\em fixed-coefficient} correlation functions were therefore given by the
diagonal elements of the rotated correlation matrix, 
which is denoted by $\tilde{C}(\tau)$:
\begin{eqnarray*}
\tilde{C}_{kk}(\tau)&\equiv& \langle \Theta_k(\tau)\overline{\Theta}_k(0)\rangle\\
&=&v_k^\dagger C(\tau)v_k\\
&=&u^\dagger_k(\tau=3a_\tau,\tau_0=a_\tau)\left[C^{-1/2}(a_\tau)C(3
a_\tau)C^{-1/2}(a_\tau)\right]u_k(\tau=3a_\tau, \tau_0=a_\tau).
\end{eqnarray*}

These fixed-coefficient correlation functions can be used to define 
fixed-coefficient effective mass functions
$$a_\tau \tilde{M}_k(\tau)=\ln\left[\frac{\tilde{C}_{kk}(\tau)}{\tilde{C}_{kk}(\tau+a_\tau)}\right]$$

\section{Fitting range}
Our next task was to find the range
$[\tau_{\mathrm{min}}, \tau_{\mathrm{max}}]$ (if any) within 
which the fixed-coefficient correlation functions $\tilde{C}_{kk}(\tau)=v_k^\dagger
C(\tau)v_k$ behave as single exponentials:
$$\tilde{C}_{kk}(\tau) \approx |\bra{k}\overline{\Theta}_k\ket{\Omega}|^2 E^{-E_k\tau},$$
and within which the off-diagonal elements of the rotated correlation matrix
$\tilde{C}_{ij}(\tau)=v_i^\dagger C(\tau)v_j$ are consistent with zero:
$$\tilde{C}_{ij}(\tau) \approx 0,\qquad i\neq j.$$
The vanishing of the off-diagonal elements tells us that any plateaus we
observe in our effective mass plots (within that range) correspond to 
orthogonal states.

\subsection{Diagonal elements}
In Chapter~\ref{chap:correlators}, we discussed the fact that correlation
functions
will receive contamination from the backward propagating states.  Because 
chiral
symmetry is spontaneously broken, the backward propagating states will have different masses
than the forward propagating states:
\begin{eqnarray*}
\tilde{C}_{kk}^{(g)}(\tau)&=& \tilde{C}_{kk}^{(u)}(T-\tau),\mbox{
but}\\
\tilde{C}^{(g)}_{kk}(\tau)&\neq& \tilde{C}_{kk}^{(g)}(T-\tau),
\end{eqnarray*}
and similarly for the odd-parity correlation functions.  $T=a_\tau\,N_\tau$
denotes the temporal extent of our lattice.

  A sophisticated fitting approach would
seek to fit the elements of $\tilde{C}(\tau)$ and $\tilde{C}(T-\tau)$ 
simultaneously using independent
parameters.  The collaboration is currently working to develop such a fitting
package to use with the production run data.  
For this work, 
we performed single-exponential fits to the diagonal elements of the
rotated correlation matrix, neglecting the backward propagating state
effects.

In order to get a rough idea of when the backward propagating state contamination
becomes significant, we can plot the effective mass function for
a {\em trial} correlation function of the form:
$$C_{\mathrm{trial}}(\tau)=e^{-m\tau}+e^{-m(T-\tau)}.$$
This simple form assumes the same mass for both the forward and backward 
propagating
states,
and a coefficient of one for both terms.

\begin{figure}[ht!] 
  \centering 
  \includegraphics[width=3in]{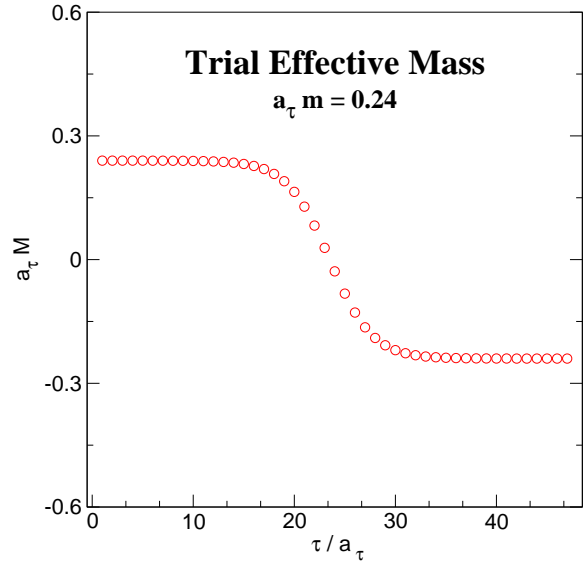}
  \caption{The effective mass function for the trial correlation function
$C(\tau)=e^{-m\tau}+e^{-m(T-\tau)}$.
The number of lattice sites in the 
temporal direction is $N_\tau=48$, and the lowest baryon state is
taken to be $a_\tau m=0.24$. The backward propagating state significantly
contaminates the effective mass function at times greater than 
$\tau \approx 20a_\tau$.}
  \label{fig:meff_trial} 
\end{figure}

The associated effective mass is:
\begin{eqnarray}
a_\tau M(\tau)&=&\ln\left[\frac{C_{\mathrm{trial}}(\tau)}{C_{\mathrm{trial}}(\tau+a_\tau)}\right],\\
&=&\ln\left[\frac{e^{-m \tau}+e^{-m(T-\tau)}}{e^{-m(\tau+a_\tau)}+e^{-m(T-(\tau+a_\tau))}}\right],\\
&=&a_\tau m+\ln\left[\frac{1+e^{-a_\tau m(N_\tau-2\tau/a_\tau)}}{1+e^{-a_\tau m(N_\tau-2(\tau/a_\tau+1)}}\right],
\end{eqnarray}
and is plotted in Figure~\ref{fig:meff_trial}.
For our lattice, $N_\tau=48$ and the lowest level seen in our preliminary
 principal effective mass plots belonged to a $G_{1g}$ state having $a_\tau m\approx 0.24$
(identified as the proton).  From Figure~\ref{fig:meff_trial} we conclude that in the (worst case) scenario 
in which our backward propagating state is the proton, we will begin to
see backward propagating state contamination at 
$$\tau_{\mathrm{max}}\approx 20a_\tau,$$
a value which is comparable to the threshold at which our signal-to-noise
ratio has degraded below acceptable levels for a fit (e.g. see
Figure~\ref{fig:fits-G1g}).  

\subsection{Off-diagonal elements}
The fixed-coefficient correlation functions $\tilde{C}_{kk}(\tau)$ differ
from the principal correlation functions $\lambda_k(\tau,\tau_0)$ because we 
are only solving the eigenvalue problem once, at $\tau^*=3a_\tau$.  Thus, we are
only guaranteed orthogonality at $\tau^*=3a_\tau$:
$$\tilde{C}_{ij}(\tau^*)=0\quad\forall\;\; i\neq j.$$
The lack of orthogonality for $\tau > \tau^*$, no matter how small, means that
our fixed-coefficient correlation functions will eventually relax to the
lowest excited state
$$\tilde{C}_{kk}(\tau) \stackrel{\tau \gg \tau^*}{\to} |c_1|^2\exp(-E_1\tau).$$

\begin{figure}[ht!] 
  \centering 
  \includegraphics[width=3in]{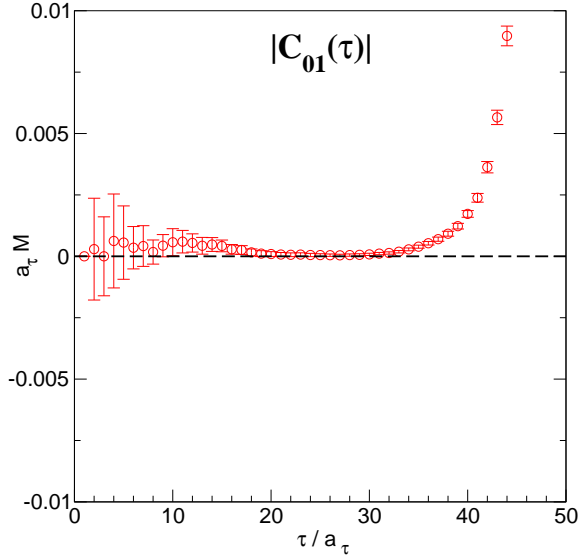}
  \caption{An example of an off-diagonal element of the rotated correlation 
matrix
$\tilde{C}_{ij}(\tau)=v^{\dagger}_iC(\tau)v_j$ having $i=0$ and $j=1$.  
Because we are using fixed 
coefficients to rotate the matrix, we are unable to maintain the orthogonality
of the states at all times $\tau$.  Here we see that we can safely work
with the fixed-coefficient correlation functions until
$\tau\approx 30a_\tau$.  After that, they all relax to the lowest excited state:
$\tilde{C}_{kk}(\tau)\to |c_1|^2\exp(-E_1\tau)$.}
  \label{fig:meff_offdiag} 
\end{figure}

Consequently, we seek a range of $\tau$ values over which
$$\vert \tilde{C}_{ij}(\tau)\vert^2=0\quad\forall\;\;i\neq j$$
within errors.  This would imply that the $v_i$ coefficient vectors are 
good estimates for the coefficients of operators which excite orthogonal
states, and would verify that we are not excluding too much information by ignoring 
the off-diagonal elements in our fits.  We found that this is indeed the
case for $\tau$ values less than $\tau\approx 30a_\tau$. An example
of one such off-diagonal element is shown in Figure~\ref{fig:meff_offdiag}.

Considering the backward propagating state contamination along with
the orthogonality of the fixed-coefficient correlation functions,
we decided to require all of our fit ranges to lie within
$$\tau_{\mathrm{min}}=a_\tau,\qquad \tau_{\mathrm{max}}=20a_\tau.$$
In performing our fits (discussed below), we found that noise was the limiting
factor for signal quality past $\tau\approx17a_\tau$.

\section{Fit method}
To estimate the value of an energy level $a_\tau E_k$, we fit directly
to the fixed-coefficient correlation function, not the principal
correlation function or any effective mass 
function\footnote{We denote our energy levels by $a_\tau E_k$ because some
may be multi-hadron states.}.
Fitting to correlation functions directly is simpler and introduces less 
noise into the process.  

\subsection{Correlated $\chi^2$ fitting}
In this study, we performed a single-exponential correlated $\chi^2$
fit to the fixed-coefficient correlation functions.  We fit each
fixed-coefficient correlation function $\tilde{C}_{kk}(\tau)$ independently.
For simplicity in notation, we consider one such fixed-coefficient
correlation function and denote it by $C(\tau)$ for the remainder of this
subsection.  We note that $C(\tau)$ is a function in this subsection, not
a matrix.

The particular form of our fitting function was
$$C_{\mathrm{fit}}(\tau,A,a_\tau E_{\mathrm{fit}})=A \exp(-a_\tau E_{\mathrm{fit}} (\tau/a_\tau) ),$$
where $A$ and $a_\tau E_{\mathrm{fit}}$ are the two fit parameters.
This simple form does not take into account the periodicity of the lattice; we
 required our fit ranges to lie within $\tau_{\mathrm{min}}=a_\tau$ and
$\tau_{\mathrm{max}}=20a_\tau$ to avoid the effects of backward propagating 
state contamination.

In our correlated $\chi^2$ fit, $\chi^2$
is given by
\begin{equation}
\chi^2(A,a_\tau E_{\mathrm{fit}})\equiv\sum_{\tau,\tau'}[C(\tau)-C_{\mathrm{fit}}(\tau,A,a_\tau 
E_{\mathrm{fit}})]\sigma^{-1}_{\tau,\tau'}[C(\tau')-C_{\mathrm{fit}}(\tau',A,
a_\tau E_{\mathrm{fit}})]
\label{eqn:chi_sq}
\end{equation}

where $\sigma^{-1}_{\tau,\tau'}$ is the inverse of the covariance
matrix:
\begin{equation}
\sigma_{\tau,\tau'}\equiv\frac{1}{N(N-1)}\sum_{n=1}^{200}[C_n(\tau)-\bar{C}(\tau)][C_n(\tau')-\bar{C}(\tau')],
\label{eqn:covar_matrix}
\end{equation}
where $C_n(\tau)$ is the value of $C(\tau)$ evaluated the $n^{th}$ 
configuration, and 
$$\bar{C}(\tau)\equiv \frac{1}{200}\sum_{n=1}^{200} C_n(\tau).$$
The covariance matrix $\sigma_{\tau,\tau'}$ measures the amount of correlation between measurements at $\tau$ 
and $\tau'$.
 
We used the Levenberg-Marquardt~\cite{press:numerical} non-linear fitting 
algorithm to find the values of $A$ and $a_\tau E_{\mathrm{fit}}$ which
minimized our $\chi^2$.  
To estimate the uncertainty in our fit values, we
first generated $2048$ bootstrap samples of size $200$ by
sampling randomly with replacement from the original set of $200$ rotated 
correlation matrices.  We repeated our single exponential fit (using the same
fit range) on each bootstrap sample, and used the standard deviation of the
resulting set of fit parameters as our uncertainty.  We also
calculated upper and lower limits for our plots using the
percentile method~\cite{efron:jackknife}.  The upper limit
of a fit parameter is chosen such that 84.1\% of the bootstrap sample values lie
below it.  Similarly, the lower limit is chosen corresponding to a 
percentile of 15.9\%. 
If the fit parameters are distributed normally across bootstrap samples, then
the upper and lower limits correspond to the estimate plus and minus
one standard deviation, respectively.  We found no significant asymmetry
in our fit error bars.

For each fixed-coefficient correlation function separately, we chose
our fit range to make the $\chi^2$ per degree of freedom 
close to $1.0$, and to make the
quality factor $Q$ as large as possible ($>0.1$ was desirable).

We examined the sensitivity of our fits as we increased or decreased
the fit range as a check of the quality of our results.
We also explored the possibility of adding a second exponential
term to model some of the high-lying contamination, but those fits
performed poorly compared to the single exponential method.

\subsection{Fitting walk-through}
In summary, the exact method we used to analyze our correlation matrices for
each irrep in $G_{1g}, H_g, G_{2g}, G_{1u}, H_u, G_{2u}$ is as follows
(using $\tau_0=\tau_{\mathrm{min}}=a_\tau,\ \tau^*=3a_\tau,\ \tau_{\mathrm{max}}= 20a_\tau,\ T=a_\tau\,N_\tau,$ and $N_\tau=48$):

\begin{enumerate}
\item Average the correlation matrix elements over all $200$ configurations
to form the best estimates for the elements of $C(\tau)$ using the improved
estimation method discussed in Chapter~\ref{chap:monte_carlo}:
$$C^{(p)}_{ab}(\tau)=\frac{1}{200}\sum_{n=1}^{200} \frac{1}{4}\left[C^{(p)}_{ab,n}(\tau)+C^{(1-p)}_{ba,n}(T-\tau)+C^{(p)*}_{ba,n}(\tau)+C^{(1-p)*}_{ab,n}(T-\tau)\right],$$
where $C^{(p)}_{ab,n}(\tau)$ is a correlation matrix element measured on
the $n^{th}$ configuration generated by the Markov chain (see Chapter
~\ref{chap:monte_carlo}), and $p=0,1$ for even and odd-parity, respectively.
\item Solve $C^{-1/2}(\tau_0)C(\tau^*)C^{-1/2}(\tau_0)\,u_i=\lambda_i\, u_i$ 
for the eigenvector $u_i$.
\item Calculate $v_i=C^{-1/2}(\tau_0)u_i$.
\item For each configuration $n=1,\cdots,200$, store the 
elements of the rotated correlation matrix $\tilde{C}$ for all $\tau$ values:
$$\tilde{C}_{ij,n}(\tau)\equiv \sum_{a,b=1}^{16}v^*_{ia} C_{ab,n}(\tau) v_{bj}.$$ 
\item Calculate the mean and jackknife error of the off-diagonal elements
of the rotated correlation matrix $\tilde{C}(\tau)$
and verify that they are consistent with zero for 
$\tau_{\mathrm{min}}\le\tau\le\tau_{\mathrm{max}}$.
\item Perform single-exponential fits to
each fixed-coefficient correlation function $\tilde{C}_{kk}(\tau)$.  In 
these fits, the covariance matrix is calculated using Eqn.~\ref{eqn:covar_matrix}
on the sample of rotated correlation matrices and then inverted for use
in the expression for $\chi^2$ given in Eqn.~\ref{eqn:chi_sq}.  The fit parameters are varied until
$\chi^2$ is minimized.
In each fit, choose
a fit range such that the $\chi^2$ per degree of freedom is near
$1.0$, the quality factor $Q$ is as large as possible ($>0.1$ is desirable), and 
the fit parameter $a_\tau E_{k,\mathrm{fit}}$ is stable under small variations
in the fit range.  Store the value of $a_\tau E_{k,\mathrm{fit}}$ as the
estimate of the level value. 
\item To estimate the uncertainty in each level estimate, generate
a large number (we used $2048$) of bootstrap samples of size $200$ by
sampling randomly with replacement from the original set of $200$ rotated 
correlation matrices.  Perform a single-exponential fit using the same
fit range used for the fit on the original sample.  Once again, the covariance
matrix is estimated using Eqn.~\ref{eqn:covar_matrix}, this time on the members
of the bootstrap sample.  Each bootstrap sample yields an estimate of the
level value $a_\tau E_{k,\mathrm{fit}}$.
\item Store the standard deviation of the bootstrap estimates of 
$a_\tau E_{k,\mathrm{fit}}$ as the uncertainty in the level estimate. 
\item Examine the fixed-coefficient effective masses, the level estimates, and
the principal effective masses to verify that the fits are reasonable and 
not adversely affected by outlying data.
\end{enumerate}

\section{Lattice nucleon spectrum results}
\begin{figure}[ht!] 
  \centering 
  \includegraphics[height=8.0in]{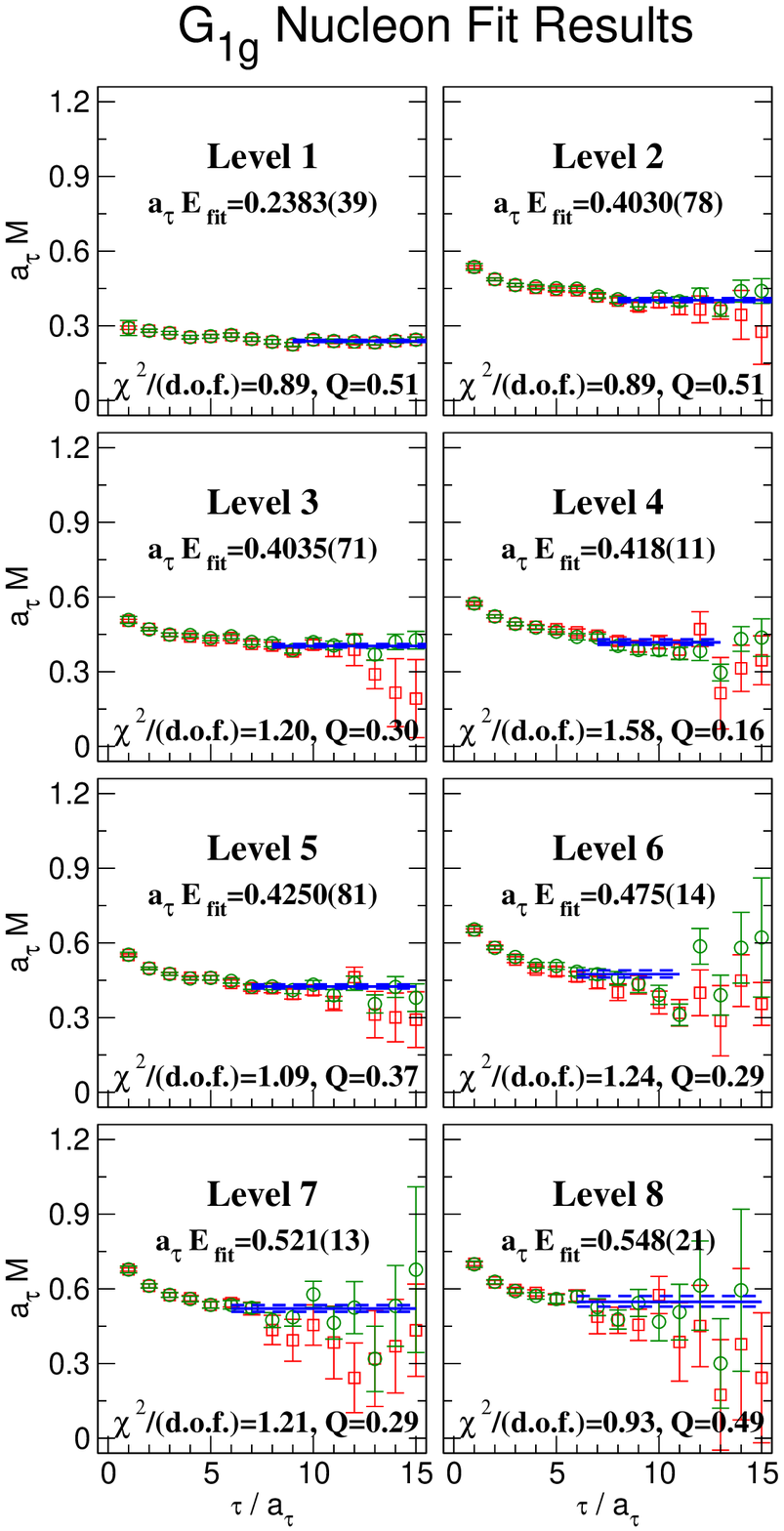}
  \caption{The lowest eight levels for the $G_{1g}$ nucleon channel. 
    The green circles are the fixed coefficient effective masses,
and the red squares are the principal effective masses.  The fits were made
to the fixed coefficient correlation functions and are denoted by the blue lines.} 
  \label{fig:fits-G1g} 
\end{figure}

\begin{figure}[ht!] 
  \centering 
  \includegraphics[height=8.0in]{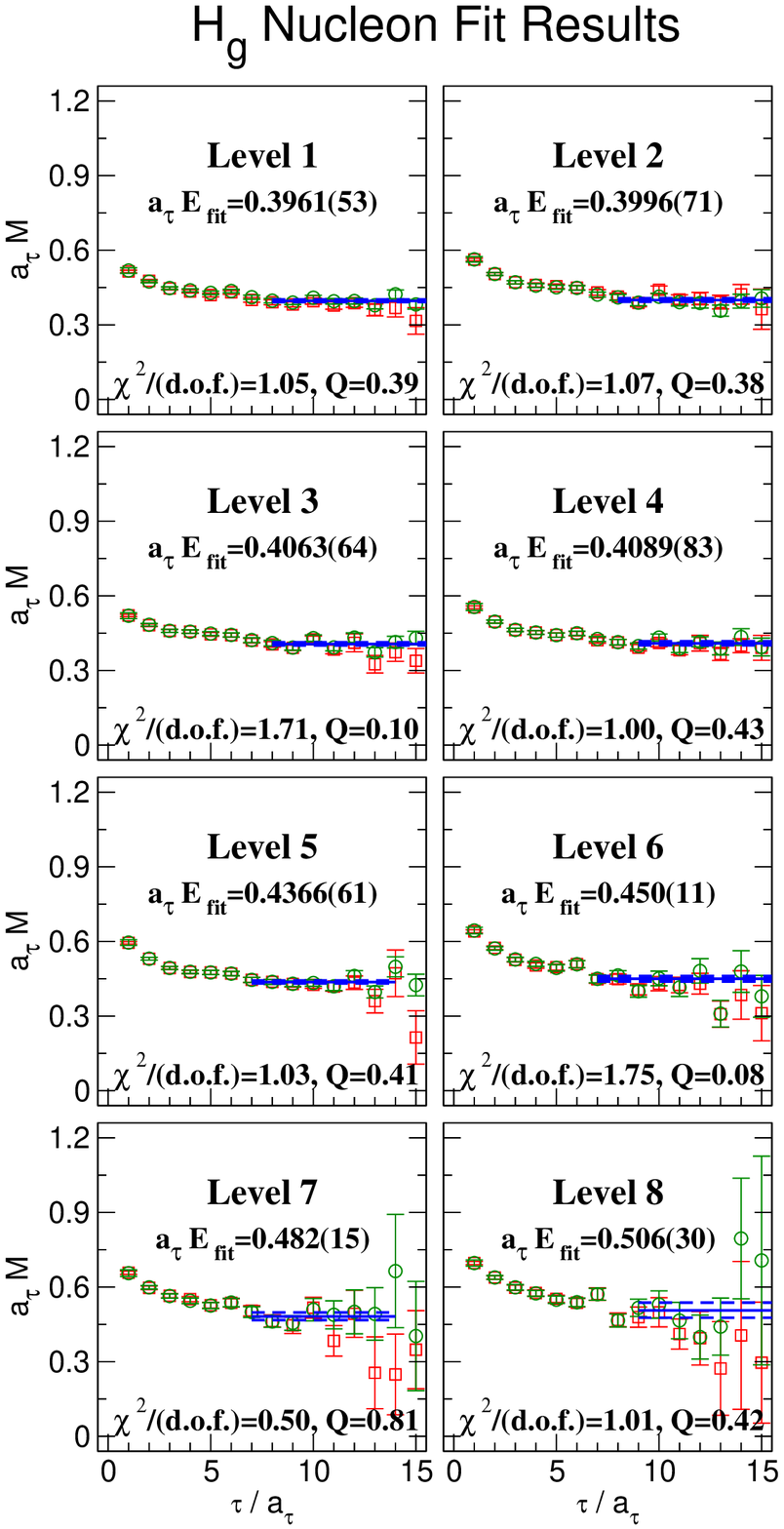}
  \caption{The lowest eight levels for the $H_{g}$ nucleon channel. 
    The green circles are the fixed coefficient effective masses,
and the red squares are the principal effective masses.  The fits were made
to the fixed coefficient correlation functions and are denoted by the blue lines.} 
  \label{fig:fits-Hg} 
\end{figure}

\begin{figure}[ht!] 
  \centering 
  \includegraphics[height=8.0in]{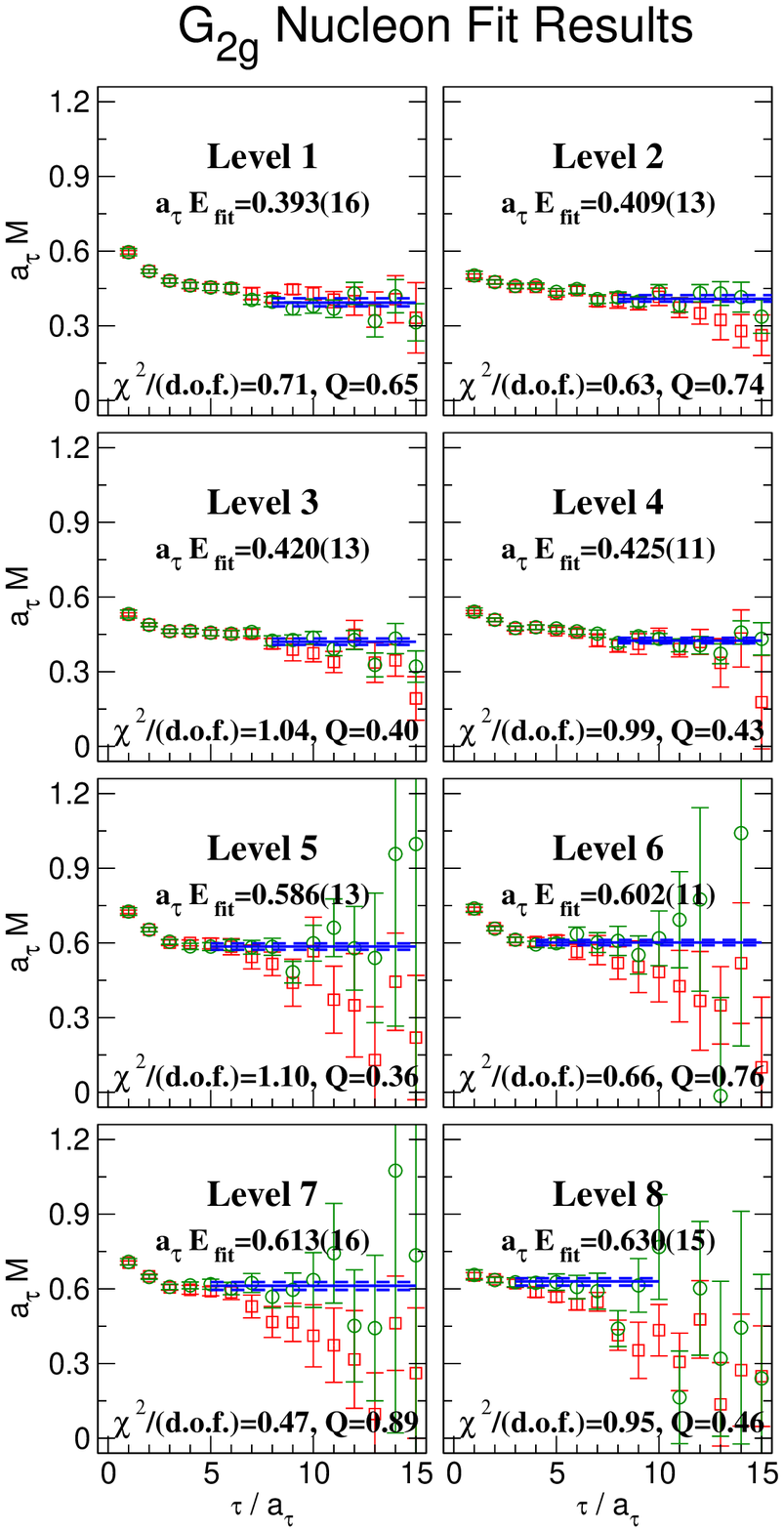}
  \caption{The lowest eight levels for the $G_{2g}$ nucleon channel. 
    The green circles are the fixed coefficient effective masses,
and the red squares are the principal effective masses.  The fits were made
to the fixed coefficient correlation functions and are denoted by the blue lines.} 
  \label{fig:fits-G2g} 
\end{figure}

\begin{figure}[ht!] 
  \centering 
  \includegraphics[height=8.0in]{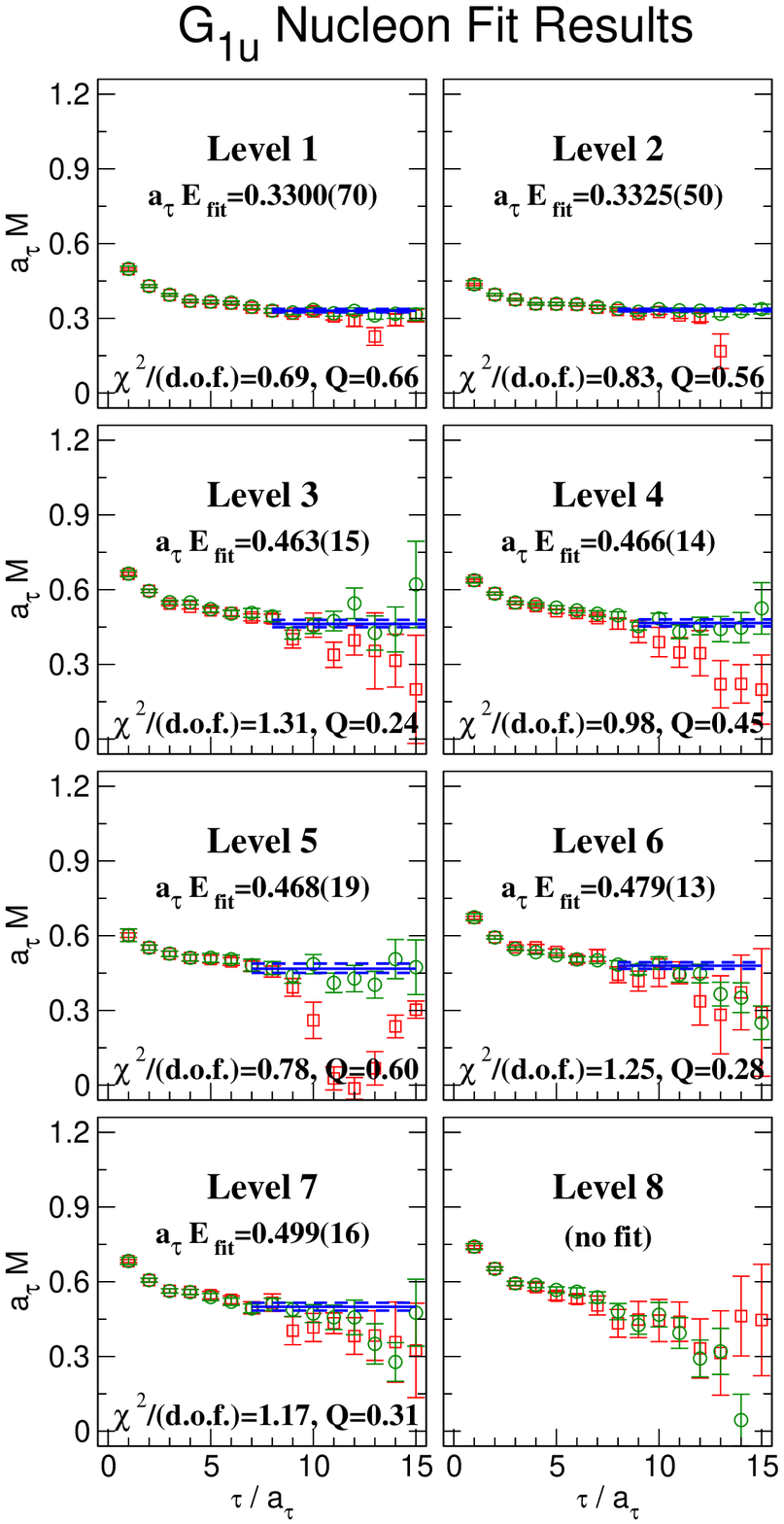}
  \caption{The lowest eight levels for the $G_{1u}$ nucleon channel. 
    The green circles are the fixed coefficient effective masses,
and the red squares are the principal effective masses.  The fits were made
to the fixed coefficient correlation functions and are denoted by the blue lines.} 
  \label{fig:fits-G1u} 
\end{figure}

\begin{figure}[ht!] 
  \centering 
  \includegraphics[height=8.0in]{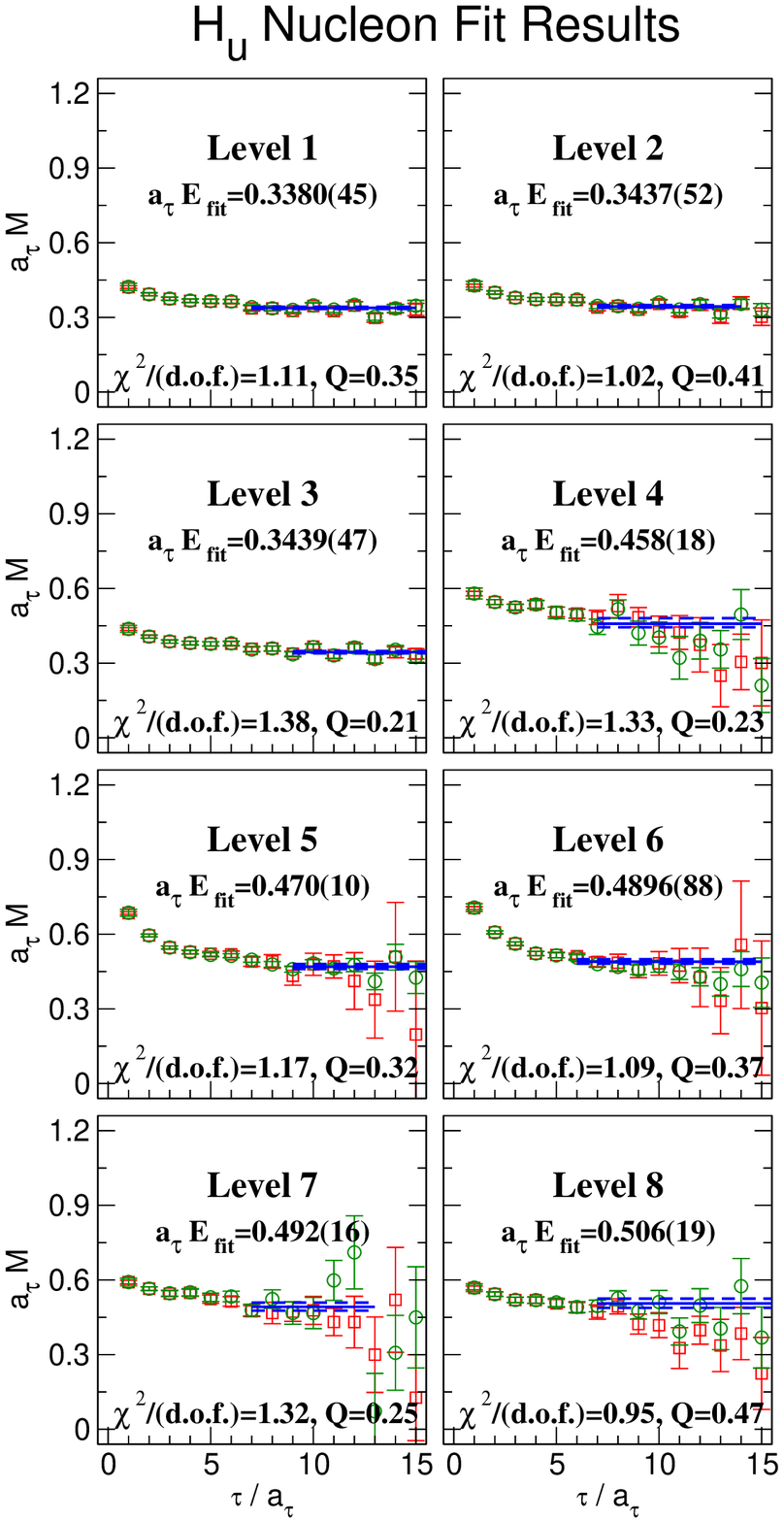}
  \caption{The lowest eight levels for the $H_{u}$ nucleon channel. 
    The green circles are the fixed coefficient effective masses,
and the red squares are the principal effective masses.  The fits were made
to the fixed coefficient correlation functions and are denoted by the blue lines.} 
  \label{fig:fits-Hu} 
\end{figure}

\begin{figure}[ht!] 
  \centering 
  \includegraphics[height=8.0in]{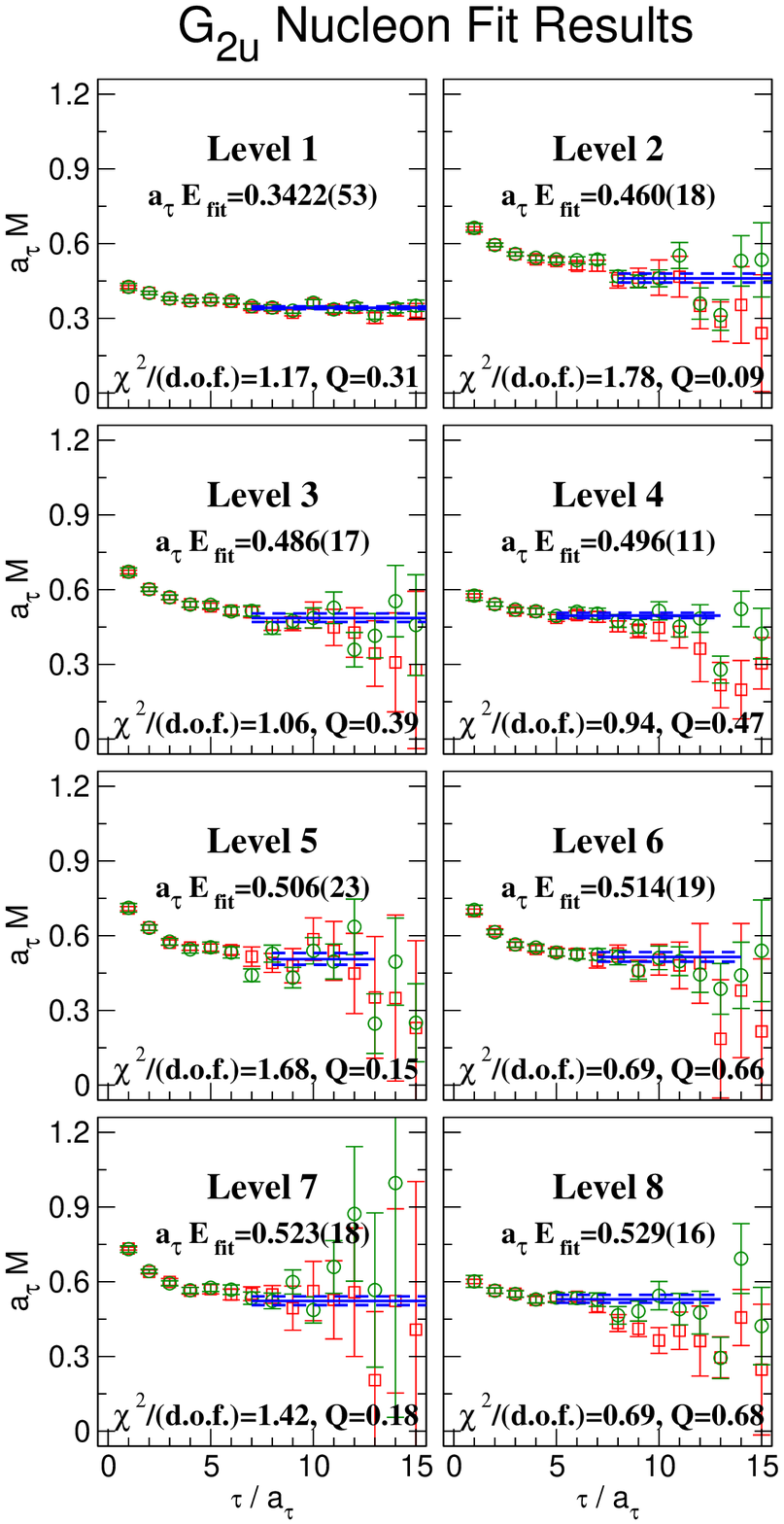}
  \caption{The lowest eight levels for the $G_{2u}$ nucleon channel. 
    The green circles are the fixed coefficient effective masses,
and the red squares are the principal effective masses.  The fits were made
to the fixed coefficient correlation functions and are denoted by the blue lines.} 
  \label{fig:fits-G2u} 
\end{figure}

\begin{figure}[ht!] 
  \centering 
  \includegraphics[width=6.0in]{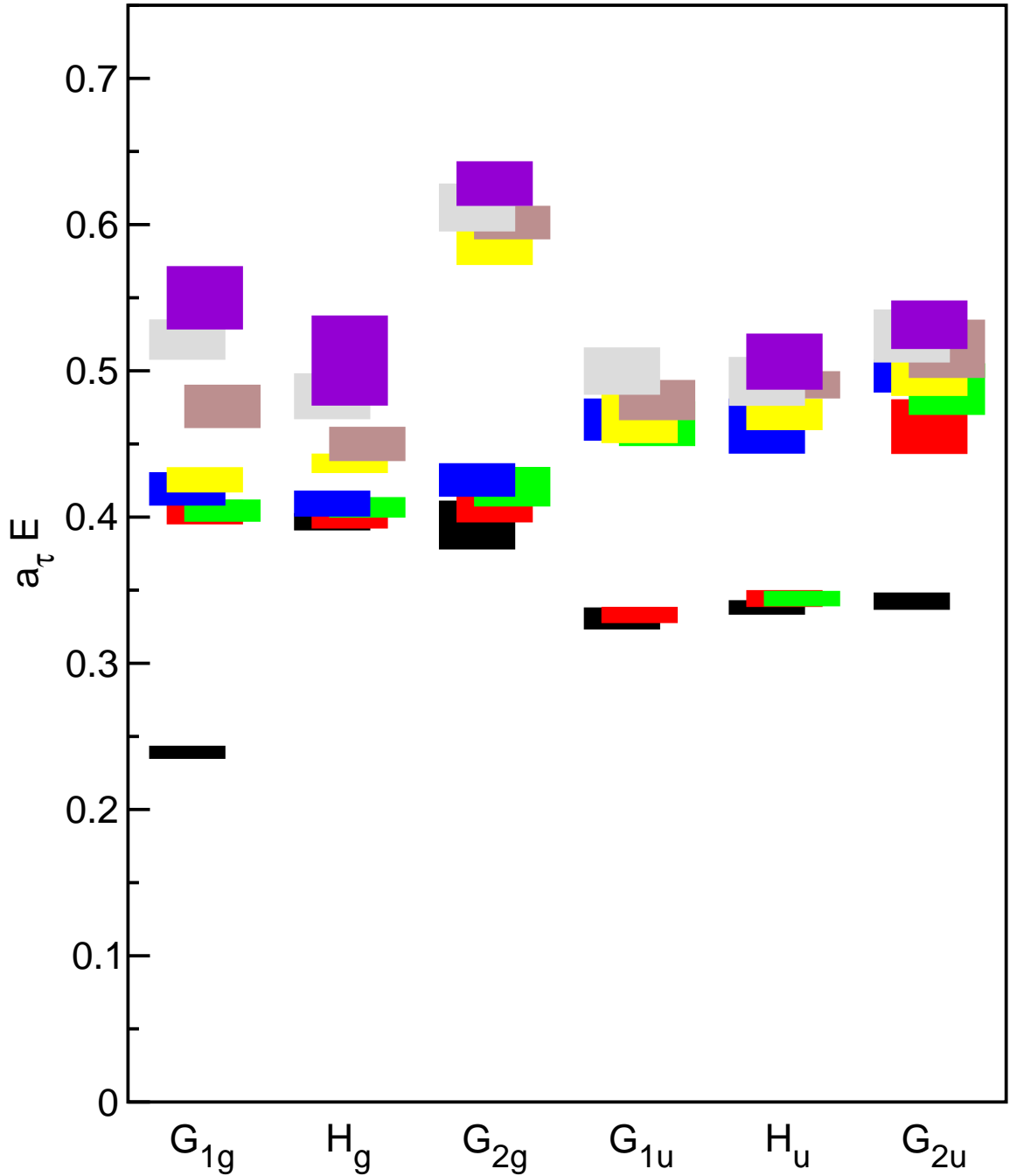}
  \caption{The low-lying $I$=1/2, $I_3$=+1/2 nucleon spectrum
 extracted from 200 quenched configurations on 
a $12^3\times 48$ anisotropic lattice using the Wilson action with 
$a_s \sim 0.1$ fm and $a_s/a_\tau \sim 3.0$.  
The vertical height of each box indicates the statistical
uncertainty in that estimate.  
Our pion mass for this study was $a_\tau M_\pi=0.1125(26)$, or approximately 
700 MeV.  A different color was chosen for each level in a symmetry channel
to help the reader discern among the different levels.}
  \label{fig:lattice_nucleon_spectrum} 
\end{figure}

This lattice study was performed using 200 quenched configurations on 
a $12^3\times 48$ anisotropic lattice using the Wilson action with 
$a_s \sim 0.1$ fm and $a_s/a_\tau \sim 3.0$.  The quark mass was such that the
mass of the pion was approximately 700 MeV.
We were able to perform satisfactory fits yielding the lowest-lying eight
states in each lattice spin-parity channel, with the exception
of the highest (eighth) $G_{1u}$ level.  Our spectrum fit results
are tabulated in Tables~\ref{table:fit_results_g} and~\ref{table:fit_results_u},
and plotted in Figures~\ref{fig:fits-G1g},~\ref{fig:fits-Hg},~\ref{fig:fits-G2g},
~\ref{fig:fits-G1u},~\ref{fig:fits-Hu}, and~\ref{fig:fits-G2u}.
The nucleon spectrum levels extracted from these fits are illustrated in
Figure~\ref{fig:lattice_nucleon_spectrum}.

\clearpage

\begin{table}
\centering
\begin{tabular}{c|ccr@{.}lcc}
\hline
\hline
Level & $\tau_{\mathrm{min}}$ & $\tau_{\mathrm{max}}$ & \multicolumn{2}{c}{$a_\tau E_{\mathrm{fit}}$} & 
$\chi^2/\mbox{(d.o.f.)}$ & $Q$\\
\hline
$G_{1g} \quad 1$ & $9$ & $17$ & $0$&$2383(39)$ & $0.89$ & $0.51$\\
$G_{1g} \quad 2$ & $8$ & $16$ & $0$&$4030(78)$ & $0.89$ & $0.51$\\
$G_{1g} \quad 3$ & $8$ & $16$ & $0$&$4035(71)$ & $1.20$ & $0.30$\\
$G_{1g} \quad 4$ & $7$ & $13$ & $0$&$418(11)$ & $1.58$ & $0.16$\\
$G_{1g} \quad 5$ & $7$ & $15$ & $0$&$4250(81)$ & $1.09$ & $0.37$\\
$G_{1g} \quad 6$ & $6$ & $11$ & $0$&$475(14)$ & $1.24$ & $0.29$\\
$G_{1g} \quad 7$ & $6$ & $15$ & $0$&$521(13)$ & $1.21$ & $0.29$\\
$G_{1g} \quad 8$ & $6$ & $15$ & $0$&$548(21)$ & $0.93$ & $0.49$\\
\hline
$H_{g} \quad 1$ & $8$ & $16$ & $0$&$3961(53)$ & $1.05$ & $0.39$\\
$H_{g} \quad 2$ & $8$ & $16$ & $0$&$3996(71)$ & $1.07$ & $0.38$\\
$H_{g} \quad 3$ & $8$ & $16$ & $0$&$4063(64)$ & $1.71$ & $0.10$\\
$H_{g} \quad 4$ & $9$ & $17$ & $0$&$4089(83)$ & $1.00$ & $0.43$\\
$H_{g} \quad 5$ & $7$ & $14$ & $0$&$4366(61)$ & $1.03$ & $0.41$\\
$H_{g} \quad 6$ & $7$ & $16$ & $0$&$450(11)$ & $1.75$ & $0.08$\\
$H_{g} \quad 7$ & $7$ & $14$ & $0$&$482(15)$ & $0.50$ & $0.81$\\
$H_{g} \quad 8$ & $9$ & $17$ & $0$&$506(30)$ & $1.01$ & $0.42$\\
\hline
$G_{2g} \quad 1$ & $8$ & $15$ & $0$&$393(16)$ & $0.71$ & $0.65$\\
$G_{2g} \quad 2$ & $8$ & $16$ & $0$&$409(13)$ & $0.63$ & $0.74$\\
$G_{2g} \quad 3$ & $8$ & $15$ & $0$&$420(13)$ & $1.04$ & $0.40$\\
$G_{2g} \quad 4$ & $8$ & $15$ & $0$&$425(11)$ & $0.99$ & $0.43$\\
$G_{2g} \quad 5$ & $5$ & $15$ & $0$&$586(13)$ & $1.10$ & $0.36$\\
$G_{2g} \quad 6$ & $4$ & $15$ & $0$&$602(11)$ & $0.66$ & $0.76$\\
$G_{2g} \quad 7$ & $5$ & $15$ & $0$&$613(16)$ & $0.47$ & $0.89$\\
$G_{2g} \quad 8$ & $3$ & $10$ & $0$&$630(15)$ & $0.95$ & $0.46$\\
\hline
\hline
\end{tabular}
\caption{The final spectrum results for the even-parity channels.  
These results are based on 200 quenched configurations on 
a $12^3\times 48$ anisotropic lattice using the Wilson action with 
$a_s \sim 0.1$ fm and $a_s/a_\tau \sim 3.0$.  
Our pion mass for this study was $a_\tau M_\pi=0.1125(26)$ 
(see Figure~\ref{fig:pion}), or approximately 700 MeV.} 
\label{table:fit_results_g}
\end{table}

\clearpage

\begin{table}
\centering
\begin{tabular}{c|ccr@{.}lcc}
\hline
\hline
Level & $\tau_{\mathrm{min}}$ & $\tau_{\mathrm{max}}$ & \multicolumn{2}{c}{$a_\tau E_{\mathrm{fit}}$} & 
$\chi^2/\mbox{(d.o.f.)}$ & $Q$\\
\hline
$G_{1u} \quad 1$ & $8$ & $15$ & $0$&$3300(70)$ & $0.69$ & $0.66$\\
$G_{1u} \quad 2$ & $8$ & $16$ & $0$&$3325(50)$ & $0.83$ & $0.56$\\
$G_{1u} \quad 3$ & $8$ & $16$ & $0$&$463(15)$ & $1.31$ & $0.24$\\
$G_{1u} \quad 4$ & $9$ & $17$ & $0$&$466(14)$ & $0.98$ & $0.45$\\
$G_{1u} \quad 5$ & $7$ & $15$ & $0$&$468(19)$ & $0.78$ & $0.60$\\
$G_{1u} \quad 6$ & $8$ & $15$ & $0$&$479(13)$ & $1.25$ & $0.28$\\
$G_{1u} \quad 7$ & $7$ & $15$ & $0$&$499(16)$ & $1.17$ & $0.31$\\
$G_{1u} \quad 8$ & $-$ & $-$ & \multicolumn{2}{c}{$-$} & $-$ & $-$\\
\hline
$H_{u} \quad 1$ & $7$ & $15$ & $0$&$3380(45)$ & $1.11$ & $0.35$\\
$H_{u} \quad 2$ & $7$ & $14$ & $0$&$3437(52)$ & $1.02$ & $0.41$\\
$H_{u} \quad 3$ & $9$ & $17$ & $0$&$3439(47)$ & $1.38$ & $0.21$\\
$H_{u} \quad 4$ & $7$ & $15$ & $0$&$458(18)$ & $1.33$ & $0.23$\\
$H_{u} \quad 5$ & $9$ & $17$ & $0$&$470(10)$ & $1.17$ & $0.32$\\
$H_{u} \quad 6$ & $6$ & $15$ & $0$&$4896(88)$ & $1.09$ & $0.37$\\
$H_{u} \quad 7$ & $7$ & $13$ & $0$&$492(16)$ & $1.32$ & $0.25$\\
$H_{u} \quad 8$ & $7$ & $16$ & $0$&$506(19)$ & $0.95$ & $0.47$\\
\hline
$G_{2u} \quad 1$ & $7$ & $16$ & $0$&$3422(53)$ & $1.17$ & $0.31$\\
$G_{2u} \quad 2$ & $8$ & $16$ & $0$&$460(18)$ & $1.78$ & $0.09$\\
$G_{2u} \quad 3$ & $7$ & $16$ & $0$&$486(17)$ & $1.06$ & $0.39$\\
$G_{2u} \quad 4$ & $5$ & $13$ & $0$&$496(11)$ & $0.94$ & $0.47$\\
$G_{2u} \quad 5$ & $8$ & $13$ & $0$&$506(23)$ & $1.68$ & $0.15$\\
$G_{2u} \quad 6$ & $7$ & $14$ & $0$&$514(19)$ & $0.69$ & $0.66$\\
$G_{2u} \quad 7$ & $7$ & $16$ & $0$&$523(18)$ & $1.42$ & $0.18$\\
$G_{2u} \quad 8$ & $5$ & $13$ & $0$&$529(16)$ & $0.69$ & $0.68$\\
\hline
\hline
\end{tabular}
\caption{The final spectrum results for the odd-parity channels.  
These results are based on 200 quenched configurations on 
a $12^3\times 48$ anisotropic lattice using the Wilson action with 
$a_s \sim 0.1$ fm and $a_s/a_\tau \sim 3.0$.  
Our pion mass for this study was $a_\tau M_\pi=0.1125(26)$ 
(see Figure~\ref{fig:pion}), or approximately 700 MeV.
We did not find a satisfactory fit range for the eighth level in the 
$G_{1u}$ channel.}
\label{table:fit_results_u}
\end{table}

\clearpage

\clearpage

\clearpage

\clearpage

\clearpage

\clearpage

\clearpage

\clearpage

\begin{figure}[ht!] 
  \centering 
  \includegraphics[width=6.0in]{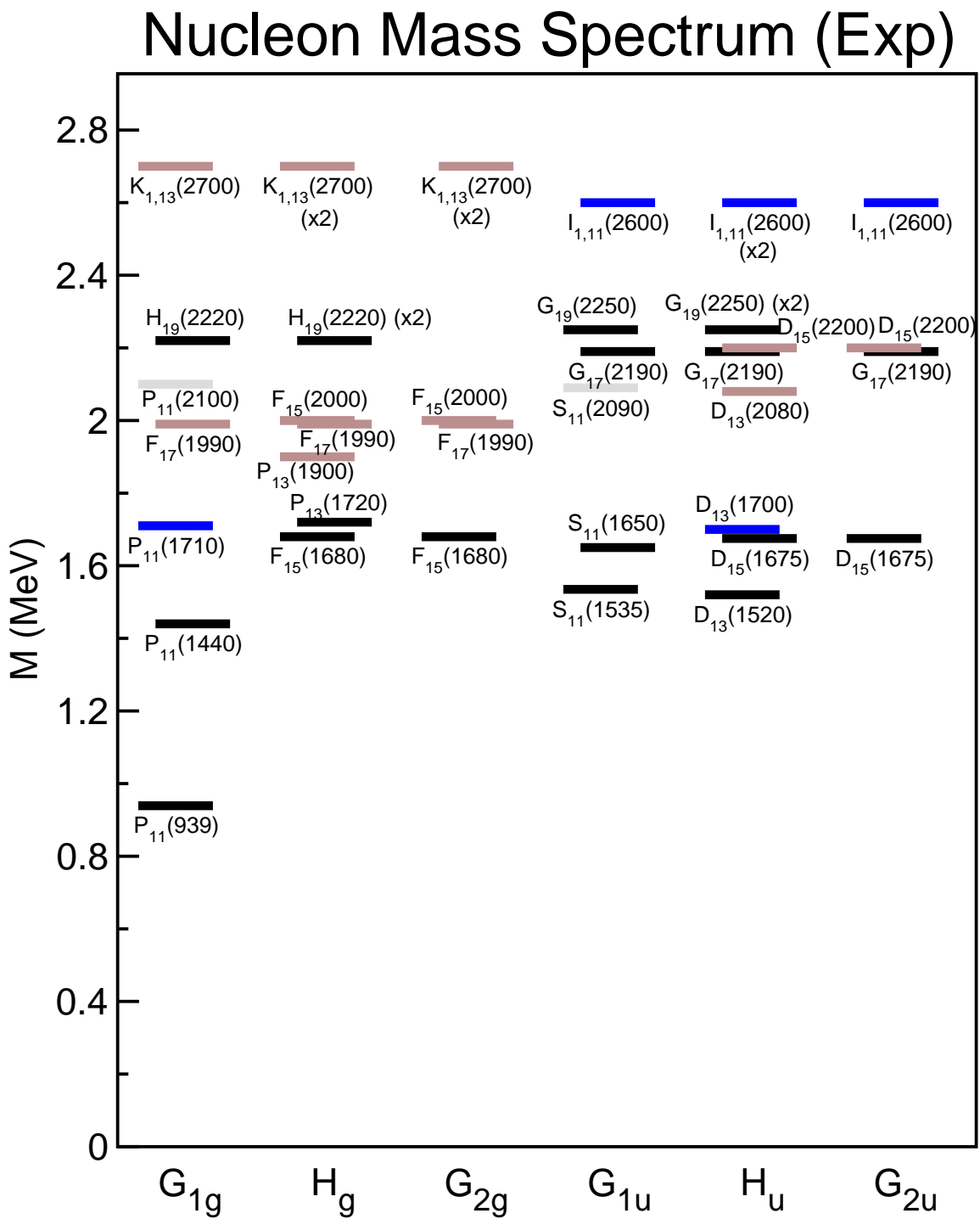}
  \caption{The $I$=1/2, $I_3$=+1/2 nucleon spectrum as determined by
experiment~\cite{yao:pdg06} and projected into the space of lattice spin-parity
 states.
Black denotes a four-star state, blue denotes a three-star state, tan denotes
a two-star state, and gray denotes a one-star state.}
  \label{fig:nucleon_spectrum_expt} 
\end{figure}

\clearpage

\begin{figure}[ht!] 
  \centering 
  \includegraphics[width=6.0in]{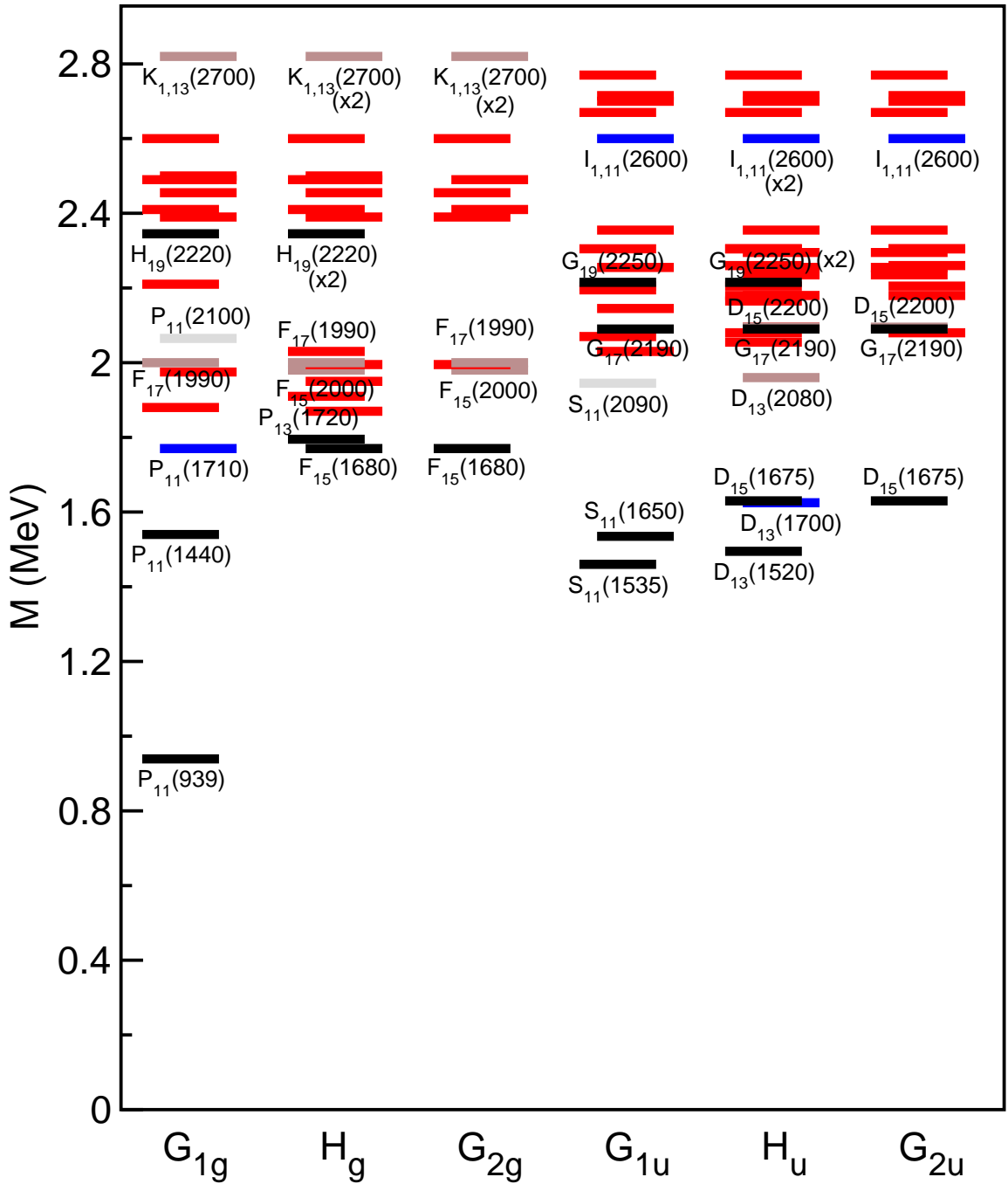}
  \caption{The low-lying $I$=1/2, $I_3$=+1/2 nucleon spectrum up to 2820 MeV
as predicted
by the relativistic quark model~\cite{capstick:quark_model2} and projected into
the space of lattice spin-parity states.  `Missing resonances' are displayed in red,
and observed resonances are labeled by their assigned state in the experimental
spectrum.  Note that the location of a labeled state is given by its
predicted energy value, and the text of the label tells its experimentally
measured energy value.  For a labeled state 
black denotes a four-star state, blue denotes a three-star state, tan denotes
a two-star state, and gray denotes a one-star state.}
\label{fig:nucleon_spectrum_rqm} 
\end{figure}

\clearpage

\section{Discussion}
Our final nucleon spectrum results for this study are tabulated in Tables~\ref{table:fit_results_g} 
and~\ref{table:fit_results_u}, and illustrated in
 Figure~\ref{fig:lattice_nucleon_spectrum}.  Comparison of these
results with experiment is not 
justified because the quenched approximation was used, an unphysically
large $u,d$ quark mass was used, and the lattice volume was too small.
Nevertheless, such a comparison is still interesting, so we
discuss our results with an eye towards the experimentally observed
spectrum, shown in Table~\ref{table:nucleon_levels_expt} and Figure~\ref{fig:nucleon_spectrum_expt}.  
It is also interesting to compare 
our results to the predictions of the relativistic constituent quark
model~\cite{capstick:quark_model, capstick:quark_model2, capstick:nucleon_levels},
shown in Tables~\ref{table:quark_model1} and ~\ref{table:quark_model2} and Figure~\ref{fig:nucleon_spectrum_rqm}. 

We emphasize that most of the
levels shown in Figure~\ref{fig:lattice_nucleon_spectrum} have never
before been calculated
using first-principles QCD.  Figure~\ref{fig:lattice_nucleon_spectrum} 
is actually a first
glimpse of the low-lying nucleon spectrum as predicted by QCD (with
the caveats mentioned above).

First, the lightest nucleon state has a mass value around $0.24~a_
\tau^{-1}$
and resides by itself in the $G_{1g}$ channel, 
indicating a spin-$\frac{1}{2}$ state.  This clearly corresponds to the
proton at 939~MeV.  

Second, there is a cluster or band of
odd-parity states just below $0.35~a_\tau^{-1}$ and well separated 
from higher-lying states.  In this band, there are two $G_{1u}$ states, 
three $H_u$ states, and one $G_{2u}$ state.  The $G_{2u}$ level most
likely
corresponds to the $D_{15}(1675)$ spin-$\frac{5}{2}$ resonance\footnote{
Experimentalists refer to a baryon resonance having isospin $I$ and total
angular momentum $J$ 
using the notation $L_{2I\ 2J}$, where
$L=S,P,D,F,G,H,I,K\cdots$ is the orbital angular momentum of 
an $N\pi$ system having the same 
$J^P$ as the state.}.  This means that one of the $H_u$
levels must correspond to this same state (see 
Table~\ref{table:nucleon_levels_expt}).  The other two $H_u$ states
most likely correspond to the spin-$\frac{3}{2}$ $D_{13}(1520)$
and $D_{13}(1700)$ resonances.  The two $G_{1u}$ levels seem to
correspond to the spin-$\frac{1}{2}$ $S_{11}(1535)$ and $S_{11}(1650)$
resonances.   There are no other experimentally-observed odd-parity 
nucleon resonances below 2.0~GeV.  So it appears that there is a
one-to-one
matching of the states in this band with experiment: every state
that should be seen is seen, and there are no extra states.
However, given that the mass of the pseudoscalar pion is about 
$0.11~a_\tau^{-1}$, we expect a two-particle $N$-$\pi$ $S$-wave state 
in the odd-parity $G_{1u}$ channel near this energy.  We
are either missing such a state, or are missing one of the
$S_{11}(1535)$ or
$S_{11}(1650)$
states.  The most likely explanation is that the two-particle state
is absent from our spectrum due to the quenched approximation
 combined with our
use of single particle operators.  Future unquenched calculations
should resolve this question.

Third, we observe another band or cluster of states in the
even-parity channels around $0.4~a_\tau^{-1}$.  There are four
states in each of the $G_{1g}, H_g, G_{2g}$ channels, although
two additional levels in $H_g$ are only slightly higher.  The most
striking
feature of this band is the lack of one $G_{1g}$ state lying just below
the previously discussed odd-parity band.  Experiments clearly show a
single
resonance near 1440~MeV, known as the Roper resonance.  The Roper
resonance in our spectrum occurs at much too high of an energy
compared with experiment.
The most likely explanation for this is our use of unphysically large
quark masses.  There is some evidence~\cite{mathur:roper} from previous
quenched lattice calculations that the Roper mass dramatically decreases
when the $u,d$ quark mass is decreased such that the pion mass falls
below 300~MeV (our pion mass is around 700~MeV).  

\begin{table}
\centering
\begin{tabular}{|c@{\quad}|@{\quad}cc|ccc|ccc|}
\hline
\hline
\raisebox{0mm}[5mm]State and& & Experimental & \multicolumn{6}{c|}{Appearance in $O^D_h$ Irrep}\\
Mass (MeV) & $J^P$ & Status & $G_{1g}$ &$H_g$ & $G_{2g}$ & $G_{1u}$ &$H_u$ & $G_{2u}$\\
\hline
\raisebox{0mm}[5mm]{$P_{11}(939)$}  & $\frac{1}{2}^+$  & $****$ & $1$ & $0$ & $0$ & $0$ & $0$ & $0$\\
\raisebox{0mm}[5mm]{$P_{11}(1440)$} & $\frac{1}{2}^+$  & $****$ & $1$ & $0$ & $0$ & $0$ & $0$ & $0$\\
\raisebox{0mm}[5mm]{$D_{13}(1520)$} & $\frac{3}{2}^-$ & $****$ & $0$ & $0$ & $0$ & $0$ & $1$ & $0$\\
\raisebox{0mm}[5mm]{$S_{11}(1535)$} & $\frac{1}{2}^-$  & $****$ & $0$ & $0$ & $0$ & $1$ & $0$ & $0$\\
\raisebox{0mm}[5mm]{$S_{11}(1650)$} & $\frac{1}{2}^-$  & $****$ & $0$ & $0$ & $0$ & $1$ & $0$ & $0$\\
\raisebox{0mm}[5mm]{$D_{15}(1675)$} & $\frac{5}{2}^-$ & $****$ & $0$ & $0$ & $0$ & $0$ & $1$ & $1$\\
\raisebox{0mm}[5mm]{$F_{15}(1680)$} & $\frac{5}{2}^+$ & $****$ & $0$ & $1$ & $1$ & $0$ & $0$ & $0$\\
\raisebox{0mm}[5mm]{$D_{13}(1700)$} & $\frac{3}{2}^-$  & $***$  & $0$ & $0$ & $0$ & $0$ & $1$ & $0$\\
\raisebox{0mm}[5mm]{$P_{11}(1710)$} & $\frac{1}{2}^+$  & $***$  & $1$ & $0$ & $0$ & $0$ & $0$ & $0$\\
\raisebox{0mm}[5mm]{$P_{13}(1720)$} & $\frac{3}{2}^+$ & $****$ & $0$ & $1$ & $0$ & $0$ & $0$ & $0$\\
\raisebox{0mm}[5mm]{$P_{13}(1900)$} & $\frac{3}{2}^+$ & $**$ & $0$ & $1$ & $0$ & $0$ & $0$ & $0$\\
\raisebox{0mm}[5mm]{$F_{17}(1990)$} & $\frac{7}{2}^+$ & $**$ & $1$ & $1$ & $1$ & $0$ & $0$ & $0$\\
\raisebox{0mm}[5mm]{$F_{15}(2000)$} & $\frac{5}{2}^+$ & $**$ & $0$ & $1$ & $1$ & $0$ & $0$ & $0$\\
\raisebox{0mm}[5mm]{$D_{13}(2080)$} & $\frac{3}{2}^-$ & $**$ & $0$ & $0$ & $0$ & $0$ & $1$ & $0$\\
\raisebox{0mm}[5mm]{$S_{11}(2090)$} & $\frac{1}{2}^-$ & $*$ & $0$ & $0$ & $0$ & $1$ & $0$ & $0$\\
\raisebox{0mm}[5mm]{$P_{11}(2100)$} & $\frac{1}{2}^+$ & $*$ & $1$ & $0$ & $0$ & $0$ & $0$ & $0$\\
\raisebox{0mm}[5mm]{$G_{17}(2190)$} & $\frac{7}{2}^-$ & $****$ & $0$ & $0$ & $0$ & $1$ & $1$ & $1$\\
\raisebox{0mm}[5mm]{$D_{15}(2200)$} & $\frac{5}{2}^-$ & $**$ & $0$ & $0$ & $0$ & $0$ & $1$ & $1$\\
\raisebox{0mm}[5mm]{$H_{19}(2220)$} & $\frac{9}{2}^+$ & $****$ & $1$ & $2$ & $0$ & $0$ & $0$ & $0$\\
\raisebox{0mm}[5mm]{$G_{19}(2250)$} & $\frac{9}{2}^-$ & $****$ & $0$ & $0$ & $0$ & $1$ & $2$ & $0$\\
\raisebox{0mm}[5mm]{$I_{1,11}(2600)$} & $\frac{11}{2}^-$ & $***$ & $0$ & $0$ & $0$ & $1$ & $2$ & $1$\\
\raisebox{0mm}[5mm][3mm]{$K_{1,13}(2700)$} & $\frac{13}{2}^+$ & $**$ & $1$ & $2$ & $2$ & $0$ & $0$ & $0$\\
\hline
\hline
\end{tabular}
\caption{The current experimental values~\cite{yao:pdg06} for all known
 nucleon resonances.
The state names are given in spectroscopic notation: $L_{2I\ 2J}$, where
$L=S,P,D,F,\cdots$ is the orbital angular momentum of an $N\pi$ system having 
the same 
$J^P$ as the state, $I$ is the isospin, and $J$ is the total angular momentum.
A four-star experimental status implies that existence is certain and that the
properties are fairly well explored.  A three-star status implies that existence
ranges from very likely to certain, but further confirmation is desirable. 
Two stars implies that the evidence for existence is only fair, 
and one star implies that
the evidence is poor.  The number of times each state is expected to appear in each 
lattice $O^D_h$ irrep (obtained from subduction) is also shown.  The experimental uncertainties 
are at the $5\%$ level or less.}
\label{table:nucleon_levels_expt}
\end{table}

\begin{table}
\centering
\begin{tabular}{|c@{\quad}|@{\quad}ccc|ccc|ccc|}
\hline
\hline
\raisebox{0mm}[5mm]{Model} & & $N\pi$ State &  Experimental & \multicolumn{6}{c|}{Appearance in $O^D_h$ Irrep}\\
State & $J^P$ & Assignment & Status & $G_{1g}$ &$H_g$ & $G_{2g}$ & $G_{1u}$ &$H_u$ & $G_{2u}$\\
\hline
\raisebox{0mm}[5mm]{$S_{1 1}^1(1460)$} & $\frac{1}{2}^-$ & $S_{1 1}(1535)$ & $****$ & $0$ & $0$ & $0$ & $1$ & $0$ & $0$\\
\raisebox{0mm}[5mm]{$D_{1 3}^1(1495)$} & $\frac{3}{2}^-$ & $D_{1 3}(1520)$ & $****$ & $0$ & $0$ & $0$ & $0$ & $1$ & $0$\\
\raisebox{0mm}[5mm]{$S_{1 1}^2(1535)$} & $\frac{1}{2}^-$ & $S_{1 1}(1650)$ & $****$ & $0$ & $0$ & $0$ & $1$ & $0$ & $0$\\
\raisebox{0mm}[5mm]{$P_{1 1}^2(1540)$} & $\frac{1}{2}^+$ & $P_{1 1}(1440)$ & $****$ & $1$ & $0$ & $0$ & $0$ & $0$ & $0$\\
\raisebox{0mm}[5mm]{$D_{1 3}^2(1625)$} & $\frac{3}{2}^-$ & $D_{1 3}(1700)$ & $***$ & $0$ & $0$ & $0$ & $0$ & $1$ & $0$\\
\raisebox{0mm}[5mm]{$D_{1 5}^1(1630)$} & $\frac{5}{2}^-$ & $D_{1 5}(1675)$ & $****$ & $0$ & $0$ & $0$ & $0$ & $1$ & $1$\\
\raisebox{0mm}[5mm]{$F_{1 5}^1(1770)$} & $\frac{5}{2}^+$ & $F_{1 5}(1680)$ & $****$ & $0$ & $1$ & $1$ & $0$ & $0$ & $0$\\
\raisebox{0mm}[5mm]{$P_{1 1}^3(1770)$} & $\frac{1}{2}^+$ & $P_{1 1}(1710)$ & $***$ & $1$ & $0$ & $0$ & $0$ & $0$ & $0$\\
\raisebox{0mm}[5mm]{$P_{1 3}^1(1795)$} & $\frac{3}{2}^+$ & $P_{1 3}(1720)$ & $****$ & $0$ & $1$ & $0$ & $0$ & $0$ & $0$\\
\raisebox{0mm}[5mm]{$P_{1 3}^2(1870)$} & $\frac{3}{2}^+$ & $-$ & $-$ & $0$ & $1$ & $0$ & $0$ & $0$ & $0$\\
\raisebox{0mm}[5mm]{$P_{1 1}^4(1880)$} & $\frac{1}{2}^+$ & $-$ & $-$ & $1$ & $0$ & $0$ & $0$ & $0$ & $0$\\
\raisebox{0mm}[5mm]{$P_{1 3}^3(1910)$} & $\frac{3}{2}^+$ & $-$ & $-$ & $0$ & $1$ & $0$ & $0$ & $0$ & $0$\\
\raisebox{0mm}[5mm]{$S_{1 1}^3(1945)$} & $\frac{1}{2}^-$ & $S_{1 1}(2090)$ & $*$ & $0$ & $0$ & $0$ & $1$ & $0$ & $0$\\
\raisebox{0mm}[5mm]{$P_{1 3}^4(1950)$} & $\frac{3}{2}^+$ & $-$ & $-$ & $0$ & $1$ & $0$ & $0$ & $0$ & $0$\\
\raisebox{0mm}[5mm]{$D_{1 3}^3(1960)$} & $\frac{3}{2}^-$ & $D_{1 3}(2080)$ & $**$ & $0$ & $0$ & $0$ & $0$ & $1$ & $0$\\
\raisebox{0mm}[5mm]{$P_{1 1}^5(1975)$} & $\frac{1}{2}^+$ & $-$ & $-$ & $1$ & $0$ & $0$ & $0$ & $0$ & $0$\\
\raisebox{0mm}[5mm]{$F_{1 5}^2(1980)$} & $\frac{5}{2}^+$ & $F_{1 5}(2000)$ & $**$ & $0$ & $1$ & $1$ & $0$ & $0$ & $0$\\
\raisebox{0mm}[5mm]{$F_{1 5}^3(1995)$} & $\frac{5}{2}^+$ & $-$ & $-$ & $0$ & $1$ & $1$ & $0$ & $0$ & $0$\\
\raisebox{0mm}[5mm]{$F_{1 7}^1(2000)$} & $\frac{7}{2}^+$ & $F_{1 7}(1990)$ & $**$ & $1$ & $1$ & $1$ & $0$ & $0$ & $0$\\
\raisebox{0mm}[5mm]{$P_{1 3}^5(2030)$} & $\frac{3}{2}^+$ & $-$ & $-$ & $0$ & $1$ & $0$ & $0$ & $0$ & $0$\\
\raisebox{0mm}[5mm]{$S_{1 1}^4(2030)$} & $\frac{1}{2}^-$ & $-$ & $-$ & $0$ & $0$ & $0$ & $1$ & $0$ & $0$\\
\raisebox{0mm}[5mm]{$D_{1 3}^4(2055)$} & $\frac{3}{2}^-$ & $-$ & $-$ & $0$ & $0$ & $0$ & $0$ & $1$ & $0$\\
\raisebox{0mm}[5mm]{$P_{1 1}^6(2065)$} & $\frac{1}{2}^+$ & $P_{1 1}(2100)$ & $*$ & $1$ & $0$ & $0$ & $0$ & $0$ & $0$\\
\raisebox{0mm}[5mm]{$S_{1 1}^5(2070)$} & $\frac{1}{2}^-$ & $-$ & $-$ & $0$ & $0$ & $0$ & $1$ & $0$ & $0$\\
\raisebox{0mm}[5mm]{$D_{1 5}^2(2080)$} & $\frac{5}{2}^-$ & $-$ & $-$ & $0$ & $0$ & $0$ & $0$ & $1$ & $1$\\
\raisebox{0mm}[5mm]{$G_{1 7}^1(2090)$} & $\frac{7}{2}^-$ & $G_{1 7}(2190)$ & $****$ & $0$ & $0$ & $0$ & $1$ & $1$ & $1$\\
\raisebox{0mm}[5mm]{$D_{1 3}^5(2095)$} & $\frac{3}{2}^-$ & $-$ & $-$ & $0$ & $0$ & $0$ & $0$ & $1$ & $0$\\
\raisebox{0mm}[5mm][3mm]{$D_{1 5}^3(2095)$} & $\frac{5}{2}^-$ & $D_{1 5}(2200)$ & $**$ & $0$ & $0$ & $0$ & $0$ & $1$ & $1$\\
\hline
\hline
\end{tabular}
\caption{The quark model predictions~\cite{capstick:quark_model2} for the 
excited nucleon spectrum below 2100 MeV.
The proton is used to set the parameters
of the model.
The state names are given in spectroscopic notation: $L_{2I\ 2J}$, where
$L=S,P,D,F,\cdots$ is the orbital angular momentum of an $N\pi$ system 
having the same 
$J^P$ as the state, $I$ is the isospin, and $J$ is the total angular momentum.
The superscripted integer on the model state name denotes the principal
quantum number in the quark model (see~\cite{capstick:quark_model2}).
The number of times each state is expected to appear in each 
lattice $O^D_h$ irrep is also shown. Dashes indicate `missing resonances.'}
\label{table:quark_model1}
\end{table}

\begin{table}
\centering
\begin{tabular}{|c@{\quad}|@{\quad}ccc|ccc|ccc|}
\hline
\hline
\raisebox{0mm}[5mm]{Model} & & $N\pi$ State &  Experimental & \multicolumn{6}{c|}{Appearance in $O^D_h$ Irrep}\\
State & $J^P$ & Assignment & Status & $G_{1g}$ &$H_g$ & $G_{2g}$ & $G_{1u}$ &$H_u$ & $G_{2u}$\\
\hline
\raisebox{0mm}[5mm]{$S_{1 1}^6(2145)$} & $\frac{1}{2}^-$ & $-$ & $-$ & $0$ & $0$ & $0$ & $1$ & $0$ & $0$\\
\raisebox{0mm}[5mm]{$D_{1 3}^6(2165)$} & $\frac{3}{2}^-$ & $-$ & $-$ & $0$ & $0$ & $0$ & $0$ & $1$ & $0$\\
\raisebox{0mm}[5mm]{$D_{1 3}^7(2180)$} & $\frac{3}{2}^-$ & $-$ & $-$ & $0$ & $0$ & $0$ & $0$ & $1$ & $0$\\
\raisebox{0mm}[5mm]{$D_{1 5}^4(2180)$} & $\frac{5}{2}^-$ & $-$ & $-$ & $0$ & $0$ & $0$ & $0$ & $1$ & $1$\\
\raisebox{0mm}[5mm]{$S_{1 1}^7(2195)$} & $\frac{1}{2}^-$ & $-$ & $-$ & $0$ & $0$ & $0$ & $1$ & $0$ & $0$\\
\raisebox{0mm}[5mm]{$G_{1 7}^2(2205)$} & $\frac{7}{2}^-$ & $-$ & $-$ & $0$ & $0$ & $0$ & $1$ & $1$ & $1$\\
\raisebox{0mm}[5mm]{$P_{1 1}^7(2210)$} & $\frac{1}{2}^+$ & $-$ & $-$ & $1$ & $0$ & $0$ & $0$ & $0$ & $0$\\
\raisebox{0mm}[5mm]{$G_{1 9}^1(2215)$} & $\frac{9}{2}^-$ & $G_{1 9}(2250)$ & $****$ & $0$ & $0$ & $0$ & $1$ & $2$ & $0$\\
\raisebox{0mm}[5mm]{$D_{1 5}^5(2235)$} & $\frac{5}{2}^-$ & $-$ & $-$ & $0$ & $0$ & $0$ & $0$ & $1$ & $1$\\
\raisebox{0mm}[5mm]{$G_{1 7}^3(2255)$} & $\frac{7}{2}^-$ & $-$ & $-$ & $0$ & $0$ & $0$ & $1$ & $1$ & $1$\\
\raisebox{0mm}[5mm]{$D_{1 5}^6(2260)$} & $\frac{5}{2}^-$ & $-$ & $-$ & $0$ & $0$ & $0$ & $0$ & $1$ & $1$\\
\raisebox{0mm}[5mm]{$D_{1 5}^7(2295)$} & $\frac{5}{2}^-$ & $-$ & $-$ & $0$ & $0$ & $0$ & $0$ & $1$ & $1$\\
\raisebox{0mm}[5mm]{$D_{1 5}^8(2305)$} & $\frac{5}{2}^-$ & $-$ & $-$ & $0$ & $0$ & $0$ & $0$ & $1$ & $1$\\
\raisebox{0mm}[5mm]{$G_{1 7}^4(2305)$} & $\frac{7}{2}^-$ & $-$ & $-$ & $0$ & $0$ & $0$ & $1$ & $1$ & $1$\\
\raisebox{0mm}[5mm]{$H_{1 9}^1(2345)$} & $\frac{9}{2}^+$ & $H_{1 9}(2220)$ & $****$ & $1$ & $2$ & $0$ & $0$ & $0$ & $0$\\
\raisebox{0mm}[5mm]{$G_{1 7}^5(2355)$} & $\frac{7}{2}^-$ & $-$ & $-$ & $0$ & $0$ & $0$ & $1$ & $1$ & $1$\\
\raisebox{0mm}[5mm]{$F_{1 7}^2(2390)$} & $\frac{7}{2}^+$ & $-$ & $-$ & $1$ & $1$ & $1$ & $0$ & $0$ & $0$\\
\raisebox{0mm}[5mm]{$F_{1 7}^3(2410)$} & $\frac{7}{2}^+$ & $-$ & $-$ & $1$ & $1$ & $1$ & $0$ & $0$ & $0$\\
\raisebox{0mm}[5mm]{$F_{1 7}^4(2455)$} & $\frac{7}{2}^+$ & $-$ & $-$ & $1$ & $1$ & $1$ & $0$ & $0$ & $0$\\
\raisebox{0mm}[5mm]{$H_{1,11}^1(2490)$} & $\frac{11}{2}^+$ & $-$ & $-$ & $1$ & $2$ & $1$ & $0$ & $0$ & $0$\\
\raisebox{0mm}[5mm]{$H_{1 9}^3(2490)$} & $\frac{9}{2}^+$ & $-$ & $-$ & $1$ & $2$ & $0$ & $0$ & $0$ & $0$\\
\raisebox{0mm}[5mm]{$H_{1 9}^2(2500)$} & $\frac{9}{2}^+$ & $-$ & $-$ & $1$ & $2$ & $0$ & $0$ & $0$ & $0$\\
\raisebox{0mm}[5mm]{$H_{1,11}^2(2600)$} & $\frac{11}{2}^+$ & $-$ & $-$ & $1$ & $2$ & $1$ & $0$ & $0$ & $0$\\
\raisebox{0mm}[5mm]{$I_{1,11}^1(2600)$} & $\frac{11}{2}^-$ & $I_{1,11}(2600)$ & $***$ & $0$ & $0$ & $0$ & $1$ & $2$ & $1$\\
\raisebox{0mm}[5mm]{$I_{1,11}^2(2670)$} & $\frac{11}{2}^-$ & $-$ & $-$ & $0$ & $0$ & $0$ & $1$ & $2$ & $1$\\
\raisebox{0mm}[5mm]{$I_{1,11}^3(2700)$} & $\frac{11}{2}^-$ & $-$ & $-$ & $0$ & $0$ & $0$ & $1$ & $2$ & $1$\\
\raisebox{0mm}[5mm]{$I_{1,13}^1(2715)$} & $\frac{13}{2}^-$ & $-$ & $-$ & $0$ & $0$ & $0$ & $1$ & $2$ & $2$\\
\raisebox{0mm}[5mm]{$I_{1,11}^4(2770)$} & $\frac{11}{2}^-$ & $-$ & $-$ & $0$ & $0$ & $0$ & $1$ & $2$ & $1$\\
\raisebox{0mm}[5mm][3mm]{$K_{1,13}^1(2820)$} & $\frac{13}{2}^+$ & $K_{1,13}(2700)$ & $**$ & $1$ & $2$ & $2$ & $0$ & $0$ & $0$\\
\hline
\hline
\end{tabular}
\caption{The quark model predictions~\cite{capstick:quark_model2} for the 
excited nucleon spectrum from 2100 MeV to 2820 MeV. Dashes indicate `missing
resonances.'}
\label{table:quark_model2}
\end{table}

In the observed spectrum (Figure~\ref{fig:nucleon_spectrum_expt})
between 1.6~GeV and 1.8~GeV, there are currently 
three well-established even-parity resonances:  the $F_{15}(1680)$, the 
 $P_{11}(1710)$, and the
$P_{13}(1720)$.  
A second band of 
four more even-parity resonances has been
tentatively 
identified near 2.0~GeV containing the $P_{13}(1900)$, 
the $F_{17}(1990)$, the $F_{15}(2000)$, and the $P_{11}(2100)$ states.
Our results do not reproduce this two-band pattern.  Rather,
we find a single band of four states in the each of the 
$G_{1g}$, $H_G$, and $G_{2g}$ channels around $0.4~a_\tau^{-1}$.

One possibility is that our
calculations have identified resonances in the 1.6 to 1.8~GeV range
which have not yet been observed
in experiments.  These states would correspond to some of the `missing
resonances' predicted by the relativistic constituent quark model
(see below and Figure~\ref{fig:nucleon_spectrum_rqm}).

However,  if the mass splittings among the experimentally observed
levels are particularly
sensitive
to the $u,d$ quark mass, one may speculate that all of those levels,
including the
Roper, might merge to a single band at large $u,d$ mass.  If so, one
would observe
four nearly degenerate levels in the $G_{1g}$ channel, five levels in
the $H_g$ 
channel, and three in the $G_{2g}$ channel.  Such a pattern would
more closely
resemble the
results of our calculation (see Figure~\ref{fig:lattice_nucleon_spectrum}).

Additionally, the single-band structure observed in the even-parity channels 
may split when we add a clover $(\sigma\cdot F)$ improvement term\footnote{The 
$\sigma$ matrices are given by $\sigma_{\mu\nu}=-\frac{i}{2}[\gamma_\mu,\gamma_\nu]$, and $F$ is a discretization of the gauge field strength.} to our 
action~\cite{sheikholeslami:clover, edwards:clover,edwards:aniso_clover_action}.  Such a term not only compensates for the $O(a)$ 
discretization errors introduced by the chiral-symmetry-breaking Wilson term,
but may also restore level splitting behavior disrupted
by the Wilson term at $O(a)$.

Increased
statistics and a 
finite lattice spacing study (involving data from several different
lattice spacings 
$a_s$) are needed in order to deduce the spin content in this even-parity band.
Note that we
also expect two-particle $\pi$-$S_{11}(1535)$ and $\pi$-$D_{13}(1520)$
states 
to occur in the even-parity spectrum just slightly above this band's
energy, 
but the quenched approximation might again make such states inaccessible
to the
operators used in this study.  Whether or not all of these levels merge
into a
single band at large $u,d$ mass will be eventually answered; in future
calculations 
we will determine the mass spectrum for various values of the $u,d$
quark mass  to resolve this issue.  

Although our odd-parity results
agree qualitatively with the quark model~\cite{capstick:quark_model2}, 
the quark model predicts more even-parity states around 2~GeV than we found.  
Quark model
calculations~\cite{capstick:quark_model2} predict a rather dense
spectrum
of four $\frac{1}{2}^+$ states, five $\frac{3}{2}^+$ states, 
three $\frac{5}{2}^+$ states, and one $\frac{7}{2}^+$ state in
the 1.65 to 2.15~GeV energy region.  If this were correct, then 
our lattice spectrum
would contain a low-lying band of five states in the 
$G_{1g}$ irrep,  nine states
in the 
$H_g$ irrep, and four states in the $G_{2g}$ irrep (see 
Figure~\ref{fig:nucleon_spectrum_rqm}).
 The number of states appearing in the lowest $H_g$ band is too small to
support 
this.  Future studies will extract more levels in the $H_g$ channel 
to shed more light on this issue.

With statistics based on only $200$ configurations, the interpretation
of
the states above $0.45~a_\tau^{-1}$ is somewhat problematic.  However,
the absence 
of states between $0.45~a_\tau^{-1}$ and $0.55~a_\tau^{-1}$ in
the $G_{2g}$ channel is an interesting feature of the
spectrum and does appear to tentatively agree with experiment.  The
quark model 
does not predict such a large gap (see
Figure~\ref{fig:nucleon_spectrum_rqm}).  On the other hand, our
calculation
also shows a large clustering of levels around $0.5~a_\tau^{-1}$ 
in the $G_{2u}$ channel, which agrees more with
 the predictions of the quark model than with the experimental data.

\clearpage

\section{Conclusion and outlook}

In this work, we have presented a systematic approach to baryon operator 
design for use in lattice QCD baryon spectrum calculations.  
We described the formalism for extracting excited states from correlation
matrix elements between quantum operators, and emphasized the importance of
using operators which couple strongly to the states of interest and
weakly to the high-lying contaminating modes.  

We discussed the evaluation of correlation matrix elements using the Monte
Carlo method on an anisotropic space-time lattice.  The anisotropy allows 
both a fine temporal resolution for the examination of exponentially
decaying correlation functions, and a coarse spatial resolution enabling
the use of
a larger spatial volume.
The signal quality associated with the diagonal correlation matrix
elements was defined in terms of the effective mass function, which provided
not only a quantitative measure of the contamination and noise, but also
a method of visualizing the presence of stationary energy states as plateaus in
effective mass plots.

The 
baryon operators used in this study were composed of
gauge-covariantly-displaced three-quark operators. 
The covariant quark displacements allowed us to incorporate radial 
structure resulting in operators which better interpolated for 
excited baryon states.  
We found it essential to use smeared gauge links in our extended operators
to suppress the noise introduced
by using gauge links to covariantly displace the quarks.  
Additionally, we found
that while quark field smearing 
had little effect on the noise, it dramatically reduced the
 coupling to high-lying contaminating modes.
By systematically exploring the smearing parameter space while monitoring
the behavior of the effective mass plots associated with three trial operators, we found values of 
the smearing parameters which reduced the noise and contamination in 
the correlation matrix elements while 
preserving the integrity of the excited state signals.

We then used group theory to construct operators which
transformed irreducibly under the symmetry group of the cubic spatial lattice.
This resulted in operators with well-defined lattice quantum
numbers, allowing us to identify the $J^P$ quantum numbers of the corresponding
continuum states.
The good quantum numbers on our lattice were the linear momentum
(chosen to be zero for our spectrum calculations), color (our operators 
were gauge-invariant, or colorless), isospin (e.g.
nucleon operators only excited
nucleon states), and the lattice spin-parity 
irreducible representation label corresponding
to the spinorial representations of $O^D_h$, the double-valued octahedral
point group of rotations and
reflections on the cubic spatial lattice.  This last label, which took the values
$G_{1g},H_g,G_{2g},G_{1u},H_u,$ and $G_{2u}$, provided a way to identify the 
continuum $J^P$ quantum
numbers of the spectral states. 

The group-theoretical approach led to an unmanageably large number of 
extended baryon operators.  
We developed a method to select sixteen operators in each lattice spin-parity
channel to be used for the
final extraction of the spectrum.  This pruning method was based on three crucial
requirements: low noise, maximal independence, and good overlap with the
excited states as judged by principal effective mass quality.
We evaluated the diagonal elements of the operator correlation matrix 
to remove noisy operators, and then selected
the sixteen operators whose renormalized correlation matrix at a fixed 
small time separation had a low condition number for both the even-
and odd-parity channels.

We then applied a variational method to rotate our 
basis set of operators into a new set which could be used to extract
the lowest seven or eight levels of the spectrum in each channel.  We found
 a range of $\tau$ values over which the temporal correlation function
for each of these new operators could be fit to a single exponential.  
After performing the fits, we interpreted the results in terms of both 
experimental data and quark model predictions.

 Although comparison
with experiment is not justified, the pattern of levels obtained
qualitatively agrees with the observed spectrum.  We also compared
our spectrum to relativistic quark model predictions;
the quark model predicts more low-lying even-parity
states than this study found, but both the quark model and
this study predict more odd-parity states near 2 GeV than
currently observed in experiments.
These results are not only interesting, but also serve 
as a powerful
validation of the operator design methodology described in this work and
provide a preview of what we expect to achieve in this long-term effort.

We focused on the nucleon channel in this work, but the methodology 
used here is readily applied to the other baryon channels ($\Delta$, $\Sigma$, 
$\Xi$, $\Lambda$, and $\Omega$).
Our approach to building operators and extracting the spectrum is also
expected
to be equally effective when adapted for the construction of meson operators,
an important next step.

An important improvement to our calculation technique will be
the use of all-to-all propagators~\cite{foley:all_to_all}, rather 
than one-to-all propagators, when evaluating correlation
matrix elements.  All-to-all propagators will allow us to treat the source
and sink operators separately, an improvement over the current method in which
we are forced to express correlation
matrix elements in terms of thousands of three-quark propagator terms.  It will
be computationally feasible to spatially average over the source as well as 
the sink to
improve
our statistics, and to evaluate correlation matrix elements between
 multi-hadron operators.
                                                                                
This research was carried out as part of the ongoing Lattice Hadron 
Physics Collaboration 
QCD spectroscopy project. 
The LHPC has recently been awarded over ten million CPU hours on the QCDOC
at Brookhaven National Laboratory, a massively parallel ASIC~\footnote{QCDOC stands for
``QCD On a Chip'' and ASIC stands for ``Application-Specific Integrated Circuit.''}
supercomputer~\cite{boyle:qcdoc}. We are currently generating 1,000 
large-volume unquenched gauge configurations.  
Because this unquenched run includes the fermion determinant, 
we are correctly including
 quark loop effects in the calculation of our propagators.  
Using clover-improved
Wilson fermions, an improved gauge action, and a pion mass around 300~MeV,
 we expect this production run data to 
resolve many of the questions raised by the preliminary spectrum
results presented here,
 and to provide us with a first {\em ab initio} look into the baryon spectrum 
as predicted by QCD.

\appendix
\chapter*{Appendix: Final operator selection}
\addcontentsline{toc}{chapter}{Appendix: Final operator selection}
\label{appx:operators}

The number of nucleon operators of each type is shown in 
Table~\ref{table:nucleon_numbers}.
For future studies, we record here the corresponding 
identification numbers for the final
sixteen nucleon operators selected from each $O^D_h$ irreducible representation in 
Tables~\ref{table:op_ids_G1},~\ref{table:op_ids_H}, and~\ref{table:op_ids_G2}.
These identification numbers are used to index the different operators in 
our \texttt{projection\_coefficients} data files.

\begin{table*}[hb!]
\centering
\begin{tabular}{l|rrr} 
\hline
\hline
 \raisebox{0mm}[5mm]{$N^+$ Operator type} &  $G_{1g}$  &   $H_g$  &  $G_{2g}$\\
\hline
 \raisebox{0mm}[5mm]{Single-Site}        &      3     &     1    &     0     \\
 \raisebox{0mm}[5mm]{Singly-Displaced}   &     24     &    32    &     8    \\
 \raisebox{0mm}[5mm]{Doubly-Displaced-I} &     24     &    32    &     8    \\
 \raisebox{0mm}[5mm]{Doubly-Displaced-L} &     64     &   128    &    64    \\
 \raisebox{0mm}[5mm][2mm]{Triply-Displaced-T} &     64     &   128    &    64    \\
\hline
\raisebox{0mm}[4mm]{\textbf{Total}} & \textbf{179} & \textbf{321} & \textbf{144}\\
\hline
\hline
\end{tabular}
\caption{The numbers of operators of each type which project
into each row of the $G_{1g}, H_g,$ and $G_{2g}$ irreps for the 
$N^+$ baryons.  The numbers
for the $G_{1u}, H_{u},$ and $G_{2u}$ irreps are the same as for the
$G_{1g}, H_g,$ and $G_{2g}$ irreps, respectively.
\label{table:nucleon_numbers}}
\end{table*}

\begin{table}[ht!]
\centering
\begin{tabular}{c|cc}
\multicolumn{3}{c}{\large \raisebox{0mm}[0mm][2mm]{$G_{1g}/G_{1u}$ Operators}}\\
\hline
\hline
\raisebox{0mm}[5mm]{Operator Number} & Operator Type &  Operator ID\\
\hline
\raisebox{0mm}[5mm]{$ 1$} & Single-Site & 2\\
\raisebox{0mm}[5mm]{$ 2$} & Singly-Displaced & 11\\
\raisebox{0mm}[5mm]{$ 3$} & Singly-Displaced & 17\\
\raisebox{0mm}[5mm]{$ 4$} & Singly-Displaced & 20\\
\raisebox{0mm}[5mm]{$ 5$} & Doubly-Displaced-I & 0\\
\raisebox{0mm}[5mm]{$ 6$} & Doubly-Displaced-I & 4\\
\raisebox{0mm}[5mm]{$ 7$} & Doubly-Displaced-I & 5\\
\raisebox{0mm}[5mm]{$ 8$} & Doubly-Displaced-I & 9\\
\raisebox{0mm}[5mm]{$ 9$} & Doubly-Displaced-I & 12\\
\raisebox{0mm}[5mm]{$10$} & Doubly-Displaced-L & 4\\
\raisebox{0mm}[5mm]{$11$} & Doubly-Displaced-L & 10\\
\raisebox{0mm}[5mm]{$12$} & Triply-Displaced-T & 3\\
\raisebox{0mm}[5mm]{$13$} & Triply-Displaced-T & 5\\
\raisebox{0mm}[5mm]{$14$} & Triply-Displaced-T & 9\\
\raisebox{0mm}[5mm]{$15$} & Triply-Displaced-T & 11\\
\raisebox{0mm}[5mm][2mm]{$16$} & Triply-Displaced-T & 25\\
\hline
\hline
\end{tabular}
\caption{The identification numbers for the final sixteen nucleon
operators selected from the $G_{1g}/G_{1u}$ channels.
The ID number corresponds to the operator number within each type (see
 Table~\ref{table:nucleon_numbers}) as indexed in our \texttt{projection\_coefficients} data files.}
\label{table:op_ids_G1}
\end{table}

\begin{table}[ht!]
\centering
\begin{tabular}{c|cc}
\multicolumn{3}{c}{\large \raisebox{0mm}[0mm][2mm]{$H_{g}/H_{u}$ Operators}}\\
\hline
\hline
\raisebox{0mm}[5mm]{Operator Number} & Operator Type &  Operator ID\\
\hline
\raisebox{0mm}[5mm]{$ 1$} & Singly-Displaced & 9\\
\raisebox{0mm}[5mm]{$ 2$} & Singly-Displaced & 10\\
\raisebox{0mm}[5mm]{$ 3$} & Singly-Displaced & 31\\
\raisebox{0mm}[5mm]{$ 4$} & Doubly-Displaced-I & 8\\
\raisebox{0mm}[5mm]{$ 5$} & Doubly-Displaced-I & 17\\
\raisebox{0mm}[5mm]{$ 6$} & Doubly-Displaced-I & 31\\
\raisebox{0mm}[5mm]{$ 7$} & Doubly-Displaced-L & 47\\
\raisebox{0mm}[5mm]{$ 8$} & Doubly-Displaced-L & 54\\
\raisebox{0mm}[5mm]{$ 9$} & Doubly-Displaced-L & 84\\
\raisebox{0mm}[5mm]{$10$} & Doubly-Displaced-L & 113\\
\raisebox{0mm}[5mm]{$11$} & Doubly-Displaced-L & 124\\
\raisebox{0mm}[5mm]{$12$} & Triply-Displaced-T & 35\\
\raisebox{0mm}[5mm]{$13$} & Triply-Displaced-T & 71\\
\raisebox{0mm}[5mm]{$14$} & Triply-Displaced-T & 86\\
\raisebox{0mm}[5mm]{$15$} & Triply-Displaced-T & 95\\
\raisebox{0mm}[5mm][2mm]{$16$} & Triply-Displaced-T & 104\\
\hline
\hline
\end{tabular}
\caption{The identification numbers for the final sixteen nucleon
operators selected from the $H_{g}/H_{u}$ channels.
The ID number corresponds to the operator number within each type (see
 Table~\ref{table:nucleon_numbers}) as indexed in our \texttt{projection\_coefficients} data files.}
\label{table:op_ids_H}

\end{table}
\begin{table}[ht!]
\centering
\begin{tabular}{c|cc}
\multicolumn{3}{c}{\large \raisebox{0mm}[0mm][2mm]{$G_{2g}/G_{2u}$ Operators}}\\
\hline
\hline
\raisebox{0mm}[5mm]{Operator Number} & Operator Type &  Operator ID\\
\hline
\raisebox{0mm}[5mm]{$ 1$} & Singly-Displaced & 0\\
\raisebox{0mm}[5mm]{$ 2$} & Singly-Displaced & 1\\
\raisebox{0mm}[5mm]{$ 3$} & Singly-Displaced & 2\\
\raisebox{0mm}[5mm]{$ 4$} & Singly-Displaced & 6\\
\raisebox{0mm}[5mm]{$ 5$} & Doubly-Displaced-I & 5\\
\raisebox{0mm}[5mm]{$ 6$} & Doubly-Displaced-I & 6\\
\raisebox{0mm}[5mm]{$ 7$} & Doubly-Displaced-I & 7\\
\raisebox{0mm}[5mm]{$ 8$} & Doubly-Displaced-L & 32\\
\raisebox{0mm}[5mm]{$ 9$} & Doubly-Displaced-L & 37\\
\raisebox{0mm}[5mm]{$10$} & Doubly-Displaced-L & 41\\
\raisebox{0mm}[5mm]{$11$} & Doubly-Displaced-L & 52\\
\raisebox{0mm}[5mm]{$12$} & Triply-Displaced-T & 1\\
\raisebox{0mm}[5mm]{$13$} & Triply-Displaced-T & 33\\
\raisebox{0mm}[5mm]{$14$} & Triply-Displaced-T & 45\\
\raisebox{0mm}[5mm]{$15$} & Triply-Displaced-T & 51\\
\raisebox{0mm}[5mm][2mm]{$16$} & Triply-Displaced-T & 61\\
\hline
\hline
\end{tabular}
\caption{The identification numbers for the final sixteen nucleon
operators selected from the $G_{2g}/G_{2u}$ channels.
The ID number corresponds to the operator number within each type (see
 Table~\ref{table:nucleon_numbers}) as indexed in our \texttt{projection\_coefficients} data files.}
\label{table:op_ids_G2}
\end{table}

\bibliography{bibliography}
\addcontentsline{toc}{chapter}{Bibliography}

\end{document}